\documentclass[12pt]{article}
\usepackage[utf8]{inputenc}
\usepackage[margin=0.87in]{geometry}
\RequirePackage{amsthm,amsmath,amsfonts,amssymb,mathtools}
\usepackage{setspace}
\usepackage{graphicx}
\usepackage{color,xcolor}
\usepackage{amsmath}
\usepackage{amsthm}
\usepackage{amssymb}
\usepackage{amsfonts}
\usepackage{titling}
\usepackage{bm}
\usepackage{mhchem}
\usepackage[inline]{enumitem}
\usepackage[compact]{titlesec}
\usepackage{times}
\usepackage{booktabs}
\usepackage{xargs}
\usepackage{subcaption}
\usepackage{accents}
\newcommand{\dbtilde}[1]{\accentset{\approx}{#1}}

\usepackage[citecolor=blue,colorlinks=true,linkcolor=blue,urlcolor=black]{hyperref}
\usepackage{varioref}
\usepackage[capitalise,noabbrev]{cleveref}
\crefformat{equation}{(#2#1#3)}
\usepackage{url} 
\usepackage{color,xcolor}
\usepackage{amsfonts}
\usepackage{bm}
\usepackage{mhchem}
\usepackage[inline]{enumitem}
\usepackage{booktabs,threeparttable}
\usepackage{xargs}
\usepackage{subcaption}
\usepackage{diagbox}
\usepackage{slashbox}
 \usepackage{multirow}
 \usepackage{float}
  \usepackage{longtable}
\usepackage{array}

\def\spacingset#1{\renewcommand{\baselinestretch}%
{#1}\small\normalsize} \spacingset{1}

\usepackage[toc,page,header]{appendix}
\usepackage{minitoc}
\def\spacingset#1{\renewcommand{\baselinestretch}%
{#1}\small\normalsize} \spacingset{1}

\newcommandx{\Emis}[1][1=ti]{E_{#1}^*}
\newcommandx{\Emiso}[1][1=ti]{E_{#1}}
\newcommandx{\Emish}[1][1=ti]{\overline{\Emis[#1]}}
\newcommandx{\Emisoh}[1][1=ti]{\overline{\Emiso[#1]}}
\newcommandx{\emis}[1][1=ti]{e_{#1}^*}
\newcommandx{\emiso}[1][1=ti]{e_{#1}}
\newcommandx{\emish}[1][1=ti]{\overline{\emis[#1]}}
\newcommandx{\emisoh}[1][1=ti]{\overline{\emiso[#1]}}

\DeclareMathAlphabet\mathbfcal{OMS}{cmsy}{b}{n}

\theoremstyle{definition}
\newtheorem{proposition}{Proposition}

\newtheorem{assumption}{Assumption}
\newtheorem{remark}{Remark}
\newtheorem{theorem}{Theorem}

\usepackage{natbib}

\begin{document}

\setstretch{1.4}
% Something related to the title: causal effect of exposure on a single unit under bipartite interference time continuity Matching methods adjusted for confounding 

\title{ \LARGE
% Natural experiments for bipartite interference due to exogenous occurring random network \\
% OR \\
%Natural experiments in time series studies with bipartite interference and a random network
% OR \\
% Estimating Single-unit Exposure Effect under Bipartite Interference by Matching with Time Continuity
% OR \\
Bipartite causal inference with interference, time series data, and a random network
}

\author{ {\large Zhaoyan Song \quad Georgia Papadogeorgou} \\} 

% \author{\large Zhaoyan Song$^1$, Lucas Henneman$^2$, Georgia Papadogeorgou$^1$
% }
% \date{\normalfont $^1$Department of Statistics, University of Florida \\
% $^2$Sid and Reva Dewberry Department of Civil, Environmental, and Infrastructure Engineering, George Mason University}
\date{\small Department of Statistics, University of Florida \\ \vspace{-20pt}}

\maketitle

\begin{abstract}
In bipartite causal inference with interference, interventional units might receive treatment or control, and they might affect the outcome of outcome units through their connections on a bipartite network. We study bipartite causal inference with interference based on observational data across time and a changing bipartite network. Under an exposure mapping framework, we define the immediate and carryover causal effects for each outcome unit, representing contrasts of potential outcomes under different values of the immediately preceding and past exposures, respectively, averaged over time. We establish unconfoundedness of the exposure received by outcome units based on unconfoundedness assumptions on the interventional units' treatment assignment and the random network, hence respecting the bipartite structure of the problem. Our results hold for binary, continuous, and multivariate exposure mappings. In the special case of binary exposure and carryover mappings, we propose algorithms for the immediate and carryover causal effects that combine matching and covariate balancing. We show that the bias of the resulting estimators is bounded. In our motivating study, we find some evidence that smoke from wildfires has an immediate impact on reducing transportation by bicycle in San Francisco.
\end{abstract}

% Something appeared in the introduction: Interventional units have covariate-related treatment assignment probability varying temporally. The network changes randomly

% In a bipartite setting, two groups are interventional units and outcome units. From time to time, interventional units are assigned treatments and through some social networks, they interfere with the outcome units. Instead of targeting the average causal effect of all the units, we are only interested in a particular unit, as to how exposure affects this unit.

\section{Introduction}
Causal inference methodology most often focuses on the scenario where units are assigned to treatment or control, and an outcome is measured on the same set of units. 
% The term interference has been used to describe the setting where a unit's potential outcomes depend on its own treatment, but also on the treatment of others. However, it is not always possible to isolate units entirely from interference, and a unit might be affected by the treatment level of others.
However, in some cases, the units that receive the treatment are distinct from the units that experience the outcome. We refer to the former as interventional units, and the latter as outcome units. The outcome units do not receive the treatment themselves. Instead, their exposure to the treatment is through their connections to potentially treated interventional units. 
The causal dependencies between interventional and outcome units can be described in a bipartite network, and, as a result, this setting has been termed bipartite interference \citep{zigler2021bipartite}.

In this manuscript, we focus on bipartite causal inference with interference from observational data measured over time.
The treatment level of an interventional unit and the bipartite network can change over time according to unknown mechanisms that depend on past covariates of the interventional units, the outcome units, and the network.
% !CUT!
% Therefore, the evolving treatment assignment of interventional units and network structure leads to temporal variations of the outcome units' exposure.
%\textcolor{blue}{
The outcomes of outcome units can be influenced by the immediately preceding and past exposures to the treatments of the interventional units through the bipartite network. 
% !CUT!
% In this setting, we define estimands representing the immediate and carryover effects of the interventional units' treatment on an outcome unit, introduce identifying assumptions on the treatment and network processes, and develop an estimation strategy that combines principles of matching and covariate balancing.
%}

Existing work in bipartite causal inference with interference is cross-sectional and mostly considers a fixed and known bipartite network. %, in the observational \citep{zigler2021bipartite,Zigler2020} and in the experimental  \citep{pouget2019variance,brennan2022cluster,Doudchenko2020,harshaw2023design} setting.
\cite{zigler2021bipartite} introduced causal estimands for bipartite causal inference, and developed weighting estimators under clustered interference.
The concept of an exposure mapping introduced in unipartite causal inference \citep[e.g.,][]{Aronow2017, forastiere2021identification} has been extended to the bipartite setting, stating that potential outcomes depend on the interventional units' treatment through known functions of the treatment and the bipartite network. 
In \cite{zigler2025bipartite},
{interventions on power plants (interventional units) can affect health outcomes across zip codes (outcome units)} and the bipartite network describes complex atmospheric and geographic dependencies among units. % Estimation and inference is based on an unconfoundedness assumption for the outcome unit's exposure.
%extends bipartite studies to a more general context where networks are characterized by spatial-varied conditions. This work also implicitly tells that covariates can be the key to driving both network and treatment, thus eventually forming complex exposure patterns.
In experimental settings and under a linear model for the potential outcomes,
\cite{harshaw2023design} designed estimators and inferential techniques for the effect of assigning all or none of the interventional units to treatment.
\cite{pouget2019variance} {developed experimentation techniques for bipartite interference that improve the efficiency of estimators with applications in marketplace experiments where discounts on Amazon listings (interventional units) can influence the behavior of customers (outcome units).} \cite{brennan2022cluster} focused on avoiding inferential bias due to interference.
\cite{Doudchenko2020} proposed using propensity scores to account for confounding 
%of a unit's exposure 
due to the network structure. 
In experimental settings, previous work has also studied bipartite causal inference from a design-based perspective \citep{chattopadhyay2023design, papadogeorgou2025causal, lu2025design}. 
All these studies have fixed causal networks, except \cite{wikle2023causal}, which considers a probabilistic bipartite network in a cross-sectional design.
%To our knowledge, only \cite{Phan2015} consider the case of non-fixed network in causal inference, focusing on how natural events impact the network of units' interconnectedness in a unipartite setting.  
% To our knowledge, bipartite causal inference has not been investigated in the case of time series data with a random bipartite network.

For time series data, most of the causal inference literature focuses on the case without interference 
\citep[see][for surveys on the topic]{abadie2018econometric,imbens2024causal}.
% (for example,
% interrupted time series \citep{mcdowall2019interrupted},
% % ITS references: \citep{penfold2013, lopez2019difference}
% difference-in-differences \citep{athey2006identification}, 
% % DiD references: \citep{card1994minimum, abadie2005semiparametric}
% synthetic controls \citep{Abadie2003},
% % SC references: \citep{ doudchenko2016balancing, ben2021augmented}
% and Bayesian methods \citep{brodersen2015inferring}).
% % Bayesian references: \citep{antonelli2023heterogeneous, papadogeorgou2023evaluating}, 
% % and others references: \citep{Du2018, Imai2021matching}).
% %
%Researchers have investigated time-series cross-sectional data extensively using these approaches for a long time, such as \cite{Du2018} and \cite{Imai2021matching}.
%, and \cite{papadogeorgou2022causal} developed a causal framework for observational data from spatio-temporal time series on a single unit where confounding adjustment is achieved via weighting. 
% In many cases with time series data, it is likely that interference occurs across the network of units. 
Recently, panel data methodology has been extended to the case with unit-to-unit interference. In \cite{cao2019estimation, grossi2020synthetic, di2020inclusive}; and \cite{menchetti2020estimating}, some units receive the treatment at some point in time and remain treated thereafter, and in \cite{clark2021approach} and \cite{agarwalnetwork} the units' treatment assignment changes over time. These methods have yet to be extended to the bipartite setting, and assume the existence of multiple units measured over time. %or under a random network.

Recent work on causal inference with data over time 
% in traditional experimental designs , including canonical difference-in-differences and two-way fixed effects models \citep{imai2019should, imai2021use, arkhangelsky2024design}. However, 
% quasi-experimental 
allows past treatments to affect future outcomes, called carryover effects.
% \citep[e.g.,][]{imai2023matching, liu2024practical}.
Relevant to our work is the literature on switchback experiments, where a single unit is followed over time, and its treatment changes according to an experimental design.
% For experiments on a single-unit, 
\cite{Bojinov2019} proposed an IPW estimator for general temporal estimands. \cite{bojinov2023design} developed minimax decision rules for optimizing the treatment assignment in switchback experiments, and \cite{hu2022switchback} used Markov models to allow for long-term carryover effects. In switchback designs the treatment assignment mechanism is known, which simplifies estimation. In contrast, in observational bipartite settings, the exposure assignment mechanism is unknown and potentially hard to model as it depends on unknown treatment and network dynamics. %}

%\textcolor{blue}{
Our motivating context is the evaluation of the effect of smoke from wildfires on population mobility using bicycles (\cref{sec:motivating}). Forested geographical areas correspond to the interventional units on which a wildfire might take place and populated geographical areas are the outcome units on which biking activity is measured. Wildfires in forested areas can lead to smoke exposure in populated areas according to weather and atmospheric processes which, in turn, can affect cycling activity. 
% More applications are discussed in discussed in \cite{chattopadhyay2023design}, also known as a special case of generalized network experiments.
%}

%The outcome units, which do not get treatment themselves, are exposed to the treatment only through their connections to potentially treated interventional units. 
%The causal dependencies across units can be described in a bipartite network, and, as a result, this setting has been termed bipartite interference \citep{zigler2021bipartite}. 

% This work develops a causal inference framework for bipartite interference with time series observational data and a random bipartite network, which allows researchers to capture the dynamic nature of causal relationships, namely direct exposure effect and indirect carryover effect, in real-world settings where networks and treatment assignments change over time. 
From a statistical perspective, our contributions are the following:
\begin{enumerate*}[label=(\alph*)]
    \item %textcolor{blue}{
    Under an exposure mapping framework, we 
    define two causal estimands for an outcome unit,
    the \textit{immediate} effect and the \textit{carryover} effect which represent the effect of the most recent and past exposures, respectively (\cref{sec:model-setup}). These estimands are defined as contrasts of potential outcomes for a given outcome unit averaged across time.%}
    \item  We introduce unconfoundedness assumptions for the treatment and network processes conditional on variables of the interventional units, outcome units and the network.  We establish the unconfoundedness for the outcome unit's most recent and carryover exposures, which implies that we can estimate the outcome-unit-specific effects while conditioning \textit{only} on temporally-varying information (\cref{sec:model-setup}).
    % We discuss that this is a strength of analyses that focus on temporally-averaged causal estimands for each unit.
    These results hold for binary, continuous, and multivariate exposures.
    \item Focusing on binary exposure mappings for the most recent and past exposures, %{\color{blue} 
    we propose estimators for the immediate and carryover effects that combine ideas from matching and covariate balancing. 
    % The proposed algorithms match time periods for an outcome unit based on their exposure values such that matched time periods occur close in time, while the matches satisfy balance constraints on time-varying covariates.
    We show that the bias of the estimators is bounded, and can be made arbitrarily small based on the choice of algorithmic parameters (\cref{sec:matching}).
%    
% \item Our approach infers the causal effect of the interventional units' treatment on each outcome unit separately. We establish how results on multiple outcome units can be combined to test a global null hypothesis of no treatment effect (\cref{sec:matching}).
%
\item In an extensive simulation study, we showcase that our approach performs well for estimating the outcome-unit effects (\cref{sec:simu_real_data}).
\item Using this approach, 
%to study whether smoke from wildfires affects outdoor physical activity in the San Francisco Bay area,
%
%The intensity of smoke is classified into three labels (light, medium, heavy) based on the visual classification of plumes using GOES-16 and GOES-17 ABI true-color imagery. We can overlay the Bay Area map with smoke polygons to distinguish which locations are exposed, thus creating a binary indicator of smoke exposure. 
%
%Our work shows that fixing time-varying confounders, the movement of air across space creates a randomization for the level of wildfire smoke exposure a bikeshare location receives. Hence, 
we find that smoke exposure from wildfires leads to an immediate reduction in the number of bike rental hours in the city of San Francisco, but does not significantly alter bike usage in nearby regions, and it does not have carryover effects (\cref{sec:real-data}). 
% Our methodology is applicable in other areas where environmental or other random processes drive how exposure from a source is transmitted across a geography of interest. 
% For example, the proposed methodology can be used to study how pollutants released in water affect ecological and biological systems, where random current patterns act as the random network, or the effect of an outcome unit's exposure under the random network of drivers and riders that are in close proximity in ride-share applications, or the random distribution of bikes or e-scooters across cities.
\end{enumerate*}
We conclude with a discussion (\cref{discussion}).

\section{Wildfire smoke and transportation by bicycle}
\label{sec:motivating}

%{\color{blue}
In recent years, climate change has led to more frequent and prolonged wildfires across North America. These events pose serious threats to human life, property, and ecological systems \citep{chen2021climate}. Wildfire smoke, a source of fine particulate matter, can be transported over long distances and poses a major public health concern. Exposure to smoke from wildfires has been associated with increased rates of respiratory infections and all-cause mortality \citep{reid2016critical}. 
Beyond direct health impacts, wildfire smoke influences human behavior in other ways as well. \cite{doubleday2021urban} found a significant decline in daily bike usage during and after wildfire smoke events, %in Seattle, WA, 
and \cite{rosenthal2020assessment} found a significant reduction in individuals' step counts with deteriorating air quality due to wildfires. %in California.
Their findings suggest that wildfire smoke can disrupt urban transportation and physical activity with broader implications for physical and mental health.

\begin{figure}[h]
\sbox0{\begin{subfigure}[b]{\dimexpr 2.6in-0.5\columnsep}% measuer height with caption
  \includegraphics[width=\linewidth, height=4in, trim=25 20 20 20, clip]{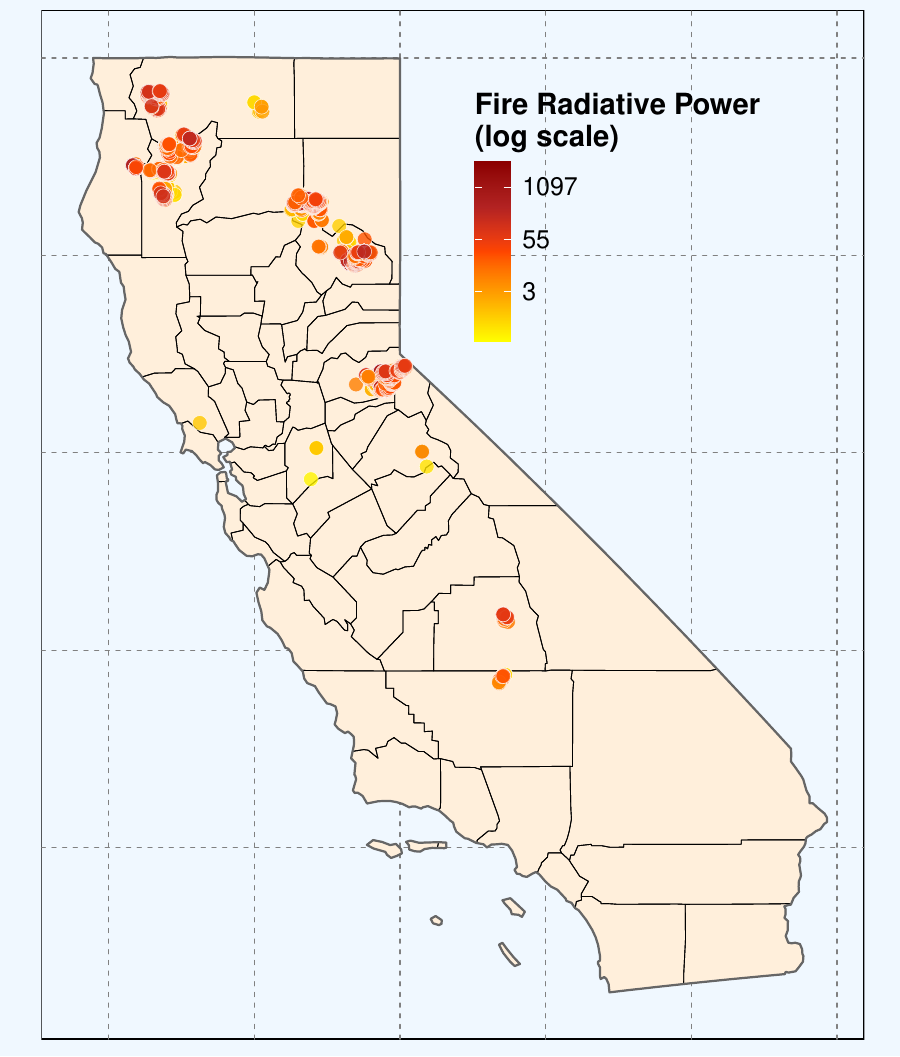}% some argitrary height
  \caption{Wildfire map in California}\label{fig:wild-fire}
\end{subfigure}}%
\usebox0\hfill\begin{minipage}[b][\ht0][s]{\wd0}% s=stretch
  \begin{subfigure}{\linewidth}
    \includegraphics[width=\linewidth]{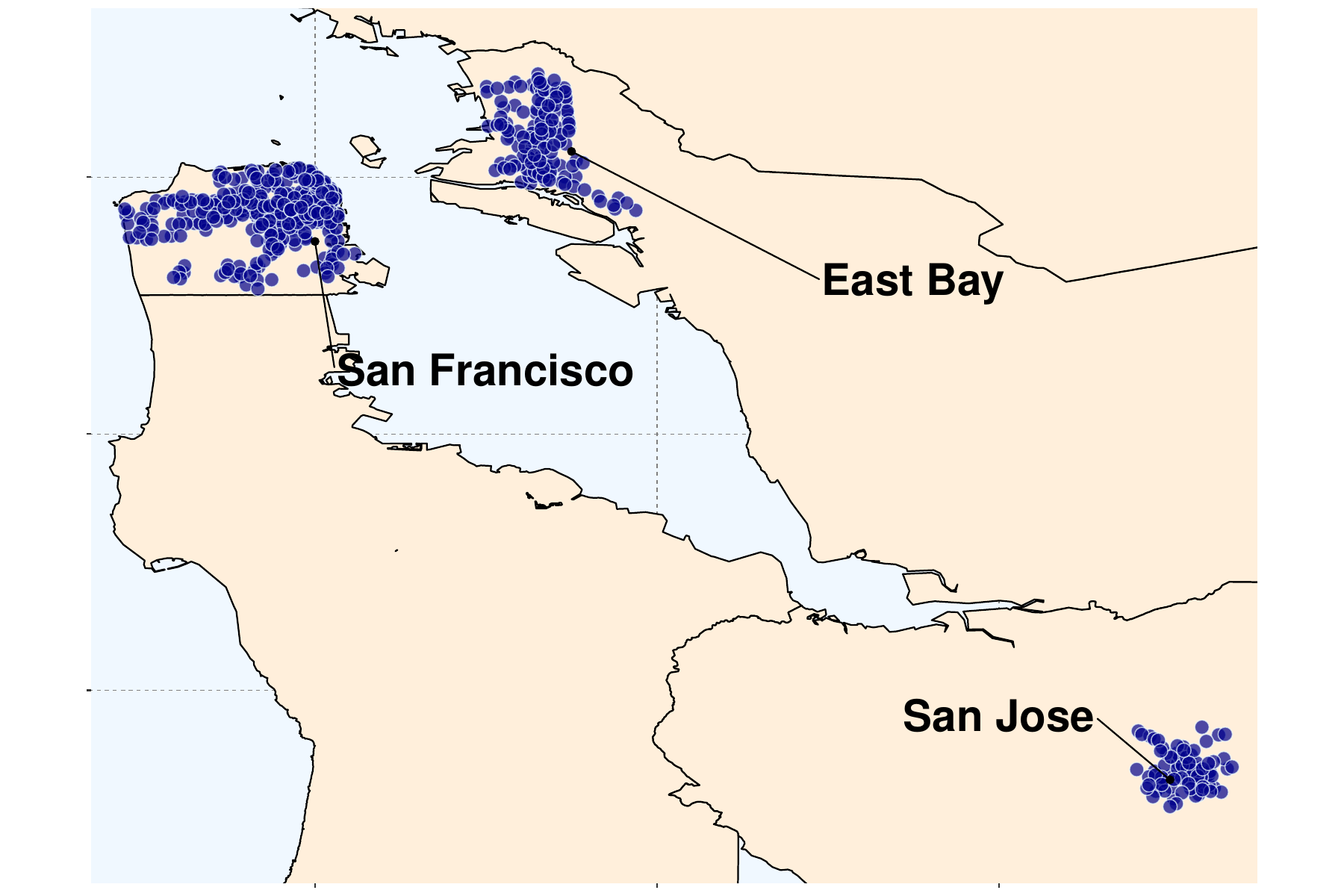}
    \caption{Bikeshare locations in Bay Area}\label{fig:bike}
  \end{subfigure}\par
  \vfill% very important
  \begin{subfigure}[b]{\linewidth}
  \includegraphics[width=\linewidth]{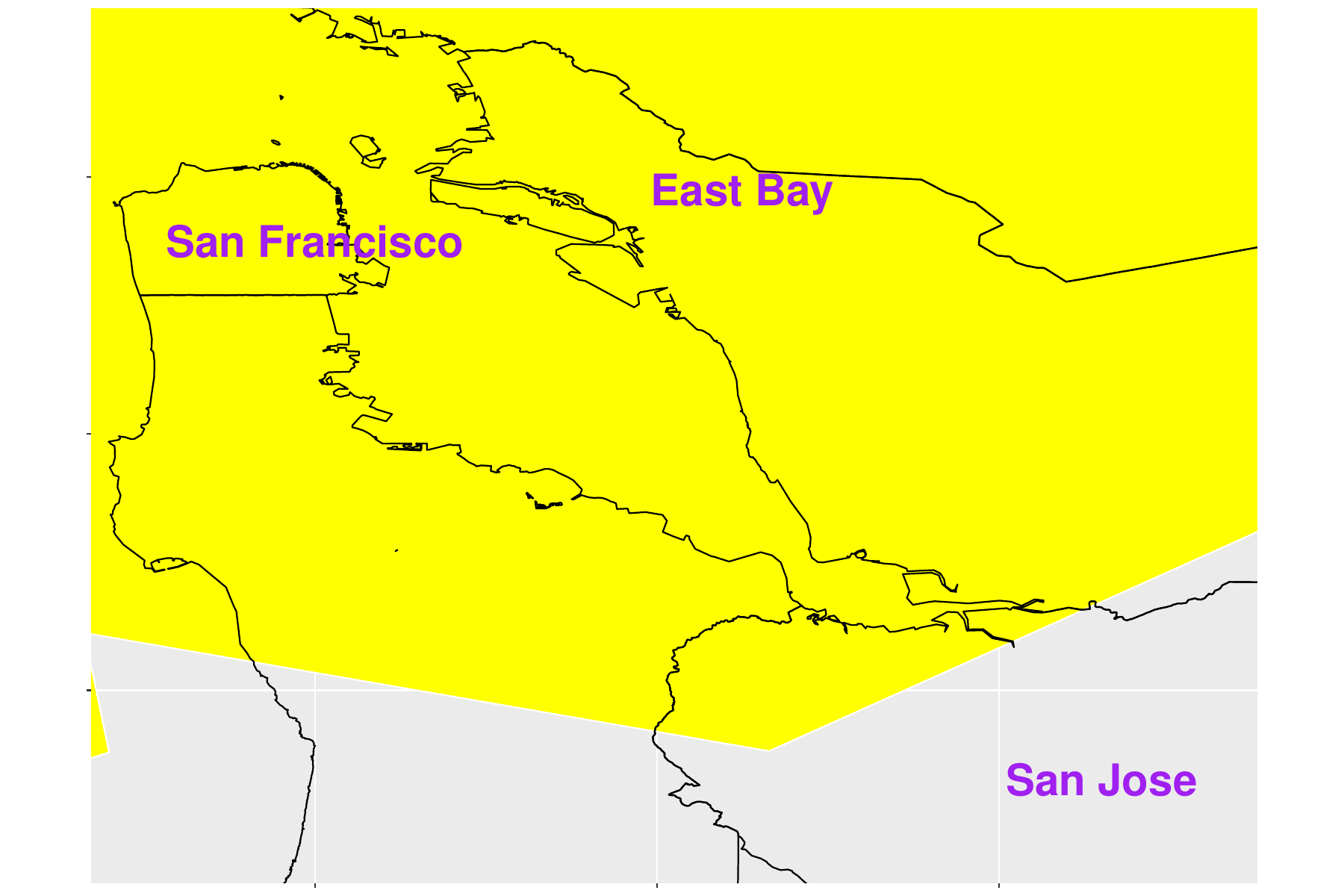}
    \caption{Wildfire exposure in Bay Area}\label{fig:exposure}
  \end{subfigure}
\end{minipage}
\caption{(a) Fire detection point data in California on August 31, 2021. Color scheme shows the fire radiative power which is rates of energy release (in log scale). (b) The locations of the bikeshare stations in the three outcome units. (c) Wildfire smoke coverage on August 31, 2021 as measured by HMS, where yellow represents light, moderate or heavy smoke.}
\end{figure}

We contribute to this literature by studying how North American wildfires affect biking activity in the San Francisco Bay Area. 
We use $\mathcal{N}=\{n_1, n_2,\cdots, n_N\}$ to denote the interventional units corresponding to forested geographical areas. Data are measured over time periods $t \in \mathcal{T} = \{1, 2, \cdots, T\}$ corresponding to daily information from January 2021 to September 2023 with $T = 1,003$.
We use $A_{ti} \in \mathcal{A}$ to denote the treatment level of interventional unit $n_i$ at time $t$, representing wildfire activity in that area. Then, $\bm{A}_t =(A_{t1}, A_{t2},\cdots, A_{tN})^\top$ is the treatment vector at time $t$ across all interventional units, with $\bm A_t \in \mathcal{A}^N$. \cref{fig:wild-fire} illustrates wildfire occurrence and intensity on August 31, 2021. The interventional units do not experience the outcome.

We study the effect of smoke from wildfires on biking activity in three populated geographical areas in northern California: San Francisco, San Jose and the East Bay.
The outcome of interest is the total daily bicycle riding time in each area recorded by Lyft’s Bay Wheels bikeshare program, which has tracked bikeshare usage since 2017 across more than 450 stations. \cref{fig:bike} shows the locations of the stations in the three areas. 
Because bikeshare usage reflects daily patterns of outdoor activity, it can serve as a proxy for how populations respond to wildfire smoke. 
% In addition, reductions in usage may imply economic losses for the bikeshare program, offering practical insights for service operators and policymakers. 
In our data, only 206 out of approximately 6 million rentals started and ended in different areas, implying minimal spatial spillover effects from one region to the other. This allows us to analyze each of the three outcome geographical areas separately.
Therefore, we focus on a single outcome unit $m$ throughout. We use $Y_{t}$ to represent the total number of bike riding hours on day $t$.
During our study period, the number of hours of bikeshare rentals was on average 1,359 in San Francisco, 105 in East Bay, and 612 in San Jose.

Smoke originating from a wildfire can travel long distances depending on weather and atmospheric conditions, with transport patterns that change over time. The vector $\bm{G}_{t} = (G_{t1},\cdots,G_{tN})^\top \in \mathcal{G}$ denotes the bipartite connectivity of all interventional units in $\mathcal N$ with the outcome unit $m$. Specifically, $G_{ti}$ represents the extent to which smoke from a wildfire in area $n_i$ is transported to the populated area of interest at time $t$. 

The treatment assignment of the interventional units, the outcome of the outcome unit, and the bipartite connectivity vector can all depend on covariates. 
We use $\bm X_{0i}^\text{int}, \bm X_0^\text{out},$ and $\bm X_{0i}^\text{net}$ to denote time-invariant covariates for the interventional unit $n_i$, the outcome unit $m$ and their network relationship, of length $p^\text{int}_0, p^\text{out}_0,$ and $p^\text{net}_0$, respectively.
We use the same notation with subscript $t$ for time-varying information, although the number of time-invariant and time-varying covariates might differ. Specifically, at time $t$, we use $\bm X_{ti}^\text{int} 
= (X_{ti1}^\text{int}, X_{ti2}^\text{int}, \cdots, X_{tip^\text{int}}^\text{int})^\top
$ to denote the $p^\text{int}$-vector of time-varying covariates associated with interventional unit $n_i$, $\bm X_{t}^\text{out}
% = (X_{t1}^\text{out}, X_{t2}^\text{out}, \cdots, X_{tp^\text{out}}^\text{out})^\top
$ to denote the $p^\text{out}$-vector of time-varying covariates for the outcome unit $m$, and $\bm X_{ti}^\text{net}
= (X_{ti1}^\text{net}, X_{ti2}^\text{net}, \cdots, X_{tip^\text{net}}^\text{net})^\top
$ to denote the $p^\text{net}$-vector of time-varying covariates characterizing the network relationship between units $n_i$ and $m$.
We use $\bm X_t^\text{int}=(\bm X_{t1}^{\text{int}\top}, \bm X_{t2}^{\text{int}\top}, \cdots, \bm X_{tN}^{\text{int}\top})^\top$ to denote the $N p^\text{int}$ time-varying covariate vector across all the interventional units. We similarly define $\bm X_{t}^\text{net}% =(X_{t1}^{\text{net}\top},X_{t2}^{\text{net}\top},\cdots, X_{tN}^{\text{net}\top})^\top
$ as the $N p^\text{net}$ covariate vector for the time-varying network covariates. Finally $\bm X_0=(\bm X_0^{\text{int}\top}, \bm X_0^{\text{net}\top}, \bm X_0^{\text{out}\top})^\top$ denotes all time-invariant covariates, while $\bm X_{t}=(\bm X_t^{\text{int}\top}, \bm X_{t}^{\text{net}\top},\bm X_{t}^{\text{out}\top})^\top$ denotes all $(p^\text{int}+p^\text{net})N+p^\text{out}$ time-varying covariates measured at time $t$. 
We use an overline notation to denote a variable's history. For example, for a variable $V_t$ measured over time, we use $\overline V_t=(V_t, V_{t-1}, \ldots, V_1)^\top$, with realization $\overline v_t$. For an integer $S$, we use $\overline V_{t, S}$ to denote the value of $V$ over the $S + 1$ time periods $t, t-1, \dots, t-S$, as $\overline V_{t, S} = (V_t, V_{t - 1}, \dots, V_{t-S})^\top.$ A glossary is included in %Supplement~\ref{supp_sec:glossary}.
Supplement~A.
% We use $\overline{\bm A}_t=({\bm A}_{t}^\top,{\bm A}_{t-1}^\top,\cdots,{\bm A}_{1}^\top)^\top$ for the vector of treatment assignments over all interventional units for all time periods up to (and including) time period $t$, with realization $\overline{\bm a}_t$. Similarly, we use $\overline{\bm G}_{t}=({\bm G}_{t}^\top,{\bm G}_{t-1}^\top,\cdots,{\bm G}_{1}^\top)^\top$ for the network history with realization $\overline{\bm g}_t$, and $\overline{\bm E}_{t}=({E}_{t},{E}_{t-1},\cdots,{E}_{1})^\top$ for the exposure history with realization $\overline{\bm e}_t \in \mathcal{E} \times \ldots \times \mathcal{E}_1$. Lastly, $\overline{\bm X}_{t}=(\bm X_{t}^\top,\bm X_{t-1}^\top, \cdots, \bm X_{1}^\top)^\top$ denotes the history of all covariates.

Time-invariant covariates may include vegetation type for the interventional units, demographic information for the outcome unit, and the geographic distance of the two for the network information. As we discuss in \cref{subsec:unconfoundedness}, time-invariant information need not be collected within our framework. Furthermore, {in \cref{sec:real-data} we discuss that, in our study, our approach does not require knowledge of $N$, nor does it rely on directly observing $\bm{A}_t$ or the covariates $\bm{X}_t^\text{int}$}. 
We acquire time-varying covariates for the outcome units, $\bm X_t^\text{out}$, representing weather information such as temperature, dew point, humidity, wind speed, and wind direction. Additional information on the data is available in Supplement~I.
%\ref{supp_sec:additional_study}. %}
%These data are obtained from \url{https://www.ncei.noaa.gov/cdo-web/datasets}.
% \textcolor{blue}{We claim that that there is no spillover outcome unit covariates that can drive the mechanism inside each outcome unit.}

% \textcolor{blue}{
% The forests are the geographical units that are plausible to take wildfire interventions but never observed with outcomes of Baywheel biking activities, whereas the biking geographical areas around Bay Areas never experience wildfires intervention, that is, they are treatment-ineligible units. This conforms to the realm of bipartite interference discussed within our content.} 

%The time-varying covariates $\bm X_{tj}^\text{out}$  We argue in \cref{sec:real-data} that no covariates beyond the above ones need to be added in our analysis.
\section{Bipartite interference with time series observational data and a random network} \label{sec:model-setup}

In this section, we formalize our causal inference framework for observational time series data with bipartite interference. We introduce exposure mappings for bipartite interference in \cref{subsec:exposure-mapping} and potential outcomes in \cref{subsec:exposure-mapping-pot-out}, define causal estimands for the immediate and carryover effect in \cref{subsec:estimands}, and establish the identifying assumptions in \cref{subsec:unconfoundedness}. 

\subsection{Exposure and carryover mappings}
\label{subsec:exposure-mapping}

The outcome unit does not receive the intervention itself, it instead experiences the treatment through its connections with potentially treated interventional units. We formalize this using exposure mappings. The function $h_{t}: \mathcal{A}^N \times \mathcal{G} \rightarrow \mathcal{E}$ maps the interventional units' treatment assignment and the outcome unit's bipartite connection vector at time $t$ to the outcome unit's exposure value at the same time, where $\mathcal{E}$ denotes the set of possible exposure values.
%Let $\bm{E}_t=\bm{h}_t(\bm{A}_t, \bm{G}_t)$  be the realized overall exposure of $N$ outcomes at time $t$.
Then, $E_{t} = h_{t}(\bm{A}_t, \bm{G}_{t})$ is the realized exposure at time $t$.
The function $h_{t}(\cdot)$ might return a scalar such as the proportion of interventional units with which $m$ is connected that are treated. It can also be completely general, specified to return the vector of treatment levels for all connected interventional units, or it can be extended to incorporate covariates.

%{\color{blue}
In our study, we acquire daily smoke exposure from the National Oceanic and Atmospheric Administration's Hazard Mapping System (HMS). HMS uses satellite imagery to track smoke plumes originating from wildfires and integrates information on fire activity, $\bm A_t$, and smoke transport, $\bm G_t$, to infer potential smoke exposure in each region on day $t$, $E_t$. This structure aligns naturally with our bipartite framework. %Satellite imagery is used to map smoke polygons, which represent the combined effect of wildfire emissions and atmospheric transport. 
HMS reports the observed smoke thickness in a given region reflecting the cumulative contribution of smoke. 
In our study, we consider a binary exposure mapping, $\mathcal{E} = \{0, 1\}$, where a populated area is considered exposed ($E_t = 1$) under light or heavier smoke thickness according to HMS, and unexposed ($E_t = 0$) otherwise.
% Three bikeshare locations might experience exposure to smoke from wildfires or not, based on random patterns of smoke transport and dispersion.
Overall, San Francisco and the East Bay experienced 147 days with smoke, while San Jose experienced 137.
% During most of our study window, the three outcome units are either simultaneously exposed or unexposed. 
\cref{fig:exposure} shows smoke exposure on August 31, 2021 when two of the areas are exposed.%}

In our time series setting, we further introduce a function of the $L$ past exposures, termed carryover mapping.
Specifically, we specify $h_t^\text{c} : \mathcal{E}^L \rightarrow \mathcal{R}$ that maps the vector of the exposures during the past $L$ time periods to a summary value. Similarly to the exposure mapping, the carryover mapping can be arbitrarily complex; it might return a scalar, or be completely general and return the vector of past exposures. We use 
$R_t = 
% h_t^\text{c}(\overline E_{t-1, L + 1}) =
h_t^\text{c}(\overline{E}_{t-1,L-1})$ 
for the observed carryover value at time $t$, with realization $r_t$. In our study, we consider a binary carryover mapping defined based on whether at least four days in the past week are exposed, $R_{t} = I(\sum_{l=1}^7 {E}_{t-l}\geq 4)$ where $I()$ is the indicator function. {Under this definition, carryover exposure occurred on 115 days in San Francisco, 166 days in the East Bay, and 104 days in San Jose.}

%Covariates for the interventional and outcome units might represent time-invariant information such as coordinates and size, or time-varying information such as daily weather conditions. A time-invariant characteristic of interest might be the geographic distance between an interventional and an outcome unit. Invariant examples can be the distances and altitude differences between two vertices while time-varying cases are daily wind intensity from interventions. 
% In a bipartite network, it is possible to have any number of interventional and outcome units, even 1 unit.
% In some studies (like ours), the number of interventional units $n$ might not even be known. 

% In our study, the interventional units are coal-powered electricity generating units. Potential covariates represent information on their operating characteristics. The outcome units are counties with covariates representing demographic and socioeconomic information, and the outcome represents to the county's all-cause mortality rate. The county exposure $E_{t}$ represents the coal exposure for county $j$ in month $t$ from all coal-burning power plants in the US, as measured using the HyADS model \citep{henneman2019characterizing, Zigler2020}. The HyADS model combines the output from a low-complexity pollution transport model that follows the trajectories of air particles based on wind fields, which represents the bipartite mapping $\bm{G}_t$, and the realized amount of emissions from the power plants, which is the interventional units' treatment $\bm{A}_t$.

\subsection{Potential outcomes under exposure mapping assumptions}
\label{subsec:exposure-mapping-pot-out}

% \textcolor{blue}{

Let $Y_{t}(\overline{\bm a}_t,\overline{\bm{g}}_{t})$ denote the potential outcome for unit $m$ at time $t$ had the treatment path of the $N$ interventional units been $\overline{\bm{a}}_t$, and under bipartite connection path  $\overline{\bm{g}}_{t}$.
The observed outcome corresponds to the potential outcome under the observed treatment and network, $Y_{t} = Y_{t}(\overline{\bm A}_t,\overline{\bm{G}}_{t})$.
We include the bipartite graph in the notation for potential outcomes to establish that the graph plays a role in how the potential outcomes vary with the interventional units' treatment, but we do not assume that the network is manipulable.
Potential outcomes are allowed to depend on the most recent and past treatments of the interventional units. 
%We make this explicit by writing $Y_{t}(\overline{\bm a}_t,\overline{\bm{g}}_{t})=Y_t(\bm a_t,\bm g_{t};\overline{\bm a}_{t-1},\overline{\bm{g}}_{t-1})$.
% We discuss this within the context of our study in \cref{sec:real-data}.
% We use $\bm{Y}_t(\bm a_t,\bm{g}_t)$ to denote the collection of the potential outcomes for all outcome units at time $t$, $\bm{Y}_t(\bm a_t,\bm{g}_t) = \big( Y_{t1}(\bm a_t,\bm{g}_{\cdot 1t}), Y_{t2}(\bm a_t,\bm{g}_{\cdot 2t}), \cdots, Y_{tM}(\bm a_t,\bm{g}_{\cdot Mt}) \big)$.
The following assumption encodes that the treatment of interventional units and the bipartite network drive the potential outcomes only through the exposure and carryover values.

\begin{assumption}\label{exp-map-cond} 
%(Exposure mapping equivalency)
For each $t \in \mathcal{T}$, the following holds. Let $\bm a_s, \bm a_s' \in \mathcal{A}^N$ and $\bm g_s, \bm g_s' \in \mathcal{G}$ be vectors with corresponding exposures 
$h_s(\bm a_s, \bm g_s) = e_s$ and 
$h_s(\bm a_s', \bm g_s') = e_s'$
for $s = t, t-1, \dots, t-L$, and carryover exposures $h_t^c(\overline{e}_{t-1,L-1}) = r_t$ and $h_t^c(\overline{e}_{t-1,L-1}') = r_t'$. %Define the path of exposures $\overline {\bm e}_{t-1} = (e_1, e_2, \dots, e_{t-1})^\top$, and similarly for $\overline {\bm e}_{t-1}'$.
If it holds that $e_t = e_t'$ and $r_t = r_t'$, then
$Y_{t}(\overline{\bm{a}}_t,\overline{\bm{g}}_{t})=Y_{t}(\overline{{\bm{a}}}_t', \overline{{\bm{g}}}_{t}')$, and the potential outcome can be denoted as $Y_t(e_t, r_t)$.
\end{assumption}

% \begin{assumption}\label{exp-map-cond} 
% %(Exposure mapping equivalency)
% Let $\bm a_s, \bm a_s' \in \mathcal{A}^N$ and $\bm g_s, \bm g_s' \in \mathcal{G}$ be vectors such that $h_s(\bm a_s, \bm g_s) = h_s(\bm a_s', \bm g_s') = e_s$ for all $s = 1, 2, \dots, t$, and $\overline{\bm{a}}_t, \overline{{\bm{a}}}_t', \overline{\bm{g}}_{t}, \overline{{\bm{g}}}_{t}'$ and $\overline {\bm e}_t$ the corresponding path vectors. 
% Then $Y_{t}(\overline{\bm{a}}_t,\overline{\bm{g}}_{t})=Y_{t}(\overline{{\bm{a}}}_t', \overline{{\bm{g}}}_{t}')$, and the potential outcome can be denoted as $Y_{t}(\overline{\bm e}_{t}) = Y_t(e_t; \overline{\bm e}_{t-1})$.
% \end{assumption}

%The following assumption codifies that past exposures drive potential outcomes only through the value of the carryover mapping.
% \begin{assumption}\label{ass:carryover-exposure}
%     For $\overline{\bm e}_{t-1}, 
%     \overline {\bm e}_{t-1}' \in \mathcal{E}_{t-1} \times \ldots \times \mathcal{E}_1$ with $h_t^\text{c}(\overline{\bm e}_{t-1}) = h_t^\text{c}(\overline{\bm e}_{t-1}') = r_t$, it holds that $Y_{t}(e_t; \overline{\bm e}_{t-1}) = Y_t(e_t; \overline{\bm e}_{t-1}')$ for any $e_t \in \mathcal{E}$, and the potential outcome can be denoted as $Y_t(e_t, r_t)$. 
% \end{assumption}

Under \cref{exp-map-cond}, potential outcomes depend on the most recent exposure value, $e_t$, and the summarized carryover exposure, $r_t$. The carryover exposure is allowed to depend on the exposure values during the most recent $L$ time periods.
%, which affects the outcome at the subsequent time point. From now on the potential outcome will be written uniquely as $Y_{t}(e_{t},r_t)$.}
The collection of all potential outcomes for outcome unit $m$ at time $t$ is the set $\mathcal{Y}_{t}(\cdot) = \{ Y_{t}(e_t, r_t), \text{ for } e_t \in \mathcal{E} \text{ and } r_t \in \mathcal{R} \}$. 
% --- Not needed based on the new assumption:
% We also define the collection of potential outcomes under a fixed value of the most recent or carryover exposure as $\mathcal{Y}_t(e_t, \cdot) = \{ Y_{t}(e_t, r_t), \text{ for } r_t \in \mathcal{R} \}$, and $\mathcal{Y}_{t}(\cdot, r_t) = \{ Y_{t}(e_t, r_t), \text{ for } e_t \in \mathcal{E}\}$, respectively.

In cross-sectional settings, assumptions like \cref{exp-map-cond} which reduce potential outcomes based on exposure mappings have been discussed in the unipartite \citep{Aronow2017, forastiere2021identification} and bipartite \citep{zigler2025bipartite, harshaw2023design, Doudchenko2020} interference literature.
\cite{savje2024causal} discusses the implications of using exposure mappings in the definition of estimands and as assumptions on potential outcomes, providing interesting distinctions between the two.

%For example, if the cardinality of $\mathcal{E}_{tj}$ is $K_{tj}$ for each $j$ and $t$, the set of potential outcomes $Y_{tj}(\cdot)$ will be $\{Y_{tj}(e_1),\cdots,Y_{tj}(e_{K_{tj}})\}.$ Given any $k_{tj} \in \{1,\cdots,K_{tj}\}$,  The equivalency of exposure at $(j,t)$ tuple leads to the same potential outcome.

\subsection{Causal estimands for the immediate and carryover effects}
\label{subsec:estimands}

We define time-specific estimands for the outcome unit $m$ that represent the immediate and carryover effect of the most recent and past exposures, respectively. Specifically, let $e, e' \in \mathcal{E}$ and $r, r' \in \mathcal{R}$ represent possible values for the exposure and carryover mappings at time $t$. Then, we define
%
% We consider estimands that are specific to each outcome unit. The contrast
% \(
% % \begin{align*}
% \tau_{t}(\overline{\bm e}_t,\overline{\bm e}_t')= Y_{t}(e_t,m(\overline{\bm e}_{t-1})) - Y_{t}(e'_t,m(\overline{\bm e}'_{t-1}))
% %\end{align*}
% \)
% represents the fundamental effect of a change in exposure from $\overline{\bm e}'_t$ to $\overline{\bm e}_t$ for the outcome unit $m$ at time $t$. We observe that \(\tau_{t}(\overline{\bm e}_t,\overline{\bm e}_t')=\tau_{t}((e_t,\overline{\bm e}_{t-1}'^\top)^\top,\overline{\bm e}'_t)+\tau_{t}(\overline{\bm e}_t,(e_t,\overline{\bm e}_{t-1}'^\top)^\top)\), an additive model with both the direct effect due to a change in $e_t$ to $e_t'$ and the indirect carryover effect due to a change in $\overline{\bm e}_{t-1}$ to $\overline{\bm e}'_{t-1}$.  Since no cross product between the current exposure and the carryover exposure, both effects are well interpreted.
% The unit- and time-specific estimands cannot be estimated without parametric assumptions. Instead, we consider target estimands that represent temporally-averaged causal effects for unit $m_j$ for a change in its exposure value. Specifically,
\begin{align*}
% \tau_t^\text{imm}(e_t, e_t' ; r_t) &= E \left[ Y_{t}(e_t, r_t) - Y_{t}(e_t', r_t) \right],
\tau_t^\text{imm}(e, e' ; r) &= Y_{t}(e, r) - Y_{t}(e', r),
% \intertext{and}
\quad \text{and} \quad
% \tau_t^\text{car}(r_t, r_t' ; e_t) &= E \left[ Y_{t}(e_t, r_t) - Y_{t}(e_t, r_t') \right],
\tau_t^\text{car}(r, r' ; e) = Y_{t}(e, r) - Y_{t}(e, r'),
\end{align*}
representing the effect at time $t$ of switching the most recent exposure from $e$ to $e'$ when the carryover exposure is set to $r$, and the effect of switching the carryover exposure from $r$ to $r'$ when the most recent exposure is equal to $e$, respectively.

The time-specific estimands cannot be estimated without parametric assumptions. Instead, we consider target estimands that represent temporally-averaged causal effects. 
% Specifically, we define the temporally-averaged immediate and carryover effects as
% \[
% \tau^\text{imm}(e, e'; r) = \frac1T \sum_{t = 1}^T \tau_t^\text{imm}(e, e'; r), \hspace{5pt} \text{and} \hspace{5pt} 
% \tau^\text{car}(r, r'; e) = \frac1T \sum_{t = 1}^T \tau_t^\text{car}(r, r'; e),
% \]
% respectively.
% Moreover, we consider these estimands averaged over only the time periods with specific values for the most recent and carryover exposure. 
Specifically, we use $\mathcal{T}_{e.}$ to denote the subset of $\mathcal{T}$ with $E_t = e$, and $\mathcal{T}_{.r}$ to denote the subset of $\mathcal{T}$ with $R_t = r$. We define 
\begin{equation}
\begin{aligned}
{\tau}^\text{imm}(e, e') & = \frac1{|\mathcal{T}_{e.}|} \sum_{t \in \mathcal{T}_{e.}} \tau_t^\text{imm}(e, e'; R_t)
= \frac1{|\mathcal{T}_{e.}|} \sum_{t \in \mathcal{T}_{e.}}
\left( Y_t - Y_t(e', R_t) \right),
\quad \text{and} \\
\tau^\text{car}(r, r') & = \frac1{|\mathcal{T}_{.r}|} \sum_{t \in \mathcal{T}_{.r}} \tau_t^\text{car}(r, r'; E_t)
= \frac1{|\mathcal{T}_{.r}|} \sum_{t \in \mathcal{T}_{.r}} \left(Y_t - Y_t(E_t, r') \right),
\end{aligned}
\label{eq:att-type-effects-no-interaction}
\end{equation}
where we have used that the observed outcome is equal to the potential outcome under the observed exposures, $Y_t = Y_t(E_t, R_t)$.
Under the following assumption, these quantities have a causal interpretation as immediate and carryover effect, respectively.
\begin{assumption}
The immediate and carryover treatment effects are constant in the carryover and most recent exposure value, respectively, i.e., for $t \in \mathcal{T}$ and any $e, e' \in \mathcal{E}$ and $r, r' \in \mathcal{R}$, it holds that $\tau_t^\text{imm}(e, e'; r) = \tau_t^\text{imm}(e, e'; r')$, and $\tau_t^\text{car}(r, r'; e) = \tau_t^\text{car}(r, r'; e')$.
\label{ass:no-interaction}
\end{assumption}
\noindent

We maintain \cref{ass:no-interaction} throughout as it allows us to gain estimation efficiency in estimating immediate effects by pulling information across time periods with different values of the carryover exposure, and similarly for estimating carryover effects. 
However, we note here that \cref{ass:no-interaction} is \textit{not} necessary. We discuss alternative estimands and estimation strategies that bypass this assumption in Supplement C.3.
%~\ref{supp_subsec:without_nointeraction_assum}.

Since these estimands represent contrasts of potential outcomes among time periods with a fixed exposure value, they resemble temporal versions of the sample average treatment effect on the treated in the cross-sectional literature without interference.
In the presence of multiple outcome units, the estimands in \cref{eq:att-type-effects-no-interaction} can be averaged across units to represent population-level effects. Alternative estimands tied to the treatment of the interventional units can also be considered, as the impact of changes in the exposure for specific changes in the treatment vector of the interventional units.

The temporally-averaged estimands defined in \cref{eq:att-type-effects-no-interaction} involve potential outcomes under exposure values that are not observed. Specifically, they involve potential outcomes of the form $Y_t(e', R_t)$ for time periods $t$ with $E_t = e$, and potential outcomes of the form $Y_t(E_t, r')$ for time periods $t$ with $R_t = r$. 
In principle, to avoid positivity violations, one can assume that all exposure levels considered in the estimands are possible for all time periods \citep{han2024population}.
Instead, if the exposure value $e'$ is impossible for some time period $t \in \mathcal{T}_{e.}$ or the carryover exposure value $r'$ is impossible for some time period $t \in \mathcal{T}_{.r}$, then we assume that this time period is excluded from the corresponding estimand. This is relevant in studies with non-stationarity in the exposure variable where the outcome units are guaranteed to experience the exposure during certain time windows. However, this is not expected to be an issue in our study, where it is always possible that the outcome units do not experience smoke from wildfires.

The estimands presented here represent contrasts of potential outcomes for a single outcome unit averaged over time. In \cref{subsec:temporal_advantages}, we discuss how focusing on temporally-average estimands might lead to weaker confounding adjustment requirements compared to estimands that average across units.

% The interference from $\mathcal{N}$ leads to different levels of exposure at $\mathcal{M}$, which is evaluated as $\bm{E}_t=(E_{t1},E_{t2},\cdots,E_{tM})$, a length $M$ binary vector, at each time $t$. Each $E_{t}$ is associated with all or partial of $\bm{A}_t$,  $\bm{G}^\top_{\cdot j}$, $W_{tj}$, and $t$, determined uniquely by a function $h(\bm{A}_t,\bm{G}^\top_{\cdot j},W_{tj},t)$.

% We assume that there exists a potential outcome at site $m_j$ only depending on the exposure. The exposure stems from the variation of both $\bm{A}_t$ and $G^\top$  at each time.  $Y_{tj}(E_{t})$ is labeled as the potential outcome for $m_j$ at $t$. The vectorized formats are $\bm{Y}_t(0) = (Y_{t1}(0),\cdots,Y_{tM}(0))^\top$ and $\bm{Y}_t(1) = (Y_{t1}(1),\cdots,Y_{tM}(1))^\top$ for nonexposed and exposed potential outcomes. The observed outcome is $Y_{tj}=(E_{t})$.

%For the observed exposed unit, we would never know the unexposed outcome simultaneously due to the fundamental problem of causal inference. However, 

%  Our interest lies in the causal effect of changes in exposure of one exposed outcome unit $m_j$; that is, $\tau_j=Y_{tj}(1)-Y_{tj}(0)$. 

\subsection{Ignorable assignments: assumptions and results}
\label{subsec:unconfoundedness}

We establish unconfoundedness assumptions that allow us to estimate the immediate and carryover effects. Our assumptions pertain to the interventional units' treatment assignment and the random bipartite network.

\begin{assumption}{(Unconfoundedness of the treatment and network assignment).}
The interventional units' treatment assignment and the network formation at times $t, t-1, \dots, t-L$ is independent of the potential outcomes $\mathcal{Y}_{t}(\cdot)$, conditional on the value of a function of time at time $t$, $f(t)$, time-invariant covariates, and time-varying covariates of the interventional units, the outcome unit, and the network at times $t, t-1, \dots, t-S$,
i.e.,
$
P \big( \overline{\bm{A}}_{t,L}, \overline{\bm{G}}_{t,L} \mid \mathcal{Y}_{t}(\cdot), f(t),\overline{\bm X}_{t, S} ,\bm X_0 \big) =
P \big( \overline{\bm{A}}_{t,L}, \overline{\bm{G}}_{t,L} \mid  f(t), \overline{\bm X}_{t, S},\bm X_0\big)$.
\label{ass:indep}
\end{assumption}

% \begin{assumption}{(Unconfoundedness of the treatment assignment).}
% The interventional units' treatment assignment at time $t$ is independent of the potential outcomes $\mathcal{Y}_{t}(\cdot, r_t)$, conditional on a function of time $f(t)$, time-invariant and historical covariates of the interventional units, the outcome unit, and the network,
% i.e.,
% $
% P \big( \bm{A}_t \mid \mathcal{Y}_{t}(\cdot,r_t), f(t),\overline{\bm X}_{t} ,\bm X_0 \big) =
% P \big( \bm{A}_t \mid  f(t), \overline{\bm X}_{t} ,\bm X_0\big)$.
% %
% Furthermore, the historical interventional units' treatment assignment path up to time $t-1$ is independent of the potential outcomes $\mathcal{Y}_{t}(e_{t},\cdot)$ given time-invariant and historical covariates, i.e., $P(\overline{\bm A}_{t-1}|\mathcal{Y}_{t}(e_{t},\cdot),\overline{\bm X}_{t-1},\bm X_0)=P(\overline{\bm A}_{t-1}|\overline{\bm X}_{t-1},\bm X_0)$.
% \label{ass:indep1}
% \end{assumption}

Under Assumption \ref{ass:indep}, the treatment level of interventional units can be driven by the characteristics of the outcome unit of interest and general temporal trends such as those that alter the overall prevalence of treatment. 
Therefore, the allowed treatment assignment mechanisms accommodate complex bipartite dependencies between interventional and outcome units. In particular, this formulation recognizes that confounding may stem not only from the covariates of the interventional units, but also from features of the network and the outcome unit that influence treatment assignment. % Additionally, Assumption \ref{ass:indep1} imposes unconfoundedness of the treatment history, conditional on all relevant time-invariant covariates and historical time-varying covariates.
%
% \begin{assumption}{(Unconfoundedness of the bipartite network).}
% The bipartite connection vector at time $t$ is independent of the potential outcomes $\mathcal{Y}_{t}(\cdot,r_t)$ given the temporal trend $f(t)$, and the treatment assignment and covariates of the interventional units, the outcome unit,   i.e.,
% \(\displaystyle P \big(\bm G_{t} |\mathcal{Y}_{t}(\cdot,r_t), f(t), \overline{\bm A}_{t},\overline{\bm X}_{t}  ,\bm X_0\big) = P \big(\bm G_{t}|f(t), \overline{\bm A}_{t},\overline{\bm X}_{t} ,\bm X_0 \big).\) 
% %
% Likewise, the historic network connectivity is independent of the potential outcomes $\mathcal{Y}_{t}(e_{t},.)$ given the historic treatment assignment, and time-invariant or historic values of the covariates, i.e., 
% $P(\overline{\bm G}_{t-1}|\mathcal{Y}_{t}(e_{t},\cdot),\overline{\bm A}_{t-1},\overline{\bm X}_{t-1},\bm X_0)=P(\overline{\bm G}_{t-1}|\overline{\bm A}_{t-1},\overline{\bm X}_{t-1},\bm X_0)$.
% \label{ass:random_graph}
% \end{assumption}
%
Furthermore, \cref{ass:indep} % \cref{ass:random_graph} 
allows the formation of connections between the outcome unit and interventional units to depend on temporal trends, as well as on past characteristics of the interventional units, the outcome unit and the network. It can also depend on the realized treatment level, which is relevant in applications where the overall treatment prevalence might lead to higher or lower outreach of the interventional units. 
The probabilistic formalization of unconfoundedness in \cref{ass:indep} % \ref{ass:random_graph} 
implicitly assumes a random network generation. That said, confounding can also arise in scenarios with a known and fixed network as illustrated in \cite{Doudchenko2020}, therefore our results are also applicable in that case.

% Assumptions \ref{ass:indep1} and \ref{ass:random_graph} allow 
% {\LARGE \color{blue} HERE}
% \cref{ass:indep} allows
% for complex dependencies of the treatment and network processes on a large class of covariates. 
Below, we establish that \cref{ass:indep} implies that the recent and carryover exposure that the outcome unit receives from the interventional units' treatment through the bipartite network is unconfounded. (The proof is in 
%Supplement B.)% 
Supplement B.)
%\ref{supp_sec:unconf_proof}.)
%
%
% \begin{proposition}{(Exposure unconfoundedness). }
% If Assumptions %\ref{exp-map-cond}, 
% \ref{ass:indep1} and \ref{ass:random_graph} hold, then 
% %and the outcome history of $m_j$, 
% the outcome unit's exposure is independent of the potential outcomes $\mathcal{Y}_{t}(\cdot, r_t)$, conditional on the temporal trend $f(t)$, and time-varying, current and historical covariate information, i.e.,
% $P({E}_{t} \mid \mathcal{Y}_{t}(\cdot,r_t),f(t),\overline{\bm X}_{t},\bm X_0) =P(E_{t} \mid f(t),   \overline{\bm X}_{t},\bm X_0)$. Furthermore, the carryover exposure is independent of the potential outcomes $\mathcal{Y}_{t}(e_t, \cdot)$, conditional on time-invariant and historical covariate information, i.e.,
% $P(R_t \mid \mathcal{Y}_{t}(e_t,\cdot),\overline{\bm X}_{t-1},\bm X_0) =P(R_t \mid   \overline{\bm X}_{t-1},\bm X_0)$.
% % i.e., $P({E}_{tj}| {Y}_{tj}(\cdot),f(t),   \bm{X}^\ast,\bm{X}_{\cdot t}, \bm{W}^\ast,\bm{W}_{tj}, \bm{P}^\ast,(\bm{P}_{t})_{\cdot j},\bm Y_{j,1:(t-1)}) =P(E_{t} |f(t),   \bm{X}^\ast,\bm{X}_{\cdot t}, \bm{W}^\ast,\bm{W}_{tj}, \bm{P}^\ast,(\bm{P}_{t})_{\cdot j},\bm Y_{j,1:(t-1)})$.
% \label{prop:as_if_randomized}
% \end{proposition}
%
\begin{proposition}{(Exposure unconfoundedness).}
If \cref{ass:indep} holds, then
the outcome unit's most recent and carryover exposure is independent of the potential outcomes $\mathcal{Y}_{t}(\cdot)$ conditional on the temporal trend at time $t$, $f(t)$, and covariate information, i.e.,
$P({E}_{t}, R_t \mid \mathcal{Y}_{t}(\cdot),f(t), \overline{\bm X}_{t, S}, \bm X_0) =P(E_{t}, R_t \mid f(t), \overline{\bm X}_{t, S}, \bm X_0)$.
\label{prop:as_if_randomized}
\end{proposition}
This result holds for exposure mappings that are arbitrarily complex. The exposure unconfoundedness statement in \cref{prop:as_if_randomized} has been evoked as an assumption in previous work on bipartite interference in cross-sectional settings \citep{zigler2025bipartite, Doudchenko2020}. Here, exposure unconfoundedness is established while acknowledging that, in a bipartite interference context, the exposure experienced by an outcome unit is governed by mechanisms operating at the treatment and network levels. This has important implications for practice. Assumption~\ref{ass:indep} yields practical insights for identifying confounders that exist in the treatment-outcome or network-outcome relationships. This is particularly relevant in bipartite interference contexts for which we have a clear grasp of the physical or mechanistic processes driving the network structure. Therefore, this assumption provides guidance that renders confounding adjustment more tangible, nuanced and actionable within the bipartite setting.

% The unconfoundedness result in \cref{prop:as_if_randomized} means that we can acquire unbiased estimators of the temporally-averaged immediate causal effect, $\tau^\text{imm}(e, e'; r)$, by comparing outcomes of time periods with similar values of the covariates in the conditioning set and carryover exposure equal to $r$, and different values of their exposure, and averaging over the covariate distribution across time \citep[see][for a related discussion]{forastiere2021identification}. We can similarly acquire unbiased estimators for the carryover effect $\tau^\text{car}(r, r'; e)$.
The unconfoundedness result in \cref{prop:as_if_randomized} means that we can acquire unbiased estimators of the temporally-averaged immediate causal effect, $\tau^\text{imm}(e, e')$, in the following manner: Compare outcomes of time periods with similar values of the carryover exposure and covariates in the conditioning set, and different values of their exposure \citep[see][for a related discussion]{forastiere2021identification}, and average over the covariate distribution among time periods with $E_{t} = e$ \citep{abadie2006large}. We can similarly acquire unbiased estimators for the carryover effect $\tau^\text{car}(r, r')$.
Importantly, the covariates $\bm X_{0}$ are constant across time. Therefore, they are implicitly {\it always} conditioned on when studying the same outcome unit across time. This implies that the time-invariant covariates that create differences in the assignment mechanism of treatment \textit{across} interventional units, or the assignment mechanism of bipartite connections \textit{across} pairs, {\it need not be measured} when focusing on estimands that average over time. Instead, we need to control for time-varying information only to estimate causal effects. We discuss this further in \cref{subsec:temporal_advantages}. Furthermore, as we discuss in \cref{sec:matching}, the temporal trend function $f(t)$ does not need to be known or specified by the analyst.

\section{Estimation}
\label{sec:matching}

Temporally-averaged causal effects can be estimated by controlling for time-varying information for exposure mappings of arbitrary complexity.
In our study of the effects of smoke from wildfires, a region's exposure is binary indicating the presence or absence of smoke exposure for the population residing in the area. Also, we consider a binary carryover exposure based on whether the region had smoke exposure during at least four out of the last seven days.
Therefore, from here onwards, we focus on the estimation of causal effects under binary exposure and carryover mappings. 

Viewing the time periods as the elementary unit of observation, in \cref{subsec:matching_algorithms} we introduce algorithms for estimating the immediate and carryover effects that match time periods with different exposure values under balance constraints for time-varying information.
% For an exposed time period, its matches act as the basis of imputing its missing counterfactual outcome, had it been unexposed.
In \cref{subsec:theory}, we define the corresponding causal effect estimators and we show, under assumptions on the potential outcome model, that the estimators' bias is bounded and can be controlled by the algorithmic tuning parameters. We discuss an inferential approach in \cref{subsec:inference}. Lastly, in \cref{subsec:temporal_advantages}, we discuss estimation advantages that arise by focusing on estimands that average over time for a specific outcome unit, over estimands that average across units.

% Unless natural disaster outbreaks occur, there is little reason to expect that there are changes in the environment that change unpredictably in a short time period, and nearby time points are expected to have similar values for the temporal trends. 

% Prior to the matching technique, the weighting mechanism is easy to put forward. For a target exposed unit, this takes the weighted average of the nearest two-sided unexposed outcome to be served as the fitted unexposed outcome had this unit not been exposed. To avoid using unexposed points multiple times, we randomly choose one exposed point for each unexposed one. The weighting approach is unbiased, and uniquely selecting unexposed units enables a good confidence level, as is indicated in Section \ref{supp_sec:simu}. However, since we delete duplicated unexposed units, it greatly reduces the number of matches. In the simulation, nearly half of the exposed time points are ignored, which renders a loss of credibility of hypothesis testing.

\subsection{Algorithms for estimating immediate and carryover effects}
\label{subsec:matching_algorithms}

%\cref{prop:as_if_randomized} indicates holding for the covariates and time equal in the condition creates randomization in exposure.
% From \cref{rm:linear_form}, we know that when outcome model equals an additive function of exposure, time, and unit-specific time-varying confounders formulated as a weighted average of interventional covariates and interactions, we only have to control for this unit-level balancing. This simplified model is common in real world, making covariate adjustment independent of other units.
 % As a result, two indicators are introduced to accommodate two different matching strategies.

% We propose three novel matching algorithms for estimating both direct and carryover causal effects in bipartite interference settings with time series observational data and a binary exposure. 
% Since we focus on {\it outcome unit specific} estimands, our matching algorithms match time periods with and without exposure for the {\it same outcome unit}. For notational simplicity, we drop notation pertaining to the unit, and 
% In what follows, we refer to a time period as exposed if $E_{t} = 1$ and unexposed if $E_t = 0$, and with carryover exposure if $R_t = 1$ and without if $R_t = 0$.
% We focus on the the immediate effect estimand among exposed time periods, and the carryover effect estimand among time periods with carryover exposure, $\widetilde \tau^\text{imm}(1, 0)$ and $\widetilde \tau^\text{car}(1, 0)$ in \cref{eq:att-type-effects-no-interaction}. 
% In the presence of multiple outcome units, the algorithms would be applied to each unit separately.

We propose algorithms for estimating the immediate and carryover effects.
We focus here on the immediate effect among time periods with exposure, $\tau^\text{imm}(1, 0)$ in \cref{eq:att-type-effects-no-interaction}.
For each exposed time period ($t \in \mathcal{T}_{1.}$ with $E_t = 1$), the algorithm ``searches'' for a corresponding unexposed time period ($t \in \mathcal{T}_{0.}$ with $E_t = 0$) in close temporal proximity to create a match, such that the matches satisfy overall balance constraints on time-varying information.
These algorithms are constructed as integer programming optimization problems with the objective of maximizing the number of matched exposed time periods, under constraints at the level of each time period, each match, and across all matches.

%\subsubsection{Matching 1-1.}\label{subsec: matching 1-1}

Specifically, let $t_e \in \mathcal{T}_{1.} =\{t: E_{t} = 1\}$ and $t_u \in \mathcal{T}_{0.} =\{t: E_{t} = 0\}$ denote an exposed and an unexposed time period, respectively. 
We introduce binary indicators $a_{t_et_u}$ for each pair of exposed and unexposed time periods such that $a_{t_et_u} = 1$ if the time periods are matched, and $a_{t_et_u} = 0$ if not.
The objective of the optimization problem is to maximize 
\begin{equation}\label{matching1-1-objective}
    \max_{\bm a} \sum_{t_e, t_u} a_{t_et_u}, %\tag{A}
\end{equation}
over all possible indicators $\bm a \in \{0, 1\}^{|\mathcal{T}_{1.}| \times |\mathcal{T}_{0.}|}$,
where $\underset{t_e, t_u}{
\Sigma}$ denotes the summation over both sets of indices,
$\displaystyle \underset{t_e \in \mathcal{T}_{1.} }{\Sigma} \underset{t_u \in \mathcal{T}_{0.} }{\Sigma} $. Each time period, exposed or unexposed, can be part of at most one match,
\begin{align*}
    &\sum_{t_u} a_{t_et_u} \leq 1, \quad \forall t_e \in \mathcal{T}_{1.} , \quad \text{and} \quad
     \sum_{t_e } a_{t_et_u} \leq 1, \quad \forall t_u \in \mathcal{T}_{0.} . 
     % \tag{\ref{matching1-1-objective}.1}
     % \label{treatment-control-constraint}
\end{align*}
%This constraint, which ensures that each time period is not overly influential over the others, improves inferential performance. 
The objective in \cref{matching1-1-objective} along with the constraint that an exposed time period can be part of at most one match 
% in \cref{treatment-control-constraint} 
implies that the optimization problem is formulated to maximize the number of matched exposed time periods.
In principle, an unexposed time period could be involved in more than one match. However, we have found that in practice % that it could have a disproportionally large influence in the causal estimator, resulting
doing so leads to inaccurate variance estimation and inference.
% By enforcing that an unexposed unit cannot be used more than once, we  avoid duplicate use of unexposed time periods, which leads to easier inference.

We impose constraints that target the balance of temporal trends, the time-varying carryover exposure, and time-varying covariates.
In reality little (if any) information is known about the temporal trends $f(t)$. We balance temporal trends indirectly by specifying that the average difference in time of exposed and unexposed time periods across all matches is at most $\delta \geq 0$,
%
% \begin{equation}
\begin{align}
\label{eq:constraint1-1:avg_time}
    \left| \sum_{t_e, t_u } 
     a_{t_et_u}(t_e-t_u)\right| \leq \delta \sum_{t_e, t_u } a_{t_et_u}.
\end{align}
%     \tag{\ref{matching1-1-objective}.2}
% \end{equation}
%
This constraint does not necessarily imply that {\it each} of the matched pairs is close in time, rather than they are close {\it on average}. 
% As we see in \cref{subsec:theory}, this constraint suffices for ensuring negligible bias of our estimator due to smooth temporal trends. 
In order to improve the balance of local temporal trends and to reduce the computationally intensive search for possible match combinations, we also impose that each matched pair of time periods is at most $\epsilon$ apart in time,
%
% \begin{equation}
% \label{single-tr-c-dif-constraint}
\begin{align*}
     \left|a_{t_et_u}(t_e-t_u) \right| \leq \epsilon, \quad \forall t_e \in \mathcal{T}_{1.} ,\ \forall t_u \in \mathcal{T}_{0.} . 
\end{align*}
% \tag{\ref{matching1-1-objective}.3}
% \end{equation}

Moreover, we specify that the carryover exposure is balanced on average across matched pairs up to some $\delta' \geq 0$,
\begin{align}
    \label{1-1-carryover}
    \left|\sum_{t_e, t_u}
     a_{t_et_u}(R_{t_e}-R_{t_u})\right| \leq \delta'\sum_{t_e, t_u} a_{t_et_u}.
% \tag{\ref{matching1-1-objective}.4}
\end{align}
We specify a balance constraint on the time-varying covariates between exposed and unexposed matched time periods up to lag $S$, as
\begin{align}
    \label{1-1-time-varying-constraint}
    \left|\sum_{t_e, t_u}
     a_{t_et_u}(\overline{\bm X}_{t_e, S} - \overline{\bm X}_{t_u, S})\right| \leq \bm 1_{(S + 1) (N p^\text{int}+ N p^\text{net} + p^\text{out}) } \cdot\delta'\sum_{t_e, t_u} a_{t_et_u},
     % \tag{\ref{matching1-1-objective}.5} % \\
     % % \mbox{~where~} t_e^\ast=t_e-s, t_u^\ast=t_u-s, \forall s = 0, 1, \dots, S. 
     % \notag
\end{align}
where $\bm 1_n$ is the $n$-vector of 1s. 
This balance constraint states that the temporal covariates are on average balanced in matched exposed and unexposed time periods. 
We discuss the choice of algorithmic parameters $\epsilon, \delta$ and $\delta'$ in \cref{subsec:theory}.
% The constant $\delta'$ is usually chosen to be 0.05 or 0.1 standard deviations of the corresponding covariate, though its choice should be part of the design phase of the study and it should be chosen without looking at outcomes \citep{Zubizarreta2015}.

%In addition to the outcome unit covariates, 
The covariate vector $\bm X_t$ includes $N p^\text{int}$ interventional unit covariates, and $N p^\text{net}$ network level covariates. If $N$ is large, the constraint in \cref{1-1-time-varying-constraint}
% The time-varying covariates, $\bm X_t^{\text{int}\top}$ and $\bm X_{t}^{\text{net}\top}$ are of dimension $N \times p^\text{int}$ and $N \times p^\text{net}$, respectively. In theory,
% balance constraints could be imposed so that matched time periods are similar with respect to all $(p^\text{int}+p^\text{net})N$ variables. However, these  balance constraints 
would be high-dimensional, which could drastically reduce the number of matches. In such cases, we propose replacing this balance constraint on the individual values of interventional and network covariates with summaries of the corresponding covariates across interventional units. To create such summaries, consider a vector $\bm q = (q_1, q_2, \cdots, q_N)^\top$. 
We define $X_{td}^\text{int,sum.} = \bm q^\top (X_{t1d}^\text{int}, X_{t2d}^\text{int}, \cdots, X_{tNd}^\text{int})^\top$ as the summary of the $d^{th}$ interventional covariate across units at time $t$, for $d = 1, 2, \dots, p^\text{int}$.
Then, ${\bm X}_{t}^\text{int,sum.}=({X}_{t1}^\text{int,sum.},{X}_{t2}^\text{int,sum.},\cdots,{X}_{tp^\text{int}}^\text{int,sum.})^\top$ denotes the $\bm q$-summaries of all the interventional units' covariates. We similarly define
${X}_{td}^\text{net,sum.} = \bm q^\top (X_{t1d}^\text{net},X_{t2d}^\text{net},\cdots,X_{tNd}^\text{net})^\top$ as a summary of the $d^{th}$ network covariate across interventional units for $d = 1, 2, \dots, p^\text{net}$, and
${\bm X}_{t}^\text{net,sum.} = ({X}_{t1}^\text{net,sum.},{X}_{t2}^\text{net,sum.},\cdots,{X}_{tp^\text{net}}^\text{net,sum.})^\top$.
Finally, we use 
${\bm X}^\text{sum.}_{t}=({\bm X}_t^{\text{int,sum.}\top}, {\bm X}_{t}^{\text{net,sum.}\top},\bm X_{t}^{\text{out}\top})^\top$ where $X^\text{sum. }_{td}$ denotes the $d^{th}$ covariate in ${\bm X}^\text{sum.}_{t}$.
The balance constraint is placed on the vector $\overline {\bm X}^\text{sum.}_{t, S}$
which includes summaries of the interventional and network covariates,  and the outcome unit covariates over the $S + 1$ time periods $t, t-1, \cdots, t-S$, as
% \begin{equation}
% \tag{\ref{matching1-1-objective}.5\textquotesingle} 
% \label{1-1-time-varying-constraint-XP}
\begin{align*}
     \left|\sum_{t_e, t_u}
     a_{t_et_u}
%     (\bm{q}^\top\bm{X}_{{t_e}}-\bm{q}^\top\bm{X}_{{t_u}})
     (\overline{{\bm X}}^\text{sum.}_{{t_e, S}} - \overline{{\bm X}}^\text{sum.}_{t_u, S})
     \right| 
     & \leq \bm 1_{(S + 1) (p^\text{int}+p^\text{net}+p^\text{out})} \cdot\delta'\sum_{t_e, t_u} a_{t_et_u}.
     % \\
     % &\mbox{~where~} t_e^\ast=t_e-s, t_u^\ast=t_u-s,  \ s  = 0, 1, \dots, S. 
\end{align*}
% \end{equation}

The vector $\bm q$ controls which covariate summary should be balanced, and its choice will be driven by the problem at hand. For example, by setting $q_i=\frac{1}{n}$ for all $i$, the algorithm balances the average covariate value across interventional units for matched time periods. Alternatively, $q_i$ could give different weights to the covariate value of interventional units based on their geographic proximity to the outcome unit, or the frequency with which they are connected. %For ease of exposition, we used the same vector $\bm q$ in the balance constraints for all covariates in \cref{1-1-time-varying-constraint-XP}. 
Different vectors $\bm q$ can be used for different covariates, and multiple summaries of the same covariate under different vectors $\bm q$ could be balanced.

\begin{figure}
    \centering
    \includegraphics[width=0.7\linewidth]{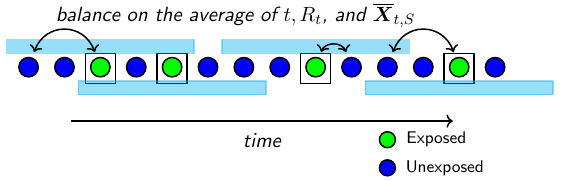}
    \caption{Illustration of the algorithm for the immediate effect. Out of 14 time periods, four are exposed (green circles) and ten are unexposed (blue circles). Available matches for each exposed time period are shaded light blue with the corresponding exposed time period in the middle. In this illustration, $\epsilon = 2$. Out of the possible available unexposed time periods, matches are acquired by maximizing the number of matches subject to balance constraints on $t$, $R_t$. and $\overline{\bm X}_{t, S}$.}
    \label{fig:matching-visualization}
\end{figure}

{A visualization of the algorithm is in \cref{fig:matching-visualization}. For each exposed time period (green circle), an available unexposed time period (blue circle) is searched within the light blue shaded area, such that the resulting set of exposed and matched unexposed time periods satisfy overall balance constraints. The arrows indicate the resulting matched pairs.}

{Our approach combines principles from matching \citep[e.g.][]{rubin1973matching, rubin1980bias, stuart2010matching, imai2023matching} and covariate balancing \citep{hainmueller2012entropy, imai2014covariate, Zubizarreta2015, li2018balancing}.
As in matching, some constraints are imposed on each pair and restrict which unexposed time periods can be paired with each exposed one. Then, as in covariate balancing methodology, some constraints specify conditions on the overall resulting data set, irrespective of who is paired with whom.}
Integer programming optimization algorithms have been previously used in the causal inference literature \citep{zubizarreta2012using, zubizarreta2013stronger, Keele2014}. Our formalization is different since the fundamental unit of observation is time (rather than physical units) and the confounders correspond to time-varying information.
By harvesting the temporal dimension of the data and by investigating the problem through its true bipartite lens, confounding adjustment for interventional, outcome, and network covariates becomes more transparent. For example, as we discuss in the context of our study in \cref{sec:real-data}, time-varying confounding due to interventional unit and network covariates might not exist. Such understanding of the needs of confounding adjustment would not be obvious if addressing the same study question by projecting it on a unipartite framework.

% The algorithms are constructed as integer programming optimization problems with the objective of maximizing the number of matches under a set of constraints. 

For the carryover effect, the proposed algorithm is defined similarly, balancing the most recent exposure in \cref{1-1-carryover} instead of the carryover exposure. The algorithm is discussed in Supplement~C.2.
%\ref{supp_subsec:alg_carryover}. 
Moreover, the algorithm discussed here finds a single unexposed time period as a match to each exposed time period. We additionally designed algorithms which match an exposed time period to two or one-or-two unexposed time periods. We defer these algorithms to Supplement C.1. %\ref{supp_subsec:matching1-2or1-12}. 
Lastly, \cref{ass:no-interaction} of constant immediate and carryover effects is not necessary. However, for estimating the immediate effect, this assumption allows us to pull information across time periods with different carryover exposures and, therefore, gain estimation efficiency, and similarly for estimating the carryover effect. In Supplement~C.3,
%\ref{supp_subsec:without_nointeraction_assum}, 
we discuss alternative estimands and estimation procedures for immediate and carryover effects that do not rely on this assumption.

% In what follows, we refer to a time period as exposed if $E_{t} = 1$ and unexposed if $E_t = 0$, and with carryover exposure if $R_t = 1$ and without if $R_t = 0$. We focus on the the immediate effect estimand among exposed time periods, and the carryover effect estimand among time periods with carryover exposure, $\widetilde \tau^\text{imm}(1, 0)$ and $\widetilde \tau^\text{car}(1, 0)$ in \cref{eq:att-type-effects-no-interaction}. 

\subsection{The matching estimators and theoretical guarantees}

\label{subsec:theory}

The matches produced by these algorithms are the basis for estimating the immediate and carryover causal effects, $\tau^\text{imm}(1, 0) $ and $\tau^\text{car}(1, 0)$. We focus here on the immediate effect, but the estimator for the carryover effect and its theoretical guarantees are similar (see Supplement C.2).
%\ref{supp_subsec:alg_carryover}).
% since they are used to impute an exposed time period's counterfactual outcome, had it been unexposed.  

Let $\mathcal{T}_{1.}^\ast = \{t: E_t = 1 \text{ and } a_{t_et_u} = 1 \text{ for some } t_u\}$ denote the set of matched exposed time periods. 
%If the exposed time period $t_e$ is matched one-to-one with the unexposed $t_u$, its counterfactual outcome is imputed as $Y_{t_e}^{\text{imp}}(0,r) = Y_{t_u}$. For $\widetilde \tau^\text{car}(1, 0) $, we impute $Y_{t_e}^{\text{imp}}(e,0) = Y_{t_u}$. 
For $t_e \in \mathcal{T}_{1.}^\ast$, $Y_{t_e}^\ast$ is the observed outcome value for its matched unexposed time period which is equal to $\sum_{t_u} a_{t_et_u} Y_{t_u}$. The estimator of the immediate effect is defined as
\begin{equation}
\widehat \tau^\text{imm}(1, 0) = \frac1{|\mathcal{T}_{1.}^\ast|} \sum_{t_e \in \mathcal{T}_{1.}^\ast} \left( Y_{t_e} - Y_{t_e}^\ast \right).
\label{eq:causal_estimator}
\end{equation}
% Specifically, for Matching 1-1, $\mathcal{T}_{e.}^\ast = \{t: E_t = 1 \text{ and } a_{t_et_u} = 1 \text{ for some } t_u\}$, and similarly for Matching 1-2 or Matching 1-1/2. We denote these estimators as $\widehat \tau_1, \widehat \tau_2$ and $\widehat \tau_{1/2}$. 
We implicitly assume that $\mathcal{T}_{1.}^\ast$ is non-empty. Otherwise, if it is not possible to satisfy the balance constraints, drawing causal inferences from such data might not be trustworthy. % In Supplement \ref{supp_subsec:alternative_matching_estimator}, we discuss an alternative way to use the design-based matching pairs to create matching estimator.

%In principle, one could average estimands and estimates across units to acquire a population-level effect.

% Let $I_1$ denote the set of 1-1 matches $(t_e,t_u)$, and $I_2$ the set of 1-2 matches $(t_e,t_{u_1},t_{u_2})$, which might differ across the algorithms. Then, the causal estimators are
% \begin{align}
%     \Hat{\tau}_{1-1} &= \frac{1}{|I_1|}\sum_{(t_e,t_u)\in I_1}
%     \left( Y_{t_e}-Y_{t_u} \right) \tag{\ref{matching1-1-objective}-sol}\\
%      \Hat{\tau}_{1-2}&=\frac{1}{|I_2|}\sum_{(t_e,t_{u_1},t_{u_2}) \in I_2} \left[ Y_{t_e}-\frac{1}{2}
%      (Y_{t_{u_1}}+Y_{t_{u_2}}) \right]
%      \tag{\ref{matching1-2-objective}-sol}\\
%      \Hat{\tau}_{1-1/2}&=\frac{1}{|I_1|+|I_2|} \left\{ \sum_{(t_e,t_u) \in I_1}\left(Y_{t_e}-Y_{t_u}\right)+\sum_{(t_e,t_{u_1},t_{u_2}) \in I_2}\left[ Y_{t_e}-\frac{1}{2}(Y_{t_{u_1}}+Y_{t_{u_2}})\right] \right\}, \tag{\ref{matching-1-1/2-objective}-sol}
% \end{align}
%for 1-1, 1-2, and 1-1/2 matching, respectively. These estimators express that the causal effect is estimated as the average over matched exposed time periods of the difference between their observed outcome and their imputed potential outcome had they been unexposed. For 1-1 matches, the imputed potential outcome is set equal to the outcome of its matched unexposed time period, whereas in 1-2 matches it is set equal to the average outcome of its two matches. 

We show that the bias of this estimator is bounded. %The form of time in the outcome model can be conceived as at least as complicated as the common temporal trends $f(t)$ in the interventional units' treatment and the potential outcomes in \cref{ass:indep1}, in that conditioning on the value for the functional form of time in the outcome model would result in conditioning on $f(t)$. 
We consider cases where the outcome is a linear or non-linear function of the exposure, time, and time-varying covariates, in line with \cite{Zubizarreta2015}. We clarify that our approach does not involve fitting an outcome model to the data, and that assuming an outcome model in the following results is to investigate the method's performance and its reliance on the algorithmic tuning parameters.  We first consider the linear case. The proofs are in Supplement D.
%\ref{supp_sec:bias_proofs}.

%
%\begin{theorem}
%\label{theorem:bias_linear}
%     If $Y_{t} = \theta+ \alpha E_{t}+\beta t +\zeta W_t+\epsilon_{tj}$ with $\operatorname{E}(\epsilon_{tj}|E_{t},t)=0$ for all $t=1,2,\cdots,T$,
%    then $|\operatorname{E}(\Hat{\tau}-\tau_j)|\leq \delta |\beta|$ for the Matching 1-1 and 1-2 estimators, and $|\operatorname{E}(\Hat{\tau}-\tau_j)|\leq (\delta+\frac{1}{2}\epsilon') |\beta|$ for the Matching 1-1/2 estimator.
%\end{theorem}
%

\begin{theorem}\label{theorem:bias_linear}
     If $Y_{t}(e,r) = \beta_0+ \beta_1 e+ \beta_2r+\beta_3 t  +\bm \beta_4^\top {\overline{\bm X}^\text{sum.}_{t, S}} +\epsilon_{t}(e,r)$ for all $t=1,2,\cdots,T$, with $\operatorname{E}(\epsilon_{t}(e,r)|E_{t},t,{\overline{\bm X}^\text{sum.}_{t, S}})=0$,
    then $|\operatorname{E}(\Hat{\tau}^\text{imm}(1, 0)-\tau^\text{imm}(1, 0))|\leq \delta |\beta_3|  +\delta' (|\beta_2| + \|\bm \beta_4\|_1)$
    % for all matching estimators on direct effect, and $|\operatorname{E}(\Hat{\tau}^\text{car}-\tau^\text{car})|\leq \delta'|\beta_1|+\delta |\beta_3|  +\delta'\|\bm \beta_4\|_1$ for all matching estimators on carryover effect,
    where $\delta$ and $\delta'$ are the balance constraints tuning parameters.
\end{theorem}

According to \cref{theorem:bias_linear}, the bias of the matching estimators is bounded by algorithmic parameters controlling how well time-varying information is balanced, and the strength of time-varying confounding in the outcome structure. We note here that we interpret the expectation operator in these derivations by a model-based perspective by seeing the potential outcomes as random variables \citep{zigler2025bipartite}. An alternative bias bound that involves the tuning parameter $\epsilon$ is provided in Supplement~D.
%\ref{supp_sec:bias_proofs}. 
We discuss the choice of algorithmic parameters at the end of this section.

{The linear temporal trend assumption in \cref{theorem:bias_linear} might not hold in many settings. However, even when the temporal trend is non-linear, a linear approximation might be relatively accurate in real-world settings where changes over time are expected to happen slowly and data are measured on a relatively short time window. In such cases, the bias of the estimator remains negligible. If temporal trends change more quickly over the study window, additional terms for time would need to be balanced as part of the proposed algorithms. We discuss these points in Supplement~D
%\ref{supp_sec:bias_proofs} 
and as part of the following extension where we allow for smooth, non-linear terms for all covariates, including the temporal trend function.}

To accommodate non-linearity, we extend the algorithm to impose balance constraints for auxiliary variables targeting higher order and localized versions of time and the measured covariates. Consider a generic variable $V_t$ measured over time. We define a localized version of this variable by breaking its support $[a, b]$ into $(b - a)/\ell$ intervals of length $\ell$. The midpoint of the $r^{th}$ interval is denoted by $\phi_r$. {Then, for each $r$, we construct the auxiliary variable $V_{tr}^\dagger$ as  $V_{tr}^\dagger = (V_t - \phi_r) I( V_t \in [\phi_r - \ell/2, \phi_r + \ell/2])$. We perform this procedure for time $t$ and for the time-varying covariates in $\bm X_t^\text{sum.}$.}%, and their higher orders $t^k$ and $(\bm X_t^\text{sum.})^k$ for $k = 2, 3, \dots, K-1$.
% for covariate $X^\text{sum. }_{td}$, as well as higher orders ${X}_{tdr}^{\text{sum.} (k)\dagger}$ for $X_{td}^{\text{sum.}(k)}$, for $k = 1, 2, \cdots, K-1$. 
We include balance constraints as in \cref{1-1-time-varying-constraint} for these auxiliary variables raised to the power of $k$ for $k = 1, 2, \dots, K-1$.

% We define a localized version of the $d^{th}$ covariate, $X^\text{sum. }_{td}$, by breaking its support $[a_{td},b_{td}]$ into $(b_{td}-a_{td})/\ell$ intervals of length $\ell$. The midpoint of the $r^{th}$ interval for $X^\text{sum. }_{td}$ is denoted by $\phi_{dr}$. Then, for each $r$, we construct the auxiliary variable $X_{tdr}^{\text{sum.}\dagger} = (X^\text{sum. }_{td}-\phi_{dr})I(X^\text{sum. }_{td}\in[\phi_{dr}-\ell/2,\phi_{dr}+\ell/2])$ for covariate $X^\text{sum. }_{td}$, as well as higher orders ${X}_{tdr}^{\text{sum.} (k)\dagger}$ for $X_{td}^{\text{sum.}(k)}$, for $k = 1, 2, \cdots, K-1$. We include balance constraints (similarly to the ones in \cref{1-1-time-varying-constraint}) for these auxiliary variables in all three matching algorithms.

We show that the bias of the causal effect estimator is still bounded, where the bound is driven by algorithmic parameters and the smoothness of functions in the outcome model. 

\begin{theorem}
\label{theorem:bias_flexible}
Suppose that the potential outcomes satisfy
\(
\displaystyle
Y_{t}(e,r) = \theta+\beta_1 e +\beta_2 r+h_0(t)+
{\scriptstyle 
{\sum\limits_{s = 0}^S
\sum\limits_{d=1}^{p^\text{int}+p^\text{net}+p^\text{out}}}} h_{sd}(X^\text{sum. }_{(t-s) d}) +
\epsilon_{t}(e,r),
\)
with $\operatorname{E}(\epsilon_{t}(e,r)|E_{t},t,{\overline{\bm X}^\text{sum.}_{t, S}})=0$ and functions $h_0, h_{01}, h_{02}, \dots, h_{S (p^\text{int}+p^\text{net}+p^\text{out})}$ that are $K$-times differentiable on their support.
% \(
% %\displaystyle
% Y_{t}(e) = \theta+\beta e +h_1(t)+\sum_{s=1}^{p_X} h_{2s}(\widetilde X_{ts})+\sum_{s=1}^{p_P} h_{3s}(\widetilde P_{ts})+\sum_{s=1}^{p_W} h_{4s}(W_{ts})+\epsilon_{t}(e)
% \)
% Here when $i=3,4,5, \ p_i=p_X, p_P, p_W$, respectively. 
% The support of the time point consists of all integer values in $[0, T]$. 
% Let $\gamma_{k,is}:=\frac{h_{is}^{(k)}(\phi_{j,is})}{k!}$ be the coefficient of the Taylor expansion of order k around $\phi_{j,s}$, which satisfies 
If $h_0^{(k)}$ and $h_{sd}^{(k)}$ represent the $k^{th}$ derivative of $h_0$ and $h_{sd}$ respectively, and $|h_0^{(k)}(t)|, |h_{sd}^{(k)}(x)| \leq c$ for some $c > 0$ for all $s =0, 1, \dots, S$, covariate $d = 1, 2, \dots, p^\text{int}+p^\text{net}+p^\text{out}$, $t, x$ in the functions' support, and $k = 1, 2, \dots, K$, then $ |\operatorname{E}(\Hat{\tau}^\text{imm}-\tau^\text{imm})| \leq  
C_T \delta + (|\beta_2|+C_{X})\delta' +C_{T X} \ell^{K-1}$, where
$C_T, C_{X}$ and $C_{T X}$ are constants proportional to $c$.  
    % Then
    % \begin{align*}
    %  |\operatorname{E}(\Hat{\tau}-\tau)| \leq \delta c_T + \delta' c_{W X P}+c_{t W X P} \ell^K.
%    |\operatorname{E}(\Hat{\tau}-\tau)| & \leq \delta\sum_{j=1}^{(b_2-a_2)/\ell}\sum_{k=1}^{K-1}|\gamma_{k,2}|+\delta' \sum_{i=3}^5\sum_{s=1}^{p_i}\sum_{j=1}^{(b_{is}-a_{is})/\ell}\sum_{k=1}^{K-1}|\gamma_{k,is}|+ \\
%    & \hspace{40pt} + 2\sum_{j=1}^{(b_2-a_2)/\ell}\frac{h_{2}^{(K)}(\phi_{j,2})}{K!}(\ell/2)^K+2\sum_{i=3}^5\sum_{s=1}^{p_i}\sum_{j=1}^{(b_{is}-a_{is})/\ell} \frac{h_{is}^{(K)}(\phi_{j,is})}{K!}(\ell/2)^K
    % \end{align*}
    % for all estimators.
    % the Matching 1-1 and Matching 1-2 estimators, and $|\operatorname{E}(\Hat{\tau}-\tau_j)|\leq\frac{3}{2}\gamma\epsilon'+ c\epsilon^2/2 $ for the Matching 1-1/2 estimator.
      
\end{theorem}
The constants $C_T, C_{X}$ and $C_{T X}$ depend on the smoothness of the functions with the corresponding indices, and their exact form is given in Supplement D.%\ref{supp_sec:bias_proofs}.
Importantly, other than smoothness assumptions, these results do not rely on restrictions on how temporal trends influence the outcome, and a non-stationary (such as monotonic) outcome poses no further challenges.
\cref{theorem:bias_flexible} establishes that, by setting the algorithms' tuning parameters $\delta, \delta'$ and $\ell$ to be small enough, the bias of the corresponding causal effect estimators can be guaranteed to be negligible. 
Since the most recent and carryover exposures are binary, the form $\beta_1 e + \beta_2 r$ in the outcome model does not impose any restriction other than the no-interaction \cref{ass:no-interaction}.
Extending our results to allow for interactions among the covariates would be theoretically straightforward. However, practically, the necessary additional balancing constraints might hinder our ability to find adequate matches.

{
The choice of $(\epsilon,\delta,\delta', \ell)$ has a direct impact on the resulting number of matches with smaller values imposing stricter constraints.
Smaller values of $\delta$ and $\delta'$ imply a smaller maximum possible bias, and they can, in principle be set arbitrarily small, even to zero. In practice, using small values for $\delta, \delta'$ might return a small number of matches and an estimated effect that is not representative of all exposed time periods.}
As with all matching procedures, the estimated immediate effect is representative of the matched exposed time periods only, $\widetilde{\tau}^\text{imm}(1, 0) = \sum_{t_e \in \mathcal{T}_{1.}^\ast} [Y_{t_e}(1,R_{t_e}) - Y_{t_e}(0,R_{t_e})] / |\mathcal{T}_{1.}^\ast|$. When the proportion of matched exposed time periods is small,
% and similarly for the carryover effect. Therefore, its interpretation 
the estimate's interpretation might be complicated, 
and might differ from the effect on all the exposed time periods $\tau^\text{imm}(1, 0) $ in the presence of heterogeneity across time 
% In these cases, and if the causal effect is heterogeneous across time, the estimated effect might differ from the effect on all the exposed time periods $\tau^\text{imm}(1, 0) $. We investigate the performance of our estimators with heterogeneous effects in the simulations of
(see Supplement H).
%\ref{supp_sec:simu}).

{
Therefore, the choice of algorithmic tuning parameters should be guided by domain knowledge and the characteristics of the dataset.
If the temporal trend is believed to change slowly over time, a larger $\epsilon$ value can be used to allow for more available unexposed time periods in forming each match. 
% However, smaller values might lead to a smaller proportion of matched exposed time periods. 
In the absence of other information, we recommend setting $\delta=\delta'=0.05$ or $0.1$ as a default for standardized covariates \citep{Zubizarreta2015} and performing sensitivity analysis to such choices.
}

\subsection{Inference}
\label{subsec:inference}

% Our inferential approach is formulated in a unified manner for the three matching estimators. 
By viewing our algorithm as part of the design phase, we construct confidence intervals conditional on the resulting data set of matches \citep{ho2007matching}.
As shown in \cref{eq:causal_estimator}, our estimators are averages of differences between a time period's observed outcome and the outcome of its match. We construct Wald-type confidence intervals using the standard deviation of the outcome differences.
Specifically, for the immediate causal effect, we define
$ \widehat{s}^2 = \sum_{t_e \in \mathcal{T}_{1.}^\ast} \big( Y_{t_e} - Y_{t_e}^\ast - \widehat \tau^\text{imm}(1, 0) \big) ^2 / \big(|\mathcal{T}_{1.}^\ast| - 1 \big)$, 
% if $\mathcal{T}_{e.}^\ast$ is the set of matched exposed time periods for estimating ${\tau}^\text{imm}(1, 0)$, 
and we
construct an $\alpha$-level confidence interval as
%To do so, we need to estimate the differences' standard error. Suppose the number of matches is $r$. We obtain r differences between the observed exposure potential outcome and the imputed unexposed potential outcome, denoted as $d_1,\cdots, d_r$. The empirical error of the estimator is $s=\sqrt{\sum_{i=1}^r (d_i-\hat{\tau})^2/(r-1)}$. We construct the confidence interval as
\(
\big[\widehat \tau^\text{imm}(1, 0) - z_{1-\alpha/2} \widehat{s} / \sqrt{|\mathcal{T}_{1.}^\ast|}, \ 
\widehat \tau^\text{imm}(1, 0) + z_{1-\alpha/2} \widehat{s} / \sqrt{|\mathcal{T}_{1.}^\ast|} \big],
\)
where
$z_{1-\alpha/2}$ is the $1-\alpha/2$ quantile of the standard normal distribution.
We similarly acquire a p-value for testing the null hypothesis of no causal effect on the outcome unit,
\(    \mathrm{H}_{0}: \tau^\text{imm}(1, 0) = 0 \text{ v.s. } \mathrm{H}_{A}: \tau^\text{imm}(1, 0) \neq 0
\)
% This $\tau_j$ is defined as the average effect for time points $t$ with $E_{t}=1$. Under the null hypothesis, the exposure received by an outcome unit does not have a causal effect on its outcome on average. Let $\widehat \tau_j$ be the estimator for the causal effect for outcome unit $m_j$ based on any of the three matching algorithms. Then, the distribution of $\hat{\tau}_j/s$ under the null hypothesis is a $t$ distribution with degrees of freedom equal to $r-1$. The p-value is therefore expressed 
as
\( \displaystyle p = P \big(|Z| >  \sqrt{|\mathcal{T}_{1.}^\ast|} \ |\widehat\tau^\text{imm}(1, 0) | / \widehat{s} \ \big),\)
where $Z \sim N(0,1)$.
P-values in one-sided hypothesis tests can be obtained similarly.
The inferential procedure for the carryover effect proceeds similarly using the matches acquired from the carryover algorithm.

% \begin{theorem}\label{theorem:conservative}
%     Under the model specifications of \cref{theorem:bias_linear} or \cref{theorem:bias_flexible}, and assuming that $E(\epsilon_t^2(e,r) \mid E_t, t, \overline{\bm X}^\text{sum.}_{t, S})$ is constant with independent errors $\epsilon_t(e,r)$, the variance estimator $\widehat{s}^2$ is conservative.
% \end{theorem}

{In Supplement~E
%\ref{supp_sec:variance estimation} 
we investigate the proposed inferential procedure. There, we show that the confidence intervals we construct are expected to cover the true value at least $(1-\alpha)100\%$ of the time, under certain conditions.
% under the potential outcome models in Theorems~\ref{theorem:bias_linear} and \ref{theorem:bias_flexible} with independent errors.
The extent of conservativeness depends on the magnitude of imbalance of time-varying information within each matched pair. As a result, the conservativeness of the inferential procedure would be alleviated if our estimator was based on matches from an alternative algorithm that balances covariates within each match. In practice, however, such alternative could return substantially fewer matches. There, we also discuss that the inferential approach performs well even with time series data that display temporal correlation. % beyond what can be explained by measured covariates. 
We illustrate these points in simulations in Supplement~H.
%\ref{supp_sec:simu}.
}

{In the presence of multiple outcome units, we discuss hypothesis testing for whether the exposure has an effect on any outcome unit in Supplement C.4.
%\ref{subsec:global_null}.
}

% The proof is in Supplement \ref{supp_sec:variance estimation}. Intuitively explaining this proof, this is because the proposed matching algorithms do not balance time-varying covariates for each match separately. As a result, the unexposed time periods that are used to impute the potential outcomes might have substantially different values for the temporal covariates compared to the corresponding exposed time periods. Therefore, the differences of observed and imputed outcomes include fluctuations in temporal predictors, leading to an estimated variance that is larger than the truth. We illustrate this slight over-coverage in the simulations of Supplement \ref{supp_subsec:simu_dif_sparsity} and \ref{appendix:new-matching}, where we observe that balancing covariates within every match could alleviate this issue at the cost of returning fewer matches.

\subsection{The potential advantages of temporal analyses in bipartite settings}
\label{subsec:temporal_advantages}

Here, we discuss the case with multiple outcome units measured over time and the potential advantages of focusing on time series data for an outcome unit, instead of the cross-sectional approach of focusing on a collection of units measured at a given time period.

In this setting, cross-sectional estimands that average across outcome units for each time period could be considered. An example of cross-sectional immediate effect is $\gamma_t(e,e'; r)=\frac{1}{M}\sum_{j=1}^M [ Y_{tj}(e,r)-Y_{tj}(e',r)]$ which averages across $M$ outcome units for a given time period, where we have extended the notation to include the subscript $j$ corresponding to outcome unit $m_j \in \{m_1, m_2, \dots, m_M\}$. 
Estimation of cross-sectional estimands requires that we measure and adjust for all meaningful differences across units that confound the exposure-outcome relationship, which can be complex and high-dimensional. For example, in the study of \cite{zigler2021bipartite}, the treatment assignment of power plants and population health can vary across the United States in intricate ways, all of which would need to be measured and adjusted. %  for estimating unit-average effects.
In our study of \cref{sec:motivating} with three outcome units, confounding adjustment for estimating cross-sectional estimands would be impossible.

In this setting, our approach would harvest the time series aspect of the data and be applied to each outcome unit separately. Estimation of the temporally-average causal effects for each outcome unit requires that we account for time-varying confounding only, which might be simpler to understand and measure, an observation that was also noted in the interrupted time series literature \citep{rockers2015inclusion} and in the causal inference literature \citep{imai2021use}. For example, consider the case where the interventional units' treatment is constant over time. Our results show that % if the treatment of interventional units is constant over time, any variation in the exposure for an outcome unit is due to the varying bipartite network, and therefore 
confounding would only be due to covariates that predict the network and the outcome. If the random network depends only on the units' time-invariant characteristics like their geographic distance, no confounding adjustment would be necessary to estimate interpretable and policy-relevant estimands. Alternatively, if the network is driven by naturally-occurring processes with temporal variation such as meteorology, one would only need to account for those for causal effect estimation, which are simpler to understand and measure. Furthermore, if confounding variables show relatively smooth temporal trends during the time window under study, such as weather variables, collecting them is unnecessary since they are indirectly balanced in our algorithms (Theorems \ref{theorem:bias_linear} and \ref{theorem:bias_flexible}).

% \begin{enumerate*}[label=(\arabic*), nosep]
%     % \item Time-varying covariates from a small temporal window in the real-world do not change too much, in the sense that we would expect a less wiggly trend of each covariate varying temporally than that of each variable varying locationally. Therefore, the measurement is much easier to manipulate.
%     \item The intrinsic bipartite aspect of the problem might make temporal confounding less instantaneous, because it takes time to learn what the other units are doing. Hence, even though we may not have up-to-date data due to limitations, it is still possible to use lagged temporal resources.
%     \item If what matters for treatment or outcome is the `average' value of the covariate during the previous $x$ time periods, then the confounder is like a moving average which is relatively smooth across time. This implies that we don't even need to measure it with full accuracy, since any bias from its independent component will be small.
% \end{enumerate*}

The inherent bipartite nature of the data suggests that temporal confounding is likely to display smoother trends compared to unipartite scenarios. 
In bipartite settings, the separation of physical units implies that decisions affecting one set of units may not immediately manifest and impact the other.
Consequently, if an interventional unit variable influences the outcome units, it might be due to its overall trend over time, such as its average over preceding time periods.
% For example, in the bipartite example of power plant interventions and ZIP code level hospitalizations of \cite{zigler2021bipartite}, time-varying information such as a power plant's operating time and capacity might drive its treatment, but they are not be immediately known at the ZIP code level to affect contemporaneous outcomes. Instead, the overall trend of the power plant's operating characteristics might be known and drive population behavior and outcomes.
In that case, this `moving average' covariate value will be relatively smooth across time. If left unmeasured, the bias occurring due to its non-temporal component is expected to be small. We illustrate this in the simulations of \cref{sec:simu_real_data}.

%Our design is akin to the interrupted time series studies, as both involve observations before and after an intervention to detect its effect. By using the same observational units at different times, bias is limited by controlling for all time-invariant characteristics. Bias due to temporal trends and time-varying confounders is minimized by focusing on outcome data observed in a short window around the time of exposure, effectively controlling bias from underlying time trends \citep{rockers2015inclusion}.

\section{Simulation Study}\label{sec:simu_real_data} 

We perform simulations to investigate the performance of our estimators and the properties of our inferential procedure.

\subsection{Simulation setup}
\label{subsec:simu_setup}

We generate data that closely resemble the data from our study described in \cref{sec:motivating} in the following ways. First, we consider a time window of $T = 1,003$ time periods, which is of equal length to the temporal window in our study. Second, we generate time-varying covariates with realizations that closely resemble the pattern of the time-varying covariates in our data. % Lastly, we simulate data such that the proportion of exposed time periods is  in our study data.
We provide an overview of our simulation setup here, with additional information on how the data are generated and the different simulation scenarios deferred to Supplement~F.
%\ref{supp_sec:real-data-simulation}.

We consider $N = 50$ interventional units. For each interventional unit $n_i$, we generate a unit-specific temporal trend, $f_i^\text{int}(t)$. The temporal trend for the outcome unit $m$, $f^\text{out}(t)$, is set equal to the trend of its geographically-closest interventional unit. For each interventional unit $n_i$, we also consider $p_0^\text{int} = 1$ time-invariant covariate $X_{0i}^\text{int}$, and $p^\text{int} = 5$ time-varying covariates $\bm X_{ti}^\text{int} = (X_{ti1}^\text{int}, X_{ti2}^\text{int}, \dots, X_{ti5}^\text{int})^\top$. Similarly, an outcome unit $m$ has $p_0^\text{out} = 1$ and $p^\text{out} = 5$ time-invariant and time-varying covariates denoted by $X_0^\text{out}$ and $\bm X_t^\text{out} = (X_{t1}^\text{out}, X_{t2}^\text{out}, \dots, X_{t5}^\text{out})^\top$, respectively. 
Time-varying covariates include location-specific variation, and hypothetical seasonality and extreme weather trends in order to capture our observed data for temperature, humidity, precipitation, wind speed, and wind direction. In Supplement~F,
%\ref{supp_sec:real-data-simulation}, 
we illustrate that the covariates in our simulated data match the observed covariates closely. In order to evaluate an approach that uses information across outcome units instead of time (see our discussion in \cref{subsec:temporal_advantages}), we generate $200$ outcome units using this setup, though our approach is only fit on one of them.

\begin{table}[!t]
    \centering
    \caption{Table of the five confounding scenarios considered in our simulations, in which the treatment $\bm{A}$, the network $\bm{G}$, and the observed outcome $\bm{Y}$ are generated based on the covariates marked with $\times$.}
    \vspace{-3pt}
    \resizebox{0.7\textwidth}{!}{%
        \begin{tabular}{llcccccccccc}
            \hline
            \\[-10pt]
            &&& \multicolumn{3}{c}{Smooth time} & \multicolumn{3}{c}{Location-varying} & \multicolumn{3}{c}{Time-varying} \\
            \cmidrule(lr){4-6} \cmidrule(lr){7-9} \cmidrule(lr){10-12}
            \multicolumn{2}{c}{Scenario}& & $t$ & $f^\text{int}$ & $f^\text{out}$ & loc & $\bm{X}^\text{int}_0$ & $\bm{X}^\text{out}_0$ & $\bm X_{ti}^\text{int}$ & $\bm X_{tj}^\text{out}$&$R_t$ \\[2pt]
            \hline
            \\[-10pt]
            \multirow{3}{*}{(a)} & \multirow{3}{*}{No confounders} 
            & $\bm{A}$ & & & & & & & & &\\
            && $\bm{G}$ & & & &  & & & & &\\
            && $\bm{Y}$ & & & & & & $\times$ & & &\\
            \hline
            \\[-10pt]
            \multirow{3}{*}{(b)} & \multirow{3}{*}{\shortstack[l]{Time-smooth\\confounders}} 
            & $\bm{A}$ & & $\times$ & & & & & & &\\
            && $\bm{G}$ & & & & $\times$ & & & & &\\
            && $\bm{Y}$ & &  &$\times$ & & & & & &\\
            \hline
            \\[-10pt]
            \multirow{3}{*}{(c)} & \multirow{3}{*}{\shortstack[l]{Location-varying\\confounders}} 
            &$\bm{A}$ & & & & & $\times$ &  & & &\\
            && $\bm{G}$ & & & & $\times$ & & & & &\\
            && $\bm{Y}$ & & & & & & $\times$ & & &\\
            \hline
            \\[-10pt]
            \multirow{3}{*}{(d)} & \multirow{3}{*}{\shortstack[l]{Time-varying\\confounders}} 
            &$\bm{A}$ & & & & & & & $\times$ &  &$\times$\\
            && $\bm{G}$ & $\times$& & &  & & & & &\\
            && $\bm{Y}$& & & & & & & & $\times$ &$\times$\\
            \hline
            \\[-10pt]
            \multirow{3}{*}{(e)} & \multirow{3}{*}{All confounders} 
            &$\bm{A}$& & $\times$ & & & $\times$ & & $\times$ & &$\times$ \\
            && $\bm{G}$& $\times$ & & & $\times$ & & & & &\\
            && $\bm{Y}$& &  &$\times$ & & & $\times$ & & $\times$ &$\times$\\
            \hline
        \end{tabular}%
    }
    \label{tab:real_sims_plan}
\end{table}

% \subsubsection{Data generative mechanisms with different confounding structure} 

We consider five data generative models corresponding to settings with different confounding structure. Under the different scenarios, the treatment assignment for interventional unit $n_i$ is allowed to depend on smooth temporal trends through $f_i^\text{int}(t)$ and on covariates through $X_{0i}^\text{int}$ and $\bm X_{ti}^\text{int}$.
The entries of the bipartite network are generated independently from Bernoulli distributions with probability that might depend on time, location, and units' spatial proximity.
The exposure of unit $m$ at time $t$ is specified as $E_{t} = I(\sum_{i=1}^N A_{ti}G_{tij}\geq d)$, {where $d$ is chosen such that the proportion of time periods with exposure is approximately 20\%}. The carryover exposure is defined as whether the majority of the previous seven time periods are exposed or not, {$R_t = I(\frac{1}{7}\sum_{s=t-7}^{t-1}E_s\geq 4)$}, as in our study.
The outcome is allowed to depend on the exposure, the carryover exposure, the smooth temporal trend $f^\text{out}(t)$, and the covariates $X_0^\text{out}$ and $\bm X_t^\text{out}$. %, and a random error term.

\cref{tab:real_sims_plan} shows the variables that are used in each data generative model component across the different scenarios. These five scenarios correspond to five different confounding structures for the exposure-outcome relationship: 
\begin{enumerate*}[label=(\alph*)]
\item no confounding, 
\item confounding by smooth temporal variables, 
\item confounding by location-varying variables, 
\item confounding by time-varying covariates, and 
\item all types of confounding. 
\end{enumerate*}
Confounding arises if a predictor of the outcome is correlated with a predictor of the treatment assignment, the bipartite network, or both. For example, in scenario (b) the temporal trend components $f_i^\text{int}(t)$ and $f^\text{out}(t)$ induce confounding due to their common smooth temporal trend, and in scenario (d) the time-varying covariates for the interventional and outcome units, which are defined based on one another, induce confounding of the exposure-outcome relationship. 
%Additional information on the simulation scenarios is provided in Supplement \ref{supp_sec:real-data-simulation}.
% Therefore, these data generative models allow for complex confounding structures in accordance to Assumption \ref{ass:indep}.

\subsection{Estimation}
\label{subsec:single-simu}

We estimate the temporally-averaged causal effect specific to an outcome unit. We fit our algorithms that match an exposed time period to one (\cref{sec:matching}), one-or-two or two (Supplement~C.1)
%(Supplement~\ref{supp_subsec:matching1-2or1-12}) 
unexposed time periods, and we denote them as \texttt{1-1}, \texttt{1-1/2}, and \texttt{1-2} in the results, respectively. We employ balance constraints on the time-varying covariates $\bm X_t^\text{out}$.
%only, shown in the last wide column of \cref{tab:sims_plan}. 
Since $f_i^\text{int}(t)$ and $f^\text{out}(t)$ represent smooth temporal trends, we do {\it not} consider balance constraints on them, illustrating that the constraints on time suffice.
For a consistent choice of tuning parameter $\delta'$ across the time-varying covariates, we standardize each one of them using the pooled standard deviation of exposed and unexposed time periods \citep{rosenbaum1985constructing}. Specifically, the entries of the $d^{th}$ covariate $\bm X_{.d}^\text{out} = (X_{1d}^\text{out}, X_{2d}^\text{out}, \dots, X_{Td}^\text{out})^\top$ are divided by $\sqrt{(\mathrm{Var}(\bm X_{(e = 1)d}^\text{out}%\mathrm{~for~ }\mathrm{exposed~} t 
) + \mathrm{Var}(\bm X_{(e=0)d}^\text{out}
%\mathrm{~for~ }\mathrm{unexposed~} t 
))/2}$, where $\mathrm{Var}(\bm X_{(e = 1)d})$ and $\mathrm{Var}(\bm X_{(e = 0)d})$ is the variance of the covariate among exposed and unexposed time periods, respectively. 
Therefore, $\delta'$ denotes the allowed covariate imbalance as the proportion of the covariate standard deviation. 
We consider three sets of tuning parameters $(\delta,\delta',\epsilon)$. The results shown here correspond to values $(2,0.05,6)$. Alternative choices for the tuning parameters are discussed in \cref{subsec:additional_simulation}.
% and shown in the Supplement \ref{appendix:tuning parameters}.
We estimate the causal effect using \cref{eq:causal_estimator}, and acquire 95\% confidence intervals as detailed in \cref{subsec:inference}.

Since there do not exist alternative approaches in the literature for estimating causal effects in bipartite time series settings, we implement three na\"ive approaches, and estimators based on linear regression, and inverse propensity weighting (IPW).
\texttt{Na\"ive-$t$} uses temporal information for the single outcome unit and estimates an effect as the difference of mean outcomes between exposed and unexposed time periods. \texttt{Na\"ive-$j$} uses information across the 200 outcome units for a single time period and estimates an effect as the difference of mean outcomes between exposed and unexposed outcome units. \texttt{Na\"ive-all} %uses all outcome units and all time periods and 
estimates an effect as the overall difference of mean outcomes in exposed and unexposed time periods across units. 
\texttt{Na\"ive-$j$} and \texttt{Na\"ive-all} are the only two approaches that use information across outcome units. 
The linear regression estimator (\texttt{Reg}) corresponds to the coefficients of the exposure and the carryover exposure in a model that incorporates all covariates linearly. For the IPW estimator (\texttt{IPW}), we consider a logistic propensity score model for the exposure on the covariates, and estimate the causal effect based on the average of the inverse-propensity weighted outcomes. Additional details on all approaches are included in Supplement G.
%\ref{appendix:sec:naive}.%\commentZ{Im not sure whether to put unadjusted ones here}

\begin{table}[!b]
    \centering

 \caption{Bias, mean squared error (MSE), coverage of 95\% intervals (\%), proportion of exposed time points matched (\%), and proportion of simulations with a solution to the optimization problem (\%). We consider the 3 na\"ive approaches denoted with `\texttt{N}', regression (\texttt{Reg}), inverse propensity weighting (\texttt{IPW}), and the three proposed estimators. The simulation scenarios correspond to (a) No confounders, (b) Time-smooth confounders, (c) Location-varying confounders, (d) Time-varying confounders, and (e) All confounders.}

    \resizebox{0.8\textwidth}{!}{%
       \begin{tabular}{cccccccccccc}

\hline
&&\multicolumn{5}{c}{Immediate effect} & \multicolumn{5}{c}{Carryover effect} \\
\cmidrule(lr){3-7} \cmidrule(lr){8-12} 
 && Bias&MSE&Cover&Prop&Success &Bias&MSE&Cover&Prop&Success\\
 \hline
\multirow{8}{*}{(a)}% $^1$}
&N-t & 0.002 & 0.008 & 94.2 & - & - & 0.032 & 0.105 & 95.5 & - & -\\
 &N-j & 0.022 & 0.306 & 91.4 & - & - & 0.072 & 0.263 & 98.2 & - & -\\
 &N-all & 0.001 & 0.000 & 96.4 & - & - & 0.005 & 0.013 & 70.2 & - & -\\
 &Reg & 0.002 & 0.008 & 94 & - & - & 0.010 & 0.095 & 96 & - & -\\
 &IPW & 0.002 & 0.008 & 93.8 & - & - & 0.009 & 0.105 & 91.8 & - & -\\
%&U1-1 & 0.009 & 0.014 & 93.6 & 100 & 100 & 0.263 & 0.269 & 89.7 & 100 & 98.8\\
 %&U1-1/2 & 0.012 & 0.011 & 94.4 & 100 & 100 & 0.260 & 0.241 & 87.4 & 100 & 98.8\\
 %&U1-2 & 0.010 & 0.010 & 95.3 & 98.3 & 100 & 0.303 & 0.234 & 84.2 & 95.8 & 98.8\\
 &1-1 & 0.011 & 0.013 & 95.8 & 100 & 100 & 0.033 & 0.160 & 93.8 & 96.5 & 87.6\\
 &1-1/2 & 0.010 & 0.013 & 95.2 & 100 & 100 & 0.042 & 0.162 & 93.4 & 96.8 & 91.2\\
 &1-2 & 0.009 & 0.010 & 95.1 & 98.3 & 100 & 0.030 & 0.141 & 95.3 & 81.5 & 85.6\\
\hline
\multirow{8}{*}{(b)\phantom{$^1$}}
&N-t & 0.031 & 0.018 & 85.6 & - & - & 0.060 & 0.123 & 73 & - & -\\
&N-j & 0.001 & 0.037 & 94 & - & - & -0.081 & 0.064 & 91.2 & - & -\\
&N-all & 0.008 & 0.002 & 26.6 & - & - & -0.041 & 0.007 & 21.6 & - & -\\
&Reg & 0.020 & 0.011 & 92.6 & - & - & 0.037 & 0.088 & 82.8 & - & -\\
&IPW & 0.019 & 0.011 & 92.8 & - & - & 0.024 & 0.074 & 82.2 & - & -\\
%&U1-1 & 0.032 & 0.013 & 93.4 & 98.7 & 100 & 0.214 & 0.152 & 75.4 & 94.1 & 92.6\\
%&U1-1/2 & 0.026 & 0.010 & 93.6 & 98.7 & 100 & 0.196 & 0.118 & 78 & 94.1 & 92.6\\
%&U1-2 & 0.034 & 0.011 & 91.6 & 87.5 & 97.4 & 0.267 & 0.148 & 68 & 69.7 & 92.6\\
&1-1 & 0.015 & 0.011 & 94.4 & 98.7 & 100 & 0.028 & 0.060 & 95 & 92.7 & 87.2\\
&1-1/2 & 0.015 & 0.011 & 94.8 & 98.7 & 99.8 & 0.039 & 0.054 & 94.1 & 92.7 & 88.4\\
&1-2 & 0.021 & 0.010 & 94.2 & 87.5 & 97 & 0.036 & 0.056 & 95.3 & 62.9 & 84.8\\
\hline
\multirow{8}{*}{(c)\phantom{$^1$}}&N-t & 0.001 & 0.008 & 96.2 & - & - & 0.022 & 0.112 & 94.5 & - & -\\
&N-j & 0.106 & 0.074 & 92.6 & - & - & 0.238 & 0.145 & 83.6 & - & -\\
&N-all & 0.113 & 0.019 & 7.4 & - & - & 0.563 & 0.354 & 0 & - & -\\
&Reg & 0.000 & 0.008 & 96.2 & - & - & -0.010 & 0.133 & 94.3 & - & -\\
&IPW & 0.000 & 0.009 & 95.4 & - & - & -0.010 & 0.145 & 87.9 & - & -\\
%&U1-1 & 0.020 & 0.015 & 94.6 & 100 & 100 & 0.251 & 0.273 & 83.2 & 99.8 & 83.4\\
%&U1-1/2 & 0.020 & 0.013 & 94.6 & 100 & 100 & 0.245 & 0.225 & 83.7 & 99.8 & 83.4\\
%&U1-2 & 0.015 & 0.012 & 94.8 & 97 & 100 & 0.264 & 0.231 & 81.1 & 91.6 & 83.4\\
&1-1 & 0.008 & 0.014 & 95.4 & 100 & 100 & 0.016 & 0.107 & 95.7 & 96.7 & 69.8\\
&1-1/2 & 0.011 & 0.014 & 95.4 & 100 & 100 & 0.026 & 0.113 & 94.6 & 96.7 & 73.6\\
&1-2 & 0.013 & 0.011 & 94.8 & 96.9 & 99.8 & 0.019 & 0.089 & 94.3 & 81.5 & 66.2\\
\hline
\multirow{8}{*}{(d)\phantom{$^1$}}&N-t & 4.782 & 23.094 & 0 & - & - & 5.028 & 25.583 & 0 & - & -\\
 &N-j & 0.022 & 0.115 & 94.2 & - & - & 0.005 & 0.163 & 96.2 & - & -\\
 &N-all & 4.841 & 23.487 & 0 & - & - & 5.086 & 25.938 & 0 & - & -\\
 &Reg & 0.211 & 0.061 & 57.4 & - & - & 0.393 & 0.183 & 26.6 & - & -\\
 &IPW & 0.422 & 0.224 & 39.6 & - & - & 0.367 & 0.207 & 62.2 & - & -\\
 %&U1-1 & 0.094 & 0.071 & 93.4 & 96.5 & 100 & 0.200 & 0.168 & 90.8 & 80.8 & 100\\ 
 %&U1-1/2 & 0.086 & 0.067 & 91.6 & 96.5 & 100 & 0.183 & 0.160 & 93.2 & 80.8 & 100\\ 
 %&U1-2 & 0.080 & 0.076 & 92.1 & 69.7 & 100 & 0.305 & 0.314 & 88.6 & 37.3 & 100\\ 
 &1-1 & 0.056 & 0.038 & 98.4 & 96.4 & 100 & 0.069 & 0.062 & 98.6 & 79.7 & 100\\ 
 &1-1/2 & 0.062 & 0.038 & 97.6 & 96.4 & 100 & 0.069 & 0.067 & 99 & 79.7 & 100\\ 
 &1-2 & 0.040 & 0.037 & 98.8 & 69.3 & 100 & 0.057 & 0.093 & 98.9 & 33.6 & 99.8\\

\hline
\multirow{8}{*}{(e)\phantom{$^1$}}&N-t & 4.342 & 19.074 & 0 & - & - & 4.694 & 22.416 & 0.2 & - & -\\
&N-j & 0.559 & 0.614 & 73.6 & - & - & 1.167 & 1.802 & 37.6 & - & -\\
&N-all & 4.670 & 21.884 & 0 & - & - & 5.108 & 26.204 & 0 & - & -\\
&Reg & 0.149 & 0.037 & 72 & - & - & 0.413 & 0.209 & 17.2 & - & -\\
&IPW & -0.480 & 0.300 & 35.8 & - & - & -0.511 & 0.497 & 56.3 & - & -\\
%&U1-1 & 0.081 & 0.051 & 94.4 & 81.4 & 100 & 0.245 & 0.178 & 87.6 & 55.3 & 99.8\\
%&U1-1/2 & 0.090 & 0.053 & 94.4 & 81.4 & 100 & 0.242 & 0.170 & 87.2 & 55.3 & 99.8\\
%&U1-2 & 0.064 & 0.056 & 93 & 53.6 & 97.8 & 0.331 & 0.357 & 87 & 22.8 & 99.8\\
&1-1 & 0.052 & 0.030 & 97.4 & 81.3 & 99.8 & 0.085 & 0.049 & 99.2 & 54.6 & 98.4\\
&1-1/2 & 0.043 & 0.032 & 97.4 & 81.4 & 100 & 0.083 & 0.048 & 99 & 54.7 & 98.6\\
&1-2 & 0.025 & 0.031 & 97.7 & 53.5 & 97 & 0.049 & 0.096 & 99.4 & 20.1 & 96\\ 
\hline

\end{tabular}\label{tab:real_sims}

    }%
% \begin{tablenotes}[flushleft]\footnotesize
%     \item $^1$The scenarios shown in \cref{tab:real_sims} correspond to 
%     \begin{enumerate*}[label=(\alph*)]
%     \item No confounders, 
%     \item Time-smooth confounders, \item Location-varying confounders, \item Time-varying confounders, and \item All confounders.
%     \end{enumerate*}
%     \end{tablenotes}
    
\end{table}

\subsection{Simulation results}

Table \ref{tab:real_sims} shows the estimation and inferential results for estimating the immediate and carryover effect for one outcome unit using the three na\"ive approaches, the estimator based on linear regression, the IPW estimator, and the three proposed estimators. For each estimator we report bias, mean squared error, and coverage of 95\% intervals. For the proposed algorithms, we further report the proportion of exposed time periods that were matched, and the success rate that the optimization problem was solved.

We focus first on the immediate causal effect.
In the absence of confounding factors (scenario (a)), all estimators are unbiased.
In the presence of location-varying confounding (scenario (c)), the \texttt{Na\"ive-$j$} and \texttt{Na\"ive-all} estimators are the only ones exhibiting bias, illustrating that estimators that focus on a specific outcome unit are robust to location-varying confounding (see \cref{subsec:temporal_advantages}). 
The \texttt{Na\"ive-$t$} estimator is biased with 95\% intervals that cover the true value less than 95\% of the time in the presence of temporal confounding (scenarios (b) and (d)). 
% \texttt{Na\"ive-all} is biased under either confounding structure. 
The regression and IPW estimators are biased when time-varying confounders exist (scenarios (d) and (e)). 
In contrast, the proposed estimators perform well in terms of bias and coverage of intervals across all confounding scenarios. Interestingly, even though the theoretical guarantees of the proposed estimator (\cref{subsec:theory}) rely on the outcome model that the regression estimator fits, the proposed approach, which combines principles of matching and calibration, is more robust to outcome misspecification. These results are in line with existing results in the literature on the relative merits of matching, calibration, and regression approaches \citep[e.g.,][]{imai2023matching}.  Lastly, the proposed algorithms successfully return matches across at least 97\% of data sets, indicating that our optimization formulation is not overly restrictive and is practical for general use.

For the carryover effect, all alternative estimation strategies show varying degrees of bias and undercoverage depending on the confounding structure, suggesting higher sensitivity to model assumptions. In comparison, the proposed estimators remain close to unbiased with coverage that is approximately or higher than nominal. However,  compared to the immediate effect algorithms, the algorithm for the carryover effect returns a solution to the optimization problem in a smaller proportion of data sets, and it maintains a smaller proportion of time periods with carryover exposure. This results from the frequency and density of the carryover exposure compared to the most recent exposure, which illustrates that different exposure structures might alter the performance of the algorithm.

Lastly, the \texttt{1-1} and \texttt{1-1/2} algorithms yield the same proportion of matched exposed time periods. This alignment is logical as the objective of the optimization problem is to maximize the number of matched exposed time periods, with the matches acquired under the \texttt{1-1} algorithm also possible under the \texttt{1-1/2} algorithm. As expected, the \texttt{1-2} algorithm yields the lowest proportion of matched exposed periods due to its requirement to find two relevant unexposed periods. Despite that, the corresponding estimator performs comparably to the estimators corresponding to the two other algorithms. We recommend that the \texttt{1-2} algorithm is preferred when it returns a similar proportion of matched exposed time periods compared to the \texttt{1-1} and \texttt{1-1/2} algorithms.

%For all three matching algorithms, the matching rate varies by the sparsity level of the exposure, with higher rates under sparser exposures. 
%As expected, Matching 1-2 returns the smallest proportion of matched exposed time periods. Combined with the fact that it has the lowest MSE in sparse scenarios, this illustrates that Matching 1-2 returns more accurate imputed potential outcomes than Matching 1-1 or 1-1/2.

\subsection{Additional simulations}
\label{subsec:additional_simulation}

We investigate the performance of the proposed algorithms with additional simulations in Supplement H,%\ref{supp_sec:simu}, 
with details on the data generative mechanism in Supplement H.1. %\ref{supp_subsec:simu_setup}.
In Supplement H.2,
%\ref{supp_subsec:simu_dif_sparsity}, 
we compare the three proposed algorithms and corresponding estimators under different sparsity levels of the exposure.
In Supplement H.3,
%\ref{subsec:multi-simu}, 
we consider estimation and inference in the presence of multiple outcome units. 
In Supplement H.4,
%\ref{appendix:unadj matching}, 
we show that the matching estimators with\textit{out} any adjustment for measured covariates are unbiased under temporally-smooth confounding, illustrating that smooth temporal trends are implicitly adjusted through the time constraints and need {\it not} be measured. 
In Supplement H.5,
%\ref{appendix:tuning parameters}, 
we assess the performance of matching methods with alternative tuning parameter values. Results are generally robust to the choice of $\delta$ and $\epsilon$, and despite some residual bias under larger $\delta'$ values in some scenarios, interval coverage is close to nominal across all cases. 
% The \texttt{1-2} algorithm often returns a substantially lower number of matches under strict tuning parameters.
In Supplement H.6,
%\ref{appendix:new-matching}, 
we illustrate that applying constraints on the covariates within {\it each} match alleviates the overcoverage of 95\% intervals, supporting the discussion in \cref{subsec:inference}.
% In Supplement \ref{appendix:network_confounding}, we illustrate that confounding can be induced by predictors of the network and the outcome, even if they are not predictors of the treatment assignment, and that the proposed estimator performs well in this scenario also. 
All the simulations are under a homogeneous treatment effect to separate the evaluation of estimation efficiency from the discussion on targeted estimand, 
and ease comparison of estimators. In Supplement H.7,%\ref{supp_subsec:simu_hetero}, 
we show that under treatment effect heterogeneity, results from the proposed estimators are representative of the matched population of exposed time periods. Lastly, in Supplement H.8,
%\ref{supp_subsec:sims_temp_corr}, 
we demonstrate that temporal correlation in the outcome has minimal effect on the performance of our inferential procedure and we compare it to an approach that uses the Newey-West variance estimator \citep{newey1986simple}.
% to compute the covariance matrix from regression on matched time periods, effectively addressing undercoverage.

%we show that temporal correlation in the outcome variable has minimal impact on the performance of our inferential procedure.

\section{The Impact of Wildfire Smoke Exposure on Bikeshare Hours}
\label{sec:real-data}

% NEED TO ADD

% Traditional causal inference to investigate riding hours change under the interference of wildfire is to consider the model as fully deterministic. We need the fire probability of every forest, the network connection between bike locations and the forest, and the function for the exposure. Then we make a regression regarding outcomes versus exposure and all confounders. To achieve that, we need every confounder available in the observation, as well as a true graph model to determine which covariates are confounders while the rest are not. It is almost impossible to achieve the goal, as we cannot prove the model is correct, nor can we even observe all covariates. For example, if no fire happens in one forest, we are not able to derive the propensity score of that location. We may also treat the context as time-series cross-sectional data. To match the exposed days to the non-exposed, we need to ensure that the covariate of interventional units is as close as possible and that the network and interaction covariate is similar. The covariate balance forces us to look for multiple-dimension covariates, thus greatly increasing the dimension of constraints. 

%However, in our study, we will not utilize the covariate data. 

Traditional unit-to-unit cross-sectional analyses would be infeasible in our study as it would be impossible to control for attributes of interventional and outcome units
%such as demographic and socio-economic information, 
with only three outcome units (see discussion in \cref{subsec:temporal_advantages}).
%The quantities of such location-specific covariates are huge, increasing the dimension of matching constraints significantly. 
However, given daily data on 1,003 days, approximately 140 of which are exposed across the three areas, 
our approach can be applied to estimate the effect of smoke exposure on each outcome area, without the need to consider location-varying covariates, and controlling for time-varying confounding only.
%These strengths enable us to estimate the effect of wildfire smoke on three regions' ride activity with more ease. 

The proposed algorithms balance daily temperature, humidity, precipitation, wind speed, and wind direction as potential time-varying confounders, while smooth seasonal trends are balanced implicitly. 
We find it plausible that no further covariates are necessary beyond weather-related data to meet the unconfoundedness assumptions.
This reasoning stems from our understanding of the bipartite structure of the problem: factors (other than weather conditions) influencing wildfire occurrence and smoke dispersion patterns are unlikely to impact biking activity in distant locations, and economic indices fluctuating over time that might affect biking activity are likely unrelated to wildfire presence in North American forests. 
Therefore, even though our analysis is performed at the level of the outcome unit without using interventional unit information explicitly, the appropriateness of the estimation approach is based on the investigation of unconfoundedness from a bipartite perspective.
%, and explicitly investigating the treatment assignment at the level of the interventional units. 
Consequently, even though implementation does not necessarily differ, there exist subtle, yet important differences in how confounding might be investigated under a bipartite lens.
\if{
\begin{figure}[!t]
\centering
%\spacingset{1}
\includegraphics[width=0.44\linewidth, trim=40 120 440 0, clip]{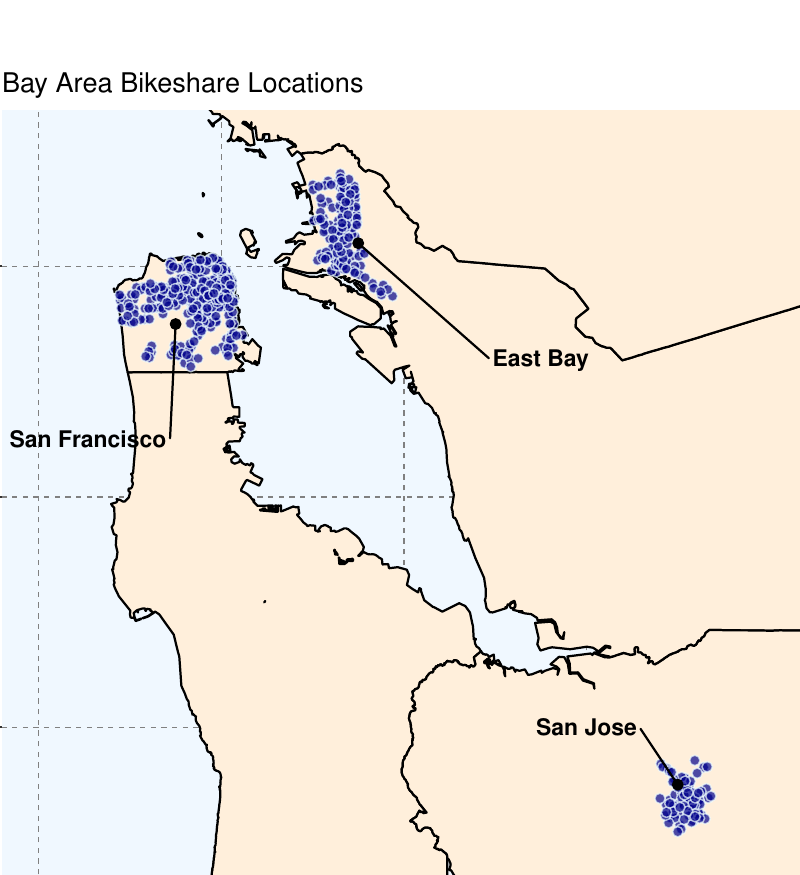}
\includegraphics[width=0.44\linewidth, trim=470 120 10 0, clip]{figures/New_real/Bay_area.pdf}
\caption{Left figure: Three major bike share locations in the Bay Area. Blue dots stand for the starting location of each trip. Right figure: The wildfire exposure on August 31, 2021, where yellow areas were affected by light exposure. }
\label{bay-area}
\end{figure}
}\fi

\if{
\begin{figure}[!t]
\includegraphics[width=\linewidth]{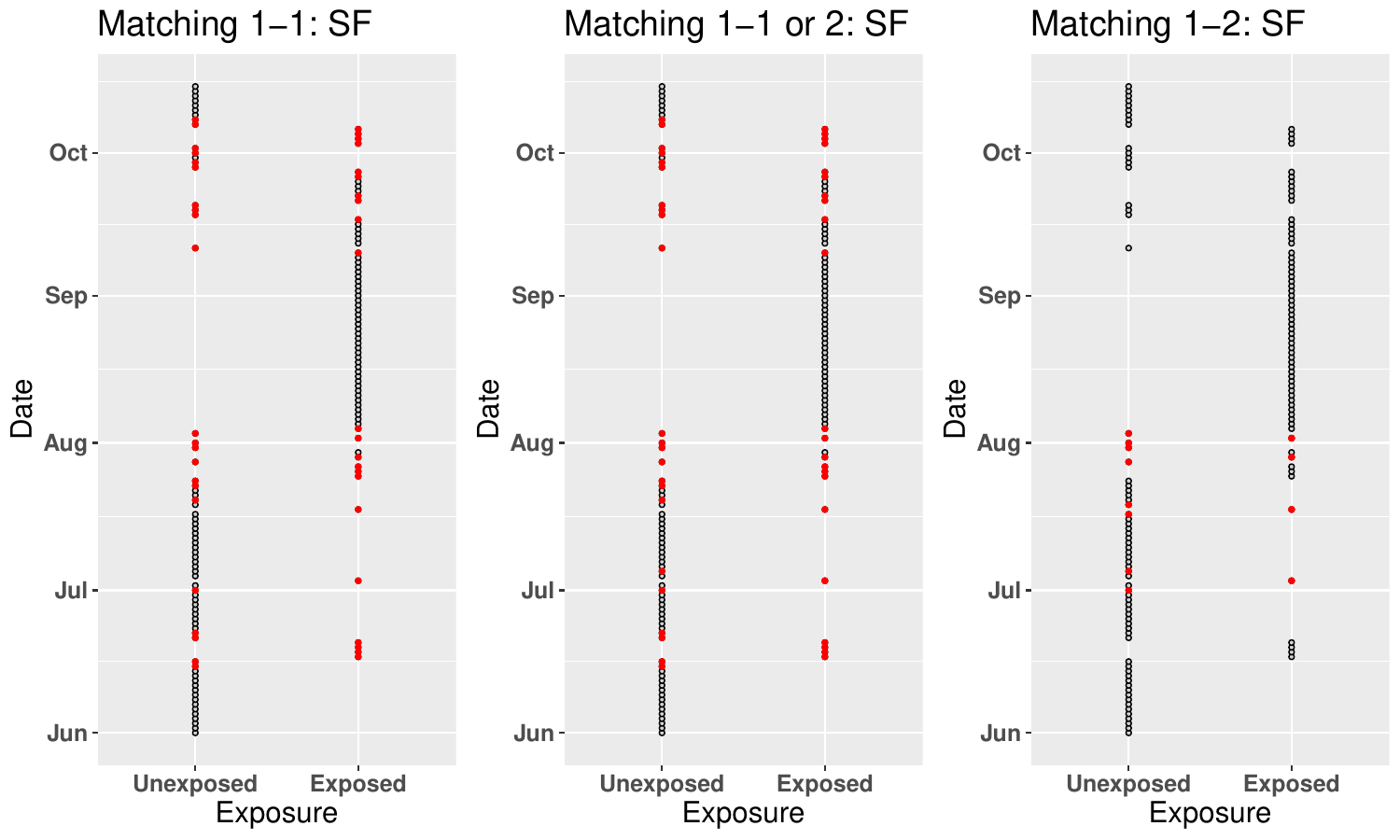}
\caption{Three matching techniques for San Francisco during the period from June 1, 2021, to October 15, 2021. The hollow dots are days without being matched while the red dots are days being matched.}
\label{SF-match}
\end{figure}
}\fi

% Selecting the maximum difference between days in a match as 6, we left out days that are too far away and irrelevant, and also give room for as many matches as possible. 
%All graphs show matches are more likely to occur on the boundary of the high-exposure and low-exposure days. Most nonexposed days are not used, since they do not constitute a reliable estimate for exposed days far from them. The middle points of high-exposure season are also rarely used, as a result of no available nonexposed time being regarded as a crucial component for the counterfactual outcome.
%Exposed days used in matching pairs differ in algorithms.  1-1 and 1-1 or 2 have no less than 100 days, 70\% of all exposed days, and much more than 55 days in the  1-2 algorithm. In comparison to \cref{SF-match1} and \cref{fig:appendix:SF_match},

\begin{table}[!b]
\caption{The immediate and carryover effect estimates of wildfire smoke on hours of bikeshare rentals in San Francisco, East Bay, and San Jose for Na\"ive-$t$ and the matching estimators. For each region, the three columns correspond to the estimate, p-value (bold font if below 0.05), and number of matches.}
\centering \small
\resizebox{0.8\textwidth}{!}{%
\begin{tabular}{cccccccccc}
\hline \\[-10pt]
& \multicolumn{3}{c}{San Francisco} & \multicolumn{3}{c}{East Bay} & \multicolumn{3}{c}{San Jose} \\
\cmidrule(lr){2-4}
\cmidrule(lr){5-7}
\cmidrule(lr){8-10}
& Est & p-value & \#Exp & Est & p-value & \#Exp & Est & p-value & \#Exp \\
\cmidrule(lr){2-2}
\cmidrule(lr){3-3}
\cmidrule(lr){4-4}
\cmidrule(lr){5-5}
\cmidrule(lr){6-6}
\cmidrule(lr){7-7}
\cmidrule(lr){8-8}
\cmidrule(lr){9-9}
\cmidrule(lr){10-10} 
\\[-10pt]
& \multicolumn{9}{c}{\underline{Immediate effect}} \\[5pt]
Na\"ive-$t$                         & 268.520 & (1.000) & {\it 147} & 24.542 & (1.000) & {\it 147} & 10.635 & (1.000) & \it {137} \\[5pt]
1-1                        & {\bf -107.457} & {\bf (0.016)} & 101 & -1.218 & (0.340) & 103 & -0.055 & (0.491) & 99 \\[5pt]
1-1/2                     & {\bf -85.293} & {\bf (0.028)} & 101 & 0.621 & (0.586) & 103 & -1.815 & (0.239) & 99 \\[5pt]
1-2                        & -58.845 & (0.117) & 56  & 2.266 & (0.736) & 59 & 0.384 & (0.514) & 59 \\
\hline \\[-5pt]
& \multicolumn{9}{c}{\underline{Carryover effect}} \\[5pt]
Na\"ive-$t$                         & 318.650 & (1.000) & {\it 115} & 25.129 & (1.000) & {\it 166} & 17.249 & (1.000) & {\it 104}  \\[5pt]
1-1                        &-14.092 & (0.441) & 71  & -4.423 & (0.102) & 72 & 3.671 & (0.869) & 65 \\[5pt]
1-1/2                     & -7.649 & (0.456) & 71  & -3.551 & (0.135) & 72 & 3.423 & (0.853) & 65 \\[5pt]
1-2                        & 39.725 & (0.970) & 31  & -0.228 & (0.476) & 32 & 1.922 & (0.856) & 31 \\
\hline
\end{tabular}}
\label{real-data-table}
\end{table}

We estimate the immediate and carryover effects for each area separately using the Na\"ive-$t$ estimator described in \cref{sec:simu_real_data} and the estimators based on the proposed algorithms that match one exposed time period to one (\cref{sec:matching}), two or one-or-two (Supplement~C.1) %(Supplement~\ref{supp_subsec:matching1-2or1-12}) 
unexposed time periods. We set the tuning parameters of the proposed algorithms as $(\delta,\delta', \epsilon)=(2,0.1,6)$. 

\cref{real-data-table} shows the causal effect estimates, p-values, and the number of exposed time periods used by each approach. Effect estimates are in hours of bikeshare rentals. The na\"ive approach, which uses all the exposed (and unexposed) time periods, estimates that both the immediate and the carryover effect is positive (albeit, not statistically significant). These results under the na\"ive approach imply that the most recent and past smoke exposure increase biking activity. This unrealistic result is likely due to temporal confounding since exposure is more common during the late summer and fall months when there is also more biking activity.

Instead, effect estimates from the proposed algorithms and estimators return more realistic results. We estimate that smoke exposure reduces biking activity in San Francisco, while it does not have an impact in the East Bay and San Jose areas.
These results are statistically significant at the 0.05 level for San Francisco for the estimator corresponding to the algorithm that matches one exposed time period to one or one-or-two unexposed time periods (1-1, and 1-1/2).
Since exposure is relatively dense during the summer and fall months, the algorithm that searches for two unexposed time periods for each exposed one returns approximately half the number of matches compared to the other two algorithms, which might explain the larger p-values.
Based on the algorithm's constraints, the proposed estimator uses unexposed time periods during the summer and fall months only (see Supplement I.2 for an illustration).%(see Supplement \ref{supp_subsec:application_matches} for an illustration).

%Compared to immediate effect estimation, the matching estimators identify 30\% fewer matched pair numbers, reflecting the fact that carryover exposure is less common than immediate exposure. 
For the carryover effect, none of the estimates are statistically significant at the 0.05 level, with point estimates generally close to 0. This suggests that smoke exposure during the previous week does not have an impact on current biking activity. % However, the corresponding p-values are large, indicating no statistically significant evidence of a carryover effect in either region. In East Bay, the carryover effect seems to be stronger than the immediate effect, but all insignificant under 95\% confidence.
% In San Jose, all matching estimators yield positive but insignificant estimates, which implies biking in San Jose is not impacted by carryover wildfire exposure. 
However, the proportions of time periods with carryover exposure that were matched are relatively low, especially for the \texttt{1-2} algorithm.
% In addition, the 1-2 matching strategy results in positive effects with large p-values across all regions, 
{This is due to the dense carryover exposure conditions in summer and fall and a small number of unexposed time periods which restricts the quantity of available matches.}

We conducted a sensitivity analysis to the definition of exposure, and performed the same analysis when an area is categorized as exposed during a given day under medium or high smoke thickness, and unexposed under no or light smoke thickness (see Supplement I.3). 
%(see Supplement \ref{supp_subsec:real-data-medium-exposure}). 
We find that all effect estimates are negative and similar or larger in magnitude in that case, even though most results are not statistically significant, likely due to the small number of days with exposure (40 or less).

\section{Discussion}\label{discussion}

In this manuscript, we introduced a causal inference framework for time series data with bipartite interference, a random network, and carryover effects. Focusing on time-averaged estimands, we showed that controlling for time-varying information allows us to attribute outcome differences to exposure's causal effects. We introduce algorithms that combine notions from matching and covariate balancing, with corresponding estimators which perform well across various scenarios. In our study, we find some evidence that smoke from wildfires reduces population mobility.

%In principle, weighting or outcome modeling estimators could be employed in our context. However, in the bipartite setting, the exposure arises as a function of the treatment and network, whose mechanisms are often unknown and would need to be modeled, adding complexity to a weighting-based estimation procedure.
Alternative estimation approaches from the causal inference literature can be adapted to our framework.
Weighting or matching on the propensity of exposure is one such alternative. However, since the exposure is defined based on the interventional units' treatments and the network, the exposure propensity score would require that we model the high-dimensional treatment and network assignment incorporating potential time-invariant confounders, which is practically difficult to do accurately.
% , especially when the number of interventional units is large.
At the same time, outcome modeling approaches like panel regression \citep{arkhangelsky2024design, imai2021use} would require correct specification of the confounding adjustment functional form, which is particularly complicated in bipartite settings where confounding relates to both the treatment and the network.
% Propensity matching and pure matching methods \citep{stuart2010matching} often suffer from lower efficiency due to a reduced number of matched time periods. 
Balancing methods \citep{hainmueller2012entropy, fong2018covariate} and methods from the econometrics literature \citep{imai2021use} may assign weights to unexposed periods that are temporally distant from the exposed periods, and therefore might be susceptible to confounding by temporal trends.
Our proposed approach which combines matching and covariate balancing bypasses these issues.
% and offer clear advantages over several benchmark approaches.
% Another benefit of matching is objectiveness in the separation between the design of putting units into blocks and the analysis of outcomes. It allows for an easier interpretation in the matched set, compared to assigning weights to each observed outcome in the propensity score weighting.
%Panel regression models (Arkhangelsky et al., 2024; Imai and Kim, 2021) may also perform poorly if important covariates are missing, nonlinearities and interactions are ignored, or temporal dependencies are not accounted for

Despite the merits of the proposed framework, several open questions remain. First, future work can extend our estimation framework to time series settings under more complicated exposure mappings.
In principle, even in the case where the exposure is not binary, matching and balancing algorithms designed for binary exposures can still be useful for estimating interpretable effects. For instance, in the setting of \cite{pouget2019variance, Doudchenko2020, brennan2022cluster}, and \cite{harshaw2023design} where an outcome unit's exposure ranges from 0 to 1, the proposed algorithms can be extended to accommodate a categorized exposure.
Second, future work can investigate the impact of misspecification of the exposure and carryover mapping functions to matching or balancing estimators, bipartite interference settings, or observational data following the work for Horvitz-Thompson estimators under exposure misspecification in the unipartite literature \citep{savje2024causal,leung2022causal,weinstein2026causal}.

{Third, in future work it is interesting to investigate how to tailor bipartite methodology to specific study designs and construct estimators with robustness properties to model misspecification. For example, in settings where the propensity score is known or can be estimated reliably, one can draw from the literature with independent data to construct doubly robust estimators by incorporating constraints on the propensity score in the algorithm of \cref{sec:matching} or by matching on both the propensity and the prognostic score \citep{antonelli2018doubly}. In shift-share study designs for observational settings with exogenous shocks, it is interesting to investigate how to design alternative, robust estimators by controlling for the expected exposures derived from multiple exposure candidate models \citep{borusyak2023nonrandom}, or by employing ensemble approaches that combine estimators with complementary failure modes \citep{gulek2024synthetic}.}
% If the size of outcome units is large, 
Lastly, it is interesting to consider extending synthetic control approaches, such as matrix completion \citep{athey2021matrix} and the augmented synthetic control method \citep{ben2021augmented}, to the bipartite setting. However, it is unclear how the methodology can incorporate spatial correlation across units \citep{grossi2025spatial}, where nearby similar areas might experience the exposure simultaneously, while distant areas may have less relevance to the target, rendering it difficult to identify a valid donor pool. Alternatively, future work can investigate how outcome autoregressive models can be employed for the estimation of causal effects in bipartite settings with time series data and carryover exposures, which would likely require additional assumptions \citep{antonelli2024autoregressive}.
Investigating, adapting and extending these approaches to the bipartite setting where interventional and outcome units are separate remains a promising area for future work.

\subsection*{Acknowledgements}
We thank Corwin Zigler, Fabrizia Mealli, Lucas Henneman, Jean Pouget-Abadie, and Jennifer Brennan for their useful comments in the preparation of this manuscript.
%They would also like to thank the anonymous referees, an Associate Editor and the Editor for their constructive comments that improved the quality of this paper.
% \end{acks}

%%%%%%%%%%%%%%%%%%%%%%%%%%%%%%%%%%%%%%%%%%%%%%
%% Funding information, if any,             %%
%% should be provided in the                %%
%% funding section.                         %%
%%%%%%%%%%%%%%%%%%%%%%%%%%%%%%%%%%%%%%%%%%%%%%
\subsection*{Funding}
The authors are grateful for support from the National Science Foundation under Grant No 2124124.

{\small
\bibliographystyle{plainnat}
\bibliography{refs}

@article{zigler2021bipartite,
  title={Bipartite causal inference with interference},
  author={Zigler, Corwin M and Papadogeorgou, Georgia},
  journal={Statistical science: a review journal of the Institute of Mathematical Statistics},
  year={2021},
  publisher={NIH Public Access}
}

@article{Aronow2017,
   author = {Peter M. Aronow and Cyrus Samii},
   issue = {4},
   journal = {Annals of Applied Statistics},
   month = {12},
   pages = {1912-1947},
   publisher = {Institute of Mathematical Statistics},
   title = {Estimating average causal effects under general interference, with application to a social network experiment},
   volume = {11},
   year = {2017},
}

@article{Doudchenko2020,
   author = {Nick Doudchenko and Minzhengxiong Zhang and Evgeni Drynkin and Edoardo Airoldi and Vahab Mirrokni and Jean Pouget-Abadie},
   month = {10},
   title = {Causal Inference with Bipartite Designs},
   year = {2020},
}

@article{zigler2025bipartite,
  title={Bipartite interference and air pollution transport: estimating health effects of power plant interventions},
  author={Zigler, Corwin and Liu, Vera and Mealli, Fabrizia and Forastiere, Laura},
  journal={Biostatistics},
  volume={26},
  number={1},
  pages={kxae051},
  year={2025},
  publisher={Oxford University Press}
}

@article{Bojinov2019,
   author = {Iavor Bojinov and Neil Shephard},
   issue = {528},
   journal = {Journal of the American Statistical Association},
   month = {10},
   pages = {1665-1682},
   publisher = {American Statistical Association},
   title = {Time Series Experiments and Causal Estimands: Exact Randomization Tests and Trading},
   volume = {114},
   year = {2019},
}

@article{Keele2014,
   author = {Luke Keele and Rocío Titiunik and José R Zubizarreta},
   journal = {Journal of the Royal Statistical Society, Series A},
   pages = {223-239},
   title = {Enhancing a geographic regression discontinuity design through matching to estimate the effect of ballot initiatives on voter turnout},
   volume = {178},
   year = {2014},
}

@article{Zubizarreta2015,
   author = {José R. Zubizarreta},
   issue = {511},
   journal = {Journal of the American Statistical Association},
   month = {7},
   pages = {910-922},
   publisher = {American Statistical Association},
   title = {Stable Weights that Balance Covariates for Estimation With Incomplete Outcome Data},
   volume = {110},
   year = {2015},
}

@article{forastiere2021identification,
  title={Identification and estimation of treatment and interference effects in observational studies on networks},
  author={Forastiere, Laura and Airoldi, Edoardo M and Mealli, Fabrizia},
  journal={Journal of the American Statistical Association},
  volume={116},
  number={534},
  pages={901--918},
  year={2021},
  publisher={Taylor \& Francis}
}

@article{ben2021augmented,
  title={The augmented synthetic control method},
  author={Ben-Michael, Eli and Feller, Avi and Rothstein, Jesse},
  journal={Journal of the American Statistical Association},
  volume={116},
  number={536},
  pages={1789--1803},
  year={2021},
  publisher={Taylor \& Francis}
}

@article{benjamini1995controlling,
  title={Controlling the false discovery rate: a practical and powerful approach to multiple testing},
  author={Benjamini, Yoav and Hochberg, Yosef},
  journal={Journal of the Royal statistical society: series B (Methodological)},
  volume={57},
  number={1},
  pages={289--300},
  year={1995},
  publisher={Wiley Online Library}
}

@article{zubizarreta2013stronger,
  title={Stronger instruments via integer programming in an observational study of late preterm birth outcomes},
  author={Zubizarreta, Jos{\'e} R and Small, Dylan S and Goyal, Neera K and Lorch, Scott and Rosenbaum, Paul R},
  journal={The Annals of Applied Statistics},
  pages={25--50},
  year={2013},
  publisher={JSTOR}
}

@article{ho2007matching,
  title={Matching as nonparametric preprocessing for reducing model dependence in parametric causal inference},
  author={Ho, Daniel E and Imai, Kosuke and King, Gary and Stuart, Elizabeth A},
  journal={Political analysis},
  volume={15},
  number={3},
  pages={199--236},
  year={2007},
  publisher={Cambridge University Press}
}

@article{doubleday2021urban,
  title={Urban bike and pedestrian activity impacts from wildfire smoke events in Seattle, WA},
  author={Doubleday, Annie and Choe, Youngjun and Isaksen, Tania M Busch and Errett, Nicole A},
  journal={Journal of Transport \& Health},
  volume={21},
  pages={101033},
  year={2021},
  publisher={Elsevier}
}

@article{rosenthal2020assessment,
  title={Assessment of accelerometer-based physical activity during the 2017-2018 California wildfire seasons},
  author={Rosenthal, David G and Vittinghoff, Eric and Tison, Geoffrey H and Pletcher, Mark J and Olgin, Jeffrey E and Grandis, Donald J and Marcus, Gregory M},
  journal={JAMA Network Open},
  volume={3},
  number={9},
  pages={e2018116--e2018116},
  year={2020},
  publisher={American Medical Association}
}

@article{clark2021approach,
  title={An Approach to Causal Inference over Stochastic Networks},
  author={Clark, Duncan A and Handcock, Mark S},
  journal={arXiv preprint arXiv:2106.14145},
  year={2021}
}

@article{harshaw2023design,
  title={Design and analysis of bipartite experiments under a linear exposure-response model},
  author={Harshaw, Christopher and S{\"a}vje, Fredrik and Eisenstat, David and Mirrokni, Vahab and Pouget-Abadie, Jean},
  journal={Electronic Journal of Statistics},
  volume={17},
  number={1},
  pages={464--518},
  year={2023},
  publisher={The Institute of Mathematical Statistics and the Bernoulli Society}
}

@article{pouget2019variance,
  title={Variance reduction in bipartite experiments through correlation clustering},
  author={Pouget-Abadie, Jean and Aydin, Kevin and Schudy, Warren and Brodersen, Kay and Mirrokni, Vahab},
  journal={Advances in Neural Information Processing Systems},
  volume={32},
  year={2019}
}

@article{brennan2022cluster,
  title={Cluster Randomized Designs for One-Sided Bipartite Experiments},
  author={Brennan, Jennifer and Mirrokni, Vahab and Pouget-Abadie, Jean},
  journal={Advances in Neural Information Processing Systems},
  volume={35},
  pages={37962--37974},
  year={2022}
}

@article{agarwalnetwork,
  title={Network Synthetic Interventions: A Causal Framework for Panel Data Under Network Interference},
  author={Agarwal, Anish and Cen, Sarah H and Shah, Devavrat and Yu, Christina Lee},
  year={2023}
}

@article{cao2019estimation,
  title={Estimation and inference for synthetic control methods with spillover effects},
  author={Cao, Jianfei and Dowd, Connor},
  journal={arXiv preprint arXiv:1902.07343},
  year={2019}
}

@article{grossi2020synthetic,
  title={Synthetic control group methods in the presence of interference: The direct and spillover effects of light rail on neighborhood retail activity},
  author={Grossi, Giulio and Lattarulo, Patrizia and Mariani, Marco and Mattei, Alessandra and Oner, O},
  journal={arXiv preprint arXiv:2004.05027},
  year={2020}
}

@article{di2020inclusive,
  title={The inclusive synthetic control method},
  author={Di Stefano, Roberta and Mellace, Giovanni},
  journal={Discussion Papers on Business and Economics, University of Southern Denmark},
  volume={14},
  year={2020}
}

@article{menchetti2020estimating,
  title={Estimating causal effects in the presence of partial interference using multivariate Bayesian structural time series models},
  author={Menchetti, Fiammetta and Bojinov, Iavor},
  journal={Harvard Business School Technology \& Operations Mgt. Unit Working Paper},
  number={21-048},
  year={2020}
}

@article{savje2024causal,
  title={Causal inference with misspecified exposure mappings: separating definitions and assumptions},
  author={S{\"a}vje, Fredrik},
  journal={Biometrika},
  volume={111},
  number={1},
  pages={1--15},
  year={2024},
  publisher={Oxford University Press}
}

@article{abadie2006large,
  title={Large sample properties of matching estimators for average treatment effects},
  author={Abadie, Alberto and Imbens, Guido W},
  journal={Econometrica},
  volume={74},
  number={1},
  pages={235--267},
  year={2006},
  publisher={Wiley Online Library}
}

@article{zubizarreta2012using,
  title={Using mixed integer programming for matching in an observational study of kidney failure after surgery},
  author={Zubizarreta, Jos{\'e} R},
  journal={Journal of the American Statistical Association},
  volume={107},
  number={500},
  pages={1360--1371},
  year={2012},
  publisher={Taylor \& Francis}
}

@article{imbens2024causal,
  title={Causal Inference in the Social Sciences},
  author={Imbens, Guido W},
  journal={Annual Review of Statistics and Its Application},
  volume={11},
  year={2024},
  publisher={Annual Reviews}
}

@article{abadie2018econometric,
  title={Econometric methods for program evaluation},
  author={Abadie, Alberto and Cattaneo, Matias D},
  journal={Annual Review of Economics},
  volume={10},
  pages={465--503},
  year={2018},
  publisher={Annual Reviews}
}

@article{wikle2023causal,
  title={Causal health impacts of power plant emission controls under modeled and uncertain physical process interference},
  author={Wikle, Nathan B and Zigler, Corwin M},
  journal={arXiv preprint arXiv:2306.05665},
  year={2023}
}

@article{rockers2015inclusion,
  title={Inclusion of quasi-experimental studies in systematic reviews of health systems research},
  author={Rockers, Peter C and R{\o}ttingen, John-Arne and Shemilt, Ian and Tugwell, Peter and B{\"a}rnighausen, Till},
  journal={Health Policy},
  volume={119},
  number={4},
  pages={511--521},
  year={2015},
  publisher={Elsevier}
}

@article{rosenbaum1985constructing,
  title={Constructing a control group using multivariate matched sampling methods that incorporate the propensity score},
  author={Rosenbaum, Paul R and Rubin, Donald B},
  journal={The American Statistician},
  volume={39},
  number={1},
  pages={33--38},
  year={1985},
  publisher={Taylor \& Francis}
}

@article{imai2023matching,
  title={Matching methods for causal inference with time-series cross-sectional data},
  author={Imai, Kosuke and Kim, In Song and Wang, Erik H},
  journal={American Journal of Political Science},
  volume={67},
  number={3},
  pages={587--605},
  year={2023},
  publisher={Wiley Online Library}
}

@article{arkhangelsky2024design,
  title={Design-robust two-way-fixed-effects regression for panel data},
  author={Arkhangelsky, Dmitry and Imbens, Guido W and Lei, Lihua and Luo, Xiaoman},
  journal={Quantitative Economics},
  volume={15},
  number={4},
  pages={999--1034},
  year={2024},
  publisher={Wiley Online Library}
}

@article{antonelli2024autoregressive,
  title={Autoregressive models for panel data causal inference with application to state-level opioid policies},
  author={Antonelli, Joseph and Rubinstein, Max and Agniel, Denis and Smart, Rosanna and Stuart, Elizabeth and Cefalu, Matthew and Schell, Terry and Eagan, Joshua and Stone, Elizabeth and Griswold, Max and others},
  journal={arXiv preprint arXiv:2408.09012},
  year={2024}
}

@article{stuart2010matching,
  title={Matching methods for causal inference: A review and a look forward},
  author={Stuart, Elizabeth A},
  journal={Statistical science: a review journal of the Institute of Mathematical Statistics},
  volume={25},
  number={1},
  pages={1},
  year={2010}
}

@article{leung2022causal,
  title={Causal inference under approximate neighborhood interference},
  author={Leung, Michael P},
  journal={Econometrica},
  volume={90},
  number={1},
  pages={267--293},
  year={2022},
  publisher={Wiley Online Library}
}

@article{hainmueller2012entropy,
  title={Entropy balancing for causal effects: A multivariate reweighting method to produce balanced samples in observational studies},
  author={Hainmueller, Jens},
  journal={Political analysis},
  volume={20},
  number={1},
  pages={25--46},
  year={2012},
  publisher={Cambridge University Press}
}

@article{antonelli2018doubly,
  title={Doubly robust matching estimators for high dimensional confounding adjustment},
  author={Antonelli, Joseph and Cefalu, Matthew and Palmer, Nathan and Agniel, Denis},
  journal={Biometrics},
  volume={74},
  number={4},
  pages={1171--1179},
  year={2018},
  publisher={Oxford University Press}
}

@article{imai2021use,
  title={On the use of two-way fixed effects regression models for causal inference with panel data},
  author={Imai, Kosuke and Kim, In Song},
  journal={Political Analysis},
  volume={29},
  number={3},
  pages={405--415},
  year={2021},
  publisher={Cambridge University Press}
}

@article{athey2021matrix,
  title={Matrix completion methods for causal panel data models},
  author={Athey, Susan and Bayati, Mohsen and Doudchenko, Nikolay and Imbens, Guido and Khosravi, Khashayar},
  journal={Journal of the American Statistical Association},
  volume={116},
  number={536},
  pages={1716--1730},
  year={2021},
  publisher={Taylor \& Francis}
}

@article{chen2021climate,
  title={Climate, fuel, and land use shaped the spatial pattern of wildfire in California’s Sierra Nevada},
  author={Chen, Bin and Jin, Yufang and Scaduto, Erica and Moritz, Max A and Goulden, Michael L and Randerson, James T},
  journal={Journal of Geophysical Research: Biogeosciences},
  volume={126},
  number={2},
  pages={e2020JG005786},
  year={2021},
  publisher={Wiley Online Library}
}

@article{fong2018covariate,
  title={Covariate balancing propensity score for a continuous treatment: Application to the efficacy of political advertisements},
  author={Fong, Christian and Hazlett, Chad and Imai, Kosuke},
  journal={The Annals of Applied Statistics},
  volume={12},
  number={1},
  pages={156--177},
  year={2018},
  publisher={JSTOR}
}

@article{chattopadhyay2023design,
  title={Design-based inference for generalized network experiments with stochastic interventions},
  author={Chattopadhyay, Ambarish and Imai, Kosuke and Zubizarreta, Jos{\'e} R},
  journal={arXiv preprint arXiv:2312.03268},
  year={2023}
}

@article{bojinov2023design,
  title={Design and analysis of switchback experiments},
  author={Bojinov, Iavor and Simchi-Levi, David and Zhao, Jinglong},
  journal={Management Science},
  volume={69},
  number={7},
  pages={3759--3777},
  year={2023},
  publisher={INFORMS}
}

@article{hu2022switchback,
  title={Switchback experiments under geometric mixing},
  author={Hu, Yuchen and Wager, Stefan},
  journal={arXiv preprint arXiv:2209.00197},
  year={2022}
}

@article{reid2016critical,
  title={Critical review of health impacts of wildfire smoke exposure},
  author={Reid, Colleen E and Brauer, Michael and Johnston, Fay H and Jerrett, Michael and Balmes, John R and Elliott, Catherine T},
  journal={Environmental health perspectives},
  volume={124},
  number={9},
  pages={1334--1343},
  year={2016},
  publisher={National Institute of Environmental Health Sciences}
}

@article{han2024population,
  title={Population interference in panel experiments},
  author={Han, Kevin and Basse, Guillaume and Bojinov, Iavor},
  journal={Journal of Econometrics},
  volume={238},
  number={1},
  pages={105565},
  year={2024},
  publisher={Elsevier}
}

@article{papadogeorgou2025causal,
  title={Causal Inference when Intervention Units and Outcome Units Differ},
  author={Papadogeorgou, Georgia and Song, Zhaoyan and Imbens, Guido and Mealli, Fabrizia},
  journal={arXiv preprint arXiv:2507.20231},
  year={2025}
}

@article{lu2025design,
  title={Design-based causal inference in bipartite experiments},
  author={Lu, Sizhu and Shi, Lei and Fang, Yue and Zhang, Wenxin and Ding, Peng},
  journal={arXiv preprint arXiv:2501.09844},
  year={2025}
}

@article{rubin1973matching,
  title={Matching to remove bias in observational studies},
  author={Rubin, Donald B},
  journal={Biometrics},
  pages={159--183},
  year={1973},
  publisher={JSTOR}
}

@article{rubin1980bias,
  title={Bias reduction using Mahalanobis-metric matching},
  author={Rubin, Donald B},
  journal={Biometrics},
  pages={293--298},
  year={1980},
  publisher={JSTOR}
}

@article{imai2014covariate,
  title={Covariate balancing propensity score},
  author={Imai, Kosuke and Ratkovic, Marc},
  journal={Journal of the Royal Statistical Society Series B: Statistical Methodology},
  volume={76},
  number={1},
  pages={243--263},
  year={2014},
  publisher={Oxford University Press}
}

@article{li2018balancing,
  title={Balancing covariates via propensity score weighting},
  author={Li, Fan and Morgan, Kari Lock and Zaslavsky, Alan M},
  journal={Journal of the American Statistical Association},
  volume={113},
  number={521},
  pages={390--400},
  year={2018},
  publisher={Taylor \& Francis}
}

@article{zhou2022psweight,
  author = {Zhou, Tianhui and Tong, Guangyu and Li, Fan and Thomas, Laine E. and Li, Fan},
  title = {PSweight: An R Package for Propensity Score Weighting Analysis},
  journal = {The R Journal},
  year = {2022},
  note = {https://doi.org/10.32614/RJ-2022-011},
  doi = {10.32614/RJ-2022-011},
  volume = {14},
  issue = {1},
  issn = {2073-4859},
  pages = {282-300}
}

@article{newey1986simple,
  title={A simple, positive semi-definite, heteroskedasticity and autocorrelationconsistent covariance matrix},
  author={Newey, Whitney K and West, Kenneth D},
  year={1986},
  publisher={National Bureau of Economic Research Cambridge, Mass., USA}
}

@article{grossi2025spatial,
  title={Spatial vertical regression for spatial panel data: Evaluating the effect of the Florentine tramway's first line on commercial vitality},
  author={Grossi, Giulio and Mattei, Alessandra and Papadogeorgou, Georgia},
  journal={arXiv preprint arXiv:2505.00450},
  year={2025}
}

@article{weinstein2026causal,
    author = {Weinstein, Bar and Nevo, Daniel},
    title = {Causal inference with misspecified network interference structure},
    journal = {Biometrics},
    volume = {82},
    number = {1},
    pages = {ujag023},
    year = {2026},
    month = {03},
    issn = {0006-341X},
    doi = {10.1093/biomtc/ujag023},
    url = {https://doi.org/10.1093/biomtc/ujag023},
    eprint = {https://academic.oup.com/biometrics/article-pdf/82/1/ujag023/67069308/ujag023.pdf},
}

@article{borusyak2023nonrandom,
  title={Nonrandom exposure to exogenous shocks},
  author={Borusyak, Kirill and Hull, Peter},
  journal={Econometrica},
  volume={91},
  number={6},
  pages={2155--2185},
  year={2023}
}

@techreport{gulek2024synthetic,
  title={Synthetic {IV} estimation in panels},
  author={Gulek, Ahmet and Vives-i-Bastida, Jaume},
  institution={MIT},
  year={2025},
  note={Job Market Paper}
}
}
\clearpage

\doparttoc % Tell to minitoc to generate a toc for the parts
\faketableofcontents % Run a fake tableofcontents command for the partocs
\part{} % Start the document part

\begin{center}
\LARGE Supplementary materials
\end{center}
\normalfont

\appendix
\setcounter{page}{1}
\setcounter{section}{0}    
\renewcommand{\thetheorem}{S.\arabic{theorem}}
\renewcommand{\theproposition}{S.\arabic{proposition}}
\renewcommand{\theassumption}{S.\arabic{assumption}}
\renewcommand{\theequation}{S.\arabic{equation}}
\renewcommand{\thetable}{S.\arabic{table}}
\renewcommand{\thefigure}{S.\arabic{figure}}

\allowdisplaybreaks

\addcontentsline{toc}{section}{Supplement} % Add the appendix text to the document TOC
\part{ } % Start the appendix part
\parttoc

\clearpage

\section{Glossary}
\label{supp_sec:glossary}

In \cref{supp_tab:notation}, we include a glossary of the notation used in the manuscript.

{
\setlength{\extrarowheight}{5pt}

\begin{longtable}{c p{4in}}
\caption{Table of notation} 
\label{supp_tab:notation} \\

\hline
\textbf{Symbol} & \textbf{Description} \\
\hline
\endfirsthead

\hline
\textbf{Symbol} & \textbf{Description} \\
\hline
\endhead

% Units
$\mathcal{N}$ & The set of interventional units, $\mathcal{N} = \{n_1, n_2, \dots, n_N\}$ where $n_i$ denotes the $i^{th}$ unit \\
$m$ & The outcome unit \\
$\mathcal{T}$ & The set of time periods, $\mathcal T = \{1, 2, \dots, T\}$ \\ 
\hline
%
% treatment and network, exposure
$A_{ti}, \bm A_t$ & $A_{ti}$ is treatment of interventional unit $n_i$ at time $t$ and $\bm A_t$ is the vector of treatments across all units at time $t$ \\
$G_{ti}, \bm G_t$ & $G_{ti}$ is the connectivity measurement of interventional unit $n_i$ with outcome unit $m$ at time $t$, and $\bm G_t$ is the vector of these measurements across interventional units \\
% $E_t$ & The exposure of the outcome unit $m$ at time $t$. It is defined based on the function $h_t$ applied on $\bm A_t, \bm G_t$ \\
% $R_t$ & The carryover exposure of outcome unit $m$ at time $t$. It is defined as the value of the function $h_t^\text{c}$ on the exposures $\overline E_{t-1, L-1} = (E_{t-1}, E_{t-2}, \dots, E_{t-L})^\top$ \\

%
% Covariates
$\bm X_{0i}^\text{int}, \bm X_0^\text{out}, \bm X_{0i}^\text{net}$ & Time-invariant covariates of dimension $p_0^\text{int}, p_0^\text{out}, p_0^\text{net}$ for the interventional unit $n_i$, the outcome unit $m$, and their relationship, respectively \\
$\bm X_{ti}^\text{int}, \bm X_t^\text{out}, \bm X_{ti}^\text{net}$ &  Time-varying covariates of dimension $p^\text{int}, p^\text{out}, p^\text{net}$ for the interventional unit $n_i$, the outcome unit $m$, and their relationship, respectively, at time $t$. Specifically, $\bm X_{ti}^\text{int} = (X_{ti1}^\text{int}, X_{ti2}^\text{int}, \dots, X_{tip^\text{int}}^\text{int})^\top$, and similarly for $\bm X_{ti}^\text{net}$ \\
$\bm X_0, \bm X_t$ & Vectors including all time-invariant and time-varying covariates, respectively \\
%
% History
$\overline V_t$ & The history of a covariate over times $1, 2, \dots, t$. Specifically, $\overline V_t = (V_t, V_{t-1}, \dots, V_1)^\top$ \\
$\overline V_{t,S}$ & The history of a covariate over times $t - S, t-S + 1, \dots, t$. Specifically, $\overline V_{t,S} = (V_t, V_{t-1}, \dots, V_{t-S})^\top$\\
\hline
%
% Exposure mappings
$h_t$ & The exposure mapping $h_t:\mathcal{A}^N \times \mathcal{G} \rightarrow \mathcal{E}$ reflects the exposure function at time $t$ that maps the interventional units' treatment and the bipartite connectivity network to the outcome unit exposure \\
$ h_t^\text{c}$ & The carryover mapping $h_t^\text{c}: \mathcal{E}^L \rightarrow \mathcal{R}$ reflects the function that summarizes the exposure values during time periods $t-L, t-L+1, \dots, t-1$ \\
$E_t$ & The exposure of the outcome unit $m$ at time $t$. It is defined based on the function $h_t$ applied on $\bm A_t, \bm G_t$ \\
$R_t$ & The carryover exposure of outcome unit $m$ at time $t$. It is defined as the value of the function $h_t^\text{c}$ on the exposures $\overline E_{t-1, L-1} = (E_{t-1}, E_{t-2}, \dots, E_{t-L})^\top$ \\
\hline
%
% Outcomes
$Y_t$ & The observed outcome for the outcome unit at time $t$ \\
$Y_t(\overline{\bm a}_t, \overline{\bm g}_t), Y_t(e_t, r_t)$ & The potential outcome of the outcome unit at time $t$ with and without the exposure mapping \cref{exp-map-cond} \\
$\mathcal Y_t(\cdot)$ & The collection of the outcome unit's all potential outcomes for time $t$ \\
\hline
%
% EStimands
$\tau_t^\text{imm}(e, e' ; r)$ & The immediate effect at time $t$ for a change in the most recent exposure, $\tau_t^\text{imm}(e, e' ; r) = Y_{t}(e, r) - Y_{t}(e', r) $ \\
% \intertext{and}
$\tau_t^\text{car}(r, r' ; e)$ & The carryover effect at time $t$ for a change in the carryover exposure, $\tau_t^\text{car}(r, r' ; e) = Y_{t}(e, r) - Y_{t}(e, r')$\\
$\mathcal{T}_{e.}, \mathcal{T}_{.r}$ & Subsets of $\mathcal T$ with exposure $E_t = e$ or carryover exposure $R_t = r$, respectively \\
${\tau}^\text{imm}(e, e')$ & Immediate effect averaged over time periods with $E_t= e$, ${\tau}^\text{imm}(e, e') = \frac1{|\mathcal{T}_{e.}|} \sum_{t \in \mathcal{T}_{e.}} \tau_t^\text{imm}(e, e'; R_t)$ \\
$ \tau^\text{car}(r, r')$ & Carryover effect averaged over time periods with $R_t = r$, $\tau^\text{car}(r, r') = \frac1{|\mathcal{T}_{.r}|} \sum_{t \in \mathcal{T}_{.r}} \tau_t^\text{car}(r, r'; E_t).$
\\
\hline
\end{longtable}
}

\section{Proof of exposure unconfoundedness}
\label{supp_sec:unconf_proof}

\begin{proof}[Proof of \cref{prop:as_if_randomized}]
First, since $\bm{X}_0$ are time-invariant, we can treat them as constants in the condition and thus omit them for simplicity. Then,
\begin{align*}
    P(&E_{t}=e, R_t=r|\mathcal{Y}_{t}(\cdot),f(t),\overline{\bm X}_{t, S}) \\
    &=\sum_{\forall \overline{\bm{g}}_{t, L},\overline{\bm{a}}_{t, L}} P(E_{t}=e,R_t=r|\overline{\bm{G}}_{t, L}=\overline{\bm{g}}_{t, L},\overline{\bm{A}}_{t, L}=\overline{\bm{a}}_{t, L}, \mathcal{Y}_{t}(\cdot),f(t),\overline{\bm X}_{t, S}) \cdot\\
    & \qquad P(\overline{\bm{G}}_{t, L}=\overline{\bm{g}}_{t, L},\overline{\bm{A}}_{t, L}=\overline{\bm{a}}_{t, L}|\mathcal{Y}_{t}(\cdot),f(t),\overline{\bm X}_{t, S})\\
    & =\sum_{\forall \overline{\bm{g}}_{t, L},\overline{\bm{a}}_{t, L}} I(h_t(\bm{a}_{t},\bm{g}_{t})=e)\cdot I(h_t^c(\bar{\bm h}_{t-1,L-1}(\overline{\bm{a}}_{t-1, L-1},\overline{\bm{g}}_{t-1, L-1}))=r)\cdot\\
    & \qquad P(\overline{\bm{G}}_{t, L}=\overline{\bm{g}}_{t, L},\overline{\bm{A}}_{t, L}=\overline{\bm{a}}_{t, L}|f(t),\overline{\bm X}_{t, S}) \\
    &=\sum_{\forall \overline{\bm{g}}_{t, L},\overline{\bm{a}}_{t, L}} P(E_{t}=e,R_t=r|\overline{\bm{G}}_{t, L}=\overline{\bm{g}}_{t, L},\overline{\bm{A}}_{t, L}=\overline{\bm{a}}_{t, L},f(t),\overline{\bm X}_{t, S}) \cdot\\
    & \qquad P(\overline{\bm{G}}_{t, L}=\overline{\bm{g}}_{t, L},\overline{\bm{A}}_{t, L}=\overline{\bm{a}}_{t, L}|f(t),\overline{\bm X}_{t, S})\\
    & = P(E_{t}=e, R_t=r|f(t),\overline{\bm X}_{t, S})
\end{align*}
where \(\overline{\bm h}_{t-1,L-1}(\overline{\bm{a}}_{t-1, L-1},\overline{\bm{g}}_{t-1, L-1})\) is equal to \[(h_{t-1}(\bm a_{t-1},\bm g_{t-1}),h_{t-2}(\bm a_{t-2},\bm g_{t-2}),\cdots,h_{t-L}(\bm a_{t-L},\bm g_{t-L}))^\top.\]

The first equation holds as a result of the law of total probability. In the second equation, the first component is simply the indicator function for whether the exposure and carryover exposure take the target values or not. In the second term, the potential outcomes $\mathcal{Y}_t(\cdot)$ drop out using \cref{ass:indep}. Then, the same expansion can be written without conditioning on potential outcomes in the third and fourth equations.
%Note when $s=L, \bm{A}_{(t-L):(t-s-1)}$ and $\bm{G}_{(t-L):(t-s-1)}$ do not exist. The first equation holds as a result of law of probability. In the second equation, the first component is simply the indicator function of the exposure vector equivalent to a known vector or not; the second term is a product, which holds due to two reasons: (1) By unconfoundedness Assumptions \textcolor{}{X and X}, the potential outcomes can be dropped on the conditioning sets, and (2) By Assumption \textcolor{}{X}, every $\overline{\bm X}_t$ in the conditioning set can be reduced to  $\overline{\bm X}_{t-s}$ for each $s$ because covariates after ${t-s}$ will not affect $\bm{G}_{t-s}$ or $\bm{A}_{t-s}$, such that these variables are constants in $\bm{G}_{t-s}$ and $\bm{A}_{t-s}$.

\end{proof}

\section{Estimation}

\subsection{Alternative algorithms with 1-2 or 1-`1 or 2' matching}
\label{supp_subsec:matching1-2or1-12}

\subsubsection{Algorithm with matching 1-2}
\label{supp_subsubsec:matching-1-2}

We propose an alternate algorithm which matches an exposed time point to {\it two} unexposed time points, one occurring temporally {\it before} and one {\it after}, while maintaining balance constraints on time-varying information across matches.
Consider indicators $a_{t_et_{u_1}t_{u_2}}$ for whether the exposed time $t_e \in \mathcal{T}_{1.}$ is matched to the unexposed timestamps $t_{u_1}, t_{u_2} \in \mathcal{T}_{0.} $.
The objective function of the optimization algorithm is  to maximize
\begin{equation}
\label{matching1-2-objective}
   \max_{\bm a} \sum_{t_e, t_{u_1}, t_{u_2}} a_{t_et_{u_1}t_{u_2}}.
%\tag{B}
\end{equation}
The constraints are similar in spirit to the ones for algorithm \cref{matching1-1-objective}, but they are adapted to accommodate the `1-2' matches. Exposed and unexposed time periods can be used in a match at most once, though if an exposed time period is matched, it is matched to two unexposed ones:
%
% \begin{equation}
\begin{align*}
\sum_{t_{u_1}, t_{u_2}} a_{t_et_{u_1}t_{u_2}} & \leq 1, \quad \forall t_e \in \mathcal{T}_{1.}  
\\
\sum_{t_e, t_{u_1}, t_{u_2}} a_{t_et_{u_1}t_{u_2}} I(t_{u_1}=t_u) +
\sum_{t_e, t_{u_1}, t_{u_2}} a_{t_et_{u_1}t_{u_2}}I(t_{u_2}=t_u) & \leq 1, \quad \forall t_u\in \mathcal{T}_{0.} 
    %\tag{\ref{matching1-2-objective}.1}
    %\label{1-2-unique-tr}
\end{align*}
% \tag{\ref{matching1-2-objective}.1}
% \label{1-2-unique-tr-c}
% \end{equation}
Therefore, the objective of the optimization problem is to maximize the number of matched exposed time periods, with a match being of the form $(t_e,t_{u_1},t_{u_2})$.

We impose constraints that balance time, the carryover exposure, and time-varying covariates. These constraints are imposed on value of the variable for the exposed unit compared to the average value of the corresponding variable for its two unexposed time periods. Specifically, the constraint in \cref{1-2-balancing-cov}
balances, {\it on average}, the time of the exposed time period compared to the average time of its unexposed matches,
\begin{align}
    \left|\sum_{t_e, t_{u_1}, t_{u_2}} a_{t_et_{u_1}t_{u_2}}\left( t_e - \frac{t_{u_1} + t_{u_2}}{2}\right) \right| \leq 
    \delta \sum_{t_e, t_{u_1}, t_{u_2}} a_{t_et_{u_1}t_{u_2}},
% \tag{\ref{matching1-2-objective}.2}
\label{1-2-balancing-cov}
\end{align}
%
% Constraint \cref{1-2-balancing-cov} restricts matching in terms of {\it average} temporal differences, 
but it does not restrict the time difference for each match. The constraint in \cref{1-2-single-t-c-constraint} restricts the temporal difference and order of time periods for each individual match, as 
\begin{equation}
\label{1-2-single-t-c-constraint}
% \tag{\ref{matching1-2-objective}.3}
\begin{aligned}
    \left|a_{t_et_{u_1}t_{u_2}} (t_e - t_{u_i})\right| & \leq \epsilon, \quad \text{for } i = 1,2 \\
    a_{t_et_{u_1}t_{u_2}} (t_e-t_{u_1}) & \geq 0, \quad \text{and} \quad 
    a_{t_et_{u_1}t_{u_2}} (t_{u_2}-t_e) \geq 0,
\end{aligned}
\end{equation}
for all $t_e \in \mathcal{T}_{1.}$, and $t_{u_1}, t_{u_2} \in \mathcal{T}_{0.}$.
The first line imposes that the time difference between an exposed time period and each of its matched unexposed ones cannot exceed $\epsilon$. The second line forces a temporal sequence within each match, requiring the exposed period to fall within the unexposed ones. 

For balancing the carryover exposure, the constraint takes the form
\begin{equation}
% \tag{\ref{matching1-2-objective}.4}
\label{1-2-carryover}
\begin{aligned}
    \left| \sum_{t_e, t_{u_1}, t_{u_2}}  a_{t_et_{u_1}t_{u_2}}
    \left(
    %\widetilde{\bm{X}}_{t_e}-
    % \frac{\bm{q}^\top\bm{X}_{t_{u_1}}+\bm{q}^\top\bm{X}_{t_{u_2}}}{2}
    R_{t_e} - 
    \frac{R_{t_{u_1}}+R_{t_{u_2}}}{2}
    \right)\right| 
    \leq
    \delta' \sum_{t_e, t_{u_1}, t_{u_2}} a_{t_et_{u_1}t_{u_2}}.
\end{aligned}
\end{equation}

In addition, the average value of time-varying covariates is balanced when comparing an exposed time periods with the average covariate value of its matches. Specifically,
\begin{align}
% \label{1-2-cov-constraint}
    \left|\sum_{t_e, t_{u_1}, t_{u_2}} a_{t_et_{u_1}t_{u_2}}
    \left( \overline{\bm X}_{t_e, S} - \frac{\overline{\bm X}_{t_{u_1}, S} + \overline{\bm X}_{t_{u_2},S}}{2}\right)\right| \leq
    \bm 1_{(S+1)(Np^\text{int}+Np^\text{net}+p^\text{out})}\cdot \delta' \sum_{t_e, t_{u_1}, t_{u_2}} a_{t_et_{u_1}t_{u_2}}.
% \tag{\ref{matching1-2-objective}.5} 
    \notag
\end{align}
Since this constraint can be high-dimensional, for a large number of interventional units $N$, we instead propose balancing summary values of the interventional and network covariates as
\begin{align}
%\label{1-2-cov-constraint'}
    \left|\sum_{t_e, t_{u_1}, t_{u_2}} a_{t_et_{u_1}t_{u_2}}
    \left( \overline{\bm X}^\text{sum.}_{t_e, S} - \frac{\overline{\bm X}^\text{sum.}_{t_{u_1}, S} + \overline{\bm X}^\text{sum.}_{t_{u_2}, S}}{2}\right)\right| \leq
    \bm 1_{(S+1)(p^\text{int}+p^\text{net}+p^\text{out})}\cdot \delta' \sum_{t_e, t_{u_1}, t_{u_2}} a_{t_et_{u_1}t_{u_2}},
% \tag{\ref{matching1-2-objective}.5\textquotesingle} 
    \notag
\end{align}

Matching algorithms seek control units that resemble the treated units in order to predict what would have happened to the treated units had they been control. Therefore, here, matching one exposed time period to two unexposed ones can improve the accuracy of imputing an exposed time period's missing potential outcome, over matching one to one. Ensuring that the exposed period falls between matched unexposed ones in \cref{1-2-single-t-c-constraint} is expected to improve balance of the temporal trends. For instance, monotonic temporal trends would be more effectively balanced when matching an exposed period with unexposed periods both before and after. Therefore, this 1-2 algorithm can be more accurate in imputing missing potential outcomes for exposed time periods, and more efficient in estimating causal effects, compared to the 1-1 algorithm.

\subsubsection{Algorithm with matching 1-1/2}
\label{subsec: matching 1-1/2}

{While the algorithm that matches one exposed time period to two unexposed time periods in Supplement~\ref{supp_subsubsec:matching-1-2} might provide more accurate imputations of missing potential outcomes in some cases, it might lead to fewer matched exposed units than the 1-1 algorithm in \cref{sec:matching}, especially in scenarios with a relatively high proportion of exposed time periods. Here, we introduce an approach that combines the previous two, and allows an exposed time period to be matched to one {\it or} two unexposed time periods. 

Specifically, we consider binary matching indicators of the form $a_{t_et_u}$ and $a_{t_et_{u_1}t_{u_2}}$ where $a_{t_e t_u} = 1$ denotes that exposed time period $t_e$ is matched only to one unexposed time period $t_u$, whereas $a_{t_et_{u_1}t_{u_2}} = 1$ denotes that $t_e$ is matched to two unexposed time periods $t_{u_1}$ and $t_{u_2}$. The target is to maximize the number of matched exposed time periods as
\begin{align*}
     %\max \sum_{t_e \in \mathcal{T}_{e.} } \sum_{t_u \in \mathcal{T}_{u.} } a_{t_et_u} + \sum_{t_e, t_{u_1}, t_{u_2}} a_{t_et_{u_1}t_{u_2}}. \tag{C}
     \max_{\bm a} \left( \sum_{t_e, t_u}a_{t_et_u} + \sum_{t_e, t_{u_1}, t_{u_2}} a_{t_et_{u_1}t_{u_2}} \right). 
% \label{matching-1-1/2-objective}
% \tag{C}
\end{align*}
The constraints we impose are similar in spirit to those in optimization problems \cref{matching1-1-objective} and \cref{matching1-2-objective}, but they are re-designed to account for the presence of two types of matches.
Each time period can be involved in either a match of type 1-1 or 1-2, and at most once,
%
% \begin{equation}
\begin{align*}
& \sum_{t_u} a_{t_et_u}  + \sum_{t_{u_1}, t_{u_2}} a_{t_et_{u_1}t_{u_2}} \leq 1,
\\
\sum_{t_e, t_u} a_{t_et_u}I(t_u=\tilde{t}_u) + \ \ & \smashoperator{\sum_{t_e, t_{u_1}, t_{u_2}}} \ \ a_{t_et_{u_1}t_{u_2}} I(t_{u_1}=\tilde{t}_u)+ \ \  \smashoperator{\sum_{t_e, t_{u_1}, t_{u_2}}} \ \  a_{t_et_{u_1}t_{u_2}}I(t_{u_2}=\tilde{t}_u) \leq 1,
    %\tag{\ref{matching1-2-objective}.1}
    %\label{1-2-unique-tr}
\end{align*}
% \tag{\ref{matching-1-1/2-objective}.1}
% \label{1-12-unique-tr-c}
% \end{equation}
for all $t_e \in \mathcal{T}_{1.}$ and $\tilde t_u \in \mathcal{T}_{0.}$. Moreover, the average time difference between an exposed time period and its one or two matches is bounded by the constant $\delta \geq 0$, as
{\small
\begin{align*}
\left|\ \smashoperator{\sum_{t_e, t_u}}
     a_{t_et_u}(t_e-t_u)+ \ \smashoperator{\sum_{t_e, t_{u_1}, t_{u_2}}} \ a_{t_et_{u_1}t_{u_2}}\Big(t_e-\frac{t_{u_1}+t_{u_2}}{2}\Big)\right| \leq \delta\Bigg(\ \smashoperator{\sum_{t_e, t_u}} a_{t_et_u}+ \ 
     \smashoperator{\sum_{t_e, t_{u_1}, t_{u_2}}} \ 
     a_{t_et_{u_1}t_{u_2}}\Bigg).
% \tag{\ref{matching-1-1/2-objective}.2}
% \label{match1-12-time-balance}
\end{align*}
}
% Similarly to Matching 1-1 and Matching 1-2, we impose restrictions on the temporal difference of individual matches. 
% Specifically, for every $t_e$ exposed time period, and $t_u, t_{u_1}$ and $t_{u_2}$ unexposed time periods, we impose
\hspace{-7pt}
Furthermore, the time difference between matched exposed and unexposed time periods is bounded by $\epsilon$ in 1-1 or 1-2 matches, and in 1-2 matches the exposed time period lies temporally between the two matched unexposed ones, as
% \begin{equation}
% \tag{\ref{matching-1-1/2-objective}.3}\label{single-tr-c-dif-constraint-2}
\begin{align*}
    & \left|a_{t_et_u}(t_e-t_u) \right| \leq \epsilon,
    \quad \quad \quad \quad \quad \quad
    \left|a_{t_et_{u_1}t_{u_2}}(t_e-t_{u_i}) \right| \leq \epsilon, \text{ for } i=1,2 \\
      & a_{t_et_{u_1}t_{u_2}} (t_e-t_{u_1}) \geq 0, \quad \quad \text{and} \quad \quad
     a_{t_et_{u_1}t_{u_2}} (t_{u_2}-t_e) \geq 0.
\end{align*}
% \end{equation} 
Finally, we impose constraints that balance the carryover exposure and time-varying covariates across the resulting data set of matches. Specifically, for the carryover exposure, we specify
\begin{equation}
% \tag{\ref{matching-1-1/2-objective}.4}
\label{1-12-carryover}
\begin{aligned}
    \Bigg|\smashoperator{\sum_{t_e, t_u}}
     a_{t_et_u}(
     R_{t_e}-
      R_{t_u}
     ) + \smashoperator{\sum_{t_e, t_{u_1}, t_{u_2}} } & a_{t_et_{u_1}t_{u_2}}  \Big(
      R_{t_e}-    \frac{ R_{t_{u_1}}+ R_{t_{u_2}}}{2}
     \Big) \Bigg|
     \\
     & \leq \delta' \Big( \smashoperator{\sum_{t_e, t_u} }a_{t_et_u}+\smashoperator{\sum_{t_e, t_{u_1}, t_{u_2}}} a_{t_et_{u_1}t_{u_2}}\Big). \\[5pt]
\end{aligned}
\end{equation}
For the time-varying covariates, we specify the balance constraint on the complete vector as
% \begin{equation}
\begin{align*}
      \Bigg| \smashoperator{\sum_{t_e, t_u}} a_{t_et_u}
      (\overline{\bm{X}}_{t_e, S}-& \overline{\bm{X}}_{t_u, S}) + \smashoperator{\sum_{t_e, t_{u_1}, t_{u_2}}} a_{t_et_{u_1}t_{u_2}} \left(\overline{\bm{X}}_{t_e, S}-\frac{\overline{\bm{X}}_{t_{u_1}, S}+\overline{\bm{X}}_{t_{u_2}, S}}{2}\right) \Bigg| \\
     &\qquad \qquad \leq \bm 1_{(S+1)(Np^\text{int}+Np^\text{net}+p^\text{out})}
     \cdot \delta'\Big(\smashoperator{\sum_{t_e, t_u}} a_{t_et_u}+\smashoperator{\sum_{t_e, t_{u_1}, t_{u_2}}} a_{t_et_{u_1}t_{u_2}}\Big),  \notag
\end{align*}
% \tag{\ref{matching-1-1/2-objective}.5}\label{match1-12-W}
% \end{equation}
or only on the summary values of covariates across interventional units as
% \begin{equation}
\begin{align*}
      \Bigg| \smashoperator{\sum_{t_e, t_u}} a_{t_et_u}
      (\overline{\bm{X}}^\text{sum.}_{t_e, S}-& \overline{\bm{X}}^\text{sum.}_{t_u, S}) + \smashoperator{\sum_{t_e, t_{u_1}, t_{u_2}}} a_{t_et_{u_1}t_{u_2}} \left(\overline{\bm{X}}^\text{sum.}_{t_e, S}-\frac{\overline{\bm{X}}^\text{sum.}_{t_{u_1}, S}+\overline{\bm{X}}^\text{sum.}_{t_{u_2}, S}}{2}\right) \Bigg| \\
     &\qquad \qquad \leq  \bm 1_{(S+1)(p^\text{int}+p^\text{net}+p^\text{out})}
     \cdot \delta'\Big(\smashoperator{\sum_{t_e, t_u}} a_{t_et_u}+\smashoperator{\sum_{t_e, t_{u_1}, t_{u_2}}} a_{t_et_{u_1}t_{u_2}}\Big).  \notag
\end{align*}
% \tag{\ref{matching-1-1/2-objective}.5\textquotesingle}\label{match1-12-W}
% \end{equation}

%In principle, we expect that Matching 1-2 is the least efficient as it requires that we find matches on both sides of the exposed time point, which can be challenging. For instance, when there are 2 points on the left and 1 point on the right satisfying the proximity constraint (A.4) or (B.4), the number of possible 1-1 matches is 3, while it is 2 for 1-2 matches.
\indent
This 1-1/2 algorithm is expected to yield more matches than the 1-2 algorithm since it allows some exposed time points to be matched to a single unexposed one. At the same time, when possible, it allows for exposed time periods to be matched to two unexposed ones, which can improve the balance of temporal trends and improve accuracy in imputing the missing potential outcomes for exposed time periods.
% If there are far more unexposed time points than the exposed ones, matching on both sides adds more reliability to the estimation. What’s more, even though there are fewer matches, the number of matched time points is twice the number of matches, allowing for greater utilization of the dataset.

%Even though these matching algorithms are developed for binary exposures, in certain settings, they can be used to estimate interpretable effects in more complicated scenarios. For example, in the bipartite settings of \cite{pouget2019variance, Doudchenko2020, brennan2022cluster} and \cite{harshaw2023design}, an outcome unit's exposure varies from 0 to 1, representing a weighted average of the connected interventional units' treatment status. In such scenarios, it is possible to estimate interpretable causal effects by matching units with two different bucket exposures, for example, less than 0.2 v.s. more than 0.8, and drop units with exposure value that is in-between. 

\subsubsection{The matching estimators for the 1-2 and 1-1/2 algorithms}\label{supp:subsec:design-phase-other-matching}

When an exposed time period $t_e$ is matched with two unexposed periods, $t_{u_1}$ and $t_{u_2}$, $Y_{t_e}^\ast$ is set as the average of the two unexposed outcomes, $Y_{t_e}^\ast = ( Y_{t_{u_1}} + Y_{t_{u_2}}) / 2$. The corresponding estimator of the immediate causal effect is the same as in \cref{eq:causal_estimator} with these values of $Y_{t_e}^\ast$ for the matched exposed time periods. 

For the 1-2 algorithm, the set of matched exposed time periods is $\mathcal{T}_{1.}^\ast = \{t: E_t = 1 \text{ and } a_{t_et_{u_1}t_{u_2}} = 1 \text{ for some } t_u\}$, and similarly for the 1-1/2 algorithm. 

\subsection{Algorithms for estimating the carryover effect}
\label{supp_subsec:alg_carryover}

We detail the three algorithms for estimating the carryover effect that correspond to the 1-1, 1-2, and 1-1/2 algorithms in \cref{sec:matching}, Supplement \ref{supp_subsubsec:matching-1-2} and  Supplement \ref{subsec: matching 1-1/2}, respectively. 

We adopt the same index notation for the binary indicators $a_{t_et_u}$ and $a_{t_et_{u_1}t_{u_2}}$ where now $t_e$ corresponds to a time period with carryover exposure, and $t_u, t_{u_1}$ and $t_{u_2}$ correspond to time periods without carryover exposure. The objective of the optimization problem for maximizing the number of matches, as well as the constraints that each time period can be part of at most one match, and the balance constraints on time and the time-varying covariates remain the same as described in Section \ref{subsec:matching_algorithms}, Supplement \ref{supp_subsubsec:matching-1-2} and  Supplement \ref{subsec: matching 1-1/2}.
We update the constraint in \cref{1-1-carryover}, \cref{1-2-carryover} and \cref{1-12-carryover} to specify balance of the most recent exposure (instead of the carryover exposure). The specific form of this constraint is
\begin{align*}
    % \label{1-1-indirect}
    \left|\sum_{t_e, t_u}
     a_{t_et_u}(E_{t_e}-E_{t_u})\right| \leq \delta\sum_{t_e, t_u} a_{t_et_u}, 
     % \tag{\ref{matching1-1-objective}.4\textquotesingle}
\end{align*}
%
% \begin{equation}
% \tag{\ref{matching1-2-objective}.4\textquotesingle}
\begin{align*}
    \left| \sum_{t_e, t_{u_1}, t_{u_2}}  a_{t_et_{u_1}t_{u_2}}
    \left(
    %\widetilde{\bm{X}}_{t_e}-
    % \frac{\bm{q}^\top\bm{X}_{t_{u_1}}+\bm{q}^\top\bm{X}_{t_{u_2}}}{2}
    E_{t_{e}} - 
    \frac{E_{t_{u_1}}+E_{t_{u_2}}}{2}
    \right)\right| 
    \leq
    \delta \sum_{t_e, t_{u_1}, t_{u_2}} a_{t_et_{u_1}t_{u_2}}.
\end{align*}
% \end{equation}
%
and
% %
% \begin{equation}
% \tag{\ref{matching-1-1/2-objective}.4\textquotesingle}
\begin{align*}
    \Bigg|\smashoperator{\sum_{t_e, t_u}}
     a_{t_et_u}(
     E_{t_{e}}-
      E_{t_{u}}
     ) + \smashoperator{\sum_{t_e, t_{u_1}, t_{u_2}} } & a_{t_et_{u_1}t_{u_2}}  \Big(
      E_{t_{e}}-    \frac{ E_{t_{u_1}}+ E_{t_{u_2}}}{2}
     \Big) \Bigg|
     \\
     & \leq \delta \Big( \smashoperator{\sum_{t_e, t_u} }a_{t_et_u}+\smashoperator{\sum_{t_e, t_{u_1}, t_{u_2}}} a_{t_et_{u_1}t_{u_2}}\Big),
\end{align*}
% \end{equation}
for the 1-1 algorithm in \cref{sec:matching}, the 1-2 algorithm in Supplement \ref{supp_subsubsec:matching-1-2}, and the 1-1/2 algorithm in Supplement~\ref{subsec: matching 1-1/2}, respectively.

Lastly, exactly like for the immediate effect, the causal estimator for the carryover effect is of the same form as in \cref{eq:causal_estimator} for $Y_{t_e}^\ast = Y_{t_u}$ for 1-1 matches, and $Y_{t_e}^\ast = (Y_{t_{u_1}} + Y_{t_{u_2}}) / 2$ for 1-2 matches. In Supplement~\ref{supp_sec:bias_proofs} we show that the causal estimators for the carryover effect have bounded bias.

\subsection{Estimands and estimation in the absence of \cref{ass:no-interaction}}
\label{supp_subsec:without_nointeraction_assum}

Under \cref{ass:no-interaction}, in the definition of the immediate effect in \cref{eq:att-type-effects-no-interaction}, we average over time periods with exposure $E_t = e$, and carryover exposure equal to its observed value $R_t$. Then, for accurate estimation of the immediate effect, the impact of the carryover exposure on the outcome is ``controlled'' by balancing the observed carryover exposure over the matched sample in the constraints \cref{1-1-carryover}, \cref{1-2-carryover} or \cref{1-12-carryover}. Therefore, this assumption allows us to use time periods with different values of observed carryover exposure.

Here, we discuss estimands and estimation of immediate and carryover effects in the absence of \cref{ass:no-interaction}.
We alter the definition of estimands in \cref{eq:att-type-effects-no-interaction} to specify the exposure and carryover value of the time periods that are averaged over. Specifically, if $\mathcal{T}_{er}$ denotes the subset of $\mathcal{T}$ for which $E_t = e$ and $R_t = r$, we consider
\begin{align}
\tau^\text{imm}(e, e'; r) & = \frac1{|\mathcal{T}_{er}|} \sum_{t \in \mathcal{T}_{er}} \tau_t^\text{imm}(e, e'; r),
\hspace{5pt} \text{and} \hspace{5pt}
\tau^\text{car}(r, r'; e) = \frac1{|\mathcal{T}_{er}|} \sum_{t \in \mathcal{T}_{er}} \tau_t^\text{car}(r, r'; e).
\label{eq:att-type-effects}
\end{align}

 Accurate estimation of the immediate causal effect requires that we match exactly on the carryover exposure and we only use time periods with the specific carryover exposure value. Specifically,
to estimate the immediate effect for carryover exposure equal to $r$, we substitute the constraints in \cref{1-1-carryover}, \cref{1-2-carryover}, or \cref{1-12-carryover} with
\begin{align*}
    R_{t_e}=R_{t_u}=r,
\end{align*}
\begin{align*}
    R_{t_e}=R_{t_{u_1}}=R_{t_{u_2}}=r, 
\end{align*}
and
\begin{align*}
    R_{t_e}=R_{t_u}=R_{t_{u_1}}=R_{t_{u_2}}=r
\end{align*}
for the algorithm in \cref{sec:matching}, Supplement \ref{supp_subsubsec:matching-1-2}, and Supplement~\ref{subsec: matching 1-1/2}, respectively.

Similar reasoning applies to the estimation of carryover effects in \cref{eq:att-type-effects}. When \cref{ass:no-interaction} is relaxed, $\tau^\text{car}(r, r'; e)$ is identified by fixing the current exposure to $e$ in matched periods in the algorithm detailed in Supplement~\ref{supp_subsec:alg_carryover}.

\subsection{Testing a null hypothesis of no causal effect with multiple outcome units}
\label{subsec:global_null}

Our matching algorithms and estimators are designed to evaluate the immediate or carryover effect of the treatment on each outcome unit separately.
In the presence of multiple outcome units, and when making general policy evaluations, we might be interested in studying whether the exposure has an effect on {\it any} of them. We use the immediate effect as an example. Suppose a collection of outcome units $\mathcal{M} = \{m_1, m_2, \dots, m_M\}$. For each outcome unit $m_j$ we use $\tau_j^{\text{imm}}$ to denote the immediate effect on this unit. We wish to test the hypothesis that
\begin{align*} % \label{multi-hypothesis}
    & \text{H}_0: \tau_j^{\text{imm}} = 0,\ \forall j=1,2,\cdots,M \quad \mathrm{v.s.} \\ &  \quad \text{H}_A: \mathrm{There~exists~at~least~one~} j \mathrm{~such~that~} \tau_j^{\text{imm}} \neq 0,
\end{align*}
We acquire a p-value for testing the null hypothesis of no effect on unit $m_j$, according to \cref{subsec:inference}, for all outcome units. 
We adjust these p-values by performing a false discovery rate (FDR) correction for multiple comparisons \citep{benjamini1995controlling}. Then, we compare the adjusted p-values to the pre-specified $\alpha$-level. If all adjusted p-values are greater than $\alpha$, we fail to reject the null hypothesis $\text{H}_0$. Otherwise, we reject the null hypothesis and identify the affected units as those with adjusted p-values less than $\alpha$.

\section{Proofs of bounded bias of causal estimators}
\label{supp_sec:bias_proofs}

% We implicitly investigate the performance of the estimators conditional on the set of matched exposed time periods $\mathcal{T}_{1.}^\ast$ being non-empty.

\subsection{Linear potential outcome model}

We re-state \cref{theorem:bias_linear} to include the statements for all three estimators and both the immediate and the carryover effect.
\begin{theorem}\label{supp_theorem:bias_linear_all}
     If $Y_{t}(e,r) = \beta_0+ \beta_1 e+ \beta_2r+\beta_3 t  +\bm \beta_4^\top {\overline{\bm X}^\text{sum.}_{t, S}} +\epsilon_{t}(e,r)$ for all $t=1,2,\cdots,T$, with $\operatorname{E}(\epsilon_{t}(e,r)|E_{t},t, \overline{\bm X}^\text{sum.}_{t, S})=0$,
    %if there exists at least one match which satisfies all the constraints, then 
    then $|\operatorname{E}(\Hat{\tau}^\text{imm}(1, 0)-\tau^\text{imm}(1, 0))|\leq \delta |\beta_3|  +\delta'(|\beta_2| + \|\bm \beta_4\|_1)$ for the estimators of the immediate effect  resulting from the 1-1, 1-2, or 1-1/2 algorithms, and $|\operatorname{E}(\Hat{\tau}^\text{car}(1, 0) - \tau^\text{car}(1, 0))|\leq \delta |\beta_3|  +\delta'(|\beta_1| + \|\bm \beta_4\|_1)$ for all estimators of the carryover effect, where $\delta$ and $\delta'$ are the balance constraints tuning parameters.
\end{theorem}

\begin{proof}[Proof of \cref{supp_theorem:bias_linear_all}]

We consider the case of bounding the bias of the matching algorithms when the outcome model has a linear form in the exposures, time, and time-varying covariates. Define $I_1$ as the set of matched pairs $(t_e,t_u)$ and $I_2$ as the set of matched triplets $(t_e,t_{u_1},t_{u_2})$. Let $\bm{E}=(E_1,E_2,\cdots, E_T)^T$ denote the observed exposure across time, and $\bm{t}=(1,2,\cdots,T)^T$ the sequence of time points. Let $\beta_{4,s, d}$ denote the coefficient for the $d$-th covariate at lag $s$, where $s \in \{0, 1, \dots, S\}$.

    \begin{itemize}[leftmargin=*]
        \item For the algorithm that matches one exposed time period to one unexposed time period:
        {\small
        \begin{align*}
        & |\operatorname{E}(\widehat \tau^\text{imm}-\tau^\text{imm})|  \\
    &=\left|\operatorname{E}\left(\operatorname{E}\left( \frac{1}{|I_1|}\sum_{(t_e,t_u) \in I_1}Y_{t_e}-Y_{t_u}\right)-\left.E\left[\left(Y_t(1,R_t)-Y_t(0,R_t)\right)I(E_t=1)\right]\right| \bm{E},\bm{t}, \overline{\bm X}^\text{sum.}_{t, S}\right)\right|\\
        &=\left|\frac{1}{|I_1|}\sum_{(t_e,t_u) \in I_1}\left(\beta_1+ \beta_2  R_{t_e}-\beta_2 
 R_{t_u}+\beta_3 t_e-\beta_3 t_u +
        \bm \beta_4^\top \overline{\bm X}^\text{sum.}_{t_e, S}-\bm \beta_4^\top \overline{\bm X}^\text{sum.}_{t_u, S}\right)-\beta_1\right|\\
        &\leq\frac{1}{|I_1|}|\beta_2|\left|\sum_{(t_e,t_u) \in I_1}(R_{t_e}-R_{t_u})\right|+\frac{1}{|I_1|}|\beta_3|\left|\sum_{(t_e,t_u) \in I_1}(t_e-t_u)\right| + \\
        & \qquad  \qquad \frac{1}{|I_1|}\left|\sum_s\sum_d\beta_{4,s,d} \sum_{(t_e,t_u) \in I_1}({ X}^\text{sum.}_{(t_e-s)d}-{ X}^\text{sum.}_{(t_u-s)d})\right|\\
        & \leq \delta' |\beta_2|+\delta |\beta_3|+\delta' \|\bm \beta_4\|_1.
    \end{align*}
    }

    \item For the algorithm that matches one exposed time period to two unexposed time periods:
    {\small
     \begin{align*}
        & |\operatorname{E}(\widehat \tau^\text{imm}-\tau^\text{imm})|  \\ &=\left|\operatorname{E}\left(\operatorname{E}\left( \frac{1}{|I_2|}\sum_{(t_e,t_{u_1},t_{u_2}) \in I_2}Y_{t_e}-\frac{1}{2}(Y_{t_{u_1}}+Y_{t_{u_2}})\right)-\left.E\left[\left(Y_t(1,R_t)-Y_t(0,R_t)\right)I(E_t=1)\right]\right|  \bm{E},\bm{t},\overline{\bm X}^\text{sum.}_{t, S}\right)\right|\\
        &= \left|\left. \frac{1}{|I_2|}\sum_{(t_e,t_{u_1},t_{u_2}) \in I_2}\left(\beta_1+\beta_2 R_{t_e}-\frac{\beta_2}{2}(R_{t_{u_1}}+R_{t_{u_2}})+\beta_3 t_e-\frac{\beta_3}{2}(t_{u_1}+t_{u_2})
        +\right.\right.\right.\\
        & \qquad \quad \left.\left. \bm \beta_4^\top \overline{\bm X}^\text{sum.}_{t_e, S}-\frac{\bm \beta_4^\top}{2} (\overline{\bm X}^\text{sum.}_{t_{u_1}, S}+\overline{\bm X}^\text{sum.}_{t_{u_2}, S})\right)-\beta_1\right|\\
        &\leq \frac{1}{|I_2|}|\beta_2|\left|\sum_{(t_e,t_{u_1},t_{u_2}) \in I_2}\left(R_{t_e}-\frac{1}{2}(R_{t_{u_1}}+R_{t_{u_2}})\right)\right|+\frac{1}{|I_2|}|\beta_3|\left|\sum_{(t_e,t_{u_1},t_{u_2}) \in I_2}\left(t_e-\frac{1}{2}(t_{u_1}+t_{u_2})\right)\right|+\\
        & \qquad \quad \frac{1}{|I_2|}\left|\sum_s \sum_d \beta_{4,s,d}\sum_{(t_e,t_{u_1},t_{u_2}) \in I_2}{ X}^\text{sum.}_{(t_e-s)d}-\frac{1}{2} ({ X}^\text{sum.}_{(t_{u_1}-s)d}+{ X}^\text{sum.}_{(t_{u_2}-s)d})\right| \\
        &  \leq \delta' |\beta_2|+\delta |\beta_3|+\delta' \|\bm \beta_4\|_1.
    \end{align*}
    }
    \item For the algorithm that matches one exposed time period to one or two unexposed time periods:
    {\small
    \begin{align*}
        & |\operatorname{E}(\widehat \tau^\text{imm}-\tau^\text{imm})| \\
        &=\left|\operatorname{E}\left(\operatorname{E}\left( \frac{1}{|I_1|+|I_2|} \sum_{(t_e,t_u) \in I_1}\left(Y_{t_e}-Y_{t_u}\right)+\sum_{(t_e,t_{u_1},t_{u_2}) \in I_2}\left(Y_{t_e}-\frac{1}{2}(Y_{t_{u_1}}+Y_{t_{u_2}})\right)\right)- \right.\right. \\
        & \left.\left. \hspace{40pt} -
        \left.E\left[\left(Y_t(1,R_t)-Y_t(0,R_t)\right)I(E_t=1)\right]\right|  \bm{E},\bm{t}, \overline{\bm X}^\text{sum.}_{t, S}\right)\right|\\
        &=\left|\frac{1}{|I_1|+|I_2|}\sum_{(t_e,t_u) \in I_1}\left(\beta_1+ \beta_2 R_{t_e}-\beta_2 R_{t_u} +\bm \beta_3 t_e-\beta_3 t_u+
        \bm \beta_4^\top \overline{\bm X}^\text{sum.}_{t_e, S}-\bm \beta_4^\top \overline{\bm X}^\text{sum.}_{t_u, S}\right)+\right.\\
       & \qquad\left.\sum_{(t_e,t_{u_1},t_{u_2}) \in I_2}\left(\beta_1+\beta_2 R_{t_e}-\frac{\beta_2}{2}(R_{t_{u_1}}+R_{t_{u_2}})+ 
        \beta_3 t_e-\frac{\beta_3}{2}(t_{u_1}+t_{u_2})+\right.\right.\\
        & \qquad \quad \left.\left. 
        \bm \beta_4^\top \overline{\bm X}^\text{sum.}_{t_e, S}-\frac{\bm \beta_4^\top}{2} (\overline{\bm X}^\text{sum.}_{t_{u_1}, S}+\overline{\bm X}^\text{sum.}_{t_{u_2}, S})\right)-\beta_1\right|\\
        &\leq\frac{1}{|I_1|+|I_2|}|\beta_2|\left|\sum_{(t_e,t_u) \in I_1}(R_{t_e}-R_{t_u})+\sum_{(t_e,t_{u_1},t_{u_2}) \in I_2}\left(R_{t_e}-\frac{1}{2}(R_{t_{u_1}}+R_{t_{u_2}})\right)\right|+\\
        &\qquad \frac{1}{|I_1|+|I_2|}|\beta_3|\left|\sum_{(t_e,t_u) \in I_1}(t_e-t_u)+\sum_{(t_e,t_{u_1},t_{u_2}) \in I_2}\left(t_e-\frac{1}{2}(t_{u_1}+t_{u_2})\right)\right|+\\
        & \qquad \frac{1}{|I_1|+|I_2|}\left|\sum_s\sum_d \beta_{4,s,d}\left(\sum_{(t_e,t_u) \in I_1} { X}^\text{sum.}_{(t_e-s)d}-  { X}^\text{sum.}_{(t_u-s)d}+ \right. \right.\\
        & \qquad \qquad \qquad \left.\left.\sum_{(t_e,t_{u_1},t_{u_2}) \in I_2} { X}^\text{sum.}_{(t_e-s)d}-\frac{1}{2} ( { X}^\text{sum.}_{(t_{u_1}-s)d}+ { X}^\text{sum.}_{(t_{u_2}-s)d})\right)\right| \\
        &  \leq  \delta'|\beta_2|+\delta |\beta_3|+\delta' \|\bm \beta_4\|_1.
    \end{align*}
    }
    \end{itemize}
\end{proof}

\begin{remark}[An alternative upper bound under linear model]
    In the proposed algorithm, Constraint \cref{eq:constraint1-1:avg_time} restricts the within-pair time difference to a maximum distance of $\epsilon$. Similar constraints apply to the 1-2 and 1-`1 or 2' algorithms discussed in Supplement \ref{supp_subsec:matching1-2or1-12}. Using the 1-1 algorithm of \cref{sec:matching} here, we derive an alternative bias bound. Specifically,
    \begin{align*}
    |E(\widehat \tau^\text{imm} - \tau^\text{imm})| &\leq \frac{1}{|I_1|}|\beta_2|\left|\sum_{(t_e,t_u) \in I_1}(R_{t_e}-R_{t_u})\right|+\frac{1}{|I_1|}|\beta_3|\left|\sum_{(t_e,t_u) \in I_1}(t_e-t_u)\right| + \\
        & \qquad  \qquad \frac{1}{|I_1|}\left|\sum_s\sum_d\beta_{4,s,d} \sum_{(t_e,t_u) \in I_1}({ X}^\text{sum.}_{(t_e-s)d}-{ X}^\text{sum.}_{(t_u-s)d})\right|\\
        & \leq  \delta'|\beta_2|+\frac{1}{|I_1|}|\beta_3|\left|\sum_{(t_e,t_u) \in I_1}(t_e-t_u)\right|+\delta'\|\bm \beta_4\|_1
        \end{align*}
        Because Constraint \cref{eq:constraint1-1:avg_time} dictates that $$\displaystyle \frac{1}{|I_1|}|\beta_3|\left|\sum_{(t_e,t_u) \in I_1}(t_e-t_u)\right| \leq \frac{1}{|I_1|}|\beta_3| \sum_{(t_e,t_u) \in I_1}|t_e-t_u| \leq |\beta_3|\epsilon,$$ an alternative upper bound for the bias is $\delta'|\beta_2|+ \epsilon|\beta_3|+\delta'\|\bm \beta_4\|_1$. 
        
        The upper bound in \cref{theorem:bias_linear} is $\delta'|\beta_2|+ \delta|\beta_3|+\delta'\|\bm \beta_4\|_1$. 
        Since 
        $\epsilon$ is a strictly positive integer, this new bound does not provide guidance for reducing the bias bound, which can instead be made small based on small values for $\delta, \delta'$. That said, this result shows that the bias bound is equal to $\delta'|\beta_2|+ \min(\epsilon, \delta)|\beta_3|+\delta'\|\bm \beta_4\|_1$.
\end{remark}

\begin{remark}[Bias of estimators in the presence of non-linear temporal trends]

The model on the potential outcomes in Theorems~\ref{theorem:bias_linear}~and~\ref{supp_theorem:bias_linear_all} specifies a linear temporal trend. This specification can be realistic and represent potential outcomes (close to) accurately in situations where temporal trends change slowly and are smooth over time, or when the observed temporal window is relatively short. In Theorems~\ref{theorem:bias_flexible}~and~\ref{supp_theorem:bias_flexible_all} we show that, when we extend the proposed algorithms to balance additional terms, our estimators have bounded bias in the presence of non-linear temporal (and covariate) terms in the potential outcome model.

Here, we address a related but different question. We consider the case where a general temporal trend function $f(t)$ is included in the model for the potential outcomes, instead of simply a term that is linear in $t$. We assume that this function is $K$-times differentiable for $K \geq 2$, and $f^{(k)}$ stands for the $k^{th}$ derivative of $f$. For this scenario, we evaluate the bias of our estimator when our balance constraints are placed only on the average time of matches as in \cref{eq:constraint1-1:avg_time}, instead of including the additional terms. 

We consider the case of the Algorithm 1-1. In the proof of the bias bound in \cref{supp_theorem:bias_linear_all}, the term
\[
\frac{1}{|I_1|}|\beta_3|\left|\sum_{(t_e,t_u) \in I_1}(t_e-t_u)\right|
\]
is now equal to
\[
\frac{1}{|I_1|}\left|\sum_{(t_e,t_u) \in I_1}(f(t_e)-f(t_u))\right|.
\]
Let $t_\text{mid}$ be the midpoint of the $[1,T]$ time window, with $t_\text{mid}=\frac{1+T}{2}.$ Using Taylor' expansion of $f$ around $t_\text{mid}$, we can write
%
%\textcolor{blue}{When the temporal trend is nonlinear, we argue in the Supplement \ref{supp_sec:bias_proofs} that the imbalance due to use $t$ for $f(t)$ in the constraint will be upper bounded by $\delta\sum_{k=1}^{K-1}\frac{|f^{(k)}(t_\text{mid})|}{k!}+2\frac{|f^{(K)}(\phi)|}{K!}(\frac{T+1}{2})^K$ where the midpoint $t_\text{mid}$ is defined as $\frac{1+T}{2}$ and $f^{(k)}$ stands for the $k$th derivative of $f$.}
%
\begin{align*}
& f(t_e) - f(t_u) = \\
& \hspace{20pt} \sum_{k=1}^{K-1}\frac{f^{(k)}(t_\text{mid})}{k!}(t_e-t_\text{mid})^k+\mathrm{RE}_{t_e,K} - \sum_{k=1}^{K-1}\frac{f^{(k)}(t_\text{mid})}{k!}(t_u-t_\text{mid})^k - \mathrm{RE}_{t_u,K}
\end{align*}
where \[\mathrm{RE}_{t_e, K} = \frac{f^{(K)}(\xi_{t_e})}{K!}(\xi_{t_e}-t_\text{mid})^K \quad \text{and} \quad \mathrm{RE}_{t_u, K} = \frac{f^{(K)}(\xi_{t_u})}{K!}(\xi_{t_u}-t_\text{mid})^K\]
%are these equal to 
%$$\frac{f^{(K)}(\xi)}{K!}? $$
%and if so I would expect $\xi$ to depend on $t_e, t_u$.
with $\xi_{t_e} \in [\min(t_e, t_\text{mid}),\max(t_e, t_\text{mid})]$ and $\xi_{t_u} \in [\min(t_u, t_\text{mid}),\max(t_u, t_\text{mid})]$.
Then, returning to the term we want to bound, using the triangle inequality we write
\begin{equation}
\begin{aligned}
   & \frac{1}{|I_1|} \left|\sum_{(t_e,t_u)\in I_1} f(t_e)-f(t_u)\right|\\
% &\left|\sum_{(t_e,t_u)\in I_1} (\sum_{k=1}^{K-1}\frac{f^{(k)}(t_\text{mid})}{k!}(t_e-t_\text{mid})^k+\mathrm{RE}_{t_e,K})-(\sum_{k=1}^{K-1}\frac{f^{(k)}(t_\text{mid})}{k!}(t_u-t_\text{mid})^k+\mathrm{RE}_{t_u,K})\right|\\
    &\leq \frac{1}{|I_1|} \sum_{k=1}^{K-1}\frac{|f^{(k)}(t_\text{mid})|}{k!}\left|\sum_{(t_e,t_u)\in I_1}\left((t_e-t_\text{mid})^k-(t_u-t_\text{mid})^k\right)\right| \\
    & \hspace{40pt} + \frac{1}{|I_1|} \sum_{(t_e,t_u)\in I_1}\frac{|f^{(K)}(\xi_{t_e})|+|f^{(K)}(\xi_{t_u})|}{K!} \left(\frac{T-1}{2} \right)^K.
\end{aligned}
\label{supp_eq:nonlinear_bound}
\end{equation}
Therefore, we notice the following that are of interest.
\begin{itemize}[itemsep=5pt]
    \item 
    For $K = 2$, the bound in \cref{supp_eq:nonlinear_bound} reduces to
\[
\frac1{|I_1|} |f'(t_\text{mid})| \left| \sum_{(t_e,t_u)\in I_1} (t_e - t_u) \right| + \frac1{|I_1|}\sum_{(t_e,t_u)\in I_1}\frac{|f''(\xi_{t_e})|+|f''(\xi_{t_u})|}2 \left(\frac{T-1}{2} \right)^2
\]
which based on the constraint on the average time periods of matched pairs in \cref{eq:constraint1-1:avg_time} also reduces to
\[
\delta |f'(t_\text{mid})| +  \frac1{|I_1|}\sum_{(t_e,t_u)\in I_1}\frac{|f''(\xi_{t_e})|+|f''(\xi_{t_u})|}2 \left(\frac{T-1}{2} \right)^2.
\]
    
   \item When the function $f(t)$ is linear and $f(t) = \beta_3 t$, this bound (for $K = 2$) reduces exactly to $|\beta_3| \delta$ as in Theorems~\ref{theorem:bias_linear}~and~\ref{supp_theorem:bias_linear_all}.
    \item When the function is non-linear, but it changes very slowly as a function of time (with small second-derivative) then the additional bias remains small. Specifically, if 
    $ |f'(t_\text{mid})| < \infty $ and the second derivative $ |f''(\xi)| \leq 
    c \left(\frac{2}{T - 1}\right)^2 
    \delta $ for all $\xi \in [1, T]$ and some $c > 0$, then the bound reduces to
    \[
    \delta |f'(t_\text{mid})| + c \delta. 
    \]

    \item In general, if the second derivative cannot be assumed small, but the $K^{th}$ derivative can for $K \geq 3$, we can extend the algorithm to balance $K - 2$ additional terms and guarantee small bias. Specifically, we would balance $(t_e - t_\text{mid})^k$ for $k = 2, 3, \dots, K-1$.
    Then the bound in \cref{supp_eq:nonlinear_bound} becomes
\begin{align*}
   % & \left|\sum_{(t_e,t_u)\in I_1} f(t_e)-f(t_u)\right|\\
   %  &\leq \sum_{k=1}^{K-1}\frac{|f^{(k)}(t_\text{mid})|}{k!}\left|\sum_{(t_e,t_u)\in I_1}\left((t_e-t_\text{mid})^k-(t_u-t_\text{mid})^k\right)\right|+\sum_{(t_e,t_u)\in I_1}\frac{|f''(\xi_{t_e})+f''(\xi_{t_u})|}{K!}\left(\frac{T-1}{2}\right)^K\\
    & %\leq
    \delta\sum_{k=1}^{K-1}\frac{|f^{(k)}(t_\text{mid})|}{k!}+ \frac1{|I_1|} \sum_{(t_e,t_u) \in I_1} \frac{|f^{(K)}(\xi_{t_e})|+|f^{(K)}(\xi_{t_u})|}{K!}\left(\frac{T-1}{2}\right)^K.
\end{align*}
Following a similar thought process to the previous bound, as long as $f^{(K)}(\xi) < c \left( \frac{T-1}2 \right)^K \delta$ for all $\xi \in [1, T]$, then the bias due to the function of time is bounded by a term that is controlled by the algorithmic parameter $\delta$.

\end{itemize}
\end{remark}

\subsection{Nonlinear potential outcome model}

We extend Theorem \ref{theorem:bias_flexible} to incorporate all three estimators derived from the 1-1, 1-2, and 1-1/2 algorithms, for both the immediate and the carryover effect.

\begin{theorem}
\label{supp_theorem:bias_flexible_all}
Suppose that the potential outcomes satisfy
\(
\displaystyle
Y_{t}(e,r) = \theta+\beta_1 e +\beta_2 r+h_0(t)+
{\scriptstyle 
{\sum\limits_{s = 0}^S
\sum\limits_{d=1}^{p^\text{int}+p^\text{net}+p^\text{out}}}} h_{sd}(X^\text{sum. }_{(t-s) d}) +
\epsilon_{t}(e,r),
\)
with $\operatorname{E}(\epsilon_{t}(e,r)|E_{t},t,{\overline{\bm X}^\text{sum.}_{t, S}})=0$ and functions $h_0, h_{01}, h_{02}, \dots, h_{S (p^\text{int}+p^\text{net}+p^\text{out})}$ that are $K$-times differentiable on their support.
If $h_{0}^{(k)}$ and  $h_{sd}^{(k)}$ represent the $k^{th}$ derivative of $h_0$ and $h_{sd}$ respectively, and $|h_{0}^{(k)}(t)|, |h_{sd}^{(k)}(x)| \leq c$ for some $c > 0$ for all $s =0, 1, \dots, S$, covariate $d = 1, 2, \dots, p^\text{int}+p^\text{net}+p^\text{out}$, $t, x$ in the function's support, and $k = 1, 2, \dots, K$, then the bias for the proposed estimator of the immediate effect  under all three algorithms satisfies $ |\operatorname{E}(\Hat{\tau}^\text{imm}(1, 0)-\tau^\text{imm}(1, 0))| \leq  
C_T \delta + (|\beta_2|+C_{X})\delta' +C_{T X} \ell^{K-1}$. Similarly, the bias for the estimator of the carryover effect satisfies $ |\operatorname{E}(\Hat{\tau}^\text{car}(1, 0)-\tau^\text{car}(1, 0))| \leq  
C_T \delta + (|\beta_1|+C_{X})\delta' +C_{T X} \ell^{K-1}$, where
$C_T, C_{X}$ and $C_{T X}$ are constants proportional to $c$ that depend on the smoothness of the functions with the corresponding indices.  
\end{theorem}

\begin{proof}[Proof of \cref{supp_theorem:bias_flexible_all}]

% \qquad
  
% We study the bias of the matching estimator based on Matching 1-1. The bias bounds of estimators based on Matching 1-2 and Matching 1-1/2 are derived similarly, and therefore are omitted here.

{Our proof proceeds similarly to the proof in \cref{supp_theorem:bias_linear_all}. We focus here on the estimator for the immediate effect based on the algorithm that matches an exposed time period to one unexposed time period. 
The proof for the bias bound for the causal estimator based on the algorithms that match one exposed time period to two, or one-or-two unexposed time periods detailed in Supplement~\ref{supp_subsec:matching1-2or1-12}, and the proof for the bias bound for the estimator of the carryover effect follow similarly to the proof below and the proof of \cref{supp_theorem:bias_linear_all}, and are therefore omitted.

% We derive a bound for the quantity $h_{sd}({ X}^\text{sum.}_{(t_e-s)d})-h_{sd}({ X}^\text{sum.}_{(t_u-s)d})$.
In non-linear settings, balance constraints are imposed on additional variables. For the variable $X_{(t-s)d}^\text{sum.}$, we partition its support $[a_{sd}, b_{sd}]$ into disjoint intervals of length $\ell$. The total number of intervals is $(b_{sd}-a_{sd})/\ell$. If $\phi_{sdr}$ is the midpoint of the $r$-th interval, our algorithm includes balance constraints on the localized auxiliary variable at the $k$-th power as
$$(X^\text{sum.}_{(t-s)dr})^{k, \dagger} = (X^\text{sum.}_{(t-s)d} - \phi_{sdr})^k I\left(X^\text{sum.}_{(t-s)d} \in [\phi_{sdr} - \ell/2, \phi_{sdr} + \ell/2]\right),$$
for $k = 1, 2, \dots, K-1$.

Then, to derive a bound for the bias, we first derive a bound for the quantity $h_{sd}({ X}^\text{sum.}_{(t_e-s)d})-h_{sd}({ X}^\text{sum.}_{(t_u-s)d})$, as follows.   %, and similar bounds can be acquired for the functions $h_s$ that correspond to $\widetilde{\bm X}$ and $\widetilde{\bm P}$.
%
% We break the support of $W_{ts} \in [a_s,b_s] \in \mathbb{R}$ into $(b_{sd}-a_{sd})/\ell$ disjoint intervals of length $\ell$. The $r$th interval has a midpoint denoted as $\xi_{sr}$. Construct the auxiliary variables ${W}_{tsr}^\dagger = (W_{ts}-\xi_{sr})I(W_{ts}\in[\xi_{sr}-\ell/2,\xi_{sr}+\ell/2])$ for $W_{ts}$ at order $k=1,\cdots,K$. And we balance the first $K-1$ powers of these transformed covariates  ${W}_{ts}$.
%
\begin{align*}
    & \left|\sum_{(t_e,t_u) \in I_1} h_{sd}(X^\text{sum.}_{(t_e-s)d}) - h_{sd}(X^\text{sum.}_{(t_u-s)d})\right|  \\
    & = \left|\sum_{r=1}^{(b_{sd}-a_{sd})/\ell} \sum_{(t_e,t_u) \in I_1} \left[ \left( \sum_{k=1}^{K-1} \gamma_{sdrk} (X^\text{sum.}_{(t_e - s)dr})^{k, \dagger} + \mathrm{RE}_{t_e-s, d,r,K} \right) - \right. \right. \\
    & \hspace{140pt} \left. \left. \left( \sum_{k=1}^{K-1} \gamma_{sdrk} (X^\text{sum.}_{(t_u - s)dr})^{k, \dagger} + \mathrm{RE}_{t_u-s, d,r,K} \right) \right] \right| \\
    & \leq \sum_{r=1}^{(b_{sd}-a_{sd})/\ell} \left( \sum_{k=1}^{K-1} |\gamma_{sdrk}| \left| \sum_{(t_e,t_u) \in I_1} (X^\text{sum.}_{(t_e - s)dr})^{k, \dagger} - (X^\text{sum.}_{(t_u - s)dr})^{k, \dagger} \right| \right. \\
    & \qquad \left. + |I_1| \frac{|h_{sd}^{(K)}(\xi_{(t_e-s)dr})| + |h_{sd}^{(K)}(\xi_{(t_u-s)dr})|}{K!} (\ell/2)^K \right) \\
    & \leq |I_1| \left( \sum_{r=1}^{(b_{sd}-a_{sd})/\ell} \left( \sum_{k=1}^{K-1} \delta' |\gamma_{sdrk}| + \frac{|h_{sd}^{(K)}(\xi_{(t_e-s)dr})| + |h_{sd}^{(K)}(\xi_{(t_u-s)dr})|}{K!} (\ell/2)^K \right) \right) \\
    & \leq |I_1| \left( (b_{sd}-a_{sd})/\ell \left( \sum_{k=1}^{K-1} \delta' c/k! + 2 \frac{c}{K!} (\ell/2)^K \right) \right)
\end{align*}}
  where 
  \begin{itemize}[leftmargin = *,itemsep=5pt]
  %\item $\phi_{sdr}$ is defined as in \cref{subsec:theory}, the mid-point of the $r$th interval on the support of the $s$-lag of the $d$th auxiliary covariate$({ X}^\text{sum.}_{(t-s)d})^{\dagger}$,
  %
  \item $\gamma_{sdrk} := \frac{h_{sd}^{(k)}(\phi_{sdr})}{k!}$ is the coefficient of the $k^{th}$ order in the Taylor expansion {for function $h_{sd}$} around $\phi_{sdr}$, where $ \frac{|h_{sd}^{(k)}(\phi_{sdr})|}{k!}\leq \frac{c}{k!}$,
  \item $\mathrm{RE}_{t-s,d,r,K}$ is the residual of the Taylor expansion {of the function $h_{sd}$} such that $|\mathrm{RE}_{t-s,d,r,K}|\leq \frac{|h_{sd}^{(K)}(\xi_{(t-s)dr})|}{K!}(\ell/2)^K\leq \frac{c}{K!}(\ell/2)^K,$ and
  \item $\xi_{(t-s)dr}$ is a value between $\min({X}^\text{sum.}_{(t-s)d}I({X}^\text{sum.}_{(t-s)d}\in [\phi_{sdr} - \ell/2, \phi_{sdr} + \ell/2]),\phi_{sdr})$ and $\max({X}^\text{sum.}_{(t-s)d}I({X}^\text{sum.}_{(t-s)d}\in [\phi_{sdr} - \ell/2, \phi_{sdr} + \ell/2]),\phi_{sdr})$.
  \end{itemize}
  {Therefore, for the causal estimator of the immediate effect we can show that
 \begin{align*}
        & |\operatorname{E}(\widehat \tau^\text{imm}(1, 0)-\tau^\text{imm}(1, 0))|  \\
    &=\left|\operatorname{E}\left(\operatorname{E}\left( \frac{1}{|I_1|}\sum_{(t_e,t_u) \in I_1}Y_{t_e}-Y_{t_u}\right)-\left.E\left[\left(Y_t(1,R_t)-Y_t(0,R_t)\right)I(E_t=1)\right]\right| \bm{E},\bm{t},  \overline{\bm X}^\text{sum.}_{t, S}\right)\right|\\
        &=\Bigg|\frac{1}{|I_1|}\sum_{(t_e,t_u) \in I_1}\Bigg(\theta+\beta_1 +\beta_2(R_{t_e}-R_{t_u}) +(h_0(t_e)-h_0(t_u))+  \\
& \hspace{120pt} \left.  {\scriptstyle \sum\limits_{s=0}^S\sum\limits_{d=1}^{p^\text{int} + p^\text{net} + p^\text{out}}} (h_{sd}({ X}^\text{sum.}_{(t_e-s)d})-h_{sd}({ X}^\text{sum.}_{(t_u-s)d}))
\right)-\beta_1\Bigg|\\
        &\leq |\beta_2|\delta' +\left(\frac{T-1}{\ell} (\sum_{k=1}^{K-1}\frac{\delta c}{k!}+2\frac{c}{K!}(\ell/2)^K)\right)+ \\
        & \hspace{120pt} \sum_s\sum_d\left(\frac{b_{sd}-a_{sd}}{\ell} (\sum_{k=1}^{K-1}\frac{\delta' c}{k!}+2\frac{c}{K!}(\ell/2)^K)\right)\\
        & \leq C_T \delta +(|\beta_2|+ C_{X})\delta' +C_{T X} \ell^{K-1}.
    \end{align*}
where the values of the constants are
% Although we match discrete time periods, we think of the smooth temporal trend as a continuous function for time over the whole interval $[0,T]$. For the time-varying covariates, their supports are indexed by their corresponding additive function, which is $[b_{sd}-a_{sd}]$ for every $h_{sd}, s=1,2,\cdots, p^\text{int} + p^\text{net} + p^\text{out}$. The constants $C_T, C_{X}$ and $C_{TX}$ are equal to
\begin{align*}
    &C_T=\sum_{k=1}^{K-1}\frac{(T-1)c}{\ell k!},\\
    &C_{X}=\sum_{s=0}^S\sum_{d=1}^{p^\text{int} + p^\text{net} + p^\text{out}}\sum_{k=1}^{K-1}\frac{(b_{sd}-a_{sd})c}{\ell k!},\\
    &C_{TX}=(\frac{1}{2})^{K-1} \left(\frac{(T-1)c}{K!} +\sum_{s=0}^S\sum_{d=1}^{p^\text{int} + p^\text{net} + p^\text{out}} \frac{(b_{sd}-a_{sd})c}{K!}\right).
\end{align*}}

\end{proof}

\section{Variance estimation}
\label{supp_sec:variance estimation}

{
We consider the case of estimating the immediate effect when we match an exposed time period to one unexposed time period (detailed in \cref{sec:matching}). The arguments follow similarly for the algorithms of estimating the immediate effect that match an exposed time period to two or one-or-two time periods (detailed in Supplement~\ref{supp_subsec:matching1-2or1-12}), or for the algorithms for estimating the carryover effect (detailed in Supplement~\ref{supp_subsec:alg_carryover}).

Here, we consider a simplified setting where all exposed time periods are matched, $|\mathcal{T}_{1.}^\ast| = |\mathcal{T}_{1.}|$, and the number of exposed time periods is fixed, so $|\mathcal{T}_{1.}^\ast|$ does not vary across data sets. This could be the case when the number of unexposed time periods is large relative to the number of exposed ones, and the number of exposed is fixed by design. Therefore, we do not consider the general case where $|\mathcal{T}_{1.}^\ast|$ has inherent variability that will have to be considered.

For the derivations below, we assume that the error terms in the potential outcome model of \cref{theorem:bias_linear} and \cref{theorem:bias_flexible} have finite second moment. For simplicity, we also assume that this variance is constant, i.e., $\text{Var}(e_t(e, r)) = \sigma^2$, for some $\sigma^2 > 0$.

\subsection{Setup and notation}

Here, it is useful to introduce some notation. For what follows we use $k$ to denote the $k^{th}$ match, where $k = 1, 2, \dots, \mathcal{T}_{1.}$, and for which time period $t_{e,k}$ is matched to time period $t_{u,k}$. This match contributes $Z_k = Y_{t_{e,k}} - Y_{t_{u,k}}$ to the estimator, which can be written as
$$\hat{\tau}^\text{imm} = \frac{1}{|\mathcal{T}_{1.}|} \sum_{k=1}^{|\mathcal{T}_{1.}|} Z_k.$$

We study the form of $Z_k$ and the linear and nonlinear models of Theorems~\ref{theorem:bias_linear} and \ref{theorem:bias_flexible}. Under the linear model, $Z_k$ is equal to
\begin{align*}
Z_k &= 
\beta_0+ \beta_1 + \beta_2 R_{t_{e,k}} + \beta_3 t_{e,k}  +\bm \beta_4^\top {\overline{\bm X}^\text{sum.}_{t_{e,k}, S}} +\epsilon_{t_{e,k}}(1, R_{t_{e,k}}) - \\
& \hspace{40pt}
\left(
\beta_0+ \beta_2 R_{t_{u,k}} +\beta_3 t_{u,k}  +\bm \beta_4^\top {\overline{\bm X}^\text{sum.}_{t_{u,k}, S}} +\epsilon_{t_{u,k}}(0, R_{t_{u,k}})
\right) \\
&= \beta_1 + \left[ \beta_2 \left(R_{t_{e,k}} - R_{t_{u,k}} \right) + \beta_3 \left(t_{e,k} - t_{u,k} \right)  + \bm \beta_4^\top \left( \overline{\bm X}^\text{sum.}_{t_{e,k}, S} - \overline{\bm X}^\text{sum.}_{t_{u,k}, S} \right) \right] + \\
& \hspace{40pt}
\epsilon_{t_{e,k}}(1, R_{t_{e,k}}) - \epsilon_{t_{u,k}}(0, R_{t_{u,k}}).
\end{align*}
We use $\text{imb}_k$ to denote the magnitude of the overall imbalance within match $k$ (weighted by each variable's importance in the outcome model) where
\[
\text{imb}_k = \beta_2 \left(R_{t_{e,k}} - R_{t_{u,k}} \right) + \beta_3 \left(t_{e,k} - t_{u,k} \right)  + \bm \beta_4^\top \left( \overline{\bm X}^\text{sum.}_{t_{e,k}, S} - \overline{\bm X}^\text{sum.}_{t_{u,k}, S} \right),
\]
and $\text{eDiff}_k$ to denote the difference of the error terms within the match defined as
\[
\text{eDiff}_k = \epsilon_{t_{e,k}}(1, R_{t_{e,k}}) - \epsilon_{t_{u,k}}(0, R_{t_{u,k}}).
\]
Then,
\[
Z_k = \beta_1 + \text{imb}_k + \text{eDiff}_k.
\]
Under the non-linear model in \cref{theorem:bias_flexible}, everything follows similarly, except the definition of the imbalance term $\text{imb}_k$ includes the term
$h_0(t_{e,k}) - h_0(t_{u,k})$ instead of $\beta_3(t_{e,k} - t_{u,k})$,
and the term
$
{\scriptstyle 
{\sum\limits_{s = 0}^S
\sum\limits_{d=1}^{p^\text{int}+p^\text{net}+p^\text{out}}}} \left[h_{sd}(X^\text{sum. }_{(t_{e,k}-s) d}) - h_{sd}(X^\text{sum. }_{(t_{u,k}-s) d}) \right] $
instead of
$\bm \beta_4^\top \left( \overline{\bm X}^\text{sum.}_{t_{e,k}, S} - \overline{\bm X}^\text{sum.}_{t_{u,k}, S} \right)$.

Under this setup, we can write the true variance of the estimator as
\begin{align*}
\text{Var}\left(\widehat \tau^\text{imm} \right) =
\text{Var}\left( \frac{1}{|\mathcal{T}_{1.}|} \sum_{k=1}^{|\mathcal{T}_{1.}|} Z_k \right) = 
\text{Var}\left( \frac{1}{|\mathcal{T}_{1.}|} \sum_{k=1}^{|\mathcal{T}_{1.}|} \left( \text{imb}_k + \text{eDiff}_k \right) \right).
% \label{supp_eq:est_true_var}
\end{align*}
When balance is specified to be exact by setting $\delta = \delta' = 0$, we have that
\(
\frac1 {|\mathcal{T}_{1.}|} \sum_{k = 1}^{|\mathcal{T}_{1.}|} \text{imb}_k = 0,
\)
and the true variance of our estimator simplifies to
\begin{align}
\text{Var}\left(\widehat \tau^\text{imm} \right) =
% \text{Var}\left( \frac{1}{|\mathcal{T}_{e.}^\ast|} \sum_{k=1}^{|\mathcal{T}_{e.}^\ast|} Z_k \right) = 
\text{Var}\left( \frac{1}{|\mathcal{T}_{1.}|} \sum_{k=1}^{|\mathcal{T}_{1.}|} \text{eDiff}_k \right) =
\sigma^2_\text{eDiff} / |\mathcal{T}_{1.}|,
\label{supp_eq:est_true_var_exact_bal}
\end{align}
since the error terms are assumed independent.
%, and $\sigma^2_\text{eDiff} = 2 \sigma^2$ with $\sigma^2$ being the constant variance of the errors. 

Finally, our estimator for the variance can be written as
\begin{align}
\widehat{\text{Var}}(\widehat \tau^\text{imm}) = 
\frac1{|\mathcal{T}_{1.}| - 1} \sum_{k = 1}^{|\mathcal{T}_{1.}|} \left(Z_k - \widehat \tau^\text{imm} \right)^2
\label{supp_eq:estimator_for_variance}.
\end{align}

\subsection{Variance estimation when matching exactly within each pair}
\label{supp_subsec:variance_match_within_pair}

We consider first an alternative algorithm that ensures that all temporal information within {\it each matched pair} is identical.  We note that this is {\it not} the proposed estimator in \cref{sec:matching}, the variance of which is considered in Supplement~\ref{supp_subsec:variance_ours}.

Under this alternative algorithm, $\text{imb}_k = 0$ for all matches $k = 1, 2, \dots, \mathcal{T}_{1.}$. Consider the resulting estimator $\widehat \tau^\text{imm}$ based on the acquired matches from such algorithm which can be written as
$$\widehat{\tau}^\text{imm} = \frac{1}{|\mathcal{T}_{1.}|} \sum_{k=1}^{|\mathcal{T}_{1.}|} Z_k =
\beta_1 + \frac{1}{|\mathcal{T}_{1.}|} \sum_{k=1}^{|\mathcal{T}_{1.}|} \text{eDiff}_k.$$
In this case, our variance estimator in \cref{supp_eq:estimator_for_variance} is equal to
\[
\frac{1}{|\mathcal{T}_{1.}| - 1} \sum_{k=1}^{|\mathcal{T}_{1.}|} \left[ \text{eDiff}_k
- 
\left(\frac{1}{|\mathcal{T}_{1.}|} \sum_{k=1}^{|\mathcal{T}_{1.}|} \text{eDiff}_k \right) \right]^2 
\]
and it is, therefore, unbiased for $\text{Var} \left(\widehat \tau^\text{imm} \right)$ in \cref{supp_eq:est_true_var_exact_bal}.

There are a few notes worth making here. If the temporal trend exists and is linear ($\beta_3 \neq 0$ in the model of \cref{theorem:bias_linear}) it is impossible to balance time exactly within each match, and $\text{imb}_k \neq 0$ necessarily. Therefore, this approach works in the absence of a temporal trend (i.e., $\beta_3 = 0$ in the linear model of \cref{theorem:bias_linear}), or when the temporal trend follows a nonlinear function (such as $h_0(t)$ in \cref{theorem:bias_flexible}) which is {\it measured} and matched on exactly. As a result, the requirements over the temporal trend term for unbiased estimation of the variance are relatively strong. If none of these hold, some imbalance within each pair will remain, which is studied in the next section. Finally, when matching exactly within each pair, unless the number of unexposed time periods far surpass the number of exposed ones, it is expected that there will be exposed time periods for which a match will not exist, and $|\mathcal{T}_{1.}| < |\mathcal{T}_{1.}|.$ This is not a case we consider here.

\subsection{Variance estimation for the causal effect estimator based on our algorithms}
\label{supp_subsec:variance_ours}

% When the matches are allowed to have imbalance of time-varying information (which implies that $\text{imb}_k \neq 0$ for some $k$), the variance of our estimator in \cref{supp_eq:est_true_var_exact_bal} reduces to
% \begin{align*}
% \text{Var}\left(\widehat\tau^\text{imm}\right) & = 
% \frac1{|\mathcal{T}_{e.}|^2} \left[ 
% \sum_{k = 1}^{\mathcal{T}_{e.}} \text{Var} \left( \text{imb}_k + \text{eDiff}_k \right)  \right. + \\
% & \hspace{40pt}
% \left.
% \sum_{1 \leq k_1 < k_2 \leq \mathcal{T}_{e.}} 2\  \text{Cov}\left(\text{imb}_{k_1} + \text{eDiff}_{k_1}, 
% \text{imb}_{k_2} + \text{eDiff}_{k_2} \right)
% \right].
% \end{align*}
% Since the outcomes are not used to inform the algorithm, and, in most situations, the covariate imbalance and the error difference can be assumed independent of one another and across time periods, the variance can be re-written as
% \begin{align*}
% \text{Var}\left(\widehat\tau^\text{imm}\right) & = \frac{\sigma^2_\text{eDiff}}{|\mathcal{T}_{e.}|} + 
% \frac1{|\mathcal{T}_{e.}|^2} \left[ 
% \sum_{k = 1}^{\mathcal{T}_{e.}} \text{Var} \left( \text{imb}_k \right) +
% \sum_{1 \leq k_1 < k_2 \leq \mathcal{T}_{e.}} 2\  \text{Cov}\left(\text{imb}_{k_1}, 
% \text{imb}_{k_2}\right)
% \right].
% \end{align*}

In this case, the algorithms do not match time-varying information exactly within each pair as in Supplement~\ref{supp_subsec:variance_match_within_pair}, rather than impose overall balance constraints (\cref{sec:matching}). The variance estimator in \cref{supp_eq:estimator_for_variance} can be re-written as
\begin{align*}
\widehat{\text{Var}}(\widehat \tau^\text{imm}) 
% &= 
% \frac1{|\mathcal{T}_{e.}| - 1} \sum_{k = 1}^{|\mathcal{T}_{e.}|} \left(Z_k - \widehat \tau^\text{imm} \right)^2 \\
% %
&= 
\frac1{|\mathcal{T}_{1.}| - 1} \sum_{k = 1}^{|\mathcal{T}_{1.}|} \left[ \beta_1 + \text{imb}_k + \text{eDiff}_k - \frac1{|\mathcal{T}_{1.}|} \sum_{k = 1}^{|\mathcal{T}_{1.}|} \left(\beta_1 + \text{imb}_k + \text{eDiff}_k \right) \right]^2 \\
&=
\frac1{|\mathcal{T}_{1.}| - 1} \sum_{k = 1}^{|\mathcal{T}_{1.}|} \left[ \text{imb}_k + \text{eDiff}_k - \frac1{|\mathcal{T}_{1.}|} \sum_{k = 1}^{|\mathcal{T}_{1.}|} \left(\text{imb}_k + \text{eDiff}_k \right) \right]^2,
\intertext{which, since the average time and time-varying covariates are balanced exactly across matched pairs with $\delta = \delta' = 0$, is also equal to}
&=
\frac1{|\mathcal{T}_{1.}| - 1} \sum_{k = 1}^{|\mathcal{T}_{1.}|} \left[ \text{imb}_k + \text{eDiff}_k - \frac1{|\mathcal{T}_{1.}|} \sum_{k = 1}^{|\mathcal{T}_{1.}|} \text{eDiff}_k  \right]^2.
\end{align*}
Then, we can write $\text{E} \left[ \widehat{\text{Var}}(\widehat \tau^\text{imm}) \right]$ as
\begin{align*}
   % \text{E}\left[ \widehat{\text{Var}}(\widehat \tau^\text{imm})\right] &=
   & \frac1{|\mathcal{T}_{1.}| - 1}\sum_{k = 1}^{|\mathcal{T}_{1.}|} \text E(\text{imb}_k^2)+\frac2{|\mathcal{T}_{1.}| - 1}\sum_{k = 1}^{|\mathcal{T}_{1.}|} \text E \left[\text{imb}_k*\left(\text{eDiff}_k - \frac1{|\mathcal{T}_{1.}|} \sum_{k = 1}^{|\mathcal{T}_{1.}|} \text{eDiff}_k \right)\right]\\
   & \qquad +\frac1{|\mathcal{T}_{1.}| - 1}\sum_{k = 1}^{|\mathcal{T}_{1.}|}\text E\left[\left(\text{eDiff}_k - \frac1{|\mathcal{T}_{1.}|} \sum_{k = 1}^{|\mathcal{T}_{1.}|} \text{eDiff}_k \right)^2\right]\\
  & = \frac1{|\mathcal{T}_{1.}| - 1}\sum_{k = 1}^{|\mathcal{T}_{1.}|} \text E(\text{imb}_k^2)+ \sigma^2_\text{eDiff}/|\mathcal{T}_{1.}|,
\end{align*}
where the equality holds because the errors (and therefore the error differences) are independent of all information including the covariates, which implies that
\begin{align*}
& \text E \left[\text{imb}_k \left(\text{eDiff}_k - \frac{1}{|\mathcal{T}_{1.}|} \sum_{k=1}^{|\mathcal{T}_{1.}|} \text{eDiff}_k \right)\right] = \\
& \text E \left[\text{imb}_k \right] \cdot \text E\left[\text{eDiff}_k - \frac{1}{|\mathcal{T}_{1.}|} \sum_{k=1}^{|\mathcal{T}_{1.}|} \text{eDiff}_k \right] = 0.
\end{align*}

This demonstrates that the variance estimator is, in expectation, larger than the true variance, by a term that depends on the expected magnitude of imbalance within each match. Therefore using our variance estimator for inference would result in $(1-\alpha)100\%$ intervals that cover the true value at least that proportion of time.

\subsection{Temporal correlation in the outcome model residuals}

In time series settings, it might be the case that outcomes are autocorrelated across time, even beyond what can be explained by measured covariates. In time series analyses of an outcome $Y_t$ that exhibits temporal autocorrelation, researchers often adopt a model on the first differences $Z_t = Y_t - Y_{t-1}$ instead of a model directly on $Y_t$. That is because, often, the first differences $Z_t$ will exhibit much less (if any) temporal autocorrelation.

In our simulations, we have found that our inferential approach is rather robust to temporal autocorrelation of the error terms of the outcome model (see Supplement~\ref{supp_subsec:sims_temp_corr}).We conjecture that this performance might have links to the fact that, in our estimator and its variance, the error terms appear only through the differences $\text{eDiff}_k$. Note that an exposed time period is required to be temporally close to its matched unexposed time period (no more than $\epsilon$-apart). Therefore, even when the errors are temporally autocorrelated, the error differences might exhibit much less autocorrelation across matches, which makes our inferential procedure robust to temporal correlation of the error terms. Future work can evaluate this connection even further.

In Supplement~\ref{supp_subsec:sims_temp_corr}, we also consider an alternative inferential procedure based on the Newey-West variance estimator \citep{newey1986simple}. In Supplement~\ref{supp_subsec:alternative_matching_estimator} we discuss how our point estimator can be acquired as the OLS estimator of a regression on the time periods matched by our algorithm, which will allow us to employ the Newey-West variance estimator.
}

\subsection{Our causal estimator within a regression framework on the matched sample}

\label{supp_subsec:alternative_matching_estimator}

In the presence of autocorrelated outcomes across time (such as in the case when the error terms are autocorrelated in the outcome models of \cref{theorem:bias_linear} or \cref{theorem:bias_flexible}), an alternative inferential strategy might employ the Newey-West variance estimator \citep{newey1986simple}. 
This estimator accounts for serial correlation in the outcomes, and often provides a more reliable, if not conservative, inference in time-series settings.

To use the Newey-West variance estimator, our first step is to rewrite our point estimator within a regression framework. Specifically, we employ regression of the outcome on the exposure on the time periods that are matched following the design phase (the optimization step in \cref{subsec:matching_algorithms}). Then, the ordinary least squares (OLS) estimator for the coefficient of the exposure is the point estimator, with regression-based standard errors to quantify uncertainty. Here, we illustrate that the OLS estimator is numerically equivalent to our proposed estimator in \cref{eq:causal_estimator}. Also, we note that the Wald-type standard errors we employ in \cref{subsec:inference} differ from the vanilla regression-based standard errors (from the regression that also assumes independent error terms) by a term of order $O(|\mathcal{T}_{1.}|^{-1})$ which diminishes with the number of matched exposed time periods.
%Framing the causal estimator as a regression is particularly advantageous for handling autoregressive errors, as it accommodates the flexible construction of a Newey-West covariance matrix. Under the independence assumption, autocorrelation can lead to underestimated variance and overly narrow confidence intervals. 
We return to our implementation of the Newey-West variance estimator in Supplement \ref{supp_subsec:sims_temp_corr} where we evaluate its performance in simulations.

\subsubsection{Equivalence of point estimates of the proposed causal estimator and the OLS estimator in the outcome on exposure regression}

%Apart from the estimator in Section \ref{subsec:theory}, we can also construct a regression-based estimator using the matched data. 
For simplicity, we focus on the algorithm and estimator for the immediate effect.
We consider the time periods that have been part of a match. Specifically, we consider the time periods $t_e \in \mathcal{T}_{1.}$ such that $a_{t_et_u} = 1$ for some $t_u \in \mathcal{T}_{0.}$, and the time periods $t_u \in \mathcal{T}_{0.}$ such that $a_{t_et_u} = 1$ for some $t_e \in \mathcal{T}_{1.}$.
Since there are $|\mathcal{T}_{1.}|$ matched exposed time periods, and each time period can be part of at most one match, the number of time periods we consider is $2|\mathcal{T}_{1.}|$. We construct the data set using these time periods only, and use $Y_k$ and $E_k$ to denote the outcome and exposure value for $k$th time period in ascending order in this data set, where $k = 1, 2, \dots, 2|\mathcal{T}_{1.}|$. On this data set, we consider a simple linear regression of $Y_k$ on $E_k$ and we use the estimated coefficient of the exposure and associated confidence interval as the estimate and confidence interval of the immediate effect.

% the regression model on 
% The core concept involves regressing the outcome on the exposure using the matched pairs. Following the notation established in \cref{subsec:theory} and Supplement \ref{supp:subsec:design-phase-other-matching}, we construct a combined dataset by interleaving the matched outcomes $Y_{t_e}$ and $Y_{t_e}^\ast$ into a single response vector $Y_k$ for $k = 1, 2, \dots, 2|\mathcal{T}_{1.}|$. The corresponding exposure indicator, $E_k$, is set to 1 for odd indices (representing $Y_{t_e}$) and 0 for even indices (representing $Y_{t_e}^\ast$).}

{First, it is easy to show that this regression estimator is numerically equivalent to the $\widehat{\tau}^\text{imm}$ introduced in Section \ref{subsec:theory}. To see this, recall from \cref{eq:causal_estimator} that the proposed estimator is given by:
\[\widehat \tau^\text{imm} = \frac1{|\mathcal{T}_{1.}|} \sum_{t \in \mathcal{T}_{1.}} \left( Y_{t_e} - Y_{t_e}^\ast \right).\]
Let $\bar{E}=\frac{1}{2|\mathcal{T}_{1.}|}\sum_{k=1}^{2|\mathcal{T}_{1.}|}E_k$ be the mean value of the exposure in the matched data set, so $\bar{E}=0.5$. The OLS estimate for the coefficient of the exposure is equal to 
\begin{align*}
\frac{\sum_{k=1}^{2|\mathcal{T}_{1.}|} (E_k-\bar{E})Y_k}{\sum_{k=1}^{2|\mathcal{T}_{1.}|^\ast} (E_k-\bar{E})^2} = \frac{1}{0.5|\mathcal{T}_{1.}|}\left(\sum_{t_e \in \mathcal{T}_{1.}} 0.5 Y_{t_e} - 0.5 Y_{t_e}^\ast\right) = \widehat{\tau}^\text{imm}.
\end{align*}
Furthermore, letting $\bar{Y} = \frac{1}{2|\mathcal{T}_{1.}|}\sum_{k=1}^{2|\mathcal{T}_{1.}|}Y_k$, the OLS estimate of the regression intercept is equal to $\bar{Y} - 0.5\widehat{\tau}^\text{imm}$.}

\subsubsection{Differences in inferences when viewing causal estimator through a regression lens}
%and the difference in estimated variances is negligible. Thus, this approach offers an alternative perspective for estimation, reinforcing that matching primarily serves as a design step in causal inference.

{Here, we compare the standard error of the regression estimator with the proposed standard error under the Wald-type construction in \cref{subsec:inference}.}

{Based on the results above about the estimated regression coefficients, the fitted values from the regression are equal to
\[\widehat{Y}_k = (\bar{Y}-0.5\widehat{\tau}^\text{imm}) + \widehat{\tau}^\text{imm}E_k.\]
Then, the sum of squared errors (SSE) can be written as
\begin{align*}
    \text{SSE}&=\sum_{t_e \in \mathcal{T}_{1.}} \left[ \left(Y_{t_e}-\widehat{Y}_{t_e}\right)^2 + \left(Y_{t_e}^\ast-\widehat{Y}_{t_e}^\ast\right)^2 \right] \\
    &=\sum_{t_e \in \mathcal{T}_{1.}} \left[ (Y_{t_e}-( \bar{Y}-0.5\widehat{\tau}^\text{imm})-\widehat{\tau}^\text{imm})^2+(Y_{t_e}^\ast-( \bar{Y}-0.5\widehat{\tau}^\text{imm}))^2 \right] \\
    &=\sum_{t_e \in \mathcal{T}_{1.}} \left[ (Y_{t_e}- \bar{Y}-0.5\widehat{\tau}^\text{imm})^2+ (Y_{t_e}^\ast- \bar{Y}+0.5\widehat{\tau}^\text{imm})^2 \right] \\
    &=\sum_{t_e \in \mathcal{T}_{1.}} \left[ \left(Y_{t_e}-\frac{\sum Y_{t_e}}{|\mathcal{T}_{1.}|} \right)^2+ \left(Y_{t_e}^\ast-\frac{\sum Y_{t_e}^\ast}{|\mathcal{T}_{1.}|}\right)^2 \right].
\end{align*}
Based on this residual variance estimate from the regression, denoted as $\text{MSE}$, is equal to
\begin{align*}
\text{MSE}= \frac{\sum_{t_e \in \mathcal{T}_{1.}} \left[ \left(Y_{t_e}-\frac{\sum Y_{t_e}}{|\mathcal{T}_{1.}|} \right)^2+ \left(Y_{t_e}^\ast-\frac{\sum Y_{t_e}^\ast}{|\mathcal{T}_{1.}|}\right)^2 \right]}{2|\mathcal{T}_{1.}|-2},
\end{align*} 
Hence the standard error for the coefficient of the exposure, denoted as $\widehat{\sigma}^2_\text{OLS}$, is $$\displaystyle \widehat{\sigma}^2_\text{OLS}=\frac{\text{MSE}}{\sum_{k=1}^{2|\mathcal{T}_{1.}|}(E_k-\bar{E})^2}=\frac{\text{MSE}}{0.5|\mathcal{T}_{1.}|}=\frac{\sum_{t_e \in \mathcal{T}_{1.}} \left[ \left(Y_{t_e}-\frac{\sum Y_{t_e}}{|\mathcal{T}_{1.}|} \right)^2+ \left(Y_{t_e}^\ast-\frac{\sum Y_{t_e}^\ast}{|\mathcal{T}_{1.}|}\right)^2 \right]}{|\mathcal{T}_{1.}|(|\mathcal{T}_{1.}|-1)},$$ 
and the resulting confidence interval for the regression-based estimator of ${\tau}^\text{imm}$ is  $[\widehat{\tau}^\text{imm}-z_{1-\alpha/2}\widehat{\sigma}_\text{OLS},\widehat{\tau}^\text{imm}+z_{1-\alpha/2}\widehat{\sigma}_\text{OLS}]$. }

We can write the variance estimator in our Wald-type inference described in \cref{subsec:inference}  as
\begin{align*}
    \widehat{s}^2&=\frac{\sum_{t_e \in \mathcal{T}_{1.}}( Y_{t_e}-Y_{t_e}^\ast-\widehat\tau^\text{imm})^2}{|\mathcal{T}_{1.}|-1}\\
    &=\frac{\sum_{t_e \in \mathcal{T}_{1.}} \left[ \left(Y_{t_e}-\frac{\sum Y_{t_e}}{|\mathcal{T}_{1.}|} \right)-\left(Y_{t_e}^\ast-\frac{\sum Y_{t_e}^\ast}{|\mathcal{T}_{1.}|} \right)\right]^2}{|\mathcal{T}_{1.}|-1} \\
    &= \frac{\sum_{t_e \in \mathcal{T}_{1.}} \left[
    \left(Y_{t_e}-\frac{\sum Y_{t_e}}{|\mathcal{T}_{1.}|}\right)^2+ \left(Y_{t_e}^\ast-\frac{\sum Y_{t_e}^\ast}{|\mathcal{T}_{1.}|}\right)^2
    -2\left(Y_{t_e}-\frac{\sum Y_{t_e}}{|\mathcal{T}_{1.}|}\right) \left(Y_{t_e}^\ast-\frac{\sum Y_{t_e}^\ast}{|\mathcal{T}_{1.}|}\right)
    \right]}{|\mathcal{T}_{1.}|-1}.
\end{align*}

Then, the difference between the standard error estimator in the regression-based confidence interval, $\widehat{\sigma}^2_\text{OLS}$, and in the proposed Wald-type confidence interval in \cref{subsec:inference}, $\widehat{s}^2/|\mathcal{T}_{1.}|$, is $\displaystyle 2 \frac{\sum_{t_e \in \mathcal{T}_{1.}} \left(Y_{t_e}-\frac{\sum Y_{t_e}}{|\mathcal{T}_{1.}|}\right)\left(Y_{t_e}^\ast-\frac{\sum Y_{t_e}^\ast}{|\mathcal{T}_{1.}|}\right)}{|\mathcal{T}_{1.}|(|\mathcal{T}_{1.}|-1)}$. Because this term represents exactly twice the sample covariance of the outcomes in the exposed and unexposed time periods in the matched pairs ($Y_{t_e}$ and $Y_{t_e}^\ast$), scaled by the sample size, this difference is of the order of $O(|\mathcal{T}_{1.}|^{-1})$, and is therefore negligible. % $O(|\mathcal{T}_{1.}^\ast|^{-1/2})$ difference between the confidence intervals produced by the two methods.}

\section{A Synthetic Dataset on Wildfire Exposure}\label{supp_sec:real-data-simulation}

We generate data over a rectangular region $[0,1] \times [0,4]$. We consider $N = 50$ interventional units, and $M = 200$ outcome units. 
We refer to the subset of our area $[0,0.5] \times [1,4]$ as the ``urban'' subregion, and we consider it to be more dense in outcome units than interventional units. The reverse is true for the remaining area which we refer to as the ``rural'' subregion. Specifically, we randomly generate the locations of 7 out of 50 interventional units and 150 out of 200 outcome units in the urban subregion, while the remaining interventional and outcome units are randomly positioned in the rural subregion. The outcome units are denoted by $\mathcal{M}=\{m_1, m_2, \dots, m_M\}$.
Since we generate $M = 200$ outcome units, in what follows we use $X_{0j}^\text{out}$ and $\bm X_{tj}^\text{out}$ to denote the time-invariant and time-varying covariates of outcome unit $m_j$, respectively, with $j = 1, 2, \dots, M$.
For our simulations in \cref{sec:simu_real_data}, one randomly chosen outcome unit in the urban area is considered, whereas we use all $M$ outcome units to evaluate our multiple-testing inferential procedure detailed in Supplement \ref{subsec:global_null}. 
To align with our real-world application, we set the total number of time periods to $T = 1003$.

\subsection{Generating the temporal trends and covariates for the interventional and outcome units}

We denote the smooth temporal trends as $f_i^{int}(t)$ and $f_j^{out}(t)$ for interventional unit $n_i$ and outcome unit $m_j$, respectively. We generate the temporal trends to include the same increasing linear trend $\mu(t) = 0.0004t$, with temporally smooth unit-specific variation drawn from a Gaussian process with a Gaussian correlation kernel.
Specifically, let $\bm f_i^\text{int}$ denote the value of the smooth temporal trend for interventional unit $n_i$ across the $T$ time periods as $ \bm f_i^\text{int} =(f_i^\text{int}(1), \cdots, f_i^\text{int}(T))^\top$. Similarly, we use $\bm \mu$ to denote the value of the overall temporal trend across the $T$ time periods, as $\bm \mu = (\mu(1), \mu(2), \dots, \mu(T))^\top$.
We draw $\bm f_i^\text{int} \sim N(\bm \mu, \Sigma)$ independently across interventional units, where the $(t_1, t_2)$ entry of the covariance matrix $\Sigma$ is $\Sigma_{t_1, t_2} = \frac{\exp\left( - (t_1 - t_2)^2 \right)}{2 \cdot 100^2}$. If $\bm f_j^\text{out}$ denotes the temporal trend for outcome unit $m_j$ at the $T$ time periods, we set $\bm f_j^\text{out}$ equal to the value of $\bm f_i^\text{int}$ for the interventional unit $n_i$ that is geographically closest to the outcome unit $m_j$.

We generate time-invariant covariates independently across units according to a different distribution for units located in the urban and rural subregions. Specifically, $X_{0i}^\text{int}$ is drawn from a $\text{Beta}(1,9)$ distribution if $n_i$ is located within the urban subregion $[0,0.5] \times [1,4]$, or a $\text{Beta}(9,1)$ distribution if it is located in the rural subregion. We generate the outcome unit time-invariant covariate in the same manner. Therefore, the time-invariant covariates of interventional and outcome units are more similar to one another when the units are located in the same subregion, with higher values in the rural compared to the urban region.

%{\color{magenta} This area comprises 50 forest zones and 200 residential zones. We partition the plane into urban locations—defined by the sub-rectangle $(0, 0.5] \times [1, 4]$—which are predominantly residential, while the remaining areas are designated as forest locations. This spatial distribution mimics the real-world clustering of urban and wilderness areas in the California region.
% 
% a species richness covariate
% In urban, X0 is lower biodiversity in urban
% For all other locations, which are more likely to be natural preserves
%we assume higher species richness 

Finally, we generate $p^\text{int} = p^\text{out} = 5$ time-varying covariates for the interventional and the outcome units. These covariates are generated to resemble the observed values for temperature, humidity, precipitation, wind speed, and wind direction in our observed data. Below, we detail how each of the five covariates are generated for an outcome unit $m_j$ with coordinates $(x_j, y_j)$, where $\bm X_{tj}^\text{out}= (X_{tj1}^\text{out}, X_{tj2}^\text{out}, \dots, X_{tj5}^\text{out})^\top$.

The first covariate $X_{tj1}^\text{out}$ represents temperature for unit $m_j$ over time. We set
\begin{align*}
X_{tj1}^\text{out} &=  \texttt{base\_temperature}_{j}+ 
\texttt{seasonal\_effect}_{t}+
\texttt{trend}_{t} + \\
& \hspace{40pt} +
\texttt{daily\_variation}_{tj},
\end{align*}
where $\texttt{base\_temperature}$, $
\texttt{seasonal\_effect}$, $
\texttt{trend},$ and $
\texttt{daily\_variation}$ are defined as follows.
We set $\texttt{base\_temperature}_{j}=20-3y_j+5x_j$ with higher values for locations towards the bottom right corner of our geography (reflecting that the ``southeast'' area of our geography has warmer conditions). 
We set the seasonal variation as
$\texttt{seasonal\_effect}_{t}=15*\sin(\frac{2\pi}{365}(t-150))$ which indicates annual temperature cycles, and $\texttt{trend}_t = t/5000$ represents a long-term warming trend. Lastly, we generate the day- and unit-specific errors as $\texttt{daily\_variation}_{tj}\sim N(0,([\log(|t-500|+2)]/1.5)^2)$, independently across units and time.

The second covariate $X_{tj2}^\text{out}$ represents humidity patterns for the outcome unit $m_j$ across time. This variable is generated based on two quantities, $\texttt{seasonal\_effect}_{tj}$ and $\texttt{noise}_{tj}$ which, in turn, are generated in two different ways depending on the outcome unit's distance from 0 on the x-axis. (This design is such that locations that are close to 0 on the x-axis can be conceived as being ``close to the coast'' with higher humidity levels).
%The synthetic humidity covariates $X_{ti2}^\text{int}$ and $X_{tj2}^\text{out}$ are primarily composed of a seasonal effect and random noise. We define two different generating processes based on a region’s distance from the coast. 
If $x_j \leq 0.5$, the location is considered as being coastal, and the seasonal effect and noise are generated in the following manner. We specify $\texttt{seasonal\_effect}_{tj} = 10 \sin(2\pi / 365(t - 170)) + 5y_j - 20x_j + 65$, reflecting higher humidity during the summer, in areas closer to the sea (small $x$ values) or in the north part of our geography (large $y$ values). The noise term is generated as $noise_{tj} \sim N(0, 5^2)$ independently across time and units.
If $x_j > 0.5$, the location is considered inland. The seasonal effect is simplified to be constant across coordinates in the inland area as $\texttt{seasonal\_effect}_{tj} = 2 \sin(2\pi / 365(t - 170)) + 10$. The noise term is generated as $\texttt{noise}_{tj} \sim N(0, 3^2)$ independently across times and units.
{Humidity is then generated from a truncated normal distribution with mean $\texttt{seasonal\_effect}_{tj}$, standard deviation equal to 15, %truncated above at the minimum of 100 and $\texttt{seasonal\_effect}_{tj}+\texttt{noise}_{tj}$. These generated values are subsequently censored to the interval $[0, 100]$, such that any values falling below 0 are set to 0, and any values exceeding 100 are set to 100.}%
truncated below at 0, and above at the minimum of 100 and $\texttt{seasonal\_effect}_{tj}+\texttt{noise}_{tj}$.}

The precipitation covariate $X^\text{out}_{tj3}$ is also generated differently based on the $x$-coordinate of each unit. For (coastal) outcome units with $x_j \leq 0.5$, we set $\texttt{seasonal\_effect}_{tj} = 0.15 + y_j / 100 - x_j / 50 + 0.1 \sin\left( \frac{2\pi(t - 180)}{365} \right)$ and we generate $\texttt{rain\_indicator}_{tj}$ independently across units and time from a $ \text{Ber}(\texttt{seasonal\_effect}_{tj})$ distribution, with $\texttt{rain\_indicator}_{tj} = 1$ reflecting rain at the location of outcome unit $m_j$ at time $t$, and $= 0$ reflecting no rain.
On days without rain, we set $X_{tj3}^\text{out} = 0$. On rainy days, we generate another Bernoulli draw as $\texttt{extreme\_rain}_{tj} \sim \text{Ber}(0.01 \times \texttt{seasonal\_effect}_{tj})$, with $\texttt{extreme\_rain}_{tj} = 1$ reflecting heavy precipitation, and $= 0$ for moderate precipitation. This specification implies that heavy precipitation events are more likely during the summer. Then, for days with moderate rain, we generate $X_{tj3}^\text{out} \sim \text{Gamma}(0.05,5)$, while on days with heavy rain we generate $X_{tj3}^\text{out} \sim \text{LogNormal}(2, 0.8^2)$. 
For (inland) outcome units with $x_j > 0.5$, we similarly generate $\texttt{rain\_indicator}_{tj}$ independently across units and time from a $ \text{Ber}(\texttt{seasonal\_effect}_{tj})$ distribution where $\texttt{seasonal\_effect}_{tj} = \max(0.1 \sin\left( \frac{2\pi(t - 180)}{365} \right),0)$. We set $X_{tj3}^\text{out} = 0$ if $\texttt{rain\_indicator}_{tj} = 0$, and draw $X_{tj3}^\text{out}$ from a Gamma$(2,5)$ distribution if $\texttt{rain\_indicator}_{tj} = 1$.

The covariate representing wind speed, $X^\text{out}_{tj4}$, is generated based on a smooth seasonal trend with heteroskedastic noise. For day $t$, the mean wind speed is specified as $\texttt{mean\_wind\_speed}_t = 10+20(1-(\frac{1}{182}t-\frac{183}{182})^2)^{0.2}$  which forms a flattened parabolic dome centered at the middle of the year. We also specify $\texttt{wind\_speed\_variance}_{t}=0.1 \texttt{mean\_wind\_speed}_t -1$. Then, the wind speed covariate is generated as $X^\text{out}_{tj4}\sim N(\texttt{mean\_wind\_speed}_{tj}, \texttt{wind\_speed\_variance}_{t})$, independently across units and time. {Lastly, to ensure sufficient variability across time, we add a random draw from the standard normal distribution $N(0, 1)$ to the $X^\text{out}_{tj4}$ values which were generated with $\texttt{wind\_speed\_variance}_{t} < 0.1$.}

Lastly, the wind direction covariate $X^\text{out}_{tj5}$ is generated to exhibit seasonal patterns. In January and December, it varies widely and is sampled uniformly on $[0, 1]$, whereas in other months, it remains relatively more constant and it is sampled uniformly on $[0.7, 0.9]$.

For an interventional unit $n_i$, their time-varying covariates are assigned to be equal to the time-varying covariates of the outcome unit $m_j$ that is geographically closest. {We scale the temporal trend variables and the time-varying covariates representing temperature, humidity, and wind speed for the interventional and outcome units to the $[0,1]$ range by subtracting the minimum and dividing by the range of each variable across units. To ensure that most (over 95\%) of values corresponding to the precipitation variable fall within the $[0,1]$ interval, we apply a $\log(1+x)$ transformation. Wind direction remains unchanged as it is naturally bounded within the $[0,1]$ range.}

The simulated covariates preserve similar fluctuation patterns compared to the observed covariates after re-scaling. As shown in \cref{supp:fig-five-panel}, the temporal patterns in simulated temperature, humidity, precipitation, wind speed, and wind direction are very similar to those in our study data.
\begin{figure}[htbp]
\centering
% --- First row with 3 figures ---
\begin{subfigure}[t]{1.8in}
  \includegraphics[width=\linewidth]{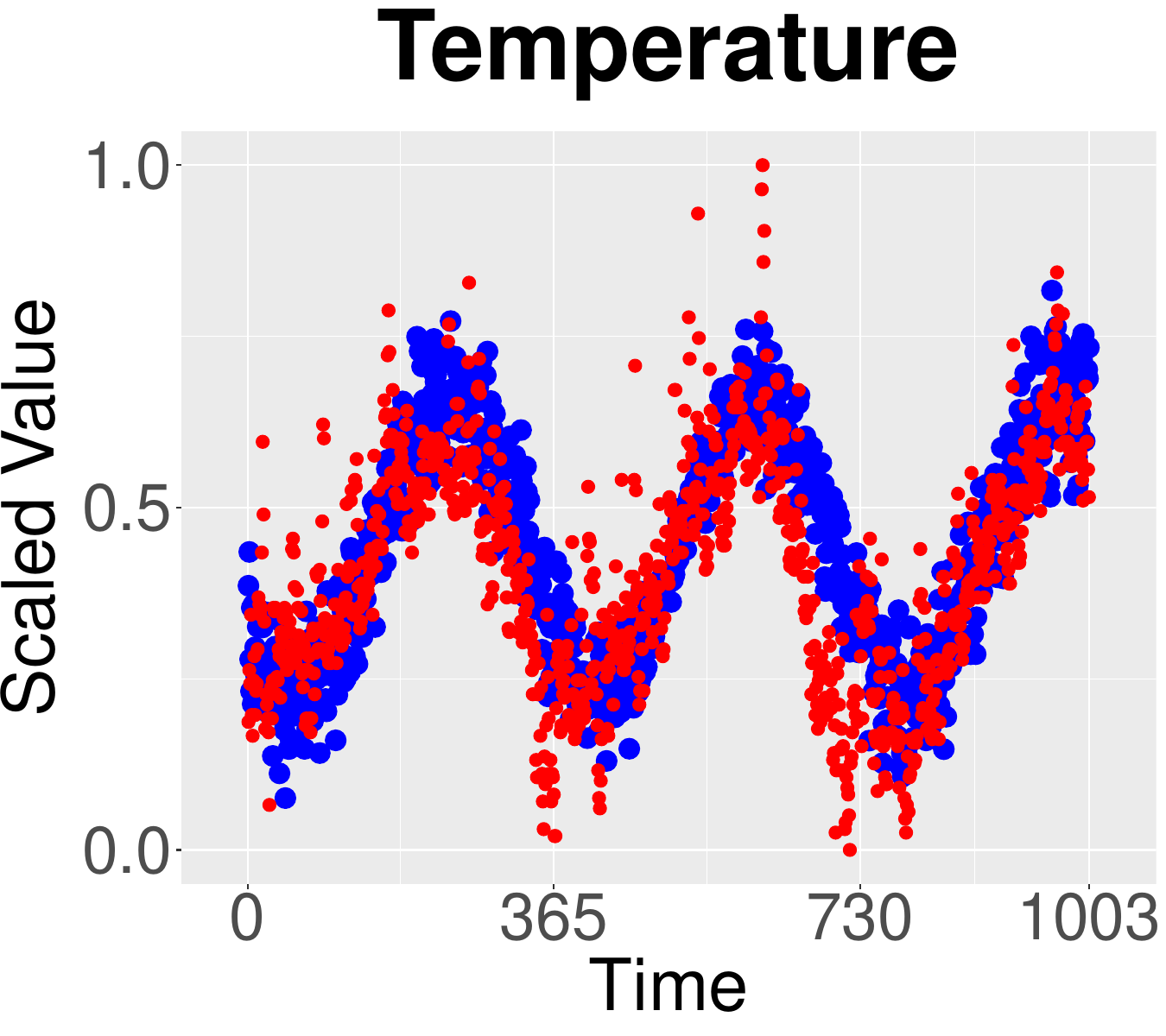}
\end{subfigure}
\hfill
\begin{subfigure}[t]{1.8in}
  \includegraphics[width=\linewidth]{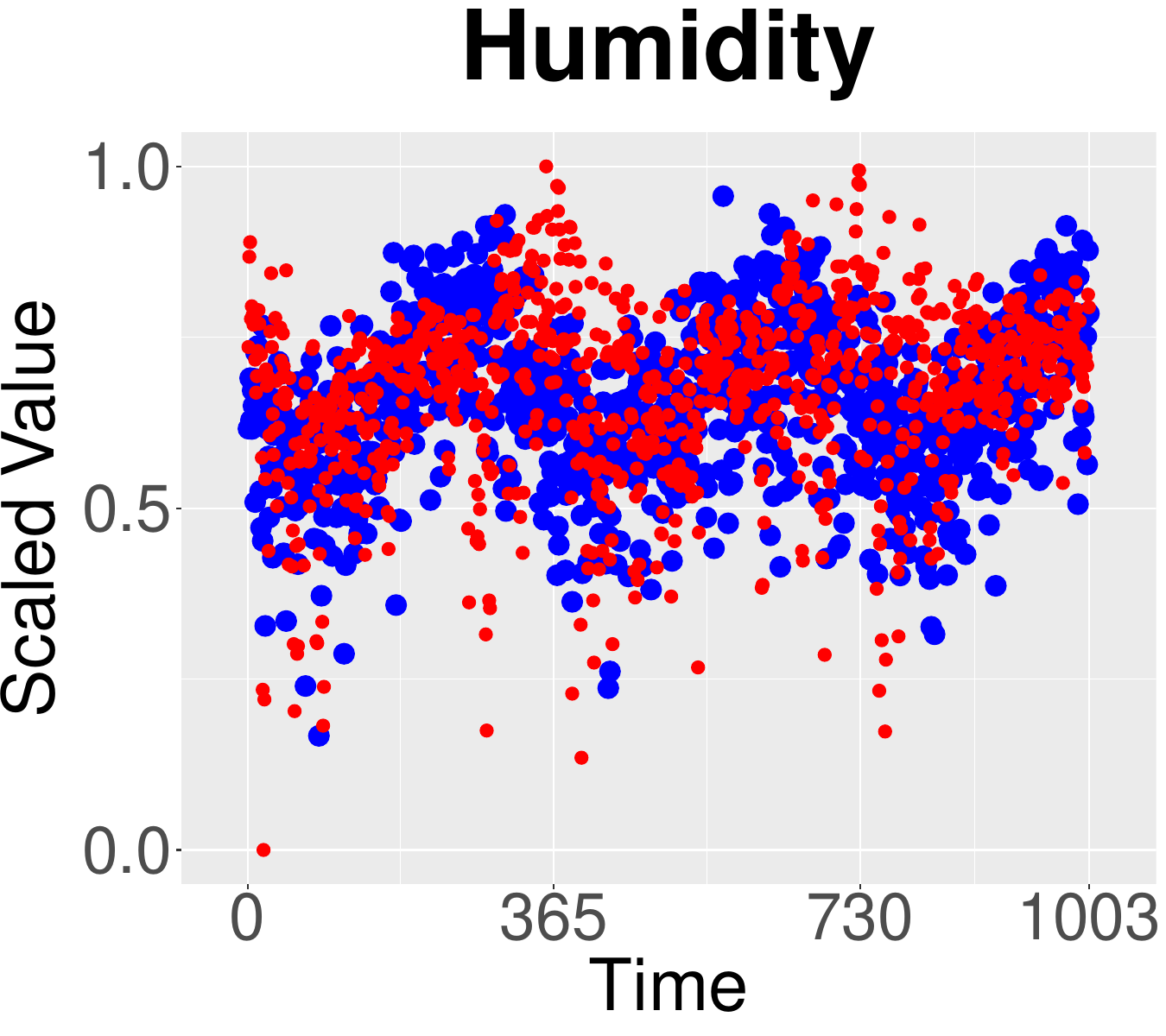}
\end{subfigure}
\hfill
\begin{subfigure}[t]{1.8in}
  \includegraphics[width=\linewidth]{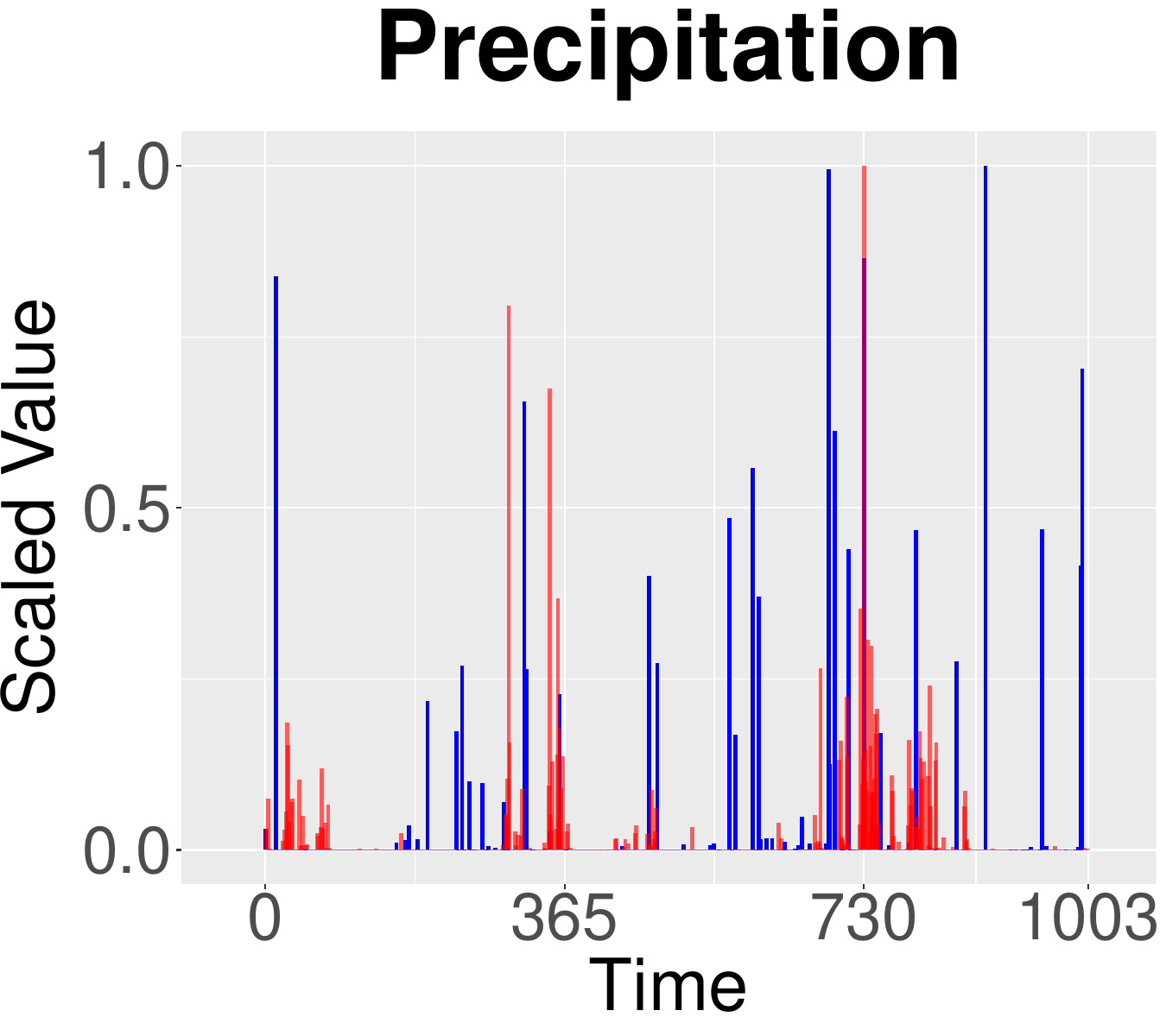}
\end{subfigure}

\vspace{1em}  % Space between rows

% --- Second row with 2 centered figures ---
\begin{subfigure}[t]{1.8in}
  \includegraphics[width=\linewidth]{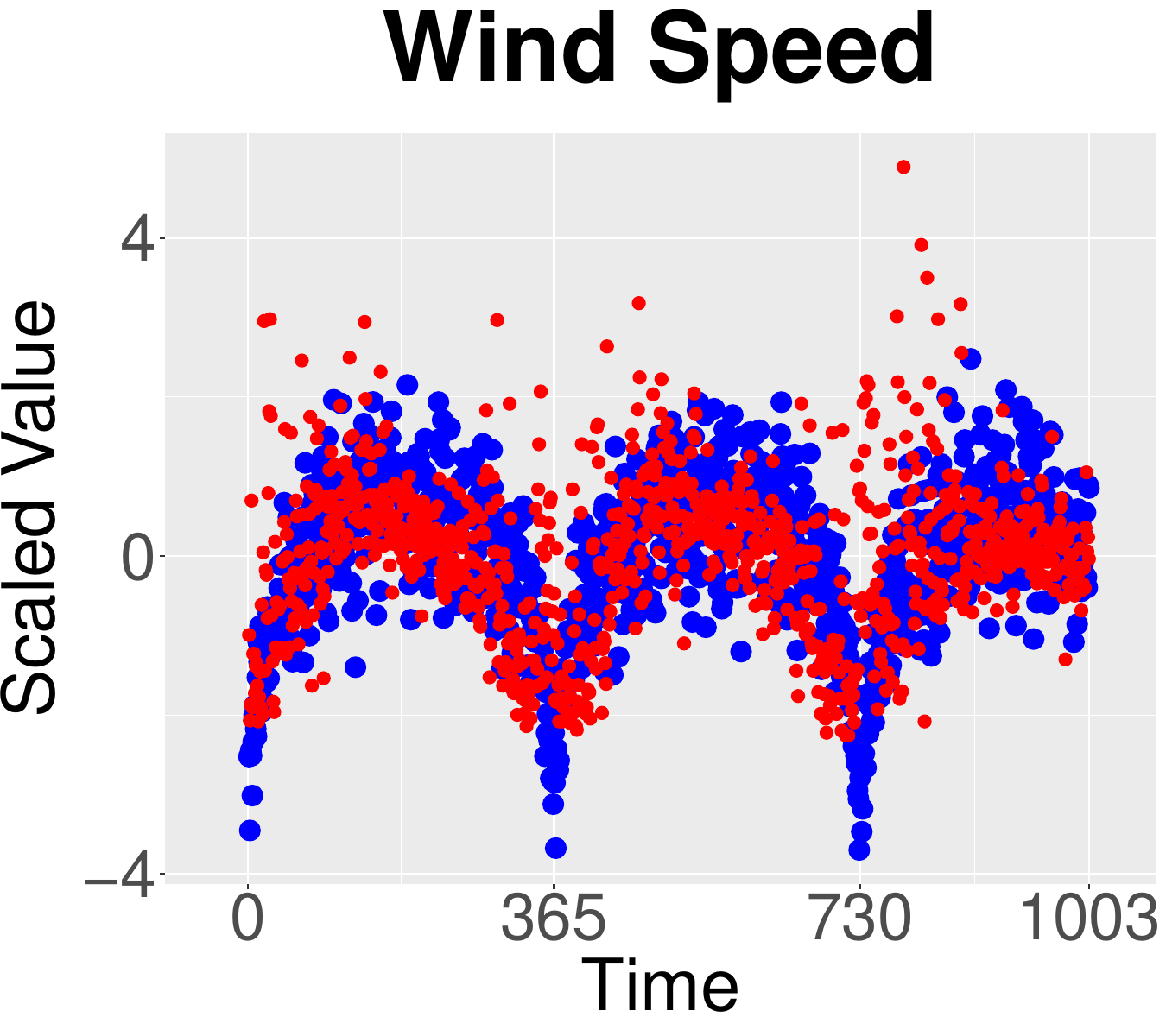}
\end{subfigure}
\hspace{0.52in}  % Center the two figures
\begin{subfigure}[t]{1.8in}
  \includegraphics[width=\linewidth]{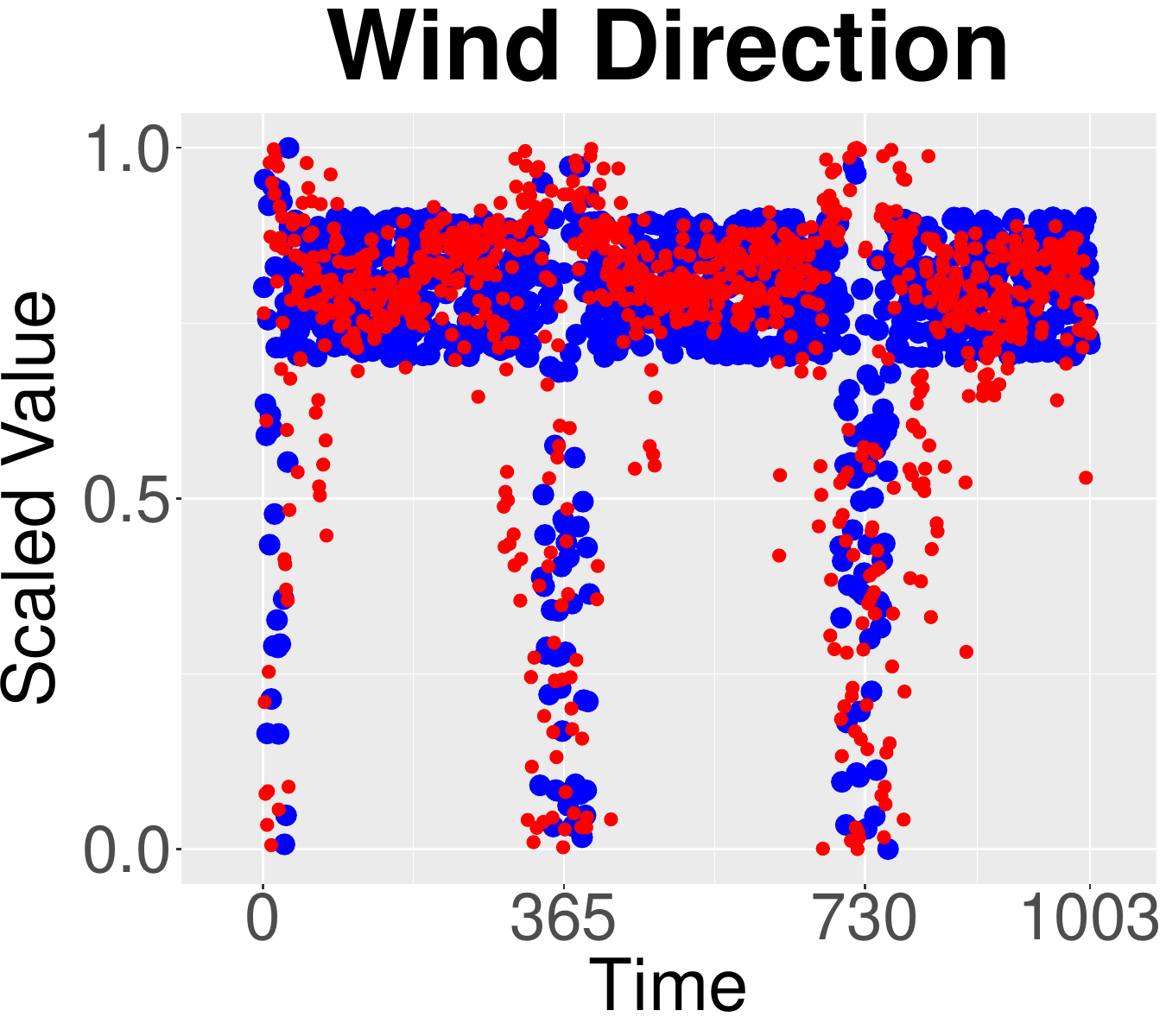}
\end{subfigure}

\caption{The comparison between real covariates (in red) and simulated covariates (blue) under the same scale.}
\label{supp:fig-five-panel}
\end{figure}
\subsection{The treatment assignment, bipartite network, and outcome}

We consider five scenarios that differ in their confounding structures. Each scenario defines a distinct way of generating treatment vectors for the interventional units over time. In the specifications below, when we write $A_{ti}, G_{tij} \sim \text{(Distribution)}$ we implicitly specify the distribution of the variable conditional on all covariate information and temporal trends. In what follows, we define {$\overline{R_t} = \frac{1}{M}\sum_{j=1}^MR_{tj}$}.

\begin{enumerate}[leftmargin=*,label=(\alph*)]
    \item No confounders: 
    \begin{align*}
    A_{ti} & \sim \text{Ber}(0.3), &\\
    G_{tij} & \sim \text{Ber}(0.3),&\\  Y_{tj} &= -E_{tj}-{0.5R_{tj}}+5X_{0j}^\text{out}+\epsilon_{tj}, & {\text{where } E_{tj} = I(\sum_{i=1}^N A_{ti}G_{tij}\geq 7) \text{ and } \epsilon_{tj} \sim N(0,1).}
    \end{align*}
    \item Only time-smooth confounders exist: 
    \begin{align*}
        A_{ti} & \sim \text{Ber}(\exp \eta_{ti}/(1+\exp \eta_{ti})), & \text{where }\eta_{ti}=4.5(f_i^\text{int}(t)-0.5),  \\
        G_{tij} & \sim \text{Ber}(\rho), & \text{where } \rho = 1/(1+\exp (dist(i,j)/4*(1+I(x_i<x_j)))),\\
        Y_{tj} &= -E_{tj}-{0.5R_{tj}}+2f_j^\text{out}(t)+\epsilon_{tj},& {\text{where } E_{tj} = I(\sum_{i=1}^N A_{ti}G_{tij}\geq 11) \text{ and } \epsilon_{tj} \sim N(0,1).}
    \end{align*}
    \item Only location-varying confounders exist: 
    \begin{align*}
        A_{ti} &\sim \text{Ber}(\exp \eta_{ti}/(1+\exp \eta_{ti})), & \text{where }\eta_{ti}=X_{0i}^\text{int}-0.5,\\
        G_{tij} & \sim \text{Ber}(\rho), & \text{where } \rho = 1/(1+\exp (dist(i,j)/4*(1+I(x_i<x_j)))),\\  
         Y_{tj} &= -E_{tj}-{0.5R_{tj}}+2X_{0j}^\text{out}+\epsilon_{tj}, & {\text{where } E_{tj} = I(\sum_{i=1}^N A_{ti}G_{tij}\geq 13) \text{ and } \epsilon_{tj} \sim N(0,1).}
    \end{align*}
    \item Only time-varying confounders exist:
    \begin{align*}
        A_{ti} & \sim \text{Ber}(\exp \eta_{ti}/(1+\exp \eta_{ti})),
        &\text{where }%\eta_{ti}=\min( \max (7.5X_{ti2}^\text{int}-0.003X_{ti3}^\text{int}-0.006X_{ti4}^\text{int}+0.0015X_{ti5}^\text{int}-0.003X_{ti6}^\text{int}-4.5,-2.3),2.3) \quad \text{when } t < 7 \\
         \eta_{ti}=2X_{ti1}^\text{int}-X_{ti2}^\text{int}-0.4X_{ti3}^\text{int}+ 1.5X_{ti4}^\text{int} \\
         & & \hspace{80pt} -0.7X_{ti5}^\text{int}+0.4 \overline{R_t}-0.6, \\%\quad \text{when } t \geq 7 \\
        G_{tij} & \sim \text{Ber}(\rho_{t}), 
        & \text{where }  \rho_{t}={1}/(1+\exp\{0.6\sin(2\pi (t+50)/365)\}),\\
        Y_{tj} &= -E_{tj}-{0.5R_{tj}}+4\exp (2X_{tj1}^\text{out}) \\
        & \hspace{20pt} +1.2X_{tj2}^\text{out} - 2X_{tj3}^\text{out}+2X_{tj4}^\text{out}\\
        & \hspace{20pt} +X_{tj5}^\text{out}+\epsilon_{tj}, & {\text{where } E_{tj} = I(\sum_{i=1}^N A_{ti}G_{tij}\geq 15) \text{ and } \epsilon_{tj} \sim N(0,1).}
    \end{align*}

    \item All confounders exist:
    {\footnotesize
    \begin{align*}
          A_{ti} % \mid f_i^\text{int}(t),X_{0i}^\text{int},\bm X_{ti}^\text{int} 
          & \sim \text{Ber}(\exp \eta_{ti}/(1+\exp \eta_{ti})), 
        &\text{where }%\eta_{ti}=\min( \max (0.75X_{ti1}^\text{int}+9X_{i0}^\text{int}+7.5X_{ti2}^\text{int}-0.003X_{ti3}^\text{int}-0.006X_{ti4}^\text{int}+0.0015X_{ti5}^\text{int}-0.003X_{ti6}^\text{int}-4.5,-2.3),2.3) \quad \text{when } t < 7 \\
        \eta_{ti}=0.4f_i^\text{int}(t)+1.6X_{0i}^\text{int}+1.6X_{ti1}^\text{int}-0.8X_{ti2}^\text{int}\\
        && -0.32X_{ti3}^\text{int}+1.2X_{ti4}^\text{int}-0.56X_{ti5}^\text{int}\\
        && +0.015 \overline{R_t}-1.92, \\%\quad \text{when } t \geq 7 \\
        % &\quad \overline{R_t} = \frac{1}{T}\sum_{j=1}^MR_{tj}\\
        G_{tij} & \sim \text{Ber}(\rho_{tij}), 
        & \text{where }  \rho_{tij}={1}/(1+\exp{0.6\sin(2\pi (t+50)/365)}+ \\
        && {dist(i,j)/4(1+I(x_i<x_j))}),\\
        Y_{tj} &= -E_{tj}-{0.5R_{tj}}+1.8f_j^\text{out}(t)+4\exp{X_{0j}^\text{out}}\\
        & \hspace{40pt} +4\exp X_{tj1}^\text{out}+1.2X_{tj2}^\text{out}-2X_{tj3}^\text{out} \\
        & \hspace{40pt} +2X_{tj4}^\text{out}+X_{tj5}^\text{out}+\epsilon_{tj},& {\text{where } E_{tj} = I(\sum_{i=1}^N A_{ti}G_{tij}\geq 16) \text{ and } \epsilon_{tj} \sim N(0,1).}
    \end{align*}
    }
\end{enumerate}

\section{Description for alternative approaches}
\label{appendix:sec:naive}

\subsection{The na\"ive approaches}

Implementing three na\"ive approaches requires a full dataset including exposure status $E_{t}$ and outcomes $Y_{tj}$ among all units $m_1,m_2,\cdots,m_M $ and all time points $t=1,2,\cdots,T$. Na\"ive-all uses the full dataset, Na\"ive-$t$ uses only the data from the first outcome unit across time, and Na\"ive-$j$ uses the data across all outcome units but only for the first time point.

\subsubsection{Estimator}

For Na\"ive-$t$, we split the temporal data for the first outcome unit into two groups: the exposed time periods, $t \in \mathcal{T}_{1.}$ if $E_{t1}=1$, and the unexposed time periods, $t \in \mathcal{T}_{0.}$ if $E_{t1} = 0$. Similarly, for Na\"ive-$j$, we split the data for the first time period in the exposed outcome units, $j \in \mathcal{J}_{1.}$ if $E_{1j}=1$, and the unexposed outcome units, $j \in \mathcal{J}_{0.}$ if $E_{1j} = 0$.
For Na\"ive-all, we consider all units and time periods, and denote $(t,j)\in \mathcal{S}_{1.}$ if the outcome unit-time period combination is exposed, $E_{tj}=1$, and  $(t,j)\in \mathcal{S}_{0.}$ otherwise. We use the estimators $\widehat{\tau}^{\text{Na\"ive}-t}, \widehat{\tau}^{\text{Na\"ive}-j}$ and $\widehat{\tau}^{\text{Na\"ive}-all}$ as follows:
\begin{align*}
    \widehat{\tau}^{\text{Na\"ive}-t} & = \frac{1}{|\mathcal{T}_{1.}|}\sum_{t\in \mathcal{T}_{1.}}Y_{t1}-\frac{1}{|\mathcal{T}_{0.}|}\sum_{t\in \mathcal{T}_{0.}}Y_{t1},\\
    \widehat{\tau}^{\text{Na\"ive}-j}& = \frac{1}{|\mathcal{J}_{1.}|}\sum_{j \in \mathcal{J}_{1.}}Y_{1j}-\frac{1}{|\mathcal{J}_{0.}|}\sum_{j \in \mathcal{J}_{0.}}Y_{1j},\\
    \widehat{\tau}^{\text{Na\"ive}-all}& = \frac{1}{|\mathcal{S}_{1.}|}\sum_{(t,j)\in \mathcal{S}_{1.}}Y_{tj}-\frac{1}{|\mathcal{S}_{0.}|}\sum_{(t,j)\in \mathcal{S}_{0.}}Y_{tj}.
\end{align*}

\subsubsection{Inference}
We adopt the classic Wald-type confidence interval construction for two independent samples. Denote $s_{1.}$ as the standard deviation of outcomes with index from $\mathcal{T}_{1.},\mathcal{J}_{1.}$ or $\mathcal{S}_{1.}$, and $s_{0.}$ as the standard deviation of outcomes with index from $\mathcal{T}_{0.},\mathcal{J}_{0.}$ or $\mathcal{S}_{0.}$ for each na\"ive approach. The confidence intervals for Na\"ive-$t$, Na\"ive-$j$, and Na\"ive-all are
\begin{align*}
    \widehat{\tau}^{\text{Na\"ive}-t} \pm Z_\alpha\sqrt{\frac{(|\mathcal{T}_{1.}|-1)s_{1.}^2+(|\mathcal{T}_{0.}|-1)s_{0.}^2}{|\mathcal{T}_{1.}|+|\mathcal{T}_{0.}|-2}}\sqrt{\frac{1}{|\mathcal{T}_{1.}|}+\frac{1}{|\mathcal{T}_{0.}|}},\\
     \widehat{\tau}^{\text{Na\"ive}-j} \pm Z_\alpha\sqrt{\frac{(|\mathcal{J}_{1.}|-1)s_{1.}^2+(|\mathcal{J}_{0.}|-1)s_{0.}^2}{|\mathcal{J}_{1.}|+|\mathcal{J}_{0.}|-2}}\sqrt{\frac{1}{|\mathcal{J}_{1.}|}+\frac{1}{|\mathcal{J}_{0.}|}},\\
\intertext{and}
      \widehat{\tau}^{\text{Na\"ive}-all} \pm Z_\alpha\sqrt{\frac{(|\mathcal{S}_{1.}|-1)s_{1.}^2+(|\mathcal{S}_{0.}|-1)s_{0.}^2}{|\mathcal{S}_{1.}|+|\mathcal{S}_{0.}|-2}}\sqrt{\frac{1}{|\mathcal{S}_{1.}|}+\frac{1}{|\mathcal{S}_{0.}|}},
\end{align*}
respectively.

\subsection{Linear regression estimator}

We consider linear regression for the first outcome unit where we regress outcomes across time on time-varying information. Specifically, the linear regression model assumes the outcome $Y_t$ is a linear function of the exposure $E_t$, the carryover effect $R_t$, time $t$, and the summarized covariates ${\overline{\bm X}^\text{sum.}_{t, S}}$:
$$Y_{t} = \beta_0+ \beta_1 E_t+ \beta_2 R_t+\beta_3 t  +\bm \beta_4^\top {\overline{\bm X}^\text{sum.}_{t, S}} +\epsilon_{t}$$
for $t=1,2,\dots,T$, where $\epsilon_t$ represents independent random errors with $E(\epsilon_t)=0$ and $\text{Var}(\epsilon_t)$ constant. We use Ordinary Least Squares (OLS) to derive the point estimators $\hat{\beta}_1$ and $\hat{\beta}_2$, alongside their respective standard errors, $\text{se}(\hat{\beta}_1)$ and $\text{se}(\hat{\beta}_2)$. These coefficients serve as the regression estimators for the immediate effect $\tau^{\text{imm}}$ and the carryover effect $\tau^{\text{car}}$, respectively. Finally, the 95\% confidence intervals are constructed as $\hat{\beta}_j \pm 1.96 \cdot \text{se}(\hat{\beta}_j)$ for $j \in \{1, 2\}$.

\subsection{IPW estimator}

Again, we consider an Inverse Probability Weighting (IPW) estimator for the first outcome unit across time.
For $\tau^{\text{imm}}$, we define the propensity score of $E_t$ as $e(R_t, t, \overline{\bm X}^\text{sum.}_{t, S}) = P(E_t = 1 \mid R_t, t, \overline{\bm X}^\text{sum.}_{t, S})$, and consider the ATT-type estimator 
$$\frac{1}{T} \sum_{t=1}^T \left( E_tY_t - \frac{e(R_t, t, \overline{\bm X}^\text{sum.}_{t, S})}{1 - e(R_t, t, \overline{\bm X}^\text{sum.}_{t, S})}(1 - E_t) Y_t \right).$$
The IPW estimator for the carryover effect is similar, switching the roles of $E_t$ and $R_t$.

We employ the R package \texttt{PSweight} \citep{zhou2022psweight} to calculate the estimators for $\tau^{\text{imm}}$ and $\tau^{\text{car}}$.
For the immediate effect, the package fits the propensity score using a logistic regression model, yielding the estimate $\widehat{e}(R_t, t, \overline{\bm X}^\text{sum.}_{t, S})$. To construct confidence intervals, \texttt{PSweight} utilizes the empirical sandwich variance estimator for propensity score weighting.%, as established by \cite{zhou2022psweight}.

\section{Additional Simulations}
\label{supp_sec:simu}

\subsection{Data generation}
%\subsection{Simulation setup}
\label{supp_subsec:simu_setup}

We consider a setting with $N=50$ interventional units and $M=200$ outcome units at randomly generated locations over the $[0,1]\times [0,1]$ square, followed over $T=400$ time periods.

\subsubsection{The locations}

The locations of the interventional and outcome units, $(x, y)$, are generated in the following manner. For the interventional units, the $x$ coordinates of units 1 to 10 and 31 to 40 are generated independently from a Uniform$(0,0.5)$ distribution, while the $x$ coordinates for the remaining units are generated from a Uniform$(0.5,1)$ distribution. The $y$ coordinates for the interventional units are generated independently from a Uniform$(0,0.5)$ distribution for units 1 to 10 and 21 to 30, and from Uniform$(0.5,1)$ distribution for the remaining units. Similarly, for the outcome units, the $x$ coordinates for units 1 to 68 and 113 to 156 are drawn from a Uniform$(0,0.5)$ distribution, while the $x$ coordinates for the remaining outcome units are drawn from a Uniform$(0.5,1)$ distribution. The $y$ coordinates are drawn from  a Uniform$(0,0.5)$ for units 1 to 50 and 101 to 150, and from a Uniform$(0.5,1)$ for the remaining units.

\subsubsection{The covariates}

We consider six covariates for the interventional units, six covariates for the outcome units, and one network covariate. We discuss how each variable is generated below. %that represent smooth temporal trends, vary across units but not time, vary across time but not units, and vary across units and time. These covariates are:
\begin{enumerate}[label=(\alph*),itemsep=5pt,parsep=5pt]
\item (Smooth temporal trends) Covariates $f_i^\text{int}(t)$ and $f_j^\text{out}(t)$ are generated independently from Gaussian processes with the same smooth function of time as the mean, and an exponential decay kernel for the covariance matrix. Therefore, these covariates represent similar but not identical smooth temporal trends. Specifically:
\begin{itemize}[label=-,itemsep=5pt]
    \item For the interventional units, we generate a variable that represents a smooth function of time in the following manner. The variable $\bm f_{i}^\text{int}=(f_{i}^\text{int}(1), f_{i}^\text{int}(2), \dots, f_{i}^\text{int}(T))$  is generated from a Gaussian process with Gaussian correlation kernel and mean representing a smooth temporal trend as $\bm f_{i}^\text{int} \sim \mathcal{N}(f(t), \Sigma)$, where $\Sigma_{t_1,t_2} = \frac{\exp (-({t_{1}}-{t_{2}})^2)}{2\cdot 100^2}$ and $f(t)=\frac{3}{400}t$. 
    
    \item For the outcome units, the vector $\bm f_{j}^\text{out}$ is generated by the same distribution as for the interventional units, as $\bm f_{j}^\text{out} \sim {N}(f(t),\Sigma)$, where $\Sigma_{t_1,t_2} = \frac{\exp (-({t_{1}}-{t_{2}})^2)}{2\cdot 100^2}$ and $f(t) = \frac{3}{400}t$.
\end{itemize}
Realizations of the smooth temporal trend $\bm f_{i}^\text{int}$ for five interventional units and $\bm f_{j}^\text{out}$ for five outcome units are shown on the top- and bottom-left panels of \cref{supp_fig:simu_covariates}, respectively. It is evident that the smooth temporal trends are similar but different. Therefore, if $\bm f_{i}^\text{int}$ is a predictor of the interventional units' treatment across time, and $\bm f_{j}^\text{out}$ is a predictor of the outcome units' outcome across time, then the common smooth temporal trend in $\bm f_{i}^\text{int}$ and $\bm f_{j}^\text{out}$ confounds the exposure-outcome relationship of interest.

\item (Location-varying covariates) 
Covariates $X_{0i1}^\text{int}$ and $X_{0j1}^\text{out}$ are constant across time,  and they are drawn independently from a scaled beta distribution with parameters that depend on the unit's location. Therefore, these covariates have similar structure across space. 
\begin{itemize}[label=-,itemsep=5pt]
    \item For the interventional units, we consider a time-invariant covariate, $X_{0i1}^\text{int}$. For units $n_i$ inside the $[0,0.5]\times [0,0.5]$ rectangle, we generate $X_{0i1}^\text{int}/8 \sim$ Beta$(9,1)$, while for the rest $n_i$'s, we draw $X_{0i1}^\text{int}/8 \sim$ Beta$(1,9)$.

     \item We generate the time-invariant covariate for the outcome units $X_{0j1}^\text{out}$ in the same manner based on outcome unit $m_j$'s coordinates.
\end{itemize}
A visualization of the time-invariant covariate for all interventional and outcome units is shown at the top- and bottom-right panels of \cref{supp_fig:simu_covariates}, respectively. It is evident that the location-varying covariates of interventional and outcome units share spatial trends. Therefore, if the former is a predictor of the interventional units' treatment, and the latter of the outcome units' outcome, then there exists location-varying confounding of the exposure-outcome relationship.
\item (Non-smooth time-varying variables) Covariates $X_{ti1}^\text{int}$, $X_{tj1}^\text{out}$ and $ X_{tij1}^\text{net}$ are time-varying covariates that are independent across units but are not smooth over time.
\begin{itemize}[label=-,itemsep=5pt]
\item For the interventional units, we generate $X_{ti1}^\text{int}\sim {N}(0,t/100), i=1,\cdots,N, t=1,\cdots,T.$ This variable varies across time without a smooth pattern. A realization is drawn on the left of \cref{supp_fig:simu_covariates_nonsmooth}.
\item For the outcome units, we generate $X_{tj1}^\text{out}/2\sim $Beta$(t/100,2), j=1,2,\cdots,M,\  t=1,2,\cdots,T.$ This variable has an {\it non}-smooth temporal trend.
\item The one network covariate has a non-smooth temporal trend. Specifically, the network covariate array $\bm X_{t..1}^\text{net}$ is an $N\times M$ matrix. %For the weights of $\bm P_t$, we make it a priori with respect to every outcome unit $j$. The weight vector is termed $\bm{q}_j$, a length $N$ vector, with constant $\frac{1}{N}$ in all entries.
    The entries of the matrix are generated independently as $X_{tij1}^\text{net} \sim$ Beta($t/50,10$), $i=1,2,\cdots,N, \ j=1,2,\cdots,M, \ t=0,1,\cdots, T$. 
\end{itemize}

\item (Bipartite covariates) We define covariates for one set of units based on the covariates of the other set. 

\begin{itemize}[label=-,itemsep=5pt]
\item For interventional units, we define location-varying covariate $X_{0i2}^\text{int}$, and the  time-varying covariates $X_{ti2}^\text{int}$ and $X_{ti3}^\text{int}$, as averages of covariates $X_{0j1}^\text{out}$, $X_{tj1}^\text{out}$, and $X_{tij1}^\text{net}$, respectively, of neighboring outcome units. Specifically, we define the $N \times M$ matrix $\bm{R}$, with entries $p_{ij} = 1$ if unit $i$ and $j$ are within distance 0.1, and $p_{ij}=0$ otherwise. We define
\begin{align*}
X_{0i2}^\text{int} &= \frac{\sum_j p_{ij} X_{0j1}^\text{out}}{\sum_j p_{ij}}, \quad
X_{ti2}^\text{int}=\frac{\sum_j p_{ij}X_{tj1}^\text{out}}{\sum_j p_{ij}}, \quad \text{ and } \quad
X_{ti3}^\text{int}=\frac{\sum_j p_{ij}X_{tij1}^\text{net}}{\sum_j p_{ij}},
\end{align*} 
for $i=1,2,\cdots,N,$ and $t=1,2,\cdots,T.$

\item Covariates $X_{0j2}^\text{out}$,  $X_{tj2}^\text{out}$ and $X_{tj3}^\text{out}$ for outcome units are similarly defined based on covariates $X_{0i1}^\text{int}$,  $X_{ti1}^\text{int}$, and $X_{tij1}^\text{net}$ of interventional units, as
\[
X_{0j2}^\text{out} = \frac{\sum_i p_{ij} X_{0i1}^\text{int}}{\sum_i p_{ij}}, \quad 
X_{tj2}^\text{out} = \frac{\sum_i p_{ij} X_{ti1}^\text{int}}{\sum_i p_{ij}}, \quad \text{ and} \quad 
X_{tj3}^\text{out} = \frac{\sum_i p_{ij} X_{tij1}^\text{net}}{\sum_i p_{ij}},
\]
for $j=1,2,\cdots,M,$ and $t=1,2,\cdots,T.$
One realization of the location-varying covariate $\bm X_{0.2}^\text{out}$ for all outcome units and one realization of the non-smooth covariate $\bm X_{t.2}^\text{out}$ are presented on the left and right panels of \cref{supp_fig:simu_covariates_outcome}, respectively.

Here, consider the matrix $\bm{Q}$ whose columns are the normalized version of the columns of $\bm{R},$ where the $(i,j)$th entry is equal to $q_{ij} = p_{ij} / \sum_i p_{ij}$. The columns of the matrix $\bm Q$ correspond to the vectors $\bm q$ we introduce in \cref{subsec:matching_algorithms} for defining the interventional and network covariate summaries. 

\end{itemize}

\item (Common non-smooth time-varying covariate) Covariates $X_{ti4}^\text{int}, X_{tj4}^\text{out}$ are equal to each other, regardless of $i$ and $j$, and represent {\it non}-smooth temporal trends. Specifically, $X_{ti4}^\text{int} = X_{tj4}^\text{out} = Z_t$, where $Z_t$ is drawn from a $\mathcal{N}(t/200, \log(t)/10)$ distribution, independently across time. A realization of this covariate is shown on the right panel of \cref{supp_fig:simu_covariates_nonsmooth}.

\end{enumerate}
% we use $\bm X_{0j}^\text{out}$ and $\bm X_{tj}^\text{out}$ to denote the time-invariant and time-varying covariates of outcome unit $m_j$, respectively, with $j = 1, 2, \dots, M$. And similarly $\bm X_{0i}^\text{int}$ and $\bm X_{ti}^\text{int}$  for interventional units $i= 1, 2, \cdots, N$.

%We generate covariates, treatments, graphs, exposures, and outcomes over $T=400$ time periods. We create scenarios with or without different confounders between exposure and outcome.

\begin{table}[!b]
    \centering
    %\spacingset{1}
    \caption{Table of five confounding scenarios, in which treatment $\bm{A}$, network graph $\bm{G}$, and observed outcome $\bm{Y}$ are associated with corresponding confounding covariates.}
    % \rowcolors{5}{}{gray!10}
    % \spacingset{1} % Assuming this command is defined elsewhere
    \vspace{-3pt}
    \resizebox{5.5in}{!}{%
        \begin{tabular}{ll*{10}{c}}
    \hline
    \\[-10pt]
    && \multicolumn{3}{c}{Smooth time} & \multicolumn{5}{c}{Location-varying} & \multicolumn{2}{c}{Time-varying} \\
    \cmidrule(lr){3-5} \cmidrule(lr){6-10} \cmidrule(lr){11-12}
    Scenario & Component & $t$ & $f_i^\text{int}(t)$ & $f_j^\text{out}(t)$ & dist & ${X}_{0i1}^\text{int}$ & ${X}_{0j1}^\text{out}$ & ${X}_{0i2}^\text{int}$& ${X}_{0j2}^\text{out}$ & $\bm{X}_{ti}^\text{int}$ & $\bm{X}_{tj}^\text{out}$\\[5pt]
    \\[-13pt]
    \hline
    \\[-10pt]
    
    \multirow{3}{*}{(a) No confounders} 
    & $\bm{A}$ & & & & & & & & & & \\
    & $\bm{G}$ & & & & & & & & & & \\
    & $\bm{Y}$ & & & & & & $\times$ & & & & \\[0pt]
    \\[-10pt]
    \hline
    \\[-10pt]
    
    \multirow{3}{*}{\shortstack[l]{(b) Time-smooth\\ confounders}} 
    & $\bm{A}$ & & $\times$ & & & & & & & & \\
    & $\bm{G}$ & & & & $\times$ & & & & & & \\
    & $\bm{Y}$ & & & $\times$ & & & & & & & \\[0pt]
    \\[-10pt]
    \hline
    \\[-10pt]
    
    \multirow{3}{*}{\shortstack[l]{(c) Location-varying\\ confounders}} 
    & $\bm{A}$ & & & & & $\times$ & & $\times$ & & & \\
    & $\bm{G}$ & & & & $\times$ & & & & & & \\
    & $\bm{Y}$ & & & & & & $\times$ & & $\times$ & & \\[0pt]
    \\[-10pt]
    \hline
    \\[-10pt]
    
    \multirow{3}{*}{\shortstack[l]{(d) Time-varying\\ confounders}} 
    & $\bm{A}$ & & & & & & & & & $\times$ & \\
    & $\bm{G}$ & $\times$ & & & $\times$ & & & & & & \\
    & $\bm{Y}$ & & & & & & & & & & $\times$ \\[0pt]
    \\[-10pt]
    \hline
    \\[-10pt]
    
    \multirow{3}{*}{(e) All confounders} 
    & $\bm{A}$ & & $\times$ & & & $\times$ & & $\times$ & & $\times$ & \\
    & $\bm{G}$ & $\times$ & & & $\times$ & & & & & & \\
    & $\bm{Y}$ & & & $\times$ & & & $\times$ & & $\times$ & & $\times$ \\[0pt]
    \\[-10pt]
    \hline
\end{tabular}%
    }%
    \label{tab:sims_plan}
\end{table}

%We generate covariates, treatments, graphs, exposures, and outcomes over $T=400$ time periods. We create scenarios with or without different confounders between exposure and outcome.

\begin{figure}[!t]
\centering
    \caption{Visualization of interventional and outcome unit covariates over time and units. The top left panel shows 5 realizations from the Gaussian process for $\bm f_i^\text{int}$. The top right figure is the $[0,1]\times[0,1]$ location square with one realization for the location of the interventional points and the corresponding location-varying covariate $\bm X_{0.1}^\text{int}$. The bottom left figure is 5 realizations from the Gaussian process for $\bm f_j^\text{out}$ and the bottom right figure is a realization from the locations of the outcome units colored by the values of one realization of $\bm X_{0.1}^\text{out}$.}
    \includegraphics[scale=0.25]{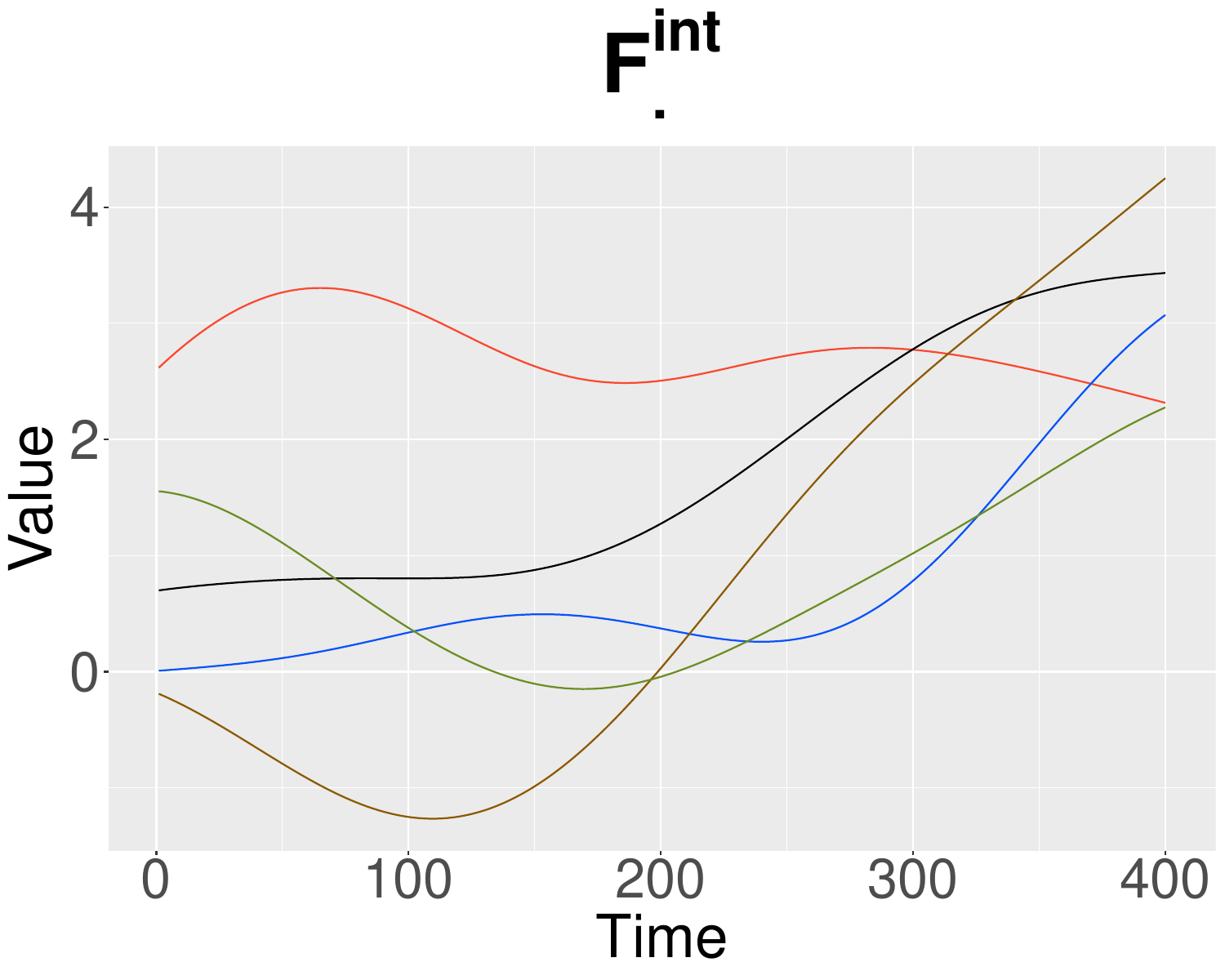}
\includegraphics[scale=0.25]{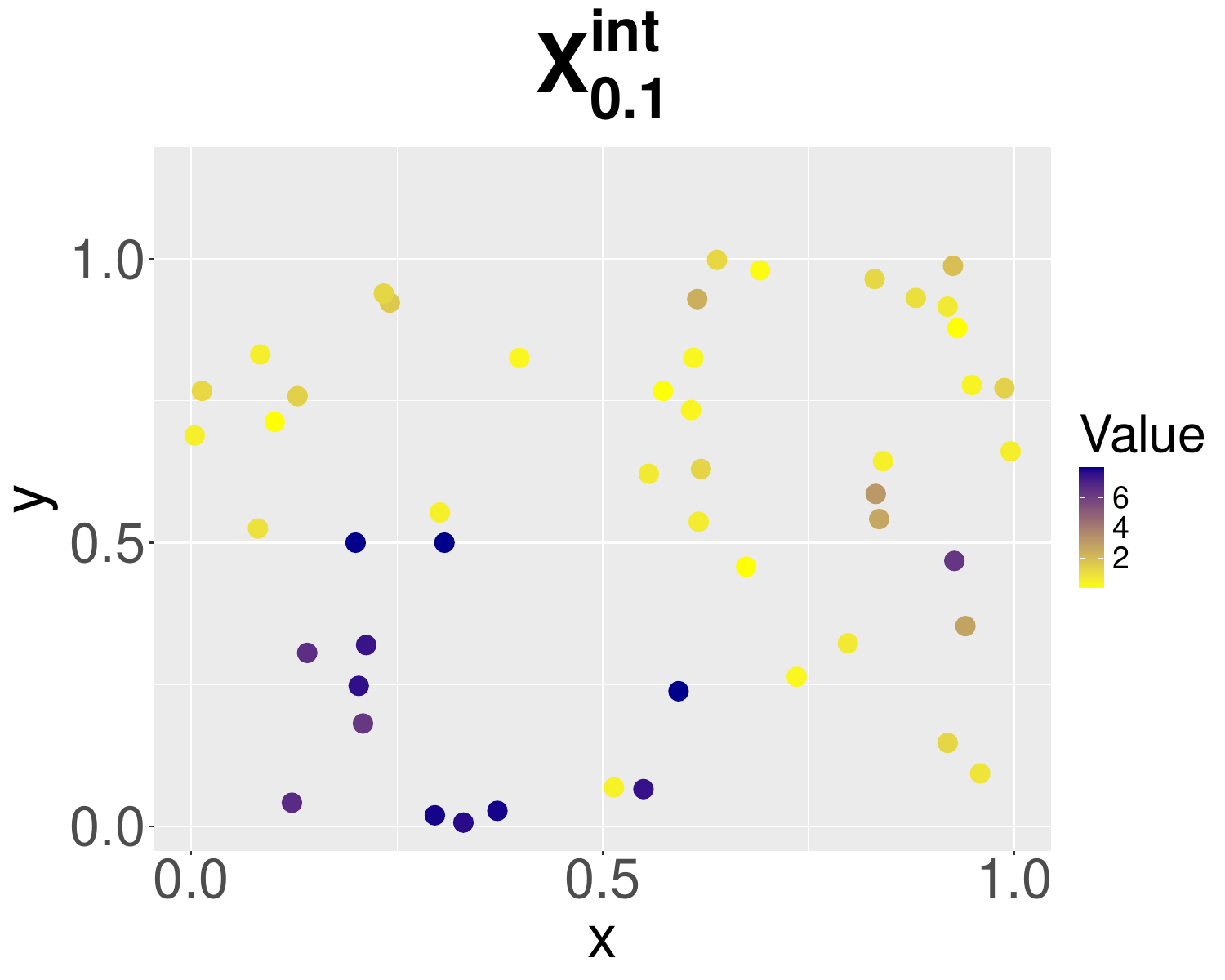}\\

\vspace{5mm}

\includegraphics[scale=0.25]{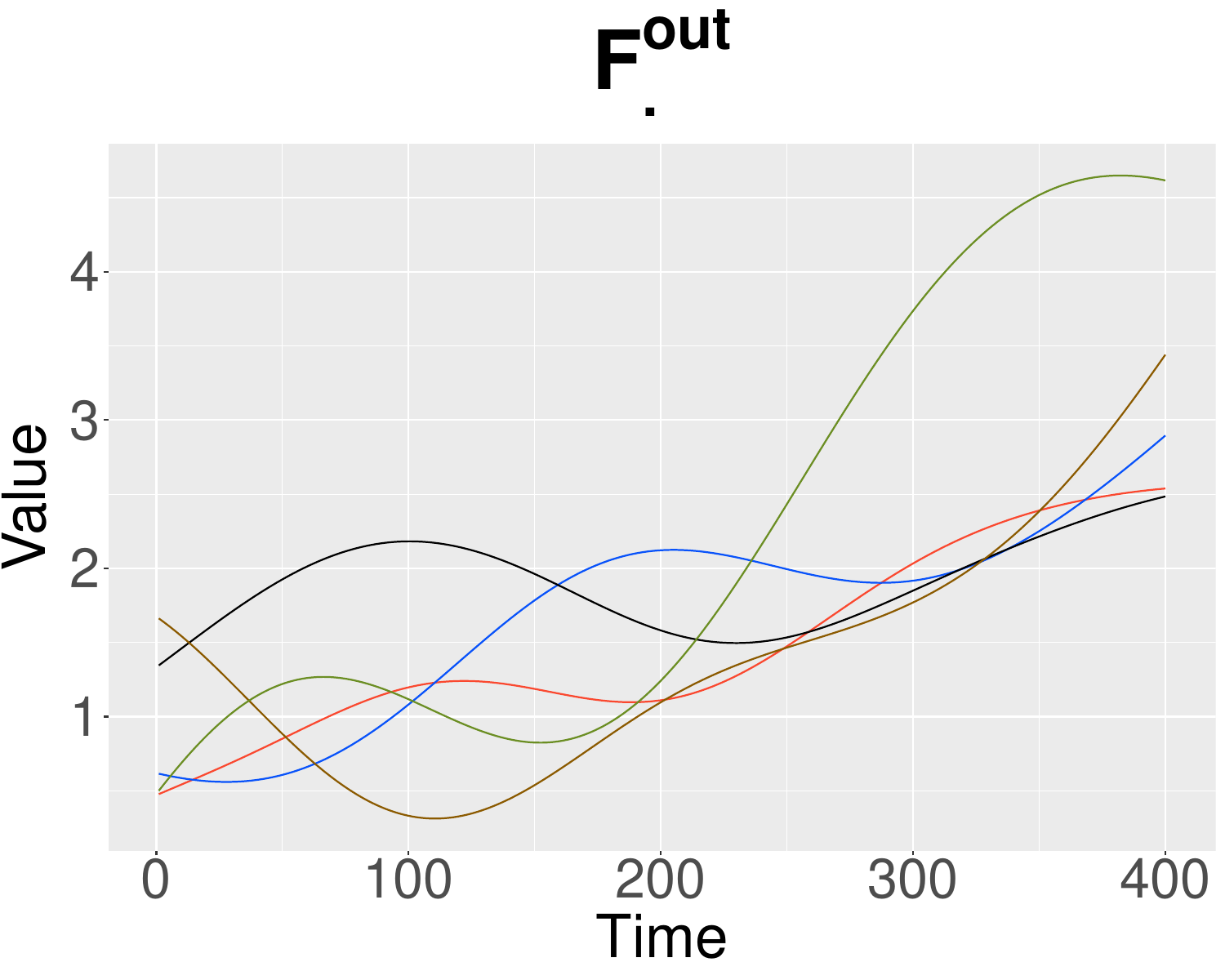}
\includegraphics[scale=0.25]{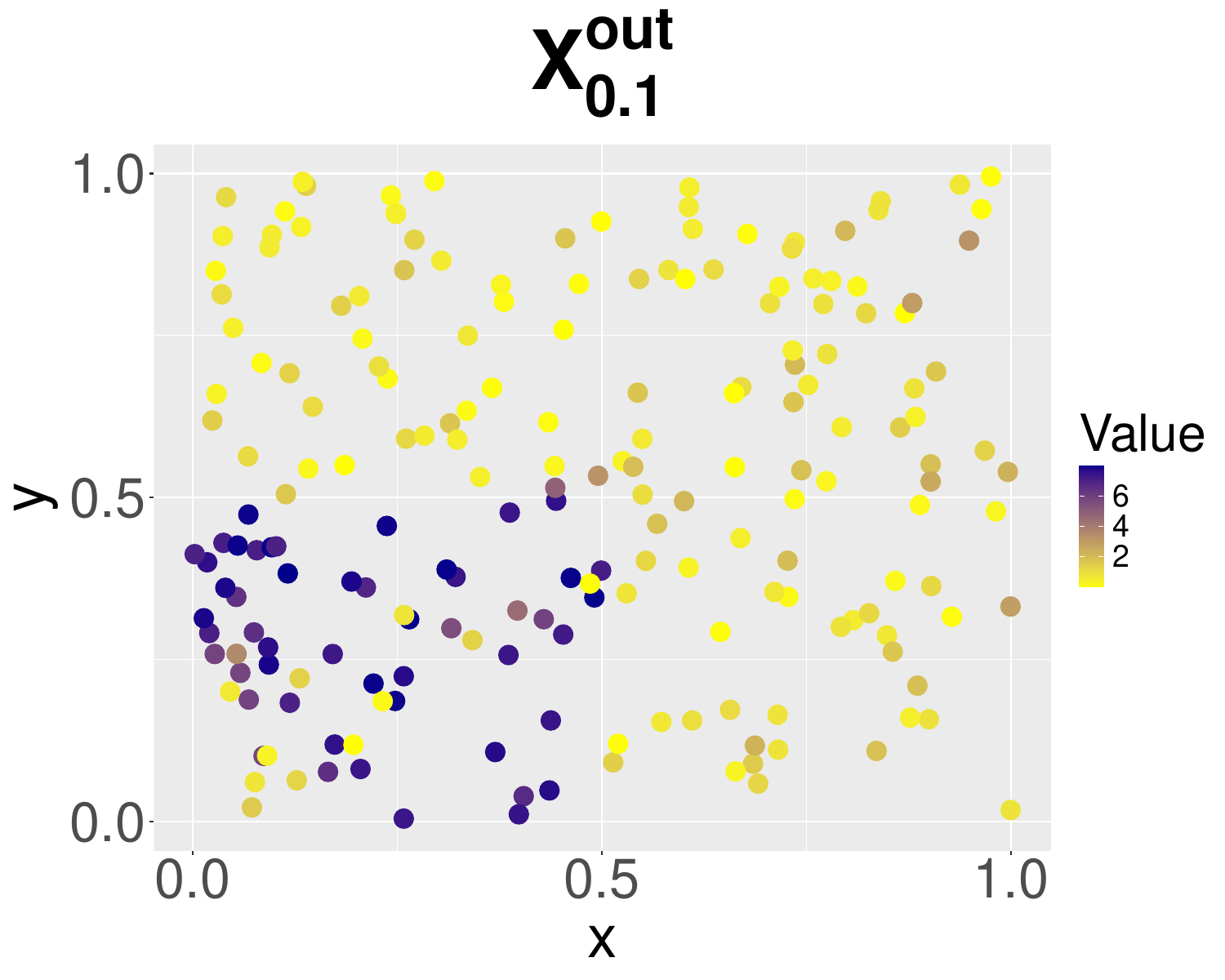}
\label{supp_fig:simu_covariates}
\end{figure}

\begin{figure}
    \centering
    \caption{Visualization of non-smooth interventional covariates $\bm X_{..1}^\text{int}$ and $\bm X_{..4}^\text{int}$. The left panel shows one realization of $\bm X_{..1}^\text{int}$, and the right panel shows one realization of $\bm X_{..4}^\text{int}$. }
       \includegraphics[scale=0.25]{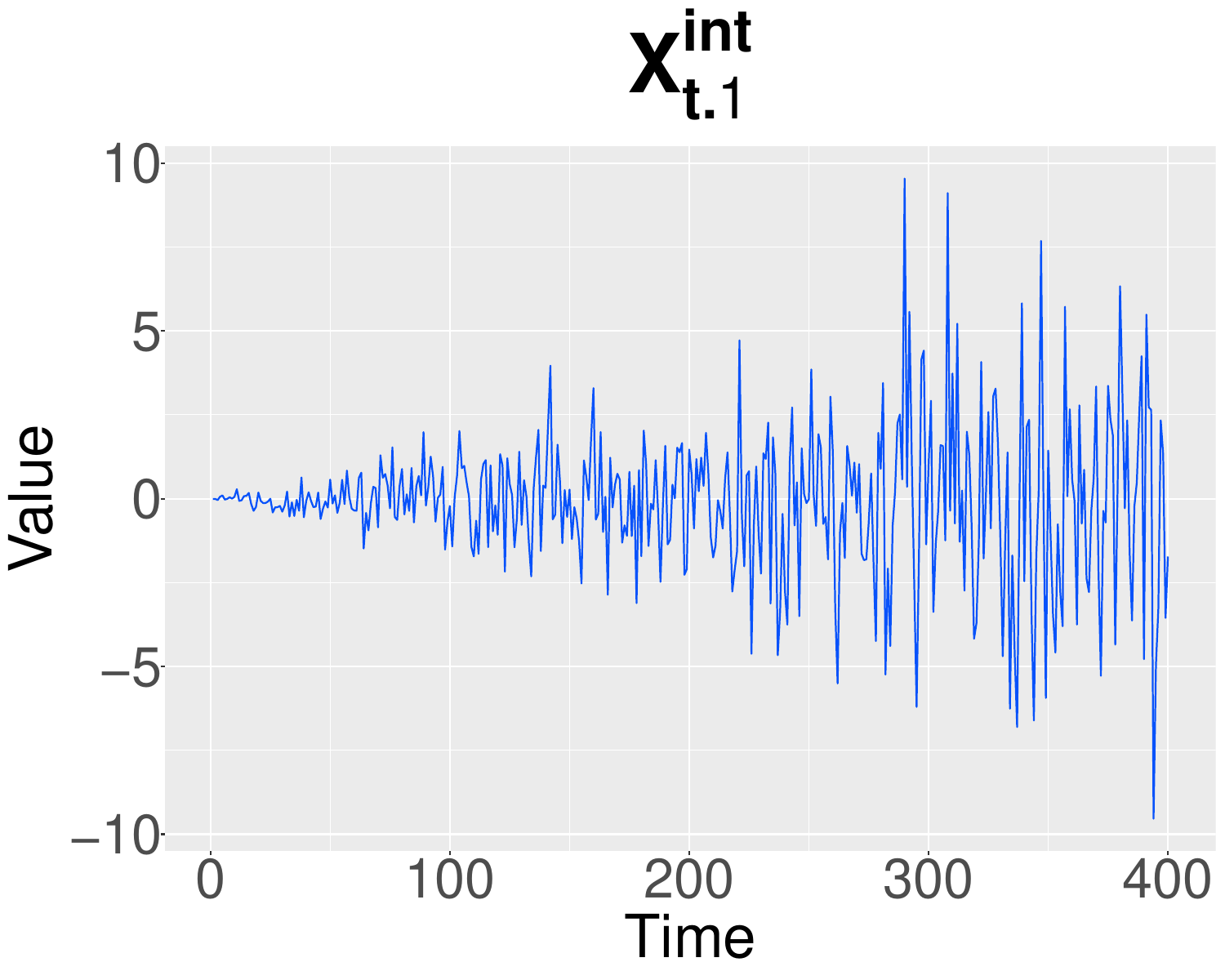}
\includegraphics[scale=0.25]{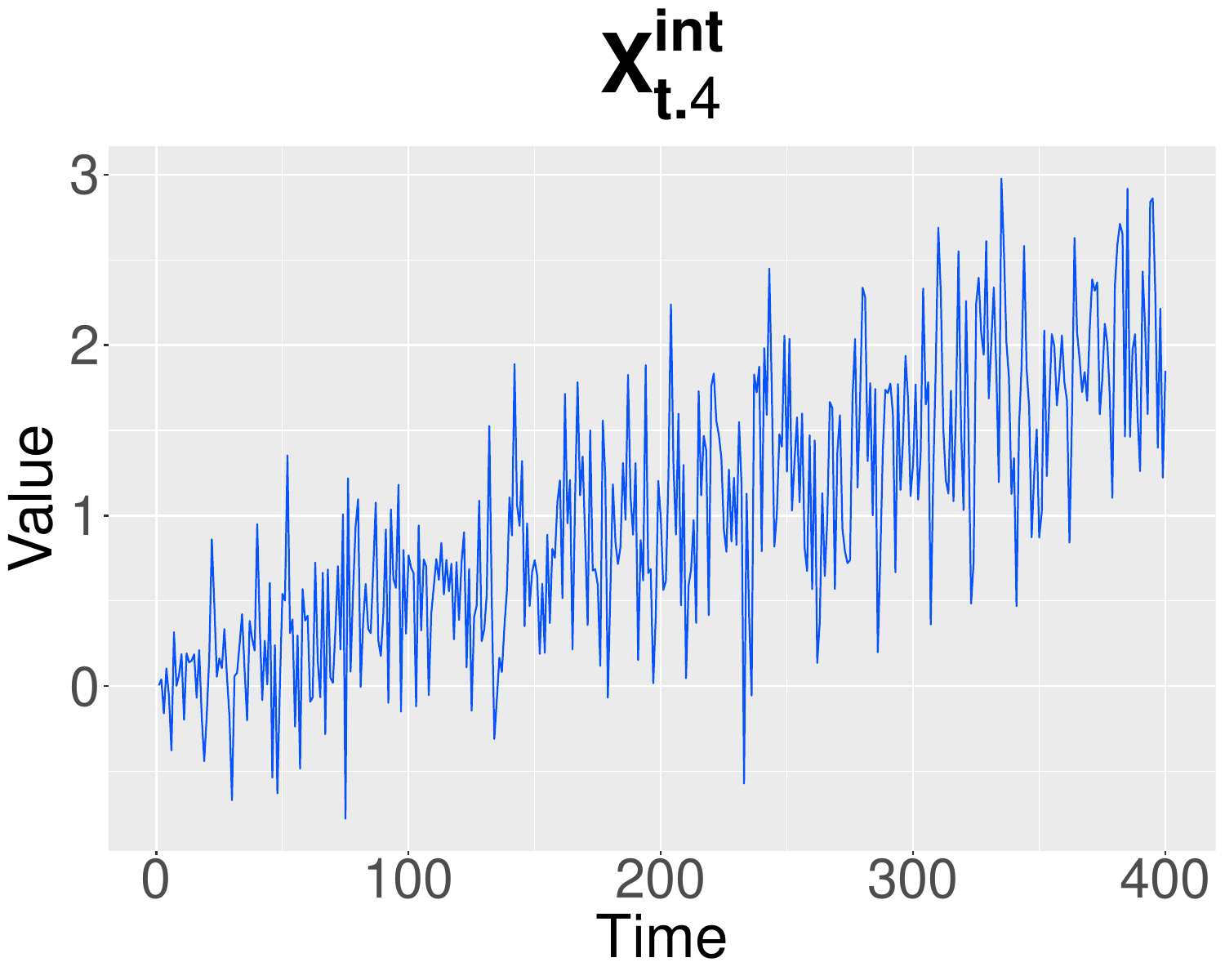}\\
    \label{supp_fig:simu_covariates_nonsmooth}
\end{figure}

\begin{figure}[ht]
    \centering
    \caption{Visualization of the location-varying outcome covariate $\bm X_{0.2}^\text{out}$ and the non-smooth covariate $\bm X_{t.2}^\text{out}$. The left panel is the $[0,1]\times[0,1]$ location square with one realization for the location of the outcome units with colors representing the value of the location-varying covariate $\bm X_{0.2}^\text{out}$. The right panel shows one realization for $\bm X_{t.2}^\text{out}$ for the outcome units which is defined based on the interventional units' covariate $\bm X_{t.1}^\text{int}$ in \cref{supp_fig:simu_covariates_nonsmooth} and their geographical distance. }
       
\includegraphics[scale=0.25]{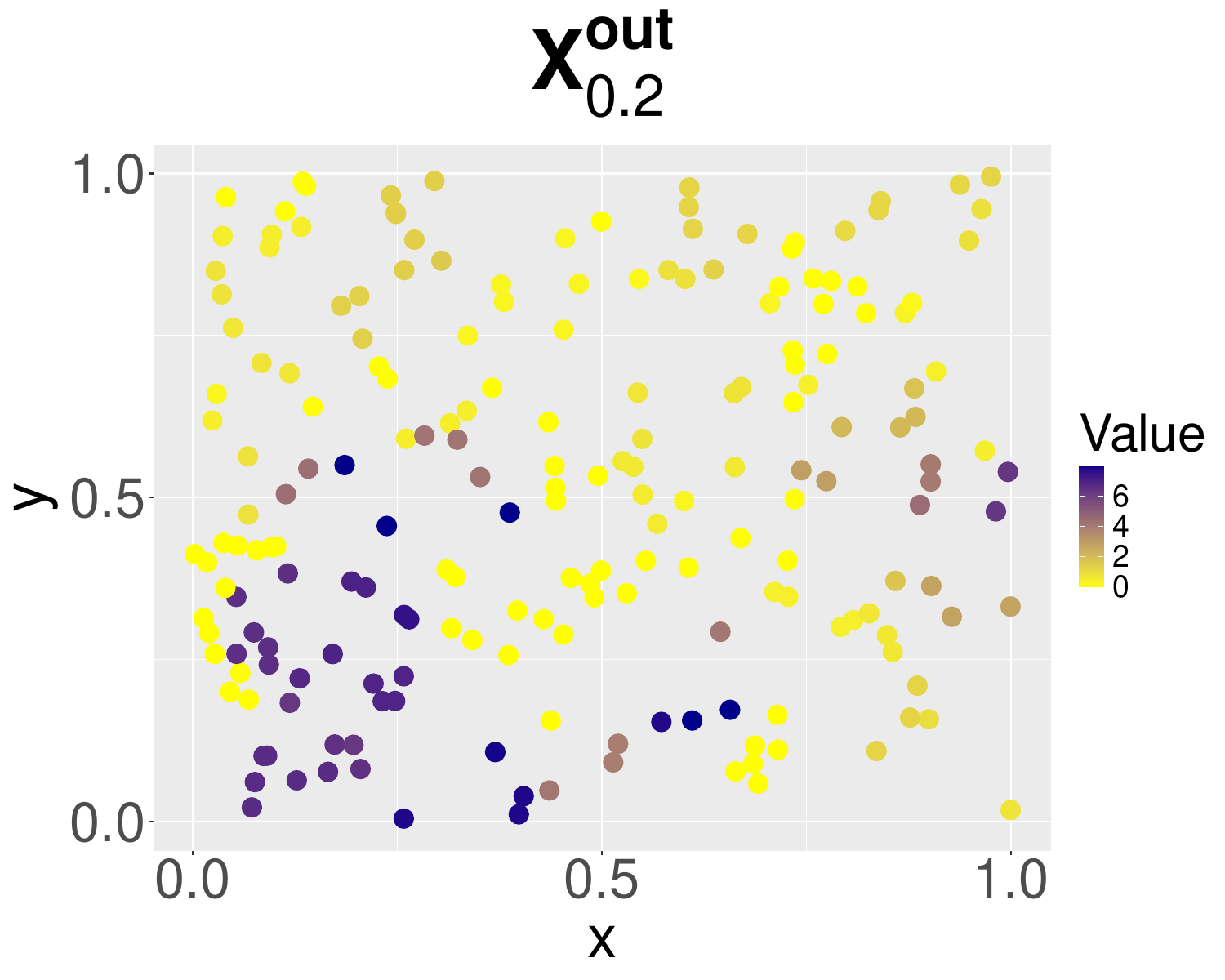}\includegraphics[scale=0.25]{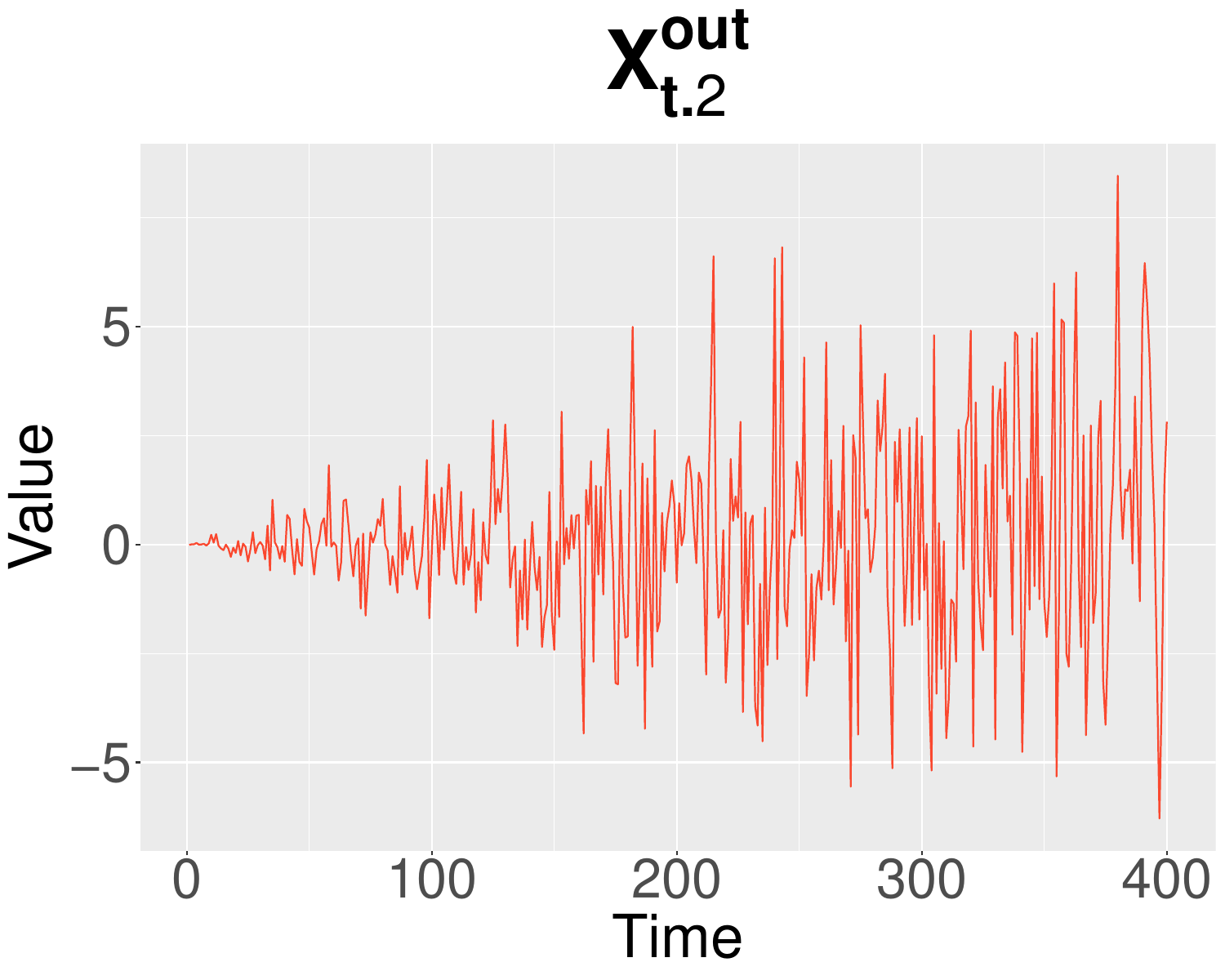}\\
    \label{supp_fig:simu_covariates_outcome}
\end{figure}

\subsubsection{The treatment assignment, bipartite network, and outcome}\label{supp_subsubsec:agy}

We consider five scenarios regarding the confounding structure. In what follows, when we write $A_{ti}, G_{tij}\sim \text{(Distribution)}$, we implicitly condition on all the variables.
\begin{enumerate}[leftmargin=*,label=(\alph*)]
    \item No confounders: 
    \begin{align*}
    A_{ti} & \sim \text{Uniform}(0,1), \\
    G_{tij} & \sim \text{Ber}(\rho), \quad \quad \text{where } \rho = 0.17,\\  Y_{tj} &= E_{t}+{0.4R_{tj}}+X_{0j1}^\text{out}+\epsilon_{tj}.
    \end{align*}
    \item Only time-smooth confounders exist: 
    \begin{align*}
        A_{ti}\mid f_i^\text{int}(t) & \sim \text{Ber}(1/(1+\exp (f_i^\text{int}(t)/1.2))\\
        G_{tij} & \sim \text{Ber}(\rho_{ij}), \quad \quad \text{where } \rho_{ij} = 1/(2(1+\exp dist(i,j)),\\
        Y_{tj} &= E_{t}+{0.4R_{tj}}+f_j^\text{out}(t)+\epsilon_{tj}
    \end{align*}
    \item Only location-varying confounders exist: 
    \begin{align*}
        A_{ti} &\sim \text{Ber}({1}/{(1+0.3\exp{(X_{0i1}^\text{int}}-X_{0i2}^\text{int}/40))})\\
        G_{tij} & \sim \text{Ber}(\rho_{ij}), \quad \quad \text{where } \rho_{ij} = 1/(1.7(1+\exp dist(i,j)))\\
        Y_{tj} & = E_{t}+{0.4R_{tj}}+4X_{0j1}^\text{out}+4X_{0j2}^\text{out}+\epsilon_{tj}
    \end{align*}
    \item Only time-varying confounders exist:
    {\small
    \begin{align*}
        A_{ti}\mid \bm X_{ti}^\text{int} &\sim \text{Ber}(\frac{1}{1+\exp (X_{ti1}^\text{int}/2 + X_{ti2}^\text{int}/2+X_{ti3}^\text{int}/10+X_{ti4}^\text{int})})\\
        G_{tij} \mid t & \sim \text{Ber}(\rho_{tij}), \quad \text{where }  \rho_{tij}={1}/(1+0.1\exp{\sin(\pi t/1000)}+\exp{dist(i,j)})\\
        Y_{tj} &= E_{t}+{0.4R_{tj}}+X_{tj1}^\text{out}+X_{tj2}^\text{out}+X_{tj3}^\text{out} +X_{tj4}^\text{out}+\epsilon_{tj}
    \end{align*}
    }
    \item All confounders exist:
    {\footnotesize
    \begin{align*}
          &A_{ti}\mid \bm f_i^\text{int},\bm{X}_{0i}^\text{int},\bm{X}_{ti}^\text{int}\sim{ \text{Ber}(\eta_{ti}+\sum_{j=1}^MR_{tj}/M)}, \quad \quad \text{where}\\
        &\eta_{ti}=\frac{1}{1+0.45\exp (f_{i}^\text{int}(t)/20+X_{0i1}^\text{int}+X_{0i2}^\text{int}+X_{ti1}^\text{int}/100+X_{ti2}^\text{int}/20+X_{ti3}^\text{int}/10+X_{ti4}^\text{int}/1.5+X_{(t-1)i4}^\text{int}/3)}\\
        &G_{tij} \mid t  \sim \text{Ber}(\rho_{tij}),\quad \quad \text{where }  \rho_{tij}={1}/(1+0.1\exp{\sin(\pi t/1000)}+\exp{dist(i,j)})\\
        &Y_{tj} = E_{t}+{0.4R_{tj}}+f_j^\text{out}(t)+2*X_{0j1}^\text{out}+2*X_{0j2}^\text{out}+X_{tj1}^\text{out}+0.1*{X_{tj1}^\text{out}}^2+X_{tj2}^\text{out}+\\
        & \qquad \qquad 1/(1+\exp(X_{tj2}^\text{out}))+ \sin \left( X_{tj3}^\text{out}\right) +2*X_{tj4}^\text{out}+X_{(t-1)j4}^\text{out}/3+\epsilon_{tj}
    \end{align*}
    }
\end{enumerate}

\cref{tab:sims_plan} shows the variables that are used in each data generative model component across the different scenarios. 
% These five scenarios correspond to five different confounding structures for the exposure-outcome relationship: 
% \begin{enumerate*}[label=(\alph*)]
% \item no confounding, 
% \item confounding by smooth temporal variables, 
% \item confounding by location-varying variables, 
% \item confounding by all time-varying information, and 
% \item all types of confounding. 
% \end{enumerate*}
Confounding arises if a predictor of the outcome is correlated with a predictor of the treatment assignment, the bipartite network, or both. For example, in scenario (b), the role of $f_i^\text{int}(t)$ and $f_j^\text{out}(t)$ induces confounding due to the variables' common smooth temporal trend, and in scenario (d), the covariate $X_{ti2}^\text{int}$ is defined based on $X_{tj1}^\text{out}$ which induce confounding in the exposure-outcome relationship.
Therefore, these data generative models allow for complex confounding structures.

For each scenario, we consider three sparsity levels for the exposure by tuning $d$, dense, medium, and sparse, corresponding to about 150-200, 80-120, and 30-60 exposed time periods, respectively. Therefore, in total, we consider 15 simulation scenarios, and generate 500 data sets for each one of them.

\subsection{Single-unit estimation and inference across different exposure sparsity levels}\label{supp_subsec:simu_dif_sparsity}

\cref{tab:sims_single} shows the estimation and inferential results for estimating the effect for one outcome unit using the three na\"ive approaches, and the three proposed estimators as also discussed in the simulations of \cref{sec:simu_real_data}. For each estimator we report bias, mean squared error and coverage of 95\% intervals. For the proposed estimators, we also report the proportion of exposed time periods that were matched.
%\textcolor{orange}{add covariate omission result, maybe only simply describe the single unit simulation?}

% \begin{enumerate*}[label=(\alph*)]
%     \item Bias: $\frac{1}{K}\sum_{k=1}^K\hat{\tau}_k - \tau$;
%     \item Mean Squared Error (MSE): $\frac{1}{K}\sum_{k=1}^K(\hat{\tau}_k - \tau)^2$;
%     \item Coverage: Build confidence intervals in every simulation as described in \cref{subsec:theory}, record the proportion of those intervals including $\tau$;
%     \item Proportion of exposed time points being matched.
%     \end{enumerate*}
% The results are shown in \cref{tab:sims_single}. 

\begin{table}[!t]
    \centering
    %\spacingset{1.12}
    \caption{Simulation results of single-unit \textbf{immediate} effects. Bias, mean squared error (MSE), coverage of 95\% intervals (\%), and proportion of exposed time points being matched (\%). We show simulation results for the 3 na\"ive approaches and 3 matching estimators over 5 confounding scenarios, and for 3 exposure levels. `N' stands for `na\"ive'. The scenarios  correspond to (a) No confounders, (b) Time-smooth confounders, (c) Location-varying confounders, (d) Time-varying confounders, and (e) All confounders.}
    % \rowcolors{5}{}{gray!10}
    %\spacingset{1}
    \resizebox{0.9\textwidth}{!}{%
       \begin{tabular}{*{14}{c}}
            \hline
            &&\multicolumn{4}{c}{Dense} & \multicolumn{4}{c}{Medium} & \multicolumn{4}{c}{Sparse} \\
            \cmidrule(lr){3-6} \cmidrule(lr){7-10} \cmidrule(lr){11-14}
            & Method & Bias & MSE & Cover & Prop & Bias & MSE & Cover & Prop & Bias & MSE & Cover & Prop \\
              \hline
             \multirow{8}{*}{(a)} &N-t & -0.00 & 0.011 & 95.2 & - & -0.01 & 0.013 & 95.4 & - & -0.01 & 0.022 & 95.2 & -\\
&N-j & -0.00 & 0.216 & 93.2 & - & -0.00 & 0.214 & 95.6 & - & -0.02 & 0.248 & 95 & -\\
&N-all & -0.00 & 0.000 & 94.8 & - & -0.00 & 0.001 & 95.4 & - & -0.00 & 0.001 & 95.8 & -\\
&Reg & -0.00 & 0.011 & 94.4 & - & -0.01 & 0.014 & 94.6 & - & -0.01 & 0.022 & 95.2 & -\\
&IPW & -0.00 & 0.011 & 94.4 & - & -0.01 & 0.014 & 95 & - & -0.01 & 0.022 & 95 & -\\
%&U1-1 & -0.04 & 0.014 & 93.2 & 97.8 & -0.03 & 0.021 & 94.4 & 100 & -0.01 & 0.039 & 95.2 & 100\\ 
%&U1-1/2 & -0.03 & 0.013 & 94.4 & 97.8 & -0.03 & 0.018 & 94.2 & 100 & -0.02 & 0.035 & 95.2 & 100\\ 
%&U1-2 & -0.07 & 0.019 & 90.8 & 62.3 & -0.03 & 0.017 & 94.2 & 90.8 & -0.02 & 0.030 & 94.4 & 98.7\\
&1-1 & -0.02 & 0.013 & 94 & 97.8 & -0.02 & 0.020 & 95.6 & 100 & -0.01 & 0.040 & 95.8 & 100\\ 
&1-1/2 & -0.02 & 0.012 & 95.4 & 97.8 & -0.03 & 0.017 & 94.6 & 100 & -0.02 & 0.032 & 95 & 100\\ 
&1-2 & -0.02 & 0.015 & 95.6 & 61.9 & -0.03 & 0.017 & 93.9 & 90.6 & -0.01 & 0.029 & 94.9 & 98.7\\ 
               \hline
\multirow{8}{*}{(b)\phantom{$^1$}} & N-t & -0.77 & 0.939 & 14.2 & - & -0.85 & 1.104 & 14.4 & - & -0.92 & 1.300 & 17.4 & -\\
&N-j & -0.00 & 0.048 & 96.2 & - & 0.01 & 0.051 & 93.2 & - & 0.00 & 0.046 & 94.2 & -\\
&N-all & -0.75 & 0.567 & 0 & - & -0.82 & 0.684 & 0 & -& -0.92 & 0.853 & 0 & -\\
&Reg & 0.01 & 0.020 & 92.8 & - & 0.01 & 0.032 & 87.4 & - & 0.03 & 0.060 & 86 & -\\
&IPW & 0.01 & 0.027 & 94.6 & - & 0.02 & 0.027 & 94.4 & - & 0.03 & 0.033 & 95 & -\\
%&U1-1 & -0.03 & 0.021 & 94.4 & 62.5 & -0.03 & 0.026 & 93.2 & 85.1 & -0.02 & 0.037 & 94.8 & 97.8\\
%&U1-1/2 & -0.02 & 0.021 & 94.2 & 62.5 & -0.02 & 0.023 & 94.4 & 85.1 & -0.02 & 0.034 & 94.8 & 97.8\\
%&U1-2 & -0.04 & 0.026 & 92.3 & 40.9 & -0.03 & 0.026 & 92.8 & 59.9 & -0.03 & 0.034 & 95.2 & 80.6\\
&1-1 & -0.02 & 0.019 & 93.8 & 62.5 & -0.02 & 0.023 & 95 & 85 & -0.00 & 0.037 & 95 & 97.6\\
&1-1/2 & -0.02 & 0.020 & 94.4 & 62.5 & -0.01 & 0.023 & 93.8 & 85 & -0.00 & 0.033 & 95 & 97.6\\
&1-2 & -0.02 & 0.023 & 93.2 & 40.8 & -0.01 & 0.025 & 94.8 & 59.7 & -0.00 & 0.035 & 93.6 & 80\\ 

               \hline
\multirow{8}{*}{(c)\phantom{$^1$}} & N-t & -0.00 & 0.013 & 94.6 & - & 0.00 & 0.023 & 94.8 & - & -0.00 & 0.048 & 94.8 & -\\
&N-j & -3.63 & 22.483 & 76 & - & -3.89 & 25.117 & 75 & - & -3.91 & 25.394 & 78.4 & -\\
&N-all & -3.31 & 11.404 & 0 & - & -3.54 & 13.006 & 0 & - & -3.88 & 15.635 & 0 & -\\
&Reg & 0.00 & 0.014 & 94.2 & - & 0.00 & 0.023 & 94.6 & - & -0.00 & 0.050 & 94.4 & -\\
&IPW & -0.00 & 0.014 & 94 & - & 0.00 & 0.024 & 94.2 & - & -0.00 & 0.050 & 92.6 & -\\
%&U1-1 & -0.03 & 0.021 & 93.6 & 99.7 & -0.01 & 0.041 & 93.8 & 100 & 0.00 & 0.086 & 94.4 & 100\\ 
%&U1-1/2 & -0.02 & 0.019 & 94.8 & 99.7 & -0.00 & 0.038 & 93.6 & 100 & -0.01 & 0.080 & 94.6 & 100\\ 
%&U1-2 & -0.03 & 0.018 & 93.5 & 85.7 & -0.01 & 0.030 & 92.6 & 97.4 & -0.00 & 0.064 & 93.8 & 99.2\\
&1-1 & -0.01 & 0.020 & 94 & 99.7 & 0.00 & 0.040 & 94.2 & 100 & 0.00 & 0.090 & 94.6 & 100\\ 
&1-1/2 & -0.01 & 0.018 & 94.8 & 99.7 & -0.01 & 0.032 & 94.2 & 100 & 0.00 & 0.065 & 94.8 & 99.9\\
&1-2 & -0.02 & 0.017 & 92.7 & 85.5 & -0.00 & 0.030 & 94.6 & 97.3 & -0.01 & 0.068 & 91.5 & 99.2\\
          \hline
\multirow{8}{*}{(d)\phantom{$^1$}} &N-t & -1.08 & 1.228 & 0.6 & - & -1.15 & 1.380 & 0.2 & - & -1.35 & 1.908 & 4.6 & -\\
&N-j & 0.01 & 0.025 & 95.6 & - & 0.01 & 0.026 & 94.4 & - & 0.01 & 0.023 & 94.4 & -\\
&N-all & -1.09 & 1.191 & 0 & - & -1.15 & 1.326 & 0 & - & -1.34 & 1.812 & 0 & -\\
&Reg & 0.00 & 0.014 & 95.8 & - & -0.00 & 0.016 & 93.2 & - & 0.00 & 0.036 & 94.4 & -\\
&IPW & -0.01 & 0.034 & 95.2 & - & 0.01 & 0.027 & 94.8 & - & 0.01 & 0.049 & 94.6 & -\\
%&U1-1 & -0.43 & 0.254 & 50 & 69.4 & -0.41 & 0.217 & 55.8 & 86.1 & -0.37 & 0.238 & 78.6 & 99.8\\
%&U1-1/2 & -0.44 & 0.251 & 46 & 69.4 & -0.40 & 0.212 & 54.8 & 86.1 & -0.37 & 0.234 & 74.6 & 99.8\\
%&U1-2 & -0.49 & 0.321 & 46.7 & 46.7 & -0.45 & 0.271 & 52 & 63 & -0.37 & 0.223 & 76 & 91.7\\
&1-1 & -0.05 & 0.024 & 98.3 & 67.5 & -0.06 & 0.026 & 99.2 & 84.7 & -0.04 & 0.068 & 95.2 & 99.6\\
&1-1/2 & -0.05 & 0.024 & 97.9 & 67.5 & -0.06 & 0.025 & 99 & 84.8 & -0.05 & 0.054 & 96.6 & 99.6\\
&1-2 & -0.06 & 0.025 & 98.5 & 43.1 & -0.06 & 0.027 & 97.8 & 58.9 & -0.05 & 0.055 & 96.5 & 86.7\\

               \hline
\multirow{8}{*}{(e)\phantom{$^1$}} & N-t & -3.13 & 10.172 & 0 & - & -3.33 & 11.519 & 0 & - & -3.58 & 13.284 & 0 & -\\
&N-j & -1.44 & 5.533 & 80.8 & - & -1.64 & 5.827 & 79.8 & - & -1.79 & 5.847 & 76.4 & -\\
&N-all & -3.91 & 15.355 & 0 & - & -4.20 & 17.698 & 0 & - & -4.59 & 21.124 & 0 & -\\
&Reg & 0.00 & 0.018 & 93.8 & - & 0.01 & 0.031 & 89.2 & - & 0.00 & 0.057 & 83.8 & -\\
&IPW & -0.12 & 0.133 & 90.8 & - & 0.00 & 0.056 & 96 & - & 0.03 & 0.053 & 95 & -\\
%&U1-1 & -0.72 & 0.601 & 24 & 55.6 & -0.68 & 0.530 & 22 & 77.8 & -0.65 & 0.508 & 35.2 & 95\\ 
%&U1-1/2 & -0.71 & 0.579 & 22.8 & 55.6 & -0.66 & 0.501 & 23 & 77.8 & -0.62 & 0.459 & 34.8 & 95\\ 
%&U1-2 & -0.72 & 0.623 & 34.1 & 34.9 & -0.72 & 0.601 & 27.6 & 54 & -0.68 & 0.549 & 31.6 & 74.5\\
&1-1 & -0.10 & 0.031 & 98.8 & 55.1 & -0.09 & 0.030 & 98.5 & 77.1 & -0.08 & 0.042 & 97.1 & 94.1\\
&1-1/2 & -0.10 & 0.029 & 98.7 & 55 & -0.09 & 0.030 & 98.5 & 77.1 & -0.08 & 0.039 & 97.1 & 94\\ 
&1-2 & -0.09 & 0.035 & 99.5 & 33.6 & -0.08 & 0.031 & 99.5 & 51.8 & -0.09 & 0.043 & 98.4 & 70.8\\

        \hline
        \end{tabular}
    }%
    \label{tab:sims_single}
    
\end{table}

\begin{table}[!t]
    \centering
    %\spacingset{1.12}
    \caption{Simulation results of single-unit  \textbf{carryover} effects. Bias, mean squared error (MSE), coverage of 95\% intervals (\%), and proportion of exposed time points being matched (\%). We show simulation results for the 3 na\"ive approaches and 3 matching estimators over 5 confounding scenarios, and for 3 exposure levels. `N' stands for `na\"ive'. The scenarios correspond to (a) No confounders, (b) Time-smooth confounders, (c) Location-varying confounders, (d) Time-varying confounders, and (e) All confounders.}
    % \rowcolors{5}{}{gray!10}
    %\spacingset{1}
    \resizebox{0.9\textwidth}{!}{%
       \begin{tabular}{*{14}{c}}
            \hline
            &&\multicolumn{4}{c}{Dense} & \multicolumn{4}{c}{Medium} & \multicolumn{4}{c}{Sparse} \\
            \cmidrule(lr){3-6} \cmidrule(lr){7-10} \cmidrule(lr){11-14}
            & Method & Bias & MSE & Cover & Prop & Bias & MSE & Cover & Prop & Bias & MSE & Cover & Prop \\
              \hline
             \multirow{8}{*}{(a)} &N-t & -8.29 & 70.166 & 0 & - & -8.06 & 66.064 & 0 & - & -7.79 & 61.750 & 0.8 & -\\
&N-j & -3.30 & 14.051 & 65.2 & - & -3.88 & 18.073 & 49.2 & - & -4.44 & 23.088 & 33.4 & -\\
&N-all & -9.20 & 85.015 & 0 & - & -9.30 & 86.723 & 0 & - & -9.62 & 92.725 & 0 & -\\
&Reg & -0.74 & 0.875 & 88.2 & - & 0.20 & 0.404 & 65.2 & - & 0.37 & 0.680 & 68.7 & -\\
&IPW & 1.50 & 3.389 & 14.6 & - & 0.04 & 0.347 & 58.8 & - & -0.47 & 0.356 & 88.4 & -\\
%&U1-1 & -0.15 & 0.042 & 84 & 86.6 & -0.21 & 0.143 & 87.4 & 99.4 & -0.32 & 0.823 & 76.6 & 99.9\\
%&U1-1/2 & -0.14 & 0.040 & 84.6 & 86.6 & -0.21 & 0.133 & 87 & 99.4 & -0.31 & 0.629 & 79.3 & 100\\ 
%&U1-2 & -0.29 & 0.117 & 68.4 & 39.1 & -0.26 & 0.159 & 82.8 & 82.4 & -0.35 & 0.572 & 77.6 & 98.2\\
&1-1 & -0.04 & 0.020 & 96 & 86.5 & -0.01 & 0.085 & 96.1 & 98.6 & -0.05 & 0.243 & 86.4 & 90.7\\
&1-1/2 & -0.04 & 0.019 & 96.4 & 86.5 & -0.03 & 0.074 & 96.2 & 98.7 & -0.00 & 0.405 & 82.2 & 87.8\\
&1-2 & -0.05 & 0.035 & 97.1 & 37.4 & -0.03 & 0.095 & 93.7 & 74.3 & -0.06 & 0.297 & 92 & 73.8\\
               \hline
\multirow{8}{*}{(b)\phantom{$^1$}} & N-t & -0.88 & 1.474 & 10.4 & - & -0.90 & 1.531 & 14.8 & - & -0.97 & 1.719 & 32.5 & -\\ 
&N-j & 0.01 & 0.064 & 94.6 & - & -0.00 & 0.069 & 93.8 & - & 0.03 & 0.054 & 94 & -\\ 
&N-all & -0.84 & 0.706 & 0 & - & -0.87 & 0.762 & 0 & - & -0.93 & 0.878 & 0 & -\\ 
&Reg & 0.01 & 0.110 & 68.8 & - & 0.02 & 0.184 & 59 & - & 0.04 & 0.261 & 72.4 & -\\
&IPW & 0.25 & 0.361 & 64.8 & - & 0.08 & 0.109 & 84.6 & - & 0.05 & 0.112 & 92.3 & -\\
%&U1-1 & -0.15 & 0.070 & 89 & 35 & -0.13 & 0.079 & 90.4 & 64.4 & -0.18 & 0.192 & 89.5 & 92.4\\
%&U1-1/2 & -0.14 & 0.069 & 87 & 35 & -0.12 & 0.072 & 90.6 & 64.4 & -0.17 & 0.164 & 90.4 & 92.4\\
%&U1-2 & -0.32 & 0.231 & 79.8 & 12.2 & -0.27 & 0.198 & 82.3 & 25.9 & -0.27 & 0.238 & 85.5 & 61.1\\
&1-1 & -0.04 & 0.042 & 95.6 & 34.7 & -0.03 & 0.053 & 94.6 & 63.1 & -0.00 & 0.126 & 96 & 87.3\\
&1-1/2 & -0.04 & 0.043 & 95.4 & 34.7 & -0.03 & 0.056 & 94.4 & 63.1 & -0.01 & 0.129 & 95.1 & 87.8\\
&1-2 & -0.06 & 0.124 & 94 & 10.7 & -0.01 & 0.122 & 94.2 & 22.7 & -0.00 & 0.159 & 93.1 & 49\\ 

               \hline
\multirow{8}{*}{(c)\phantom{$^1$}} & N-t & -0.03 & 0.059 & 95.2 & - & -0.06 & 0.205 & 95 & - & -0.06 & 0.333 & 94.4 & -\\
&N-j & -6.93 & 56.281 & 42.9 & - & -7.42 & 62.293 & 43.2 & - & - & - & - & -\\
&N-all & -6.62 & 45.379 & 0 & - & -7.39 & 56.350 & 0 & - & -8.43 & 74.005 & 0.6 & -\\
&Reg & -0.01 & 0.053 & 94 & - & -0.02 & 0.304 & 93.6 & - & -0.10 & 0.442 & 91.2 & -\\
&IPW & -0.01 & 0.063 & 92.4 & - & -0.00 & 0.302 & 78 & - & -0.04 & 0.448 & 67 & -\\
%&U1-1 & -0.22 & 0.154 & 84.4 & 97.4 & -0.35 & 0.523 & 84.5 & 99.9 & -0.30 & 0.664 & 84.5 & 100\\ 
%&U1-1/2 & -0.20 & 0.125 & 85.9 & 97.4 & -0.28 & 0.402 & 86.1 & 99.9 & -0.29 & 0.564 & 83.1 & 100\\ 
%&U1-2 & -0.26 & 0.153 & 82 & 74.6 & -0.30 & 0.365 & 87.9 & 94.6 & -0.33 & 0.686 & 74.3 & 96.1\\
&1-1 & -0.02 & 0.074 & 95.1 & 96.4 & 0.00 & 0.152 & 95.3 & 94.8 & 0.06 & 0.139 & 93.8 & 91.5\\
&1-1/2 & -0.05 & 0.058 & 96.4 & 96.6 & -0.02 & 0.205 & 94.4 & 95.3 & -0.03 & 0.261 & 95.8 & 91.5\\
&1-2 & -0.03 & 0.076 & 95.2 & 67.4 & -0.03 & 0.231 & 86.8 & 76.8 & -0.02 & 0.092 & 100 & 75.7\\

          \hline
\multirow{8}{*}{(d)\phantom{$^1$}} &N-t & -1.05 & 1.158 & 0.2 & - & -1.13 & 1.354 & 0.4 & - & -1.34 & 2.099 & 55.4 & -\\
&N-j & 0.03 & 0.041 & 93.4 & - & 0.02 & 0.036 & 93.8 & - & 0.04 & 0.031 & 94 & -\\
&N-all & -1.02 & 1.035 & 0 & - & -1.11 & 1.241 & 0 & - & -1.28 & 1.638 & 0 & -\\
&Reg & -0.01 & 0.019 & 93.4 & - & -0.00 & 0.024 & 95 & - & -0.00 & 0.186 & 95.7 & -\\
&IPW & 0.23 & 0.134 & 80.4 & - & 0.11 & 0.068 & 90.2 & - & -0.00 & 0.231 & 85.9 & -\\
%&U1-1 & -0.11 & 0.106 & 93.2 & 42.2 & -0.10 & 0.093 & 92.2 & 64.1 & -0.22 & 0.463 & 89.5 & 98.9\\
%&U1-1/2 & -0.11 & 0.108 & 92.2 & 42.2 & -0.09 & 0.085 & 93.4 & 64.1 & -0.19 & 0.386 & 88.2 & 99\\ 
%&U1-2 & -0.19 & 0.298 & 91 & 16.3 & -0.18 & 0.225 & 90.4 & 26.8 & -0.22 & 0.385 & 85.3 & 83.8\\
&1-1 & -0.05 & 0.041 & 98.2 & 42 & -0.04 & 0.048 & 98.4 & 63.1 & -0.04 & 0.204 & 93 & 89.6\\
&1-1/2 & -0.05 & 0.040 & 98.2 & 41.9 & -0.04 & 0.046 & 98.8 & 63.2 & -0.05 & 0.226 & 93.2 & 91.3\\
&1-2 & -0.05 & 0.076 & 98.8 & 14.8 & -0.03 & 0.103 & 98.1 & 23.7 & -0.03 & 0.251 & 91.4 & 62.1\\

               \hline
\multirow{8}{*}{(e)\phantom{$^1$}} & N-t & -3.49 & 12.943  & 0&- & -3.70 & 14.497 & 0 & - & -3.78 & 15.149 & 0.6 & -\\
&N-j & -2.70 & 10.445 & 65.2 & - & -3.28 & 13.783 & 49.2 & - & -3.84 & 18.119 & 33.4 & -\\
&N-all & -4.34 & 18.866 & 0 & - & -4.84 & 23.566 & 0 & - & -5.54 & 30.895 & 0 & -\\
&Reg & 0.01 & 0.078 & 74.2 & - & 0.02 & 0.166 & 61.8 & - & 0.03 & 0.263 & 69.9 & -\\
&IPW & 1.49 & 2.935 & 18 & - & 0.52 & 0.503 & 60.8 & - & 0.16 & 0.127 & 88.2 & -\\
%&U1-1 & -0.04 & 0.139 & 94 & 31 & -0.04 & 0.113 & 93.2 & 53.7 & -0.09 & 0.184 & 92.3 & 86.6\\
%&U1-1/2 & -0.05 & 0.128 & 95.4 & 31 & -0.04 & 0.104 & 93 & 53.7 & -0.07 & 0.169 & 93.3 & 86.6\\
%&U1-2 & -0.04 & 0.415 & 90 & 10.8 & -0.12 & 0.279 & 92.2 & 20.9 & -0.13 & 0.304 & 89.2 & 48.8\\
&1-1 & -0.02 & 0.041 & 99.4 & 30.7 & -0.02 & 0.049 & 98.8 & 53 & -0.01 & 0.095 & 98.3 & 83.1\\
&1-1/2 & -0.02 & 0.039 & 99.6 & 30.7 & -0.02 & 0.051 & 97.6 & 53.1 & -0.01 & 0.097 & 97.5 & 83.3\\
&1-2 & -0.01 & 0.100 & 100 & 9.6 & -0.03 & 0.117 & 98.3 & 19 & -0.00 & 0.135 & 98.6 & 42.2\\

        \hline
        \end{tabular}
    }%
    \label{tab:sims_single_carryover}
    
\end{table}

The performance across different confounding scenarios closely mirrors the results from the simulation study in \cref{sec:simu_real_data}. For that reason, we omit a detailed summary here on the relative performance of the methods. We instead focus on comparing the performance of the proposed estimators under the different levels of exposure density.

In terms of MSE, the \texttt{1-1} estimator performs best or close to best in the dense scenarios, and the \texttt{1-2} estimator performs best or close to best in the sparse scenarios. The \texttt{1-1/2} estimator performs as well, or close to as-well as the best estimator across all scenarios considered, while maintaining a high proportion of matched exposed time periods.

As in the simulations of \cref{sec:simu_real_data}, the 1-1 and 1-1/2 algorithms yield the same proportion of matched exposed time periods, as the matches generated under the 1-1 algorithm are also possible under the 1-1/2 algorithm.
For all three algorithms, the proportion of matched exposed time periods varies with the sparsity level of the exposure, with higher rates of matched time periods under sparser exposures. 
As expected, the \texttt{1-2} algorithm returns the smallest proportion of matched exposed time periods. Combined with the fact that it has {close to} the lowest MSE in sparse scenarios, this illustrates that the 1-2 estimator returns more accurate predictions for the missing potential outcomes compared to the 1-1 and 1-1/2 algorithms.
The coverage of 95\% intervals for all three estimators  
is close to or above nominal across all scenarios.

The overcoverage observed in some scenarios for all estimators can likely be explained by the fact that our variance estimator is biased upwards (see Supplement~\ref{supp_sec:variance estimation}). In a few settings, we find that coverage of the immediate effect in \cref{tab:sims_single} is below the nominal level. This is more often the case in sparse scenarios. Based on Q-Q plots of our estimates across data sets, we believe that this undercoverage might be explained by violations of the normality approximation, possibly due to the small number of observations used in estimation in the sparse scenarios.

\if{
In the case of no confounding, all estimators are unbiased. In the presence of location-varying confounding (scenario (c)), the Na\"ive-$j$ estimator that is based on comparing outcomes across units is biased. In contrast, under temporal confounding (scenarios (b) and (d)), the Na\"ive-$t$  estimator is biased. Na\"ive-all is biased under both confounding structures. Unadjusted matching estimators are unbiased with nominal coverage in the absence of {\it non-smooth} time-varying confounders, even in the presence of confounding temporal trends, in cases (a), (b) and (c). However, in the scenarios with non-smooth time-varying confounding, unadjusted estimators will be biased. The proportion of exposed time points that are matched using unadjusted approaches is only slightly higher than the proportion for adjusted approaches in some scenarios. In contrast, the matching estimators have minimal bias in all cases.

}\fi

\subsection{Testing the global null hypothesis} \label{subsec:multi-simu}

For the scenarios with temporal confounding (b, d, and e), we consider simulations where all immediate effects are set to zero under a dense frequency for the exposure, and evaluate the properties of the inferential technique of Supplement~\ref{subsec:global_null} for testing the global null hypothesis of no immediate effect in the presence of multiple outcome units at the 0.05 level.
% as in (a) and (c) na\"ive-$t$ should also be valid for multiple hypothesis testing. 
We alter the simulations to impose that the global null holds, and impose that  $\tau_j^\text{imm}=0$ for all outcome units. %In other words, the coefficient for $E_{t}$ should be 0 in the outcome models.
We generate $500$ data sets for each one of the three scenarios. We acquire point estimates and p-values for the exposure effect on each of the 200 outcome units, and adjust the p-values using the procedure detailed in Supplement~\ref{subsec:global_null}.

The optimizer returned a solution for approximately 85\% of the outcome units for each matching method.
The results are in \cref{tab:sims_multiple}. We report the average estimated effect across outcome units with matches and across data sets, the proportion of the available p-values across the outcome units and data sets that are below 0.05, the proportion of data sets where any of the available p-values is below 0.05, and the proportion of data sets where any of the FDR-adjusted p-values is below 0.05. 
\begin{table}[!t]
%\spacingset{1}
\centering

\caption{Simulation results for testing the global null for the {immediate} effect using the Na\"ive-$t$ and the estimator based on the three proposed algorithms for scenarios (b), (d) and  (e), and under a dense exposure level.}
    \label{tab:sims_multiple}
    % \rowcolors{5}{}{gray!10}
    % \spacingset{1} % Assuming this command is defined elsewhere
    
    \resizebox{0.9\textwidth}{!}{%
        \begin{tabular}{cccccc}
& & \begin{tabular}[c]{@{}c@{}}Average  \\ estimator mean\end{tabular} & \begin{tabular}[c]{@{}c@{}}Average rate of \\ p-value $\leq$ 0.05\end{tabular} & \begin{tabular}[c]{@{}c@{}}Rate of \\ min(p-value) $\leq$ 0.05\end{tabular} & \begin{tabular}[c]{@{}c@{}}Rate of FDR \\ min (adj.p-value) $\leq$ 0.05\end{tabular} \\ 
\cmidrule(lr){3-3} \cmidrule(lr){4-4} \cmidrule(lr){5-5} \cmidrule(lr){6-6}
\multirow{9}{*}{(b)}          
&Na"ive-$t$ & -0.903 & 0.904 & 1.000 & 1.000 \\ 
&  Reg & -0.000 & 0.063 & 1.000 & 0.106 \\ 
&  IPW & 0.008 & 0.059 & 1.000 & 0.149 \\ 
&  1-1 & -0.005 & 0.052 & 1.000 & 0.078 \\ 
&  1-1/2 & -0.005 & 0.053 & 1.000 & 0.074 \\ 
 & 1-2 & -0.000 & 0.054 & 1.000 & 0.102 \\ 
 & NW1-1 & -0.005 & 0.004 & 0.524 & 0.000 \\ 
&  NW1-1/2 & -0.005 & 0.004 & 0.524 & 0.004 \\ 
&  NW1-2 & -0.000 & 0.004 & 0.538 & 0.010 \\ \hline
\multirow{6}{*}{(d)}          
&Na\"ive-$t$ & -1.199 & 0.999 & 1.000 & 1.000 \\ 
&  Reg & -0.000 & 0.049 & 1.000 & 0.058 \\ 
&  IPW & -0.038 & 0.066 & 1.000 & 0.204 \\  
&  1-1 & -0.045 & 0.014 & 0.900 & 0.010 \\ 
&  1-1/2 & -0.046 & 0.014 & 0.922 & 0.008 \\ 
&  1-2 & -0.044 & 0.013 & 0.854 & 0.002 \\ 
 \hline 
\multirow{6}{*}{(e)}          
 &Na"ive-$t$ & -3.321 & 1.000 & 1.000 & 1.000 \\ 
 & Reg & 0.001 & 0.055 & 1.000 & 0.068 \\ 
 & IPW & -0.191 & 0.127 & 1.000 & 0.746 \\ 
 & 1-1 & -0.082 & 0.000 & 0.000 & 0.000 \\ 
 & 1-1/2 & -0.083 & 0.000 & 0.000 & 0.000 \\ 
 & 1-2 & -0.075 & 0.000 & 0.002 & 0.000 \\ 
    \hline 
\end{tabular}%
    }%
\end{table}

Since Na\"ive-$t$ is biased in the presence of temporal confounding, its inferential performance suffers, and using this estimator would mistakenly reject the global null hypothesis every time, using FDR-corrected p-values or not. 
%%Additionally, in \cref{tab:sims_multiple_unadj}, we showed that matching without adjustment does not guarantee an unbiased estimator when the scenarios involve time-varying confounders.
For the proposed estimators, up to 6 \% of the p-values across outcome units and data sets are below 0.05. % Since 5\% of outcome units have p-values below 0.05, 
%Therefore, it is not surprising that the minimum (unadjusted) p-value across outcome units is very often below 0.05. This illustrates that controlling for multiple comparisons is necessary in order to maintain the level of the test, especially with a large number of outcome units.
Hence, it is not surprising that the minimum (unadjusted) p-value across outcome units is below 0.05 in most data sets, emphasizing the necessity of controlling for multiple comparisons to maintain the level of the test, especially with a large number of outcome units.
%%For case (b), the second column of \cref{tab:sims_multiple} indicates that for all matching methods the null hypothesis is true since, on average, $5\%$ of outcome units have p-values below 0.05 in each data set. 
% When using the FDR-adjusted p-values, the inferential approach for all matching algorithms respects the target level of the hypothesis test, and maintains that the rate with which the minimum p-value is below 0.05 is controlled. 
With FDR-adjusted p-values, the rate of rejection of the null hypothesis is much closer to the target level of the test. 
%%The application of FDR adjustment warrants that only 5\% of the multiple hypothesis tests report false rejection. Such a low possibility makes us 95\% confident regarding which hypotheses should be rejected. Hence, whenever a less than 0.05 adjusted p-value is observed, its corresponding null hypothesis is very likely to be rejected.
%In cases (d) and (e), matching methods are relatively conservative in rejecting the global null hypothesis, aligned with the fact that hypothesis tests for individual outcome units are conservative. 
Matching methods in cases (d) and (e) are conservative in rejecting the global null hypothesis, consistent with the conservative nature of individual outcome unit hypothesis tests (also discussed in Supplement~\ref{supp_sec:variance estimation}).
%% This can be well solved by balancing within-match covariates, though efficiency may be reduced thereby. After all, this can bring enough confidence in rejecting any of the hypotheses as long as its FDR-adjusted p-value is below 0.05. On the contrary, na\"ive-$t$ and matching without adjustment do not offer an unbiased estimator of true effect, and consequently their adjusted p-values are not reliable.

% The FDR control shows slight undercoverage, as indicated by the rate of minimum adjusted p-values below 0.05. 
In scenario (b), we also considered the Newey-West variance estimator in acquiring the p-values for each outcome unit before applying the FDR correction in testing the global null hypothesis. This is shown in \cref{tab:sims_multiple} as \texttt{NW1-1}, \texttt{NW1-1/2}, and \texttt{NW1-2}. Using the Newey-West variance estimator yields fewer p-values below 0.05, resulting in substantially lower rates of both minimum unadjusted and adjusted p-values falling below the threshold. We investigate the performance of the Newey-West variance estimator in simulations in more detail in Supplement~\ref{supp_subsec:sims_temp_corr}.

\subsection{Algorithms balancing time but not the time-varying covariates}
\label{appendix:unadj matching}

We consider the three algorithms, 1-1, 1-2, and 1-1/2, with balance constraint on the time of matches, but not on the time-varying covariates. We call these algorithms and the corresponding estimators as {\it unadjusted}. We consider these approaches in order to investigate the impact of ignoring time-varying confounding factors within the proposed framework, and the extent to which the proposed framework can adequately control for smooth temporal trends without explicitly specifying them. We evaluate the performance of the unadjusted estimators in simulations under the generative scenarios described in Supplement~\ref{supp_subsec:simu_setup}. For brevity, we focus here on the performance of the estimator for the immediate effect.

In \cref{tab:sims_single_unadj} we show the bias, mean squared error, coverage, and proportion of matched exposed time periods when applying the three unadjusted matching algorithms to estimate the immediate effect. We find that, as expected, the estimators are unbiased in the absence of confounding (scenario (a)), in the presence of smooth temporal trends only (scenario (b)), or in the presence of location-varying confounding only (scenario (c)).
We also find that, in these simulations, these estimators which do not adjust for any measured covariates have smaller bias compared to the na\"ive approaches, likely thanks to the implicit adjustment for the smooth temporal trend.

\begin{table}[!t]
    \centering
    %\spacingset{1.12}
    \caption{Bias, mean squared error (MSE), coverage of 95\% intervals (\%), and proportion of exposed time points being matched (\%). We show simulation results for the three estimators of the immediate effect based on the unadjusted algorithms that balance time but not time-varying covariates. `U' stands for `unadjusted'. The scenarios correspond to (a) No confounders, 
    (b) Time-smooth confounders, (c) Location-varying confounders, (d) Time-varying confounders, and (e) All confounders.}
    % \rowcolors{5}{}{gray!10}
    %\spacingset{1}
     \resizebox{0.9\textwidth}{!}{%
       \begin{tabular}{*{14}{c}}
            \hline
            &&\multicolumn{4}{c}{Dense} & \multicolumn{4}{c}{Medium} & \multicolumn{4}{c}{Sparse} \\
            \cmidrule(lr){3-6} \cmidrule(lr){7-10} \cmidrule(lr){11-14}
            & Method & Bias & MSE & Cover & Prop & Bias & MSE & Cover & Prop & Bias & MSE & Cover & Prop \\
              \hline
             \multirow{3}{*}{(a)} 
&U1-1 & -0.04 & 0.014 & 93.2 & 97.8 & -0.03 & 0.021 & 94.4 & 100 & -0.01 & 0.039 & 95.2 & 100\\ 
&U1-1/2 & -0.03 & 0.013 & 94.4 & 97.8 & -0.03 & 0.018 & 94.2 & 100 & -0.02 & 0.035 & 95.2 & 100\\ 
&U1-2 & -0.07 & 0.019 & 90.8 & 62.3 & -0.03 & 0.017 & 94.2 & 90.8 & -0.02 & 0.030 & 94.4 & 98.7\\ 
               \hline
\multirow{3}{*}{(b)\phantom{$^1$}} 
&U1-1 & -0.03 & 0.021 & 94.4 & 62.5 & -0.03 & 0.026 & 93.2 & 85.1 & -0.02 & 0.037 & 94.8 & 97.8\\
&U1-1/2 & -0.02 & 0.021 & 94.2 & 62.5 & -0.02 & 0.023 & 94.4 & 85.1 & -0.02 & 0.034 & 94.8 & 97.8\\
&U1-2 & -0.04 & 0.026 & 92.3 & 40.9 & -0.03 & 0.026 & 92.8 & 59.9 & -0.03 & 0.034 & 95.2 & 80.6\\

               \hline
\multirow{3}{*}{(c)\phantom{$^1$}} 
&U1-1 & -0.03 & 0.021 & 93.6 & 99.7 & -0.01 & 0.041 & 93.8 & 100 & 0.00 & 0.086 & 94.4 & 100\\ 
&U1-1/2 & -0.02 & 0.019 & 94.8 & 99.7 & -0.00 & 0.038 & 93.6 & 100 & -0.01 & 0.080 & 94.6 & 100\\ 
&U1-2 & -0.03 & 0.018 & 93.5 & 85.7 & -0.01 & 0.030 & 92.6 & 97.4 & -0.00 & 0.064 & 93.8 & 99.2\\
          \hline
\multirow{3}{*}{(d)\phantom{$^1$}} 
&U1-1 & -0.43 & 0.254 & 50 & 69.4 & -0.41 & 0.217 & 55.8 & 86.1 & -0.37 & 0.238 & 78.6 & 99.8\\
&U1-1/2 & -0.44 & 0.251 & 46 & 69.4 & -0.40 & 0.212 & 54.8 & 86.1 & -0.37 & 0.234 & 74.6 & 99.8\\
&U1-2 & -0.49 & 0.321 & 46.7 & 46.7 & -0.45 & 0.271 & 52 & 63 & -0.37 & 0.223 & 76 & 91.7\\
               \hline
\multirow{3}{*}{(e)\phantom{$^1$}} 
&U1-1 & -0.72 & 0.601 & 24 & 55.6 & -0.68 & 0.530 & 22 & 77.8 & -0.65 & 0.508 & 35.2 & 95\\ 
&U1-1/2 & -0.71 & 0.579 & 22.8 & 55.6 & -0.66 & 0.501 & 23 & 77.8 & -0.62 & 0.459 & 34.8 & 95\\ 
&U1-2 & -0.72 & 0.623 & 34.1 & 34.9 & -0.72 & 0.601 & 27.6 & 54 & -0.68 & 0.549 & 31.6 & 74.5\\
        \hline
        \end{tabular}
    }%
    \label{tab:sims_single_unadj}

\end{table}

We also evaluated the performance of the inferential procedure for testing the global null of no immediate effect with unadjusted matching in the presence of multiple outcome units, similarly to the simulations for the proposed estimators in Supplement~\ref{subsec:multi-simu}. The results are shown in \cref{tab:sims_multiple_unadj}. The unadjusted methods give reliable inferences for testing the global null in the presence of only time-smooth confounders. In the presence of non-smooth temporal confounding, the unadjusted estimators' bias will lead to identifying statistically significant causal effects too often.

\begin{table}[!t]
%\spacingset{1}
\centering
\caption{Simulation results for using the three unadjusted algorithms and estimators to test the global null hypothesis of no immediate causal effect for scenarios (b), (d) and  (e), and under a medium exposure level. `U' stands for `unadjusted' and `NW' stands for `Newey-West'. The scenarios correspond to 
    (b) Time-smooth confounders, (d) Time-varying confounders, and (e) All confounders.}
    \label{tab:sims_multiple_unadj}
    % \rowcolors{5}{}{gray!10}
    % \spacingset{1} % Assuming this command is defined elsewhere
    \resizebox{0.9\textwidth}{!}{%
        \begin{tabular}{cccccc}
& & \begin{tabular}[c]{@{}c@{}}Average  \\ estimator mean\end{tabular} & \begin{tabular}[c]{@{}c@{}}Average rate of \\ p-value $\leq$ 0.05\end{tabular} & \begin{tabular}[c]{@{}c@{}}Rate of \\ min(p-value) $\leq$ 0.05\end{tabular} & \begin{tabular}[c]{@{}c@{}}Rate of FDR \\ min (adj.p-value) $\leq$ 0.05\end{tabular} \\ 
\cmidrule(lr){3-3} \cmidrule(lr){4-4} \cmidrule(lr){5-5} \cmidrule(lr){6-6}
\multirow{3}{*}{(b)}    
 &  U1-1 & -0.004 & 0.053 & 1.000 & 0.078 \\ 
&  U1-1/2 & -0.004 & 0.052 & 1.000 & 0.068 \\ 
&  U1-2 & -0.001 & 0.054 & 1.000 & 0.098 \\ 
& NWU1-1 & -0.004 & 0.004 & 0.512 & 0.002 \\ 
&  NWU1-1/2 & -0.003 & 0.004 & 0.526 & 0.008 \\ 
 & NWU1-2 & -0.001 & 0.004 & 0.544 & 0.004 \\  \hline 
\multirow{3}{*}{(d)}          
&  U1-1 & -0.422 & 0.486 & 1.000 & 1.000 \\ 
&  U1-1/2 & -0.421 & 0.504 & 1.000 & 1.000 \\ 
&  U1-2 & -0.447 & 0.449 & 1.000 & 0.996 \\ \hline 
\multirow{3}{*}{(e)}          
 & U1-1 & -1.073 & 0.998 & 1.000 & 1.000 \\ 
 & U1-1/2 & -1.044 & 0.998 & 1.000 & 1.000 \\ 
 & U1-2 & -1.098 & 0.995 & 1.000 & 1.000 \\ \hline 
\end{tabular}%
    }%
\end{table}

\subsection{Analysis of tuning parameters}\label{appendix:tuning parameters}

Following the simulation study in this section, we consider two additional choices of the algorithmic parameters $(\delta,\delta',\epsilon)$ per simulation. The ``strict'' choice of algorithmic parameters corresponds to $(\delta,\delta',\epsilon) = (0,0.05,2)$ imposing tight constraints on time and covariates, and the  ``loose'' choice of algorithmic parameters $(\delta,\delta',\epsilon) = (2,0.1,6)$ corresponds to looser constraints.

The results for the immediate effect are shown in \cref{tab:sims_tuning}, along with the results based on the default setting $(\delta,\delta',\epsilon)=(2,0.05,6)$ (the results for which are also shown in \cref{tab:sims_single}). Comparing the results under the strict and default values allows us to investigate the impact of $(\delta,\epsilon)$ that correspond to tuning parameters of time, while comparing the results under the default and loose values allows us to investigate the impact of $\delta'$ that is a tuning parameter for the measured covariates.  For brevity, we focus on the immediate effect estimates.

Some conclusions from this comparison are discussed in \cref{subsec:additional_simulation}. Here, we also point out that, even though the estimator based on the \texttt{1-2} algorithm performed best under sparse scenarios in \cref{tab:sims_single}, when $(\delta,\epsilon)$ are small under strict tuning parameters, this approach no longer performs best. That is because of the large variability for the corresponding \texttt{1-2} estimator since the number of available unexposed time periods to form matches is smaller and the proportion of matched exposed time points is almost half of that based on the \texttt{1-1} or \texttt{1-1/2} algorithms. % Matching 1-2 is also the worst in the scenario (e), since almost no periods are matched, resulting in the information loss.

\begin{table}[htbp!]
    \centering
    %\spacingset{1.12}
    \caption{Simulation results for estimating the immediate effect under different values of the tuning parameters. The tuning parameters in this table correspond to 
    (1) strict: $(\delta, \delta', \epsilon) = (0, 0.05, 2)$, 
    (2) medium: $(\delta, \delta', \epsilon) = (2, 0.05, 6)$ (the results in \cref{tab:sims_single} correspond to this choice of tuning parameters),
    (3) loose: $(\delta, \delta', \epsilon) = (2, 0.1, 6)$.}
    % \label{tab:sims_tuning}
    % \rowcolors{5}{}{gray!10}
    %\spacingset{0.9}
    \resizebox{0.8\textwidth}{!}{%
       \begin{tabular}{cl*{12}{c}}
            \hline
            &&\multicolumn{4}{c}{Dense} & \multicolumn{4}{c}{Medium} & \multicolumn{4}{c}{Sparse} \\
            \cmidrule(lr){3-6} \cmidrule(lr){7-10} \cmidrule(lr){11-14}
            Method&  & Bias & MSE & Cover & Prop & Bias & MSE & Cover & Prop & Bias & MSE & Cover & Prop \\
              \hline
              \hline
         \\[-10pt]
        \multicolumn{14}{c}{(a) No confounders} \\[0pt]
        \\[-10pt]
        \hline
        \multirow{3}{*}{1-1} & (1) & -0.02 & 0.013 & 96.0 & 83.5 & -0.02 & 0.021 & 94.2 & 97.3 & -0.02 & 0.039 & 94.3 & 99.8 \\
& (2) & -0.02 & 0.013 & 94.0 & 97.8 & -0.02 & 0.020 & 95.6 & 100.0 & -0.01 & 0.040 & 95.8 & 100 \\
& (3) & -0.03 & 0.013 & 94.2 & 97.8 & -0.03 & 0.021 & 95.6 & 100.0 & -0.02 & 0.037 & 95.8 & 100 \\
\hline
\multirow{3}{*}{1-1/2} & (1) & -0.02 & 0.014 & 94.3 & 83.5 & -0.03 & 0.020 & 94.4 & 97.3 & -0.02 & 0.037 & 94.3 & 99.8 \\
& (2) & -0.02 & 0.012 & 95.4 & 97.8 & -0.03 & 0.017 & 94.6 & 100.0 & -0.02 & 0.032 & 95 & 100 \\
& (3) & -0.03 & 0.013 & 94.0 & 97.8 & -0.03 & 0.019 & 94 & 100.0 & -0.02 & 0.032 & 93.7 & 100 \\
\hline
\multirow{3}{*}{1-2} & (1) & -0.03 & 0.023 & 94.6 & 39.7 & -0.02 & 0.023 & 94.6 & 63.8 & -0.01 & 0.037 & 93.7 & 81.9 \\
& (2) & -0.02 & 0.015 & 95.6 & 61.9 & -0.03 & 0.017 & 93.9 & 90.6 & -0.01 & 0.029 & 94.9 & 98.7 \\
& (3) & -0.04 & 0.017 & 92.6 & 62.3 & -0.03 & 0.018 & 95.2 & 90.8 & -0.02 & 0.028 & 94.9 & 98.7 \\
               \hline
               \hline
                 \\[-10pt]
        \multicolumn{14}{c}{(b) Time-smooth confounders} \\[0pt]
        \\[-10pt]
        \hline
\multirow{3}{*}{1-1} & (1) & -0.01 & 0.021 & 95.0 & 57.1 & -0.01 & 0.028 & 94.6 & 76.9 & -0.00 & 0.039 & 95.7 & 91.1 \\
& (2) & -0.02 & 0.019 & 93.8 & 62.5 & -0.02 & 0.023 & 95 & 85.0 & -0.00 & 0.037 & 95 & 97.6 \\
& (3) & -0.03 & 0.021 & 94.4 & 62.5 & -0.02 & 0.025 & 94.2 & 85.1 & -0.02 & 0.039 & 94.0 & 97.8 \\
\hline
\multirow{3}{*}{1-1/2} & (1) & -0.01 & 0.021 & 95.0 & 57.1 & -0.01 & 0.026 & 95 & 76.9 & 0.00 & 0.037 & 94.3 & 91.1 \\
& (2) & -0.02 & 0.020 & 94.4 & 62.5 & -0.01 & 0.023 & 93.8 & 85.0 & -0.00 & 0.033 & 95 & 97.6 \\
& (3) & -0.02 & 0.019 & 94.2 & 62.5 & -0.03 & 0.024 & 93.4 & 85.1 & -0.01 & 0.035 & 95.7 & 97.7 \\
\hline
\multirow{3}{*}{1-2} & (1) & -0.00 & 0.034 & 94.2 & 27.9 & -0.01 & 0.038 & 94.4 & 42.3 & 0.01 & 0.048 & 95.1 & 58.5 \\
& (2) & -0.02 & 0.023 & 93.2 & 40.8 & -0.01 & 0.025 & 94.8 & 59.7 & -0.00 & 0.035 & 93.6 & 80.0 \\
& (3) & -0.03 & 0.023 & 94.6 & 40.9 & -0.03 & 0.026 & 94 & 59.9 & -0.02 & 0.034 & 93.7 & 80.4 \\
\hline
        \hline
          \\[-10pt]
        \multicolumn{14}{c}{(c) Location-varying confounders} \\[0pt]
        \\[-10pt]
        \hline
   \multirow{3}{*}{1-1}&     (1) &-0.02 & 0.023 & 94.8 & 95 & -0.01 & 0.041 & 94.8 & 99.4 & 0.01 & 0.075 & 95.3 & 99.4\\
&(2) & -0.01 & 0.020 & 94 & 99.7 & 0.00 & 0.040 & 94.2 & 100 & 0.00 & 0.090 & 94.6 & 100\\ 
&(3) & -0.02 & 0.020 & 95.2 & 99.7 & 0.00 & 0.037 & 95.6 & 100 & -0.01 & 0.079 & 93.7 & 100\\ 
 \hline
  \multirow{3}{*}{1-1/2}&(1) & -0.02 & 0.021 & 93.5 & 95 & 0.00 & 0.036 & 94.6 & 99.4 & -0.01 & 0.085 & 94.9 & 99.5\\
& (2) & -0.01 & 0.018 & 94.8 & 99.7 & -0.01 & 0.032 & 94.2 & 100 & 0.00 & 0.065 & 94.8 & 99.9\\
& (3) & -0.01 & 0.020 & 94.4 & 99.7 & -0.00 & 0.032 & 94.6 & 100 & -0.01 & 0.070 & 94.3 & 99.9\\
  \hline
  \multirow{3}{*}{1-2}&(1) & -0.01 & 0.026 & 93.5 & 59.3 & 0.00 & 0.037 & 94.4 & 78.9 & -0.00 & 0.074 & 94.2 & 87.6\\
& (2) & -0.02 & 0.017 & 92.7 & 85.5 & -0.00 & 0.030 & 94.6 & 97.3 & -0.01 & 0.068 & 91.5 & 99.2\\
& (3) & -0.03 & 0.017 & 92.7 & 85.6 & -0.01 & 0.030 & 93.6 & 97.4 & -0.01 & 0.071 & 93.1 & 99.1\\
\hline
\hline
  \\[-10pt]
        \multicolumn{14}{c}{(d) Time-varying confounders} \\[0pt]
        \\[-10pt]
        \hline
 \multirow{3}{*}{1-1}&     (1) &-0.06 & 0.026 & 98.8 & 57.7 & -0.06 & 0.030 & 98.4 & 72.9 & -0.05 & 0.071 & 96.3 & 90.7\\
& (2) & -0.05 & 0.024 & 98.3 & 67.5 & -0.06 & 0.026 & 99.2 & 84.7 & -0.04 & 0.068 & 95.2 & 99.6\\
& (3) & -0.11 & 0.042 & 96.2 & 68.4 & -0.10 & 0.040 & 94.8 & 85.6 & -0.06 & 0.069 & 94.2 & 99.7\\ 
 \hline
  \multirow{3}{*}{1-1/2}&(1) & -0.06 & 0.025 & 98.8 & 57.7 & -0.06 & 0.030 & 99 & 73 & -0.04 & 0.068 & 97 & 90.9\\
& (2) & -0.05 & 0.024 & 97.9 & 67.5 & -0.06 & 0.025 & 99 & 84.8 & -0.05 & 0.054 & 96.6 & 99.6\\
& (3) & -0.11 & 0.045 & 95.6 & 68.5 & -0.11 & 0.042 & 95.8 & 85.6 & -0.07 & 0.062 & 95.1 & 99.8
 \\ 
  \hline
  \multirow{3}{*}{1-2}&(1) & -0.05 & 0.037 & 98.2 & 26.5 & -0.06 & 0.042 & 98.8 & 37.2 & -0.03 & 0.074 & 97.5 & 55.6\\
& (2) & -0.06 & 0.025 & 98.5 & 43.1 & -0.06 & 0.027 & 97.8 & 58.9 & -0.05 & 0.055 & 96.5 & 86.7\\
& (3) & -0.12 & 0.047 & 96.8 & 44.6 & -0.11 & 0.045 & 96.9 & 60.7 & -0.09 & 0.065 & 94.4 & 89\\ 
\hline
\hline
  \\[-10pt]
            \multicolumn{14}{c}{(e) All confounders} \\[0pt]
        \\[-10pt]
        \hline
\multirow{3}{*}{1-1}&(1) & -0.11 & 0.035 & 99.2 & 48.3 & -0.09 & 0.033 & 98.6 & 67.3 & -0.09 & 0.047 & 98.2 & 83.1\\
&(2) & -0.10 & 0.031 & 98.8 & 55.1 & -0.09 & 0.030 & 98.5 & 77.1 & -0.08 & 0.042 & 97.1 & 94.1\\
&(3) & -0.20 & 0.067 & 94 & 55.4 & -0.18 & 0.061 & 93.2 & 77.6 & -0.16 & 0.062 & 95.2 & 94.7\\
 \hline
 \multirow{3}{*}{1-1/2}& (1) & -0.10 & 0.035 & 99.4 & 48.3 & -0.09 & 0.035 & 99.2 & 67.4 & -0.09 & 0.049 & 98 & 83.3\\
&(2) & -0.10 & 0.029 & 98.7 & 55 & -0.09 & 0.030 & 98.5 & 77.1 & -0.08 & 0.039 & 97.1 & 94\\ 
&(3) & -0.19 & 0.068 & 91.7 & 55.4 & -0.18 & 0.062 & 92.2 & 77.5 & -0.16 & 0.063 & 93.2 & 94.7\\
 \hline
 \multirow{3}{*}{1-2}& (1)& -0.08 & 0.044 & 99.4 & 20.9 & -0.09 & 0.048 & 99.6 & 33.3 & -0.07 & 0.053 & 99 & 45.8\\
&(2) & -0.09 & 0.035 & 99.5 & 33.6 & -0.08 & 0.031 & 99.5 & 51.8 & -0.09 & 0.043 & 98.4 & 70.8\\
&(3) & -0.20 & 0.073 & 96.4 & 34.4 & -0.18 & 0.066 & 94.8 & 53.2 & -0.18 & 0.074 & 95.1 & 72.7\\
\hline
\hline
        \end{tabular}
        
    }%
    \label{tab:sims_tuning}
\end{table}

\if{
\begin{table}[!t]
    \centering
    %\spacingset{1.12}
    \caption{Simulation outputs when $(\delta,\delta',\epsilon)$ is $(0,0.01,2)$ or $(2,0.05,6)$}
    % \rowcolors{5}{}{gray!10}
    %\spacingset{1}
    \resizebox{5.5in}{!}{%
        \begin{tabular}{*{16}{c}}
            \hline
            \\[-10pt]
             &\multicolumn{3}{c}{No Confounder} & \multicolumn{3}{c}{Time-smmoth Confounders} & \multicolumn{3}{c}{Location-varying confounders} & \multicolumn{3}{c}{Time-varying confounders} & \multicolumn{3}{c}{All confounders} \\
            \cmidrule(lr){2-4} \cmidrule(lr){5-7} \cmidrule(lr){8-10} \cmidrule(lr){11-13}\cmidrule(lr){14-16}
            Method & Bias (MSE) & Cover & Prop  & Bias (MSE) & Cover & Prop & Bias (MSE) & Cover & Prop & Bias (MSE) & Cover & Prop & Bias (MSE) & Cover & Prop\\[5pt]
        \hline
        \hline
         \\[-10pt]
        \multicolumn{16}{c}{Dense exposure} \\[0pt]
        \\[-10pt]
        \hline
        \\[-8pt]
        \multicolumn{3}{c}{$\delta=0,\delta'=0.01,\epsilon=2$}
        \\[2pt]
        \cmidrule{1-4}
           U1-1 & -0.004 (0.013) & 0.956 & 0.871 & 0.005 (0.021) & 0.946 & 0.582 & -0.002 (0.014) & 0.944 & 0.811 & -0.395 (0.218) & 0.564 & 0.647 & -0.63 (0.459) & 0.27 & 0.701 \\ 
   U1-1/2 & -0.005 (0.013) & 0.948 & 0.871 & 0.005 (0.02) & 0.954 & 0.582 & -0.004 (0.013) & 0.952 & 0.811 & -0.4 (0.218) & 0.526 & 0.647 & -0.632 (0.459) & 0.256 & 0.701 \\ 
   U1-2 & -0.003 (0.023) & 0.942 & 0.408 & 0.002 (0.036) & 0.942 & 0.284 & -0.003 (0.025) & 0.948 & 0.368 & -0.431 (0.297) & 0.678 & 0.318 & -0.672 (0.577) & 0.418 & 0.349 \\ 
   1-1 & -0.001 (0.014) & 0.938 & 0.871 & 0.006 (0.021) & 0.954 & 0.582 & -0.006 (0.013) & 0.942 & 0.81 & -0.005 (0.019) & 0.988 & 0.577 & -0.019 (0.019) & 0.992 & 0.648 \\ 
   1-1/2 & -0.003 (0.013) & 0.95 & 0.871 & 0.005 (0.02) & 0.956 & 0.582 & -0.004 (0.013) & 0.946 & 0.81 & -0.004 (0.019) & 0.988 & 0.577 & -0.019 (0.02) & 0.99 & 0.648 \\ 
   1-2 & -0.006 (0.02) & 0.964 & 0.408 & 0.006 (0.033) & 0.948 & 0.283 & 0.001 (0.025) & 0.934 & 0.368 & 0.001 (0.033) & 0.99 & 0.263 & -0.016 (0.035) & 0.994 & 0.305 \\ 
                 \hline
        \\[-8pt]
        \multicolumn{3}{c}{$\delta=2,\delta'=0.05,\epsilon=6$}
        \\[2pt]
        \cmidrule{1-4}
       U1-1 & 0 (0.012) & 0.958 & 0.978 & -0.006 (0.019) & 0.952 & 0.626 & -0.007 (0.013) & 0.944 & 0.918 & -0.41 (0.238) & 0.544 & 0.699 & -0.647 (0.484) & 0.264 & 0.766 \\ 
   U1-1/2 & 0.004 (0.012) & 0.944 & 0.978 & -0.001 (0.02) & 0.952 & 0.626 & -0.007 (0.013) & 0.948 & 0.918 & -0.422 (0.237) & 0.5 & 0.699 & -0.632 (0.459) & 0.218 & 0.766 \\ 
   U1-2 & -0.006 (0.015) & 0.932 & 0.624 & -0.006 (0.023) & 0.944 & 0.41 & -0.011 (0.016) & 0.942 & 0.571 & -0.458 (0.296) & 0.534 & 0.472 & -0.675 (0.533) & 0.286 & 0.515 \\ 
   1-1 & 0.001 (0.012) & 0.948 & 0.978 & -0.007 (0.02) & 0.942 & 0.626 & -0.007 (0.013) & 0.944 & 0.918 & -0.06 (0.022) & 0.982 & 0.687 & -0.104 (0.03) & 0.972 & 0.76 \\ 
   1-1/2 & 0.001 (0.012) & 0.946 & 0.978 & -0.009 (0.019) & 0.932 & 0.626 & -0.008 (0.013) & 0.946 & 0.918 & -0.061 (0.022) & 0.978 & 0.687 & -0.107 (0.031) & 0.972 & 0.76 \\ 
   1-2 & -0.004 (0.014) & 0.948 & 0.624 & -0.008 (0.022) & 0.954 & 0.41 & -0.006 (0.016) & 0.948 & 0.571 & -0.058 (0.025) & 0.98 & 0.445 & -0.103 (0.033) & 0.984 & 0.5 \\  [3pt]
        \hline
        \hline
         \\[-10pt]
        \multicolumn{16}{c}{Medium exposure} \\[0pt]
        \\[-10pt]
        \hline
        \\[-8pt]
        \multicolumn{3}{c}{$\delta=0,\delta'=0.01,\epsilon=2$}
        \\[2pt]
        \cmidrule{1-4}
      U1-1 & -0.007 (0.021) & 0.948 & 0.983 & 0.008 (0.028) & 0.948 & 0.786 & 0 (0.021) & 0.938 & 0.963 & -0.367 (0.192) & 0.658 & 0.806 & -0.614 (0.451) & 0.338 & 0.892 \\ 
  U1-1/2 & -0.007 (0.019) & 0.952 & 0.983 & 0.009 (0.024) & 0.946 & 0.786 & 0.001 (0.019) & 0.944 & 0.963 & -0.37 (0.19) & 0.598 & 0.806 & -0.611 (0.437) & 0.324 & 0.892 \\ 
  U1-2 & -0.005 (0.024) & 0.946 & 0.639 & 0.011 (0.038) & 0.938 & 0.431 & 0.01 (0.025) & 0.942 & 0.598 & -0.42 (0.276) & 0.694 & 0.456 & -0.669 (0.55) & 0.438 & 0.542 \\ 
   1-1 & -0.008 (0.021) & 0.946 & 0.983 & 0.003 (0.026) & 0.956 & 0.786 & 0 (0.021) & 0.944 & 0.963 & -0.015 (0.025) & 0.99 & 0.732 & -0.012 (0.027) & 0.996 & 0.837 \\ 
   1-1/2 & -0.007 (0.02) & 0.95 & 0.983 & 0.006 (0.025) & 0.956 & 0.786 & 0 (0.02) & 0.942 & 0.963 & -0.015 (0.025) & 0.99 & 0.732 & -0.015 (0.027) & 0.994 & 0.837 \\ 
   1-2 & -0.001 (0.023) & 0.946 & 0.639 & 0.012 (0.039) & 0.932 & 0.429 & 0.011 (0.026) & 0.938 & 0.597 & -0.009 (0.038) & 0.99 & 0.371 & -0.004 (0.04) & 0.996 & 0.461 \\          \hline
        \\[-8pt]
        \multicolumn{3}{c}{$\delta=2,\delta'=0.05,\epsilon=6$}
        \\[2pt]
        \cmidrule{1-4}
   U1-1 & -0.004 (0.02) & 0.948 & 1 & -0.001 (0.024) & 0.946 & 0.853 & 0.007 (0.019) & 0.942 & 0.997 & -0.376 (0.196) & 0.612 & 0.864 & -0.595 (0.426) & 0.336 & 0.948 \\ 
   U1-1/2 & -0.003 (0.018) & 0.944 & 1 & -0.002 (0.023) & 0.944 & 0.853 & 0.003 (0.017) & 0.948 & 0.997 & -0.374 (0.193) & 0.602 & 0.864 & -0.589 (0.415) & 0.372 & 0.948 \\ 
   U1-2 & -0.01 (0.016) & 0.956 & 0.909 & 0.001 (0.026) & 0.942 & 0.6 & 0.004 (0.017) & 0.936 & 0.859 & -0.417 (0.239) & 0.592 & 0.633 & -0.639 (0.492) & 0.334 & 0.741 \\ 
   1-1 & -0.011 (0.02) & 0.95 & 1 & -0.001 (0.025) & 0.942 & 0.853 & 0.007 (0.019) & 0.958 & 0.997 & -0.06 (0.025) & 0.978 & 0.857 & -0.088 (0.036) & 0.964 & 0.945 \\ 
   1-1/2 & -0.003 (0.017) & 0.96 & 1 & -0.005 (0.022) & 0.944 & 0.853 & 0.004 (0.018) & 0.956 & 0.997 & -0.061 (0.026) & 0.97 & 0.857 & -0.093 (0.036) & 0.964 & 0.945 \\ 
   1-2 & -0.007 (0.017) & 0.954 & 0.909 & 0 (0.026) & 0.95 & 0.6 & 0.004 (0.016) & 0.948 & 0.859 & -0.061 (0.026) & 0.97 & 0.604 & -0.096 (0.036) & 0.978 & 0.721 \\  [3pt]
        \hline
        \hline
         \\[-10pt]
        \multicolumn{16}{c}{Sparse exposure} \\[0pt]
        \\[-10pt]
        \hline
        
        %\multicolumn{13}{c}{$\beta_{UZ} = 0$ and $\beta_U = \beta_{\bar U} = 0$}
        %\\[0pt]
        \\[-8pt]
        \multicolumn{3}{c}{$\delta=0,\delta'=0.01,\epsilon=2$}
        \\[2pt]
        \cmidrule{1-4}
     U1-1 & -0.008 (0.041) & 0.942 & 0.999 & 0.009 (0.039) & 0.952 & 0.932 & 0.004 (0.035) & 0.956 & 0.996 & -0.34 (0.22) & 0.808 & 0.979 & -0.587 (0.46) & 0.57 & 0.979 \\ 
   U1-1/2 & -0.007 (0.032) & 0.956 & 0.999 & 0.01 (0.035) & 0.944 & 0.932 & 0.005 (0.032) & 0.946 & 0.996 & -0.35 (0.216) & 0.786 & 0.979 & -0.604 (0.459) & 0.478 & 0.979 \\ 
   U1-2 & -0.015 (0.036) & 0.958 & 0.823 & 0.027 (0.052) & 0.948 & 0.6 & 0.008 (0.035) & 0.95 & 0.795 & -0.388 (0.28) & 0.812 & 0.728 & -0.673 (0.582) & 0.52 & 0.719 \\ 
   1-1 & -0.003 (0.039) & 0.954 & 0.999 & 0.015 (0.04) & 0.934 & 0.932 & 0.01 (0.041) & 0.954 & 0.996 & -0.012 (0.064) & 0.976 & 0.903 & 0.003 (0.046) & 0.992 & 0.924 \\ 
   1-1/2 & -0.011 (0.034) & 0.958 & 0.999 & 0.016 (0.037) & 0.948 & 0.932 & 0.009 (0.039) & 0.944 & 0.996 & -0.016 (0.062) & 0.984 & 0.905 & 0.012 (0.047) & 0.988 & 0.925 \\ 
   1-2 & -0.009 (0.038) & 0.942 & 0.818 & 0.02 (0.049) & 0.958 & 0.591 & 0.01 (0.037) & 0.952 & 0.788 & -0.009 (0.079) & 0.986 & 0.547 & 0.004 (0.058) & 0.998 & 0.579 \\ 
                 \hline
        \\[-8pt]
        \multicolumn{3}{c}{$\delta=2,\delta'=0.05,\epsilon=6$}
        \\[2pt]
        \cmidrule{1-4}
    U1-1 & -0.009 (0.04) & 0.954 & 1 & 0.01 (0.036) & 0.958 & 0.978 & 0.005 (0.041) & 0.938 & 1 & -0.349 (0.224) & 0.826 & 0.998 & -0.615 (0.488) & 0.568 & 0.997 \\ 
   U1-1/2 & -0.014 (0.035) & 0.96 & 1 & 0.007 (0.034) & 0.944 & 0.978 & 0.005 (0.037) & 0.948 & 1 & -0.352 (0.223) & 0.778 & 0.998 & -0.591 (0.46) & 0.518 & 0.997 \\ 
   U1-2 & -0.016 (0.03) & 0.944 & 0.987 & 0.004 (0.034) & 0.96 & 0.806 & 0.001 (0.03) & 0.928 & 0.974 & -0.351 (0.208) & 0.804 & 0.917 & -0.612 (0.479) & 0.502 & 0.913 \\ 
   1-1 & -0.011 (0.039) & 0.948 & 1 & 0.015 (0.038) & 0.962 & 0.978 & 0.008 (0.038) & 0.956 & 1 & -0.044 (0.062) & 0.956 & 0.998 & -0.064 (0.047) & 0.992 & 0.997 \\ 
   1-1/2 & -0.011 (0.033) & 0.946 & 1 & 0.009 (0.035) & 0.948 & 0.978 & 0.01 (0.03) & 0.964 & 1 & -0.05 (0.054) & 0.952 & 0.998 & -0.074 (0.044) & 0.982 & 0.997 \\ 
   1-2 & -0.015 (0.029) & 0.962 & 0.987 & 0.003 (0.035) & 0.954 & 0.806 & 0.004 (0.028) & 0.944 & 0.974 & -0.055 (0.056) & 0.964 & 0.889 & -0.077 (0.045) & 0.972 & 0.893 \\  [3pt]
        \hline
        \\[-10pt]

        \end{tabular}
    }%
    \label{tab:appendix}
\end{table}
}\fi

\subsection{Modification of the algorithms to impose constraints within each match}\label{appendix:new-matching}

Here, we investigate the performance of our inferential procedure when applied on the estimator derived from an alternative matching algorithm that imposes constraints on time-varying covariates within each match, instead of overall balance constraints. We consider this approach in accordance to the observations made in Supplement~\ref{supp_sec:variance estimation}, and particularly Supplement~\ref{supp_subsec:variance_match_within_pair}.
Specifically, here, we illustrate that this issue of over-coverage that is sometimes observed based on our inferential procedure and for our algorithm and estimator is alleviated when covariate balance is imposed for {\it every} match.

To do so, we consider the following extension to our algorithms in \cref{sec:matching} and Supplement~\ref{supp_subsec:matching1-2or1-12} to include additional constraints. Given a small value $\delta''$ when matching on all covariates, the 1-1 algorithm in \cref{sec:matching} is extended to impose
\begin{align*}
    &|a_{t_et_u}(\overline{\bm X}_{t_e, S}-\overline{\bm X}_{t_u, S})|\leq \bm 1_{Np^\text{int}+Np^\text{net}+p^\text{out}}\cdot\delta'',
    % \tag{\ref{matching1-1-objective}.8}\\
\intertext{and similarly for the 1-2 algorithm in Supplement~\ref{supp_subsubsec:matching-1-2}}
&\left|a_{t_et_{u_1}t_{u_2}}\left( \overline{\bm X}_{t_e, S}-\frac{\overline{\bm X}_{t_{u_1}, S}+ \overline{\bm X}_{t_{u_2}, S}}{2}\right)\right| \leq \bm 1_{Np^\text{int}+Np^\text{net}+p^\text{out}}\cdot\delta''. 
% S.% \tag{\ref{matching1-2-objective}.10}   
\end{align*}
For the 1-1/2 algorithm in Supplement~\ref{subsec: matching 1-1/2} we impose both constraints. If constraints are imposed on summarized covariates only, we replace $\overline{\bm X}_{t, S}$ with $\overline{\bm X}^\text{sum.}_{t, S}$ for all constraints above. 
We refer to these algorithms and the corresponding estimators as {\it extended} algorithms and estimators. 

We evaluated the performance of these extended algorithms and estimators against the standard algorithms and estimators introduced in \cref{sec:matching} (which impose only constraints on the overall matched populations, and not within each match).
We consider the confounding scenario (d) with time-varying confounder and medium sparsity level. We set $\delta'' = 0.25$, assuming the covariates have been standardized as discussed in \cref{subsec:single-simu}. We set the remaining tuning parameters equal to the values $(\delta, \delta', \epsilon) = (2, 0.05, 6)$.

%As a comparison, the standard deviation of time-varying covariates $\bm{W}_{..3}$, $\bm{W}_{..5}$, $\bm{W}_{..6}$ and $\widetilde{\bm P}_{.}$ varies from 0.3 to 1.7. 

In \cref{fig:appendix-new1-2}, we see that the three extended matching estimators are close to unbiased for the true causal effect (which is equal to 1). 
The bias of the three estimators based on the extended algorithms is $ 0$ for the 1-1 algorithm, $-0.04$ for the 1-1/2 algorithm, and $-0.05$ for the 1-2 algorithm.
Any residual bias is most likely due to residual covariate imbalance.  
{Since bias can affect coverage rates, we compare coverage of 95\% intervals of the extended matching estimators, with the results of the standard estimators under tuning parameters $(\delta, \delta', \epsilon)= (2, 0.05, 6)$ which show similar bias (see \cref{tab:sims_tuning}, bias equal to $-0.06$ for all three estimators). }
The coverage rate for the extended 1-1, 1-1/2, and 1-2 estimators is equal to 97.2\%, 95.4\%, and 95.0\%, respectively, closer to the nominal level compared to the corresponding coverage rates of the standard 1-1, 1-1/2, and 1-2 estimators (included in \cref{tab:sims_tuning}) which are equal to 99.2\%, 99\% and 97.8\%, respectively. }
%We suspect that the extended estimators have coverage rate below 95\% due to the minor amount of bias, and we expect that they will have closer to nominal coverage compared to the standard matching estimators in unbiased scenarios.

However, even though inference might be closer to the nominal level (rather than being conservative) when the extended estimators are employed, they might suffer due to small number of matches depending on the value of $\delta''$. Even with $\delta'' = 0.25$, the proportion of matched exposed time periods decreases significantly (compared to the algorithms that do not impose constraints within each match) to 10.8\%, 21.3\%, and 12.6\% for the Extended 1-1, Extended 1-1/2, and Extended 1-2 algorithm, respectively {(compared to 63.1\%, 63.2\% and 23.7\% for the standard 1-1, 1-1/2, and 1-2 algorithm, respectively)}. 

We conclude that incorporating the within-match adjustment has the benefit of coverage that is closer to the nominal level, but suffers from a lower proportion of matched exposed time periods, which might lead to higher overall uncertainty and lower estimation efficiency.

\begin{figure}[!t]
    \centering
    \includegraphics[width=0.6\linewidth]{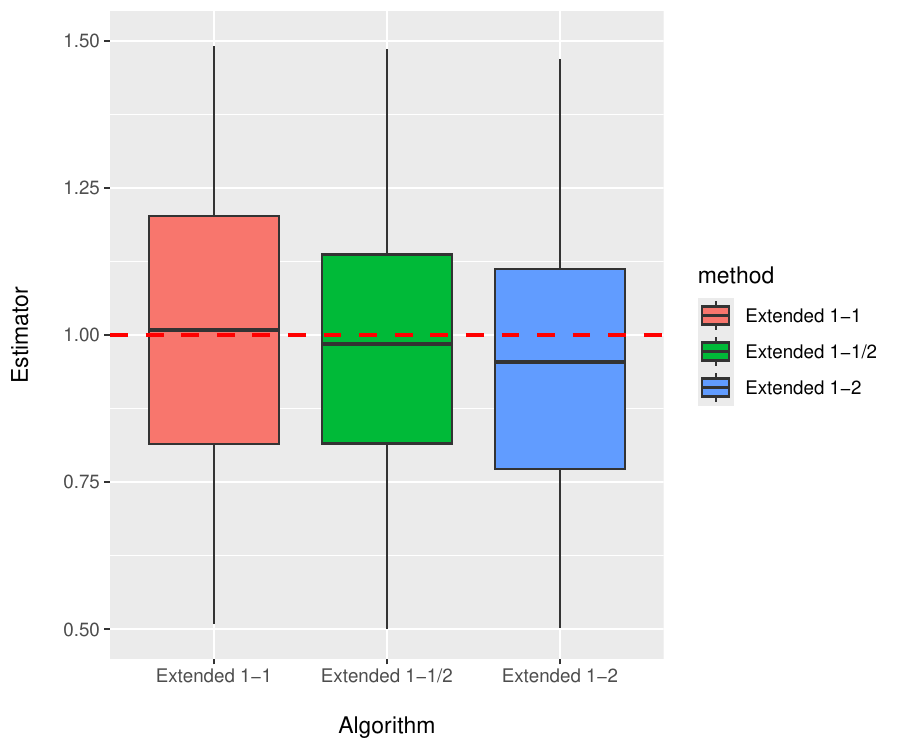}
    \caption{{Boxplots of estimated causal effects over 500 data sets based on the extended estimators. The horizontal line in the boxplot corresponds to the median estimate.}}% `N' stands for `New' compared to adjusting for only average covariates.}
    \label{fig:appendix-new1-2}
\end{figure}

\subsection{Simulations under heterogeneous treatment effects}
\label{supp_subsec:simu_hetero}

We performed a simulation where exposure effects are no longer homogeneous over time for an individual outcome unit. %Now denoted as $\tau_{jt}$, the exposure effect of unit $m_j$ at $t$ is related to both its own characteristics and time period. 
We %define the average in the potential outcomes for all time periods as $\widetilde{\tau}^\text{imm}(1, 0)  = \frac1{T} \sum_{t = 1}^T \tau_t^\text{imm}(1, 0; R_t),$ and 
define three new quantities, denoted as $\widetilde{\tau}^{\text{imm},\mathrm{1-1}}(1, 0), \widetilde{\tau}^{\text{imm},\mathrm{1-1/2}}(1, 0)$ and $\widetilde{\tau}^{\text{imm},\mathrm{1-2}}(1, 0)$, as the average difference in the potential outcomes under exposures $0$ and $1$ over the exposed time periods {\it that are matched} according to the 1-1, 1-1/2, and 1-2 algorithm, respectively. For example, $\widetilde{\tau}^{\text{imm},\mathrm{1-1}}(1, 0)$ is defined as
\begin{align*}
& \widetilde{\tau}^{\text{imm},\mathrm{1-1}}(1, 0)= \\
& \hspace{20pt} \dfrac{1}{\sum_{t:\ \mathrm{matched\ in \ 1-1}}I(E_{t}=1)}\sum_{t:\ \mathrm{matched\ in \ 1-1}} (Y_{t}(1,R_t)-Y_{t}(0,R_t))I(E_{t}=1).
\end{align*}
These quantities resemble the estimand $\tau^{\text{imm}}(1, 0)$ defined in \cref{sec:model-setup}, but they average over the {\it matched} exposed time periods only.
For ease of notation, we denote them as $\widetilde{\tau}^{\text{imm},\mathrm{1-1}}, \widetilde{\tau}^{\text{imm},\mathrm{1-1/2}}$ and $\widetilde{\tau}^{\text{imm},\mathrm{1-2}}$.

The data generating models follow the scenarios (b) and (d) with medium exposure, where we alter the outcome model to specify heterogeneous treatment effects. Specifically, we generate outcomes according to
\begin{align*}
Y_{tj}=&(1+\epsilon'_{tj})E_{t} + 0.005 (400-t) E_{t} + (0.4+\epsilon_{tj}'')R_t-0.0005tR_t\\
&+X_{tj3}^\text{out}+X_{tj5}^\text{out}+\sum_i q_{ij}X_{tij}^\text{net} +X_{tj6}^\text{out}+\epsilon_{tj},
\end{align*}
where $\epsilon'_{tj}$ and $\epsilon''_{tj}$ are generated independently from  $N(0,1).$ We simulate 500 data sets and consider the estimation of effects for the first outcome unit.

We compare the matching approaches to Na\"ive-$t$. We exclude Na\"ive-$j$ and Na\"ive-all from simulations under heterogeneity since the estimands they target average across units, whereas Na\"ive-$t$ and the matching estimators target unit-specific effects that average across time. %For every simulated data set, we have an estimation for each individual outcome unit. $\tau$ and $\Tilde{\tau}$ are fixed while $\Tilde{\tau}^{\mathrm{1-1}}, \Tilde{\tau}^{\mathrm{1-1/2}}$ and $\Tilde{\tau}^{\mathrm{1-2}}$ vary based on matching. For the latter three statistics, we only compare the performance of Na\"ive-$t$ and the estimation under their own matching methods. We simulated 500 cases, with the average of MSE and coverage written below.

The bias, MSE, Coverage and proportion of exposed units that are matched are shown in \cref{appendix:tab:hetero}. We evaluate the performance of the three proposed estimators and Na\"ive-$t$ for estimating %the average effect over all time points $\widetilde \tau^\text{imm}$, and 
the average effect over all exposed time points $\tau^\text{imm}$.
%In the presence of heterogeneous effects across time, all of the estimators are biased for $\tau^\text{imm}$.the effect over all time periods, $\widetilde \tau^\text{imm}$. 
The proposed estimators perform relatively well for estimating ${\tau}^\text{imm}$, and substantially better than the Na\"ive-$t$ approach that suffers from confounding bias.
Particularly, the estimators based on the 1-1 and 1-1/2 algorithms which find matches for 85\% of the exposed time points are close to unbiased with appropriate coverage of confidence intervals.

We also evaluate the performance of each estimator against the causal effect over the set of exposed time periods that are in fact matched, in that we evaluate the estimator based on the 1-1 algorithm for estimating $\widetilde\tau^{\text{imm}, 1-1}$, the estimator based on the 1-1/2 algorithm for estimating $\widetilde\tau^{\text{imm},1-1/2}$ and the estimator based on the 1-2 algorithm for estimating $\widetilde\tau^{\text{imm}, 1-2}$.
When compared against the effect over the matched exposed population, the estimators are unbiased with appropriate coverage. This suggests that even though the effect of exposure is not constant at each time point, matching approaches can still serve as a useful tool for inferring the average effect of the exposure over the population that was in fact matched. 

% \usepackage{multirow}
\if{
\begin{table}[hbtp!]
\centering
\begin{tabular}{|l|llll|llll|}
\hline
        & \multicolumn{4}{c|}{AEA$_{j}$}                           & \multicolumn{4}{c|}{AEE$_{j}$}                                                  \\ \hline
         & Na\"ive & 1-1                        & 1-1/2 & 1-2   & Na\"ive & 1-1                        & 1-1/2              & 1-2               \\ \hline
MSE      & 8.38  & 0.05                       & 0.04  & 0.03  & 7.19  & 0.02                       & 0.03               & 0.04              \\ \hline
Coverage & 0.003 & 0.776                      & 0.853 & 0.936 & 0.005 & 0.954                      & 0.922              & 0.858             \\ \hline \hline
         & \multicolumn{6}{c|}{AEM$_{j}$}                                                        & \multicolumn{2}{l|}{\multirow{4}{*}{}} \\ \cline{1-7}
         & Na\"ive & \multicolumn{1}{l|}{1-1}   & Na\"ive & 1-1/2 & Na\"ive & \multicolumn{1}{l|}{1-2}   & \multicolumn{2}{l|}{}                  \\ \cline{1-7}
MSE      & 7.38  & \multicolumn{1}{l|}{0.02}  & 7.86  & 0.02  & 7.98  & \multicolumn{1}{l|}{0.02}  & \multicolumn{2}{l|}{}                  \\ \cline{1-7}
Coverage & 0.005 & \multicolumn{1}{l|}{0.965} & 0.004 & 0.956 & 0.003 & \multicolumn{1}{l|}{0.942} & \multicolumn{2}{l|}{}                  \\ \hline

\end{tabular}
\caption{Comparison of the average heterogeneous effect among the na\"ive and matching approaches}
\end{table}

\begin{table}[!t]
    \centering
    %\spacingset{1.12}
    \caption{Simulation results of single-unit estimation under heterogeneous effect. Mean squared error (MSE) and coverage of 95\% intervals (\%) of the Na\"ive-$t$ method and matching methods.}
     %\spacingset{1.05}
    \resizebox{0.76\textwidth}{!}{%
    \begin{tabular}{*{9}{c}}
        \hline
        & \multicolumn{4}{c}{$\tau$} & \multicolumn{4}{c}{${\tau}$} \\
        \cmidrule(lr){2-5} \cmidrule(lr){6-9} 
        & Na\"ive & 1-1 & 1-1/2 & 1-2 & Na\"ive & 1-1 & 1-1/2 & 1-2  \\[2pt]
        \hline
        \\[-10pt]
        MSE & 0.78 &0.11 &0.11 &0.06& 1.53& 0.03& 0.04& 0.07 \\ 
        Cover & 2.6 & 81.2 & 83 & 95.4 & 0 & 97.4 & 97.6 & 93.6 \\[3pt]
        \hline
        \hline
         \\[-10pt]
        
        & \multicolumn{2}{c}{$\tilde{\tau}^{\mathrm{1-1}}$} & \multicolumn{2}{c}{$\tilde{\tau}^{\mathrm{1-1/2}}$} & \multicolumn{2}{c}{$\tilde{\tau}^{\mathrm{1-2}}$} \\
        \cmidrule(lr){2-3} \cmidrule(lr){4-5}\cmidrule(lr){6-7}
        & Na\"ive & 1-1 & Na\"ive & 1-1/2 & Na\"ive & 1-2 \\[2pt]
        \hline
        \\[-10pt]
        MSE & 1.33 &0.02 &1.33 &0.02 &1.09& 0.02 \\
        Cover& 0.2 & 99.2 & 0.2 & 99.4 & 1 & 99.8 \\
        \hline
        \hline
    \end{tabular}
    }%
    \label{appendix:tab:hetero}
\end{table}
}\fi

% time-smooth confounder
\if{
\begin{table}[!t]
\caption{Simulation results under heterogeneous effects for a single outcome unit. Bias, mean squared error (MSE), coverage of 95\% intervals (\%) of the Na\"ive-$t$ method and matching methods, and proportion of exposed units being matched of matching methods.}
%\spacingset{1}
    \centering
    \begin{tabular}{cccccc}
Estimand & Method & Bias & MSE & Coverage & Proportion \\ \hline\hline
 \multirow{4}{*}{$\tau$}  & Na\"ive-$t$ &  -0.62 &0.45&29.2&-\\
         & 1-1 & 0.41 &0.21&58.6&85.1\\
         & 1-1/2 & 0.41& 0.21&56.4&85.1\\
         & 1-2 & 0.30 &0.14&83&59.9\\ \hline
\multirow{4}{*}{$\widetilde \tau$} & Na\"ive-$t$ &  -1.09 &1.26&1.4&-\\
         & 1-1 &  -0.07 &0.04&97.6&85.1\\
         & 1-1/2 & -0.06&0.03&97&85.1\\
         & 1-2 &-0.16&0.07&94.6&59.9\\ \hline
\multirow{2}{*}{$\dbtilde \tau_{1-1}$} & Na\"ive-$t$ &   -1.02&1.12&2&-\\
         & 1-1 & 0.00&0.03&98.2&85.1\\
         % & 1-2 & \\
         % & 1-1/2\\
         \hline
\multirow{2}{*}{$\dbtilde \tau_{1-2}$} & Na\"ive-$t$ &   -1.03&1.12&2&-\\
         % & 1-1 & \\
         & 1-2 & 0.01&0.03&98.8&85.1\\
         % & 1-1/2\\ 
         \hline
\multirow{2}{*}{$\dbtilde \tau_{1-1/2}$} & Na\"ive-$t$ &  -0.92&0.92 &4.8&-\\
         % & 1-1 & \\
         % & 1-2 & \\
         & 1-1/2 & 0.01&0.03&99.4&59.9\\ \hline
    \end{tabular}
    \label{appendix:tab:hetero}
\end{table}
}\fi

\begin{table}[!t]
\caption{Simulation results under heterogeneous {immediate} effect over time. Bias, mean squared error (MSE), coverage of 95\% intervals (\%) of the Na\"ive-$t$ and the proposed estimator for the causal effect over the exposed time periods and over the matched exposed time periods. The proportion of exposed time periods that are matched on average when employing each matching method is reported once.}
%\spacingset{1}
    \centering
    \begin{tabular}{cccccc}
Estimand & Method & Bias & MSE & Coverage & Proportion \\ \hline\hline 
         \\[-5pt]
\multicolumn{6}{c}{(b) Time-smooth confounders} \\[0pt]
\hline
%\multirow{4}{*}{$\widetilde \tau^\text{imm}$}  & Na\"ive-$t$ &  -0.77 &0.65 & 8.99&-\\
%         & 1-1 & 0.24& 0.09 &87.76&84.82\\
%         & 1-1/2 &0.23& 0.09 &88.53&84.82\\
%         & 1-2 & 0.30 &0.14&83&58.84\\
%         \hline
\multirow{4}{*}{$ \tau^\text{imm}$} & Na\"ive-$t$ &  -1.14& 1.35 & 0.00& -\\
         & 1-1 &  -0.13& 0.05 &98.09& 84.82\\
         & 1-1/2 & -0.13 &0.05 &97.51& 84.82\\
         & 1-2 &-0.26& 0.11 &91.78& 58.84\\
          \hline
\multirow{2}{*}{$\widetilde{\tau}_{1-1}^\text{imm}$} & Na\"ive-$t$ &   -1.05 &1.17 & 0.76&\\
         & 1-1 & -0.05& 0.03 &99.81&\\
         % & 1-2 & \\
         % & 1-1/2\\
         \hline
\multirow{2}{*}{$\widetilde{\tau}_{1-2}^\text{imm}$} & Na\"ive-$t$ &   -1.05 &1.17 & 0.96&\\
         % & 1-1 & \\
         & 1-2 & -0.05 &0.03 &99.43&\\
         % & 1-1/2\\ 
         \hline
\multirow{2}{*}{$\widetilde{\tau}_{1-1/2}^\text{imm}$} & Na\"ive-$t$ &  -0.93 &0.94 & 2.29&\\
         % & 1-1 & \\
         % & 1-2 & \\
         & 1-1/2 & -0.05 &0.03 &99.43&\\
\hline \hline  \\[-5pt]
\multicolumn{6}{c}{(d) Time-varying confounders} \\[0pt] \hline
% \multirow{4}{*}{$\widetilde \tau^\text{imm}$}  & Na\"ive-$t$ &  -0.52 &0.33 &46.46&-\\
 %        & 1-1 & 0.40& 0.20 &67.85&84.96\\
  %       & 1-1/2 &0.39& 0.20 &67.51&84.96\\
  %       & 1-2 & 0.28 &0.13 &87.54&59.59\\ \hline
\multirow{4}{*}{$ \tau^\text{imm}$} & Na\"ive-$t$ &  -0.99 &1.04  &2.36&-\\
         & 1-1 & -0.08& 0.04 &97.98&84.96\\
         & 1-1/2 & -0.08& 0.04 &97.47&84.96\\
         & 1-2 &-0.19& 0.08 &94.95&59.59\\ \hline
\multirow{2}{*}{$\widetilde{\tau}_{1-1}^\text{imm}$} & Na\"ive-$t$ &   -0.92& 0.91 & 3.20&\\
         & 1-1 &0.00 &0.03& 99.49&\\ 
         % & 1-2 & \\
         % & 1-1/2\\
         \hline
\multirow{2}{*}{$\widetilde{\tau}_{1-2}^\text{imm}$} & Na\"ive-$t$ &   -0.92& 0.92 & 3.54&\\
         % & 1-1 & \\
         & 1-2 & -0.01& 0.03 &99.16&\\
         % & 1-1/2\\ 
         \hline
\multirow{2}{*}{$\widetilde{\tau}_{1-1/2}^\text{imm}$} & Na\"ive-$t$ &  -0.82& 0.74 & 8.75&\\
         % & 1-1 & \\
         % & 1-2 & \\
         & 1-1/2 & -0.02 &0.03 &99.83& \\ \hline
    \end{tabular}
    \label{appendix:tab:hetero}
\end{table}

\subsection{The impact of outcome temporal correlation on inference}
\label{supp_subsec:sims_temp_corr}

When outcomes exhibit temporal autocorrelation, conventional Wald-type intervals in a regression analysis generally underestimate the true variance of the coefficients' OLS estimator. In these situations, the Newey-West variance estimator provides robust inference for regression coefficients.

Here, we perform simulations to investigate the impact of outcome autocorrelation on inference. First, we investigate the extent to which this is an issue within our framework. Second, we consider an alternative inference strategy that uses the Newey-West variance estimator.

We describe first the alternative inference strategy. In Supplement \ref{supp_subsec:alternative_matching_estimator} we showed that our point estimator is equal to the OLS estimator of an outcome on exposure regression on the matched sample. Then, standard errors are acquired by replacing the standard regression covariance matrix with the Newey-West matrix.
% The Newey-West estimator replaces the standard OLS variance assumption with a sandwich covariance estimator. 
This approach incorporates squared residuals and distance-weighted autocovariances to robustly correct for both heteroskedasticity and serial correlation \citep{newey1986simple}. We compute the Newey-West variance-covariance matrix using the \texttt{NeweyWest()} function from the \texttt{sandwich} R package. We then performed revised statistical tests using the \texttt{coeftest()} function from the \texttt{lmtest} package to ensure the significance levels accounted for the adjusted standard errors.

Our simulation setup is as follows. We adopt scenario (a) without any confounders to eliminate any impact of confounding on simulation results. We consider the scenario under dense exposure. Importantly, we alter how the outcome error terms $\epsilon_t$ are generated across time to impose temporal autocorrelation. Specifically, we set  $\epsilon_t=\rho\epsilon_{t-1}+ \sqrt{1 - \rho^2} e_t$, where $e_t \sim N(0, 1),$ and $\rho$ is a tuning parameter adjusting the amount of autocorrelation of the error terms. We simulate 500 data sets under different levels of autocorrelation, $\rho=0.2, 0.4, 0.6$ and $0.8$, and record whether the true value of the causal effect is included in the 95\% confidence intervals created based on each of the three estimators, and based on the Wald-type intervals in \cref{subsec:inference} and the intervals employing the Newey-West estimated variance. 

The results for the immediate effect are shown in \cref{appendix:tab:tempporal_corr}. Temporal correlation in the outcome variable has a minimal impact in confidence interval coverage for the approach described in \cref{subsec:inference}. Even in the presence of high outcome variable autocorrelation ($\rho = 0.8$) coverage is above 93\%. Implementing the Newey-West estimator yields more conservative inference across all degrees of temporal correlation, illustrating that it might be more robust, albeit too conservative when autoregression is suspected.

\begin{table}[!t]
    \centering
    %\spacingset{1.12}
    \caption{Simulation results for the immediate effect in the presence of outcome temporal correlation. Coverage of 95\% intervals (\%) of the proposed estimators based on Wald-type and Newey-West intervals for estimating the immediate effect. We consider autocorrelation values  $\rho=0.2, 0.4, 0.6$ and $0.8$ in the random error term for the outcome variable.}
     %\spacingset{1.05}
    \resizebox{0.6\textwidth}{!}{%
    \begin{tabular}{cccc|ccc}
        \hline \\
        [-10pt]
        &\multicolumn{3}{c|}{Standard Method}&\multicolumn{3}{c}{Newey-West Method}\\[2pt]
        \cline{2-7}\noalign{\vskip 2pt}
        $\rho$& 1-1 & 1-1/2 & 1-2  & 1-1 & 1-1/2 & 1-2\\[2pt]
        \hline
        \\[-10pt]
        0.2 & 96.34 &96.99 &95.91 &97.85 &98.06& 95.91 \\ 
        0.4 & 95.70 &95.48& 95.70& 98.92& 98.28 &98.49  \\
        0.6 & 95.73 & 94.87 & 95.09&  99.79 &100.00  &99.57  \\
        0.8 & 93.36 & 94.22 & 93.58& 100.00 &100.00 &100.00 \\[3pt]
        \hline
    \end{tabular}
    }%
    \label{appendix:tab:tempporal_corr}
\end{table}

\section{Additional study information}
\label{supp_sec:additional_study}

\subsection{Additional information on the study data set}
\label{appendix:data availability}

\subsubsection{Additional information on the wildfire data}

Satellite instruments scanning the Earth’s surface provide the primary evidence for wildfire activity. The National Oceanic and Atmospheric Administration’s Hazard Mapping System (HMS) integrates automated detection with human verification to delineate fire perimeters, producing a spatio-temporal dataset that includes Fire Radiative Power (FRP). These FRP values identify the most intense segments of a fire, characterized by high rates of energy release. The wildfire occurrence and intensity data in \cref{fig:wild-fire} are acquired from \url{https://www.ospo.noaa.gov/Products/land/hms.html} and plotted for August 31, 2021.

\subsubsection{Additional information on regions' exposure}

Hazardous smoke produced by wildfires can travel long distances, and affect population exposure and behavior across different areas.
% The National Oceanic and Atmospheric Administration's Hazard Mapping System (HMS)
HMS combines data from polar and geostationary satellites in real time, enabling experts to accurately identify and track the dispersion of smoke. The HMS smoke data are acquired from \url{https://www.ospo.noaa.gov/Products/land/hms.html}. 
% The system filters out persistent sources like gas flaring and urban fires to avoid false detections. 
Smoke plumes are labeled qualitatively as light, medium, or heavy based on their apparent opacity in satellite imagery. 
Each smoke polygon includes metadata such as the time window of observation and the satellite used.

\subsubsection{Additional information on the outcome}

The outcome represents the daily total number of bikeshare hours in the San Francisco, East Bay, and San Jose areas in California, US, as measured through the publicly-available Bay Wheels data provided by Lyft at \url{https://www.lyft.com/bikes/bay-wheels/system-data}. 

We find that over 97\% of rides last less than one hour, indicating that long-lasting rentals are rare. Additionally, we find that only 206 out of approximately 6 million rentals started in one area and ended in a different one.
This suggests that our data include very minimal commuting by bicycle that crosses area boundaries, which supports the assumption of minimal spillover effects across outcome units, and allows us to analyze the three areas separately. % For instance, bike rentals in San Jose are unlikely to affect outcomes in East Bay, and vice versa. To eliminate any remaining possibility of spillover, w
We exclude these rare cross-region trips from our analysis.

\subsubsection{Additional information on the collected covariates}

The time-varying covariates are regional temperature, precipitation, humidity, wind speed and wind direction among three bike locations, which are retrieved from \url{https://www.ncei.noaa.gov/cdo-web/datasets}.
Due to high temperatures and the dry season, there are more wildfires and gusty winds in late summer than in the other seasons. As a result, the exposed time periods concentrate around the summer and fall months (see Supplement \ref{supp_subsec:application_matches}). 

\subsection{Illustrations of resulting data for San Francisco based on the proposed algorithms}
\label{supp_subsec:application_matches}

\begin{figure}[!t]
    \centering
    \includegraphics[width=3.3in, trim=4 8 0 5.2, clip]{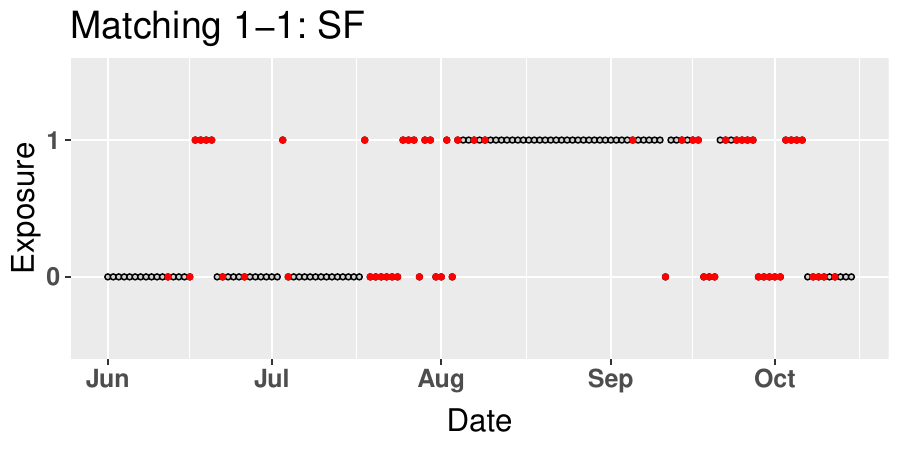} \\
    \includegraphics[width=3.3in]{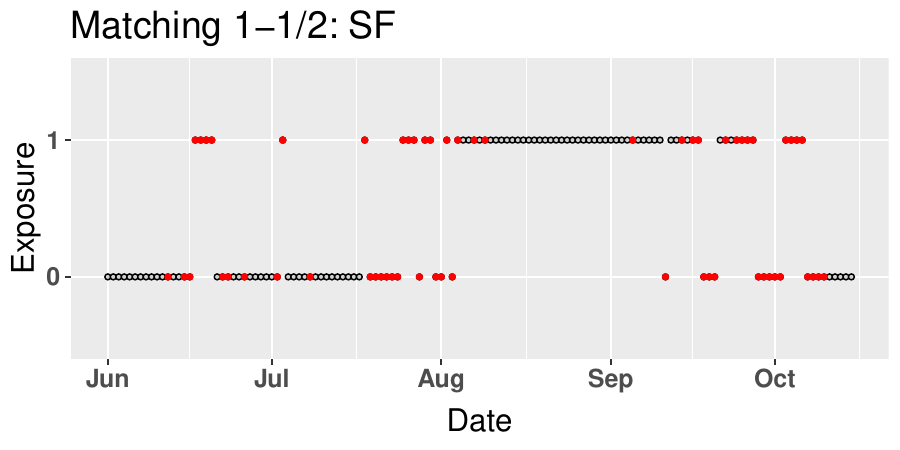} \\
    \includegraphics[width=3.3in]{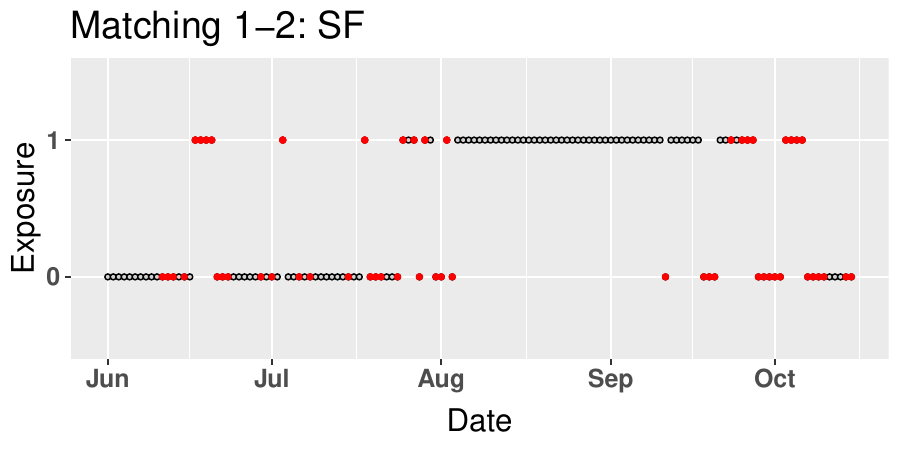}
    \caption{Exposure and matching information for the {1-1}, {1-1/2}, and {1-2} algorithms detailed in \cref{sec:matching} and Supplement~\ref{supp_subsec:matching1-2or1-12} for San Francisco during the period from June 1, 2021, to October 15, 2021. The red and hollow points correspond to time periods that are or are not part of the resulting data set of matched time periods, respectively.}
    \label{fig:appendix:SF_match}
\end{figure}

\begin{figure}[!t]
    \centering
    \includegraphics[width=3.3in, trim=4 8 0 5.2, clip]{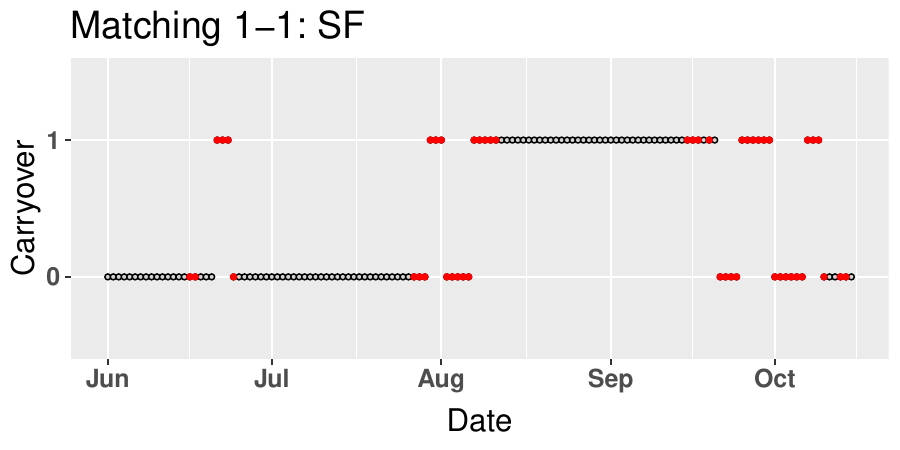} \\
    \includegraphics[width=3.3in]{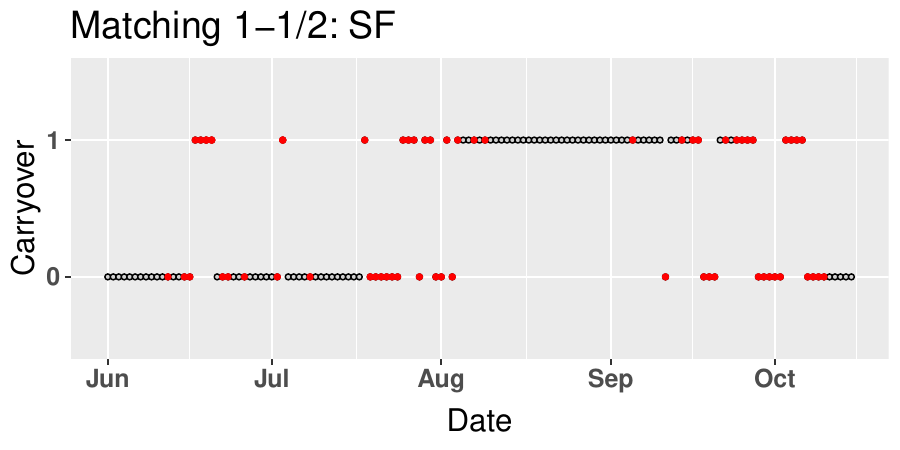} \\
    \includegraphics[width=3.3in]{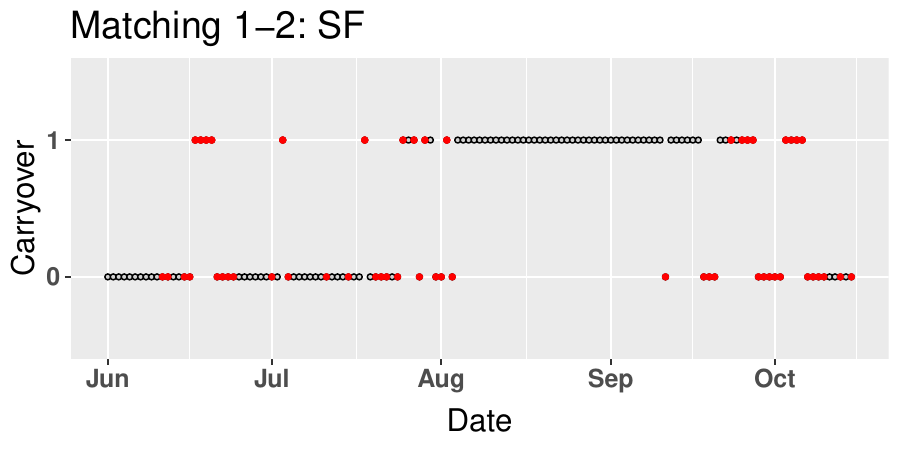}
    \caption{{Carryover exposure and matching information for the {1-1}, {1-1/2}, and {1-2} algorithms in Supplement~\ref{supp_subsec:alg_carryover} for San Francisco during the period from June 1, 2021, to October 15, 2021. The red and hollow points correspond to time periods that are or are not part of the resulting data set of matched time periods, respectively.}}
    \label{fig:appendix:SF_carryover_match}
\end{figure}
Our data set includes information from January 2021 until September 2023. Most exposed time periods, and as a result most matches for the immediate and the carryover effect, occur during the summer and fall months.
As an illustration, the exposed and unexposed time periods for a {\it subset} of our time window and for San Francisco are shown in \cref{fig:appendix:SF_match}, with the red color denoting whether the time period was used in a match according to the 1-1, 1-1/2, or 1-2 algorithms for the immediate effect.
Under $\epsilon=6$, the  1-1 and 1-1/2 algorithms return similar but not identical matching patterns, while the 1-2 algorithm matches fewer exposed time periods. 

\cref{fig:appendix:SF_carryover_match} is similar figure for the algorithms for the carryover effect. Here, we see a strong temporal trend in the carryover exposure with almost almost all of the time periods in the late summer and early fall, and almost in the rest of the time window  have carryover exposure. This explains why the algorithm matching one time period with carryover exposure to two time periods without carryover exposure manages to find such few number of matches (\cref{real-data-table}).

\subsection{Results under alternative definition of exposure}
\label{supp_subsec:real-data-medium-exposure}

As discussed in Supplement \ref{appendix:data availability}, the HMS categorizes smoke exposure as no exposure, light, medium, or high exposure. In our analysis of \cref{sec:real-data}, we considered an area at a given time period as exposed if the HMS classification was light or higher, and unexposed otherwise. Here, we evaluate the sensitivity of our conclusions when an area is considered exposed under medium or high smoke exposure, and unexposed under no smoke exposure or light smoke. Under this definition, out of the 1003 total number of days, San Francisco was exposed during 39 days, the East Bay during 40 days, and San Jose during 37 days. {The number of days with carryover exposure is 16 in all three areas.}

\begin{table}[h]
\caption{{The immediate and carryover effect estimates of wildfire smoke on bikeshare hours in San Francisco, East Bay, and San Jose for Na\"ive-$t$ and the matching estimators. For each region, the three columns correspond to the estimate, p-value, and number of matches. In this analysis, an area is considered exposed at a given time period if it is classified to have medium or high smoke exposure according to HMS, and unexposed if it is classified to have no smoke or light smoke exposure.}}
\centering \small
\resizebox{0.9\textwidth}{!}{%
\begin{tabular}{cccccccccc}
\hline \\[-10pt]
& \multicolumn{3}{c}{San Francisco} & \multicolumn{3}{c}{East Bay} & \multicolumn{3}{c}{San Jose} \\
\cmidrule(lr){2-4}
\cmidrule(lr){5-7}
\cmidrule(lr){8-10}
& Est & p-value & \#Exp & Est & p-value & \#Exp & Est & p-value & \#Exp \\
\cmidrule(lr){2-2}
\cmidrule(lr){3-3}
\cmidrule(lr){4-4}
\cmidrule(lr){5-5}
\cmidrule(lr){6-6}
\cmidrule(lr){7-7}
\cmidrule(lr){8-8}
\cmidrule(lr){9-9}
\cmidrule(lr){10-10} 
\\[-10pt]
& \multicolumn{9}{c}{\underline{Immediate effect}} \\[5pt]
Na\"ive-$t$                         & \phantom{*}159.337 & (0.983) & {\it 39}& 25.773 & (1.000) & {\it 40}& 37.640 & (1.000)  &{\it 37} \\[5pt]
1-1    & { -26.349} & { (0.354)} &  34    & -2.757 & (0.319) & 35 & -4.633 & (0.233) & 32  \\[5pt]
1-1/2 & { -35.746} & { (0.289)} &  34    & -1.698 & (0.382) & 35 & 2.190 & (0.783) & 32  \\[5pt]
1-2   & -60.263 & (0.169) &  25    & -5.908& (0.086) & 26 & -5.302 & (0.129) & 23 \\
\hline \\[-5pt]
& \multicolumn{9}{c}{\underline{Carryover effect}} \\[5pt]
 Na\"ive-$t$                         & -6.389 & (0.478) &{\it 16} & 31.904 & (1.000) &{\it 16} & 47.135 & (1.000)  &{\it 16} \\[5pt]
1-1                        &-132.589 & (0.139) & 15  & -12.256 & (0.110) & 10 & -1.993 & (0.586) & 12 \\[5pt]
1-1/2                     & -102.387 & (0.163) & 15  & -7.918 & (0.208) & 10 & -2.818 & (0.588) & 12 \\[5pt]
1-2                        & -179.933 & (0.223) & 4  & - & - & 0 & -12.256 & (0.003) & 3 \\
\hline
\end{tabular}}
\label{supp_tab:real-data-medium-smoke}
\end{table}

\cref{supp_tab:real-data-medium-smoke} shows the causal effect estimates under this alternative specification of exposure. We focus on the immediate effect first. We find that estimates are either negative or very close to zero, indicating that, if smoke exposure has an effect, it leads to a reduction in bikeshare hours. 
The effect estimates here, where a time period is defined as exposed under medium or high smoke and unexposed otherwise, are similar or larger in magnitude compared to the effect estimates reported in \cref{real-data-table}, where a time period is defined as exposed under light, medium or high smoke and unexposed otherwise. Since, here, a time period is classified as exposed under heavier smoke conditions, we believe that this comparison might be because heavier smoke has a larger impact on bikeshare hours compared to lighter smoke exposure.
However, the effect estimates in \cref{supp_tab:real-data-medium-smoke} are not statistically significant. We believe that this is largely because of the small number of matches (ranging from 23 to 35), that is partially explained by the small number of days with medium or high smoke exposure.

For the analysis of the carryover there are at most 15 matched time periods with carryover exposure. As a result, there is insufficient statistical support to draw meaningful conclusions when categorizing medium and heavy smoke as exposed.

A similar analysis that considers time periods to be exposed under heavy smoke exposure only would not be feasible in our data set. That is because the number of days with heavy smoke thickness was small across our time window. Specifically, in the three areas there are approximately 17 days of heavy exposure, 22 days of medium exposure, 70 days of light exposure, and more than 800 days of no exposure. Therefore, analyzing this exposure as categorical would be largely infeasible.

\end{document}